\documentclass[runningheads, 10pt,table]{llncs}
\usepackage[T1]{fontenc}

\usepackage{algorithm,algorithmic}
\usepackage{bm}
\usepackage{pgfplots}
\usepackage{soul}
\usepackage{stmaryrd}
\usepackage{graphicx}
\usepackage{float}
\usepackage{amsmath, amssymb, amsfonts}
\usepackage[nottoc]{tocbibind}
\usepackage{latexsym}
\usepackage{mathtools}
\usepackage{pstricks-add}
\usepackage[all,pdf]{xy}
\usepackage{url}
\usepackage{verbatim}
\usepackage{xcolor}
\usepackage{arydshln}
\usepackage{threeparttable}
\usepackage[numbers,sectionbib]{natbib}
\usepackage[colorlinks, linkcolor=blue, anchorcolor=blue, citecolor=blue]{hyperref}
\usepackage{enumitem}
\usepackage[figuresright]{rotating}
\usepackage{framed}
\usepackage{appendix}
\usepackage{marvosym}

\usepackage{listings}
\usepackage{makecell}

\usepackage{multirow}
\usepackage[normalem]{ulem}
\useunder{\uline}{\ul}{}

\usepackage{tikz, pgfplots}
\usetikzlibrary{matrix, fit, calc, matrix, patterns, backgrounds, arrows, 
shapes, chains, fit, decorations, intersections, positioning}

\tikzset{
    shapenode/.style = {draw, rectangle, fill=none, minimum size=0.5cm, minimum height=0.6cm, minimum width=0.6cm, 
    auto, node distance=0em, font=\normalsize}, 
    roundnode/.style = {draw, rectangle, rounded corners=0.5em, inner sep = 0.2em, 
      node distance=0em, minimum height=1.5em},
    rectanglenode/.style = {draw, rectangle, inner sep = 0em, node distance=0em},
    textnode/.style  = {draw=none, fill=none, rectangle, minimum size=0.5cm, auto, node distance=0em, font=\normalsize},
    smalltextnode/.style  = {draw=none, fill=none, rectangle, minimum size=0.5cm, auto, node distance=0em, font=\small},
    circlenode/.style = {draw, circle, auto, inner sep = 0em, node distance=0em, minimum size=1.5em}, 
    dotnode/.style = {draw, fill=black, circle, auto, minimum size=0.1em, inner sep = 0em, node distance=0em},  
    connect/.style = {black,->}, 
    every text node part/.style={align=center}
}

\newcommand{\FuncOT}{\mathcal{F}_\mathsf{OT}}

\newcommand{\Funcpeqt}{\mathcal{F}_\mathsf{PEQT}}

\newcommand{\Funcopprf}{\mathcal{F}_\mathsf{OPPRF}}

\newcommand{\rs}{\mathsf{r}_{\mathcal{S}}}
\newcommand{\rr}{\mathsf{r}_{\mathcal{R}}}
\newcommand{\ids}{\mathsf{IDGen}_{\mathcal{S}}}
\newcommand{\idr}{\mathsf{IDGen}_{\mathcal{R}}}
\newcommand{\id}{\mathsf{ID}}
\newcommand{\dist}{\mathsf{dist}}

\newcommand{\hint}{\mathsf{hint}}
\newcommand{\Funcdidg}{\mathcal{F}_{\mathsf{dIDG}}}
\newcommand{\Protocoldidg}{\Pi_{\mathsf{dIDG}}}

\newcommand{\rand}{\mathsf{Rand}}
\newcommand{\rands}{\mathsf{Rand}^{\mathcal{S}}}
\newcommand{\randr}{\mathsf{Rand}^{\mathcal{R}}}
\newcommand{\pids}{\mathsf{pid}^{\mathcal{S}}}
\newcommand{\pidr}{\mathsf{pid}^{\mathcal{R}}}

\newcommand{\setup}{\mathsf{Setup}}
\newcommand{\sta}{\mathsf{st}}

\newcommand{\Funcsoprf}{\mathcal{F}_{\mathsf{sOPRF}}}

\newcommand{\Funcfpsi}{\mathcal{F}_{\mathsf{FPSI}}}
\newcommand{\Funcfpsica}{\mathcal{F}_{\mathsf{FPSI\text{-}CA}}}
\newcommand{\Funcfpsisp}{\mathcal{F}_{\mathsf{FPSI\text{-}SP}}}
\newcommand{\Funclfpsi}{\mathcal{F}_{\mathsf{LFPSI}}}
\newcommand{\Funcifmat}{\mathcal{F}_{\mathsf{IFmat}}}
\newcommand{\Funcssifmat}{\mathcal{F}_{\mathsf{ssIFmat}}}
\newcommand{\Protocolfpsi}{\Pi_{\mathsf{FPSI}}}

\newcommand{\Funcps}{\mathcal{F}_{\mathsf{PS}}}
\newcommand{\Funcand}{\mathcal{F}_{\mathsf{AND}}}

\newcommand{\vecq}{\mathbf{q}}
\newcommand{\vecw}{\mathbf{w}}

\newcommand{\Funcecss}{\mathcal{F}_{\mathsf{ECSS}}}

\newcommand{\Funcecs}{\mathcal{F}_{\mathsf{ECS}}}

\newcommand{\Funcecips}{\mathcal{F}_{\mathsf{ECIPS}}}

\newcommand{\Funcsspeqt}{\mathcal{F}_\mathsf{ssPEQT}}
\newcommand{\Funcor}{\mathcal{F}_\mathsf{OR}}
\newcommand{\Funcip}{\mathcal{F}_\mathsf{IP}}

\newcommand{\veca}{\mathbf{a}}
\newcommand{\vecb}{\mathbf{b}}

\begin{document}

\mainmatter  

\title{Fuzzy PSI from Symmetric Primitives with Exact Logarithmic Dependence on Distance Threshold}


%
%
\author{Cong Zhang\inst{1}
\and Yang Cao\inst{2}
\and Yujie Bai\inst{2}
\and Shuaishuai Li\inst{3}
\and Juntong Lin\inst{2}
\and Yu Chen\inst{2}
\and Anyu Wang\inst{1}
\and Xiaoyun Wang\inst{1}
}

\institute{Tsinghua University\\
\email{\{zhangcong,anyuwang,xiaoyunwang\}@tsinghua.edu.cn}, 
\and
Shandong University\\
\email{yuchen@sdu.edu.cn, \{202437063,juntonglin\}@mail.sdu.edu.cn, byj8348560@163.com}, 
\and 
Zhongguancun Laboratory\\
\email{liss@zgclab.edu.cn}, 
}

%
%


\maketitle

\begin{abstract}
Fuzzy private set intersection (FPSI) is a protocol that enables a sender with a set of $d$-dimensional vectors $Q$ and a receiver with a set of $d$-dimensional vectors $W$ to learn items $\vecq \in Q$ for which there exists some $\vecw \in W$ satisfying $\dist(\vecq, \vecw) \leq \delta$ under a given distance metric. 
Previous FPSI works have demonstrated a linear scaling with the distance threshold $\delta$, while some recent works have achieved a poly-logarithmic dependence on $\delta$. However, these protocols either support only the $L_\infty$ distance, or they support general $L_{p\in[1,\infty]}$ distances but rely on expensive additive homomorphic encryption (AHE). Achieving exact logarithmic dependence on $\delta$ for general $L_{p\in[1,\infty]}$ distances without relying on costly AHE would constitute a theoretical breakthrough in optimal threshold scaling and a practical advance toward scalable FPSI applications.

In this work, we present new FPSI protocols for $L_{p\in[1,\infty]}$ distances that are entirely built from oblivious transfer (OT) and symmetric-key primitives. We propose FPSI protocols based on both the apart and the separate assumptions, which are applicable to low- and high-dimensional settings, respectively. Our constructions achieve strictly logarithmic complexity in $\delta$, which is optimal in the sense that distinguishing all values in an interval of length $O(\delta)$ necessarily requires $\Omega(\log \delta)$ bits of information. Our core idea is to perform fuzzy matching via prefix representation and interactively determine the correct prefix using equality conditions. To this end, we propose a suite of new components that can be implemented efficiently using only OT and symmetric-key operations.

We implement our FPSI protocols and compare them with the state-of-the-art FPSI protocols for $L_{p\in[1,\infty]}$ distance. 
Experiments show that our protocols outperform the prior state-of-the-art by up to $43.7\times$ in runtime and $31.3\times$ in communication.

\end{abstract}

\setcounter{footnote}{0}

\section{Introduction}\label{sec:intro}

Fuzzy private set intersection (FPSI) allows a receiver to identify elements in a sender's set that lie within a certain distance from its own elements, without revealing any additional information. This functionality is essential for applications involving noisy or approximate data, including genomic sequence matching \cite{DBLP:conf/ccs/WangHZTWB15}, similar image retrieval \cite{DBLP:conf/uss/KulshresthaM21}, compromised credential detection \cite{DBLP:journals/corr/abs-2109-14490}, illegal content screening \cite{DBLP:conf/eurocrypt/BaarsenP24}, and weak password identification \cite{DBLP:conf/uss/ChakrabortiFR23}.

While standard private set intersection (PSI) has matured over decades of research \cite{DBLP:conf/uss/Pinkas0Z14,DBLP:conf/ccs/KolesnikovKRT16,DBLP:conf/crypto/PinkasRTY19,DBLP:conf/crypto/ChaseM20,DBLP:conf/eurocrypt/PinkasRTY20,DBLP:conf/crypto/GarimellaPRTY21,DBLP:conf/pkc/GarimellaMRSS21,DBLP:conf/ccs/RosulekT21,DBLP:conf/eurocrypt/RindalS21,DBLP:conf/ccs/RaghuramanR22,DBLP:conf/pkc/BuiC23,DBLP:conf/pkc/ChenZZDL24} and achieved high practical efficiency, FPSI remains in a comparatively early research phase. Although introduced twenty years ago by Freedman et al. \cite{DBLP:conf/eurocrypt/FreedmanNP04}, FPSI has only attracted sustained research interest in recent years \cite{DBLP:conf/crypto/GarimellaRS22,DBLP:conf/crypto/GarimellaRS23,DBLP:conf/uss/ChakrabortiFR23,DBLP:conf/crypto/GarimellaGM24,DBLP:conf/eurocrypt/BaarsenP24,DBLP:conf/asiacrypt/GaoQLLW24,DBLP:journals/iacr/ChongchitmateLO24,cryptoeprint:2024/1642,cryptoeprint:2025/217,cryptoeprint:2025/885,AC/BGMP25,AC/BP25,DBLP:conf/ccs/piske+25,DBLP:conf/ccs/DZL25}. Conceptually related to PSI, FPSI entails greater design challenge due to its support for fuzzy matching rather than exact matching. As a result, current FPSI protocols still face unresolved theoretical issues and substantial efficiency bottlenecks, which hinder their practical deployment. Neither the asymptotic complexity nor the concrete efficiency of existing solutions has yet reached a satisfactory level. 

\subsection{Related Work}
\label{subsec:related-work}
In this work, we focus on the FPSI protocols for general Minkowski ($L_{p\in[1,\infty]}$) distance metrics in the semi-honest setting. Unlike standard PSI, the overhead of FPSI is related to more parameters, including set size $m,n$, dimension $d$, and distance threshold $\delta$. 
An overview of existing FPSI constructions is presented below, with a focus on their parameter complexity and cryptographic operations used.

\begin{trivlist}
    \item \textbf{Set size and dimension.} FPSI typically builds upon a fuzzy matching sub-protocol that allows a receiver to verify whether the distance between its element and the sender's element is within a threshold $\delta$. This sub-protocol can be viewed as a special case of FPSI where both sets contain a single element. A straightforward FPSI construction would perform fuzzy matching for every sender-receiver element pair, resulting in quadratic complexity $O(mn)$, where $m$ and $n$ denote the set sizes of the sender and receiver, respectively.

    To reduce the number of fuzzy matching operations, existing FPSI protocols typically rely on distributional assumptions about set elements. The two primary assumptions adopted in the literature are the \textit{apart} assumption and the \textit{separate} assumption. The apart assumption requires that the distance between any two elements within a party's set exceeds $2\delta d^{1/p}$ (or $2\delta(d^{1/p}+1)$), which simplifies to $2\delta$ and $4\delta$ for $L_\infty$ distance. This represents a relatively mild constraint. Under this assumption, spatial hashing techniques \cite{DBLP:conf/crypto/GarimellaRS22,DBLP:conf/eurocrypt/BaarsenP24} can reduce the number of fuzzy matching operations from $O(mn)$ to $O(\min\{m,n\} \cdot 2^d)$, making it suitable for low-dimensional setting.

    For high-dimensional setting, the stronger \textit{separate} assumption is adopted, which assumes that for each element, there exists at least one dimension where its distance from all other elements exceeds $2\delta$. This assumption implies the apart assumption. As shown by van Baarsen and Pu \cite{DBLP:conf/eurocrypt/BaarsenP24}, randomly generated sets satisfy the separate assumption with overwhelming probability in dimension $d$. FPSI protocols based on this assumption typically achieve polynomial dependence on $d$, making them suitable for high-dimensional settings. Notably, under this assumption, Gao et al. \cite{DBLP:conf/asiacrypt/GaoQLLW24} achieved linear complexity in $d$ for the first time, i.e., $O(d(m+n))$.

    \item \textbf{Distance threshold.} Fuzzy matching protocols typically rely on an interval testing sub-protocol. For $L_\infty$ distance, verifying whether the distance between $d$-dimensional elements $\vecq = (q_1,\ldots,q_d)$ and $\vecw = (w_1,\ldots,w_d)$ is within $\delta$ requires checking that $|w_k - q_k| \leq \delta$ for every $k \in [d]$, which is equivalent to testing whether each $w_k$ lies in the interval $[q_k - \delta, q_k + \delta]$. A straightforward method enumerates all $t \in [q_k - \delta, q_k + \delta]$ and tests whether $w_k = t$, resulting in a linear cost of $O(\delta)$ in the threshold. This approach has been employed in several recent works \cite{DBLP:conf/asiacrypt/GaoQLLW24,DBLP:conf/eurocrypt/BaarsenP24,cryptoeprint:2025/885,DBLP:conf/ccs/piske+25}, resulting in $O(d\delta)$ complexity for fuzzy matching.

    To improve efficiency, Chakraborti et al. \cite{DBLP:conf/uss/ChakrabortiFR23} introduced a prefix-based method that represents the interval $[q_k - \delta, q_k+\delta]$ using only $O(\log \delta)$ prefixes. Interval testing then reduces to checking whether any prefix of $w_k$ matches the prefix set of the interval, achieving an optimal logarithmic complexity of $O(\log \delta)$ per dimension. While this reduces the interval testing cost from $O(\delta)$ to $O(\log\delta)$, its application to fuzzy matching typically introduces an additional factor of $O((\log\delta)^d)$ \cite{DBLP:conf/crypto/GarimellaGM24,AC/BGMP25,AC/BP25,DBLP:conf/ccs/DZL25}, leading to an overall complexity of $O(d\log\delta+(\log\delta)^d)$ for $L_\infty$.

    Moreover, unlike $L_\infty$ distance, general Minkowski distances $L_{p\in[1,\infty)}$ do not support independent per-dimension interval testing, making prefix techniques considerably more challenging to apply. Consequently, nearly all FPSI protocols for $L_{p\in[1,\infty)}$ distance achieve only linear complexity $O(\delta)$ in the threshold \cite{DBLP:conf/eurocrypt/BaarsenP24,DBLP:conf/asiacrypt/GaoQLLW24,AC/BP25,cryptoeprint:2025/885,DBLP:conf/ccs/piske+25}. A recent exception is the work by Dang et al. \cite{DBLP:conf/ccs/DZL25}, which proposed a prefix-based fuzzy matching protocol for $L_{p\in[1,\infty)}$ distances with complexity $O(dp\log\delta+(2\log\delta)^d)$. However, their construction still retains an exponential term $O((2\log\delta)^d)$ and, more critically, contains a security flaw that we have identified (See Appendix \ref{app:flaw} for more details). Therefore, constructing a secure fuzzy matching protocol for general $L_{p\in[1,\infty)}$ distances with complexity $O(dp\log\delta)$ remains an open problem.

    \item \textbf{Public-key vs. symmetric-key.} Early private set intersection (PSI) protocols relied on public-key primitives such as DDH \cite{DBLP:conf/sp/Meadows86} or additively homomorphic encryption (AHE) \cite{DBLP:conf/eurocrypt/FreedmanNP04}, thus suffered from limited efficiency. The emergence of OT extension \cite{DBLP:conf/crypto/IshaiKNP03,DBLP:conf/ccs/BoyleCGIKRS19} enabled efficient generation of numerous OT instances using only lightweight symmetric-key operations and a small number of base OTs. This breakthrough shifted the design paradigm, leading to modern PSI protocols \cite{DBLP:conf/eurocrypt/RindalS21,DBLP:conf/ccs/RaghuramanR22} that achieve orders of magnitude improvement over public-key-based approaches. A similar transition is now underway in FPSI, where recent work aims to replace public-key operations with OT and symmetric-key techniques. 

    However, many current FPSI protocols still heavily rely on inefficient DDH \cite{DBLP:conf/eurocrypt/BaarsenP24} or AHE \cite{DBLP:conf/asiacrypt/GaoQLLW24,DBLP:conf/ccs/DZL25}. Recently, several concurrent works \cite{DBLP:conf/ccs/piske+25,AC/BGMP25,cryptoeprint:2025/885,AC/BP25} have employed OT and symmetric-key operations to enhance FPSI efficiency. Despite this progress, these constructions have not achieved optimal asymptotic efficiency. They either lack support for general $L_{p\in[1,\infty)}$ distance \cite{AC/BGMP25,DBLP:conf/ccs/piske+25}, or exhibit linear dependency on the threshold $\delta$ for $L_{p\in[1,\infty)}$ distance \cite{DBLP:conf/ccs/piske+25,cryptoeprint:2025/885}.

\end{trivlist}

\begin{table}[t]
\renewcommand\arraystretch{1}
	\centering
 \resizebox{1\linewidth}{!}{
\begin{tabular}{|c|c|c|c|c|c|}
\hline
\multirow{2}{*}{Metric} &
  \multirow{2}{*}{Protocol} &
  \multirow{2}{*}{Assumption} &
  \multirow{2}{*}{Communication} &
  \multirow{2}{*}{Computation} &
  \multirow{2}{*}{\begin{tabular}[c]{@{}c@{}}Symmetric\\ Crypto\end{tabular}} \\
 &
   &
   &
   &
   &
   \\ \hline
\multirow{13}{*}{$L_\infty$} &
  \multirow{2}{*}{\cite{DBLP:conf/eurocrypt/BaarsenP24}} &
  Apart &
  $O(\delta d n+2^dm)$ &
  $O(2^ddm+\delta d n)$ &
  \multirow{2}{*}{$\times$} \\ \cline{3-5}
 &
   &
  Separate &
  $O((\delta d)^2n+m)$ &
  $O((\delta d)^2)n+d^2m)$ &
   \\ \cline{2-6} 
 &
  \cite{DBLP:conf/crypto/GarimellaGM24} &
  Apart &
  $O(dn\log\delta+m(2\log\delta)^d)$ &
  $O(m2^d((\log\delta)^d+\kappa d \log\delta)+(\log\delta)^dn)$ &
  $\checkmark$ \\ \cline{2-6} 
 &
  \cite{DBLP:conf/asiacrypt/GaoQLLW24} &
  Separate* &
  $O(\delta d (m+ n))$ &
  $O(\delta d (m+n))$ &
  $\times$ \\ \cline{2-6} 
 &
  \cite{AC/BGMP25} &
  Apart &
  $O(dn\log\delta+m(2\log\delta)^d)$ &
  $O(m2^d((\log\delta)^d+ d \log\delta)+(\log\delta)^dn)$ &
  $\checkmark$ \\ \cline{2-6} 
 &
  \cite{cryptoeprint:2025/885} &
  Separate* &
  $O(\delta^s\frac{d}{s}(m+n))$ &
  $O(\delta^s\frac{d}{s}(m+n))$ &
  $\checkmark$ \\ \cline{2-6} 
 &
  \multirow{2}{*}{\cite{AC/BP25}} &
  Apart* &
  $O(d2^dn+\delta d m)$ &
  $O(d2^dn+\delta d m)$ &
  \multirow{2}{*}{$\checkmark$} \\ \cline{3-5}
 &
   &
  Apart* &
  $O((m2^d+n)d\log\delta)$ &
  $O((m2^d+n)d\log\delta+ n(\log\delta)^{d/2})$ &
   \\ \cline{2-6} 
 &
  \cite{DBLP:conf/ccs/piske+25} &
  Apart* &
  $O(d2^dn+\delta d m)$ &
  $O(2^ddn+dm\delta)$ &
  $\checkmark$ \\ \cline{2-6} 
 &
  \multirow{2}{*}{\cite{DBLP:conf/ccs/DZL25}} &
  Apart &
  $O((m2^d+n)d\log\delta)$ &
  $O((m2^d+n)d\log\delta+ n(\log\delta)^{d})$ &
  \multirow{2}{*}{$\times$} \\ \cline{3-5}
 &
   &
  Separate &
  $O(d\log\delta(m+n))$ &
  $O(d\log\delta(m+n))$ &
   \\ \cline{2-6} 
 &
  \multirow{2}{*}{Ours} &
  Apart* &
  $O((m2^d+n)d\log\delta)$ &
  $O((m2^d+n)d\log\delta)$ &
  \multirow{2}{*}{$\checkmark$} \\ \cline{3-5}
 &
   &
  Separate* &
  $O(d\log\delta(m+n))$ &
  $O(d\log\delta(m+n))$ &
   \\ \hline
\multirow{8}{*}{$L_{p\in[1,\infty)}$} &
  \cite{DBLP:conf/eurocrypt/BaarsenP24} &
  Apart &
  $O(\delta 2^ddn+\delta^pm)$ &
  $O(\delta 2^ddn+\delta^pm+dm)$ &
  $\times$ \\ \cline{2-6} 
 &
  \cite{DBLP:conf/asiacrypt/GaoQLLW24} &
  Separate &
  $O(\delta d (m+n) + pm\log \delta )$ &
  $O(\delta d(m+n)+p\log \delta m)$ &
  $\times$ \\ \cline{2-6} 
 &
  \cite{cryptoeprint:2025/885} &
  Separate* &
  $O(\delta d (m+n) + pm\log \delta )$ &
  $O(\delta d(m+n)+p\log \delta m)$ &
  $\checkmark$ \\ \cline{2-6} 
 &
  \cite{AC/BP25} &
  Apart &
  $O(d2^dn+\delta d m+(m+2^dn)p\log\delta)$ &
  $O(d2^dn+\delta d m+(m+2^dn)p\log\delta)$ &
  $\checkmark$ \\ \cline{2-6} 
 &
  \multirow{2}{*}{\cite{DBLP:conf/ccs/DZL25}} &
  Apart &
  $O(d2^dp\log\delta m +(d\log\delta+(2\log\delta)^dp)n)$ &
  $O(d2^dp\log\delta m +(dp\log\delta+(2\log\delta)^d)n)$ &
  \multirow{2}{*}{$\times$} \\ \cline{3-5}
 &
   &
  Separate &
  $O(d\log\delta(pm+n))$ &
  $O(dp\log\delta(m+n))$ &
   \\ \cline{2-6} 
 &
  \multirow{2}{*}{Ours} &
  Apart* &
  $O(d\log\delta(pm+2^dn))$ &
  $O(d\log\delta(pm+2^dn))$ &
  \multirow{2}{*}{$\checkmark$} \\ \cline{3-5}
 &
   &
  Separate* &
  $O(d\log\delta(pm+n))$ &
  $O(dp\log\delta(m+n))$ &
   \\ \hline
\end{tabular}
}
\caption{Asymptotic complexities of existing FPSI protocols for $L_{p\in[1,\infty]}$ distance, where the sender holds $m$ elements and the receiver holds $n$ elements in a $d$-dimensional space. $\delta$ is the threshold of FPSI. We ignore multiplicative factors of the computational security parameter $\kappa$ and statistical parameter $\lambda$. 
The apart assumption means that elements are either $2\delta d^{1/p}$ or $2\delta(d^{1/p}+1)$-apart, which simplifies to $2\delta$ and $4\delta$ for $L_\infty$ distance. The separate assumption means that at least one dimension of each element has a distance greater than $2\delta$ from all other elements. The $*$ symbol indicates that both the sets of senders and receivers satisfy the assumption.}
\label{tab:compare}
\end{table}

\subsection{Our Contribution}
\label{subsec:our-contribution}

We summarize our main contributions as follows, with a detailed comparison of asymptotic complexities against previous works provided in Table \ref{tab:compare}.


\begin{trivlist}
    \item \textbf{Equality Conditional Sum/Inner-Product Share.} 
    We introduce two new cryptographic protocols as the core building blocks of our construction: Equality Conditional Sum Share (ECSS) and Equality Conditional Inner-Product Share (ECIPS). The ECSS protocol enables efficient computation of $L_\infty$ distance, while ECIPS supports $L_{p\in[1,\infty)}$ distance. Both protocols can be efficiently implemented using OT and symmetric-key operations and may be of independent interest.
    
    \item \textbf{New Fuzzy Matching for Both $L_\infty$ and $L_{p\in[1,\infty)}$ Distances.} 
    Built on ECSS and ECIPS, we propose the first fuzzy matching protocol based on OT and symmetric-key operations that supports both $L_\infty$ and $L_{p\in[1,\infty)}$ distances while achieving exact logarithmic dependence on the distance threshold $\delta$. Existing solutions either exhibit strictly worse asymptotic complexities \cite{DBLP:conf/ccs/piske+25,AC/BP25,cryptoeprint:2025/885} or rely on expensive AHE \cite{DBLP:conf/ccs/DZL25}.

    \item \textbf{Fuzzy PSI Protocols for Low-Dimensional Spaces.} 
    By combining our fuzzy matching protocol with spatial hashing and cuckoo hashing techniques, we construct FPSI protocols for both $L_\infty$ and $L_{p\in[1,\infty)}$ distances in low-dimensional settings. Our protocols rely on apart assumption \cite{DBLP:conf/eurocrypt/BaarsenP24}, where the distance between any two elements within a party's set must exceed $2\delta d^{1/p}$, or alternatively $2\delta(d^{1/p}+1)$. For $L_\infty$ distance, these bounds simplify to $2\delta$ and $4\delta$, respectively. The corresponding asymptotic complexities are provided in Table \ref{tab:compare}.

    \item \textbf{Fuzzy PSI Protocols for High-Dimensional Spaces.} 
    For high-dimensional setting, we avoid the exponential $O(2^d)$ cost of spatial hashing by employing an interactive fuzzy mapping method, specifically a distributed ID generation (dIDG) protocol. Consistent with prior works \cite{DBLP:conf/asiacrypt/GaoQLLW24,DBLP:conf/eurocrypt/BaarsenP24,DBLP:conf/ccs/DZL25}, our high-dimensional protocols rely on the separate assumption. Asymptotic complexities are summarized in Table \ref{tab:compare}.
   
    \item \textbf{Experimental Evaluation.} 
    We implement our FPSI protocols for $L_{p\in[1,\infty]}$ distance and conduct comprehensive comparisons with state-of-the-art FPSI protocols \cite{DBLP:conf/asiacrypt/GaoQLLW24,DBLP:conf/ccs/DZL25,AC/BGMP25}. Experimental results show that our protocols achieve significant performance improvements, with runtime up to $43.7\times$ faster and communication reduced by up to $31.3\times$.

\end{trivlist}

\subsection{Overview of Our Techniques}
\label{subsec:overview}
Before presenting a high-level overview of our FPSI protocols, we first fix the notations. The sender $\mathcal{S}$ holds set $Q = \{\vecq_j\}_{j\in[m]} \subseteq \mathbb{U}^{d}$ and the receiver $\mathcal{R}$ holds set $W = \{\vecw_i\}_{i\in[n]} \subseteq \mathbb{U}^{d}$, where $d$ is the element dimension and $\delta$ the distance threshold. The $k$-th component of $\vecq_j$ ($\vecw_i$) is denoted $q_{j,k}$ ($w_{i,k}$) for $k\in[d]$.


\begin{trivlist}

    \item \textbf{Fuzzy Matching for $L_\infty$.} 
    An intuitive idea for an \(L_\infty\) fuzzy matching protocol is described below. Here, a sender holds $\mathbf{q}$ and a receiver holds $\mathbf{w}$, and the receiver will learn whether $\dist_\infty(\mathbf{q},\mathbf{w})\leq\delta$. 
    The protocol employs AHE and an Oblivious Key-Value Store (OKVS) scheme~\cite{DBLP:conf/crypto/GarimellaPRTY21}. The basic idea is for the receiver to encode a ciphertext of $0$ for every integer within $\delta$ in each dimension of its element, namely $E_k:=\mathsf{Encode}(\{(w_k+v,\mathsf{Enc}_{pk}(0))\}_{v\in[-\delta,\delta]})$ for $k\in[d]$. The sender then decodes using its element $\mathbf{q}$ and sums the results as $c:=\sum_{k\in[d]}\mathsf{Decode}(E_k,q_k)$. If $\dist_\infty(\mathbf{q},\mathbf{w})\leq\delta$, the sum $c$ is a ciphertext of $0$; otherwise, it is random. Next, the sender adds a random plaintext $r$ to $c$ using additive homomorphism, sends the result, and the receiver decrypts to verify if it matches $r$. This basic method, adopted in prior works~\cite{DBLP:conf/eurocrypt/BaarsenP24,DBLP:conf/asiacrypt/GaoQLLW24}, incurs communication and computation costs linear in $\delta$.
    
    To reduce the overhead, the protocol can be optimized via prefix representation, which constitutes the core contribution of \cite{DBLP:conf/ccs/DZL25}. Instead of enumerating all integers in the interval, the receiver encodes only the $l=O(\log\delta)$ \textit{prefixes} representing the interval. Correspondingly, the sender decodes using the \textit{prefixes} of its own element. Concretely, the receiver first computes the prefix representation $\{w_{k,1}^*,\dots,w_{k,l}^*\}$ of each interval $[w_k-\delta,w_k+\delta]$ and constructs an OKVS $E_k:=\mathsf{Encode}(\{(w_{k,v}^*,\mathsf{Enc}_{pk}(0))\}_{v\in[l]})$ for each dimension $k\in[d]$. The sender computes its own prefixes $\{q_{k,1}^*,\dots,q_{k,l'}^*\}$, picks a random value $a_k$ per dimension, and computes for each prefix the value $u_{k,v}:=\mathsf{Decode}(E_k,q_{k,v}^*)+\mathsf{Enc}_{pk}(a_k)$. Note that if $q_{k,v}^*$ matches some $w_{k,v'}^*$, we have $|q_k-w_k|\leq \delta$ and $u_{k,v}$ is a ciphertext of $a_k$. Then, the sender sends $\{u_{k,v}\}_{k\in[d],v\in[l']}$ along with $r=H(\sum_{k\in[d]} a_k)$, where $H$ is a universal hash function. Finally, the receiver exhaustively checks all combinations $\mathbf{v}=(v_1,\dots,v_d)\in[l']^d$ by testing whether $r = H(\sum_{k} \mathsf{Dec}_{sk}(u_{k,v_k}))$. 

    The avove protocol suffers from three main issues. First, it relies on expensive AHE operations, which impacts efficiency. Second, the receiver must traverse all $O((\log\delta)^d)$ possible prefix combinations, introducing significant computational overhead. Third and most critically, the protocol is insecure. The security flaw originates in the sender's masking step. For each dimension $k$, the same random mask $a_k$ is added to all prefix decoding results $\{\mathsf{Decode}(E_k,q_{k,v}^*)\}_{v\in[l']}$. This leaks information about the sender's input. A detailed attack demonstrating this vulnerability is provided in Appendix~\ref{app:flaw}.


    We now describe our approach to address these issues. For the first problem, our key observation is that we can achieve the same functionality using additive secret sharing (ASS) instead of AHE. 
    In this approach, the receiver encodes \(\{(w_{k,v}^*, r_k)\}_{v \in [l]}\) into an OKVS, where $r_k$ is a random value picked by the receiver. The sender subsequently decodes it using its own prefixes \(\{q_{k,v}^*\}_{v\in[l']}\) to obtain the values \(r_{k,v}'\). If \(q_k\) lies within the interval \([w_k - \delta, w_k + \delta]\), there exists some \(v \in [l']\) such that \(r_{k,v}' = r_k\); otherwise, with overwhelming probability, \(r_{k,v}' \neq r_k\) for every \(v \in [l']\). Thus, the pair \((r_{k,v}', -r_k)\) forms a conditional zero-sharing: it yields shares of zero exactly when the interval test succeeds, and shares of a random value otherwise. 
    A challenge arises because OKVS guarantees obliviousness only when encoded values are uniformly random, so we cannot directly send an OKVS encoding a fixed value. We resolve this by employing an Oblivious Programmable PRF (OPPRF) \cite{DBLP:conf/ccs/KolesnikovMPRT17} instead of OKVS. This leverages the standard technique of using an OPRF to generate a one-time pad, thereby ensuring the necessary randomness for the encoded values. To tackle the remaining efficiency and security concerns, we introduce a new primitive named Equality Conditional Sum (ECS). In the ECS functionality, both the sender and receiver input sets of two-dimensional vectors, denoted $A = \{ \veca_{k,v} = (a_{k,v}^0, a_{k,v}^1) \}_{k \in [d], v \in [l]}$ and $B = \{ \vecb_{k,v} = (b_{k,v}^0, b_{k,v}^1) \}_{k \in [d], v \in [l]}$, where $k$ indexes dimensions and $v$ indexes prefixes. The component with superscript $1$ is used for comparison, while the component with superscript $0$ is used for summation. The functionality identifies, for each dimension $k$, an index $v_k^* \in [l]$ such that $a_{k,v_k^*}^1 = b_{k,v_k^*}^1$\footnote{The prefix property ensures at most one such $v_k^*$ exists per dimension.}; if no such index exists, a random $v_k^*$ is selected. It then computes the sum $\mathsf{Sum} := \sum_{k \in [d]} (a_{k, v_k^*}^0 + b_{k, v_k^*}^0)$ and delivers $\mathsf{Sum}$ to the receiver.

    Afger the OPPRF evaluation, both parties proceed to invoke the ECS functionality. In this phase, the sender acts as the ECS receiver, providing input vectors in which both components are set to its OPPRF outputs, that is, \(b_{k,v}^0 = b_{k,v}^1 := r_{k,v}'\). Meanwhile, the receiver participates as the ECS sender, constructing input vectors with the first component set to zero and the second component containing the random values generated during OPPRF encoding: \(a_{k,v}^0 := 0\), \(a_{k,v}^1 := r_k\). This construction ensures that the sender learns \(\sum_{k \in [d]} r_k\) if and only if a zero-sharing condition is satisfied in every dimension, which is equivalent to \(|w_k - q_k| \leq \delta\) for all \(k \in [d]\). To complete the verification, the receiver also computes \(\mathsf{Rand} := \sum_{k \in [d]} r_k\), and the parties check whether the sender’s output from ECS equals \(\mathsf{Rand}\). Therefore, a secure ECS protocol based on OT and symmetric cryptographic primitives, with an overhead of only \(O(d \log \delta)\), is sufficient to address all the challenges mentioned above.

 \item \textbf{Equality Conditional Sum (Share).} We now present the construction of our ECS protocol, along with a variant called ECS Share (ECSS) that outputs secret shares of $\mathsf{Sum}$ rather than delivering the result directly to the receiver. This ECSS variant is useful in our high-dimensional FPSI construction. 

    The general idea of our ECS protocol is to let the receiver obtain \(v_k^*\) when \(a_{k,v_k^*}^1 = b_{k,v_k^*}^1\); otherwise, the receiver gets a random \(v_k^*\). The receiver subsequently calculates and sums over these \(v_k^*\) values. Specifically, the protocol proceeds through the following steps. First, both parties execute a secret-shared Private Equality Test (ssPEQT) on all pairs $(a_{k,v}^1, b_{k,v}^1)$, obtaining secret-shared bits $e_{k,v}^\mathcal{S}$ and $e_{k,v}^\mathcal{R}$ such that $e_{k,v}^\mathcal{S} \oplus e_{k,v}^\mathcal{R} = 1$ if and only if $a_{k,v}^1 = b_{k,v}^1$. Next, each party locally XORs its bits across all $v$, yielding $\tilde{e}_{k,0}^\mathcal{S} := \bigoplus_{v\in[l]} e_{k,v}^\mathcal{S}$ and $\tilde{e}_{k,0}^\mathcal{R} := \bigoplus_{v\in[l]} e_{k,v}^\mathcal{R}$. Consequently, $\tilde{e}_{k,0}^\mathcal{S} \oplus \tilde{e}_{k,0}^\mathcal{R} = 1$ exactly when there exists a $v_k^* \in [l]$ satisfying $a_{k,v_k^*}^1 = b_{k,v_k^*}^1$. 
    If such $v_k^*$ exists, let $\eta=\log l$ and define $S_\alpha:=\{x_\eta\dots x_{\alpha+1} 1 x_{\alpha-1}\dots x_1:x_1,\dots,x_{\eta}\in\{0,1\}\}$, the parties compute $\tilde{e}_{k,\alpha}^\mathcal{S} := \bigoplus_{i\in S_\alpha} e_{k,i}^\mathcal{S}$ and $\tilde{e}_{k,\alpha}^\mathcal{R} := \bigoplus_{i\in S_\alpha} e_{k,i}^\mathcal{R}$. It can be simply verified that $(\tilde{e}_{k,\alpha}^\mathcal{S},\tilde{e}_{k,\alpha}^\mathcal{R})$ forms the secret shares of the $\alpha$-th bit of $v_k^*$. By assembling all $\eta$ bits, the parties learn exactly the shares of $v_k^*$.
    Specifically, the sender learns $v_k^\mathcal{S}$ and the receiver learns $v_k^\mathcal{R}$, with $v_k^\mathcal{S} \oplus v_k^\mathcal{R} = v_k^*$. Now, the goal is for the receiver to learn $v_k^*$ only if such an index exists; otherwise it should receive a random index. This is achieved via a 2-choice-bit OT: the sender samples a random index $r_k \leftarrow [l]$ and inputs $(v_k^\mathcal{S}, r_k)$ if $\tilde{e}_{k,0}^\mathcal{S}=1$, and $(r_k, v_k^\mathcal{S})$ otherwise. The receiver inputs $\tilde{e}_{k,0}^\mathcal{R}$. This ensures the receiver reconstructs the correct $v_k^\mathcal{S}$ precisely when $\tilde{e}_{k,0}^\mathcal{S} \oplus \tilde{e}_{k,0}^\mathcal{R} = 1$.

    After the receiver obtains $v_k^*$, the parties invoke a 1-out-of-$l$ OT. In this step, the sender selects random masks $\{\mathsf{mask}_k\}_{k\in[d]}$ subject to $\sum_{k\in[d]} \mathsf{mask}_k = 0$, and inputs $\{a_{k,v}^0 + \mathsf{mask}_k\}_{v\in[l]}$. The receiver inputs $v_k^*$ and receives $\gamma_k = a_{k,v_k^*}^0 + \mathsf{mask}_k$. The receiver then computes $\mathsf{Sum} := \sum_{k\in[d]} (\gamma_k + b_{k,v_k^*}^0)$. For ECSS, the sender instead picks all masks arbitrarily and sets its own share as $-\sum_{k\in[d]} \mathsf{mask}_k$.

    The steps described above correctly implement the ECS functionality, but they introduce a security issue that could compromise sender privacy. Specifically, allowing the receiver to learn each \(v_k^*\) directly could reveal information about the sender's set. To address this issue, we propose to let the receiver learn a permuted index \(\pi(v_k^*)\) instead, where \(\pi\) is a random permutation over \([l]\) chosen by the sender and kept hidden from the receiver. This is achieved using a Permute + Share (PS) protocol \cite{DBLP:conf/asiacrypt/ChaseGP20,DBLP:conf/pkc/PecenyRRS25}. In the PS protocol, the sender inputs a permutation \(\pi\), the receiver inputs a vector \(\vecb_k = (b_{k,1}^1,\dots,b_{k,l}^1)\), and both parties receive additive shares of the permuted vector. Concretely, the sender obtains \(\vecb^\mathcal{S} = (b_{k,1}^\mathcal{S},\dots,b_{k,l}^\mathcal{S})\) and the receiver obtains \(\vecb^\mathcal{R} = (b_{k,1}^\mathcal{R},\dots,b_{k,l}^\mathcal{R})\), such that \(b_{k,v}^\mathcal{S} + b_{k,v}^\mathcal{R} = b_{k,\pi(v)}^1\). The sender then computes \(\tilde{a}_{k,v}^1 := a_{k,\pi(v)}^1 - b_{k,v}^\mathcal{S}\). At this stage, the parties can invoke ssPEQT on \(\tilde{a}_{k,v}^1\) and \(b_{k,v}^\mathcal{R}\); one can verify that \(\tilde{a}_{k,v}^1 = b_{k,v}^\mathcal{R}\) if and only if \(a_{k,\pi(v)}^1 = b_{k,\pi(v)}^1\). Subsequently, following the steps previously outlined, the parties obtain secret shares of \(\pi(v_k^*)\). 

    Combining the above techniques yields the ECS (and ECSS) protocol. For $L_{p\in[1,\infty)}$ distance, we introduce a similar component called Equality Conditional Inner-Product Share (ECIPS), which is constructed analogously to ECSS but replaces the local summation with an inner product protocol. Details are provided in Section \ref{subsec:ecips}.
    

    \item \textbf{Fuzzy Matching for $L_{p\in[1,\infty)}$.} Previous works \cite{DBLP:conf/eurocrypt/BaarsenP24,DBLP:conf/asiacrypt/GaoQLLW24} encode $\{(w_k + v, \mathsf{Enc}_{pk}(|v|^p))\}_{v\in[-\delta,\delta]}$ into an OKVS. If $q_k \in [w_k-\delta, w_k+\delta]$, the sender decodes precisely the ciphertext of $|w_k - q_k|^p$. Using additive homomorphism, a ciphertext of $\dist_p(\vecq,\vecw)^p = \sum_{k\in[d]}|w_k - q_k|^p$ can then be obtained. This method, however, requires enumerating all values in the interval. Dang et al. \cite{DBLP:conf/ccs/DZL25} proposed a new approach leveraging prefix representation. Assuming $q_k \leq w_k$, their core idea is to find a public midpoint $q_k \leq w' \leq w_k$ such that $|w_k - q_k| = |w_k - w'| + |q_k - w'|$. The receiver encodes ciphertext of $|w_k - w'|$ into the OKVS, and the sender locally adds $|q_k - w'|$ after decoding. They observed that the right endpoint of the prefix-generated interval satisfies this property\footnote{When $q_k > w_k$, the left endpoint is used.}. More concretely, let $\{w_{k,1}^*, \dots, w_{k,l}^*\}$ be the prefix representation of $[w_k-\delta, w_k]$, and let $q_{k,v'}^*$ be a prefix of $q_k$ that matches some $w_{k,v}^*$. The receiver encodes $\{w_{k,v}^*, \mathsf{Enc}(|w_k - \mathsf{UpBound}(w_{k,v}^*)|)\}_{v\in[l]}$, where $\mathsf{UpBound}(w_{k,v}^*)$ denotes the right endpoint of the interval corresponding to prefix $w_{k,v}^*$. The sender computes the ciphertext of $|w_k - q_k|$ as $\mathsf{Decode}(E, q_{k,v'}^*) + |q_k - \mathsf{UpBound}(q_{k,v'}^*)|$. To obtain the ciphertext of $|w_k - q_k|^p$, the receiver encodes for each $t\in[p]$ the values $\mathsf{Enc}(|w_k - \mathsf{UpBound}(w_{k,v}^*)|^t)$. The sender then uses the binomial formula to derive the ciphertext of $|w_k - q_k|^p$ as $\sum_{t\in[0,p]}\binom{p}{t}\mathsf{Enc}(|w_k - \mathsf{UpBound}(w_{k,v}^*)|^t)\cdot |q_k - \mathsf{UpBound}(q_{k,v'}^*)|^{p-t}$.


    The prefix technique expands the sender's decoding attempts per dimension by a multiplicative factor $l = O(\log \delta)$. Dang et al.~\cite{DBLP:conf/ccs/DZL25} employed a method analogous to their $L_\infty$ construction to sum across dimensions, but it inherits the same shortcomings of insecurity and an additional $O((\log \delta)^d)$ factor. To resolve these issues, our initial step is again to adopt ASS in place of AHE. In particular, the parties obtain secret shares of $|w_k - w'|$, denoted $\llbracket |w_k - w'| \rrbracket_\mathcal{S}$ and $\llbracket |w_k - w'| \rrbracket_\mathcal{R}$, rather than a ciphertext. Consequently, both parties should locally apply the binomial formula to their shares to compute shares of $|w_k - q_k|^p$, instead of having the sender perform this computation alone. This presents a challenge because the sender alone knows the binomial coefficients $\binom{p}{t}|q_k - w'|^{p-t}$, and revealing them would leak information about its input.

    We address this by reframing the computation as an inner product problem. Define the vectors
    \begin{align*}\label{aaa}
    \veca^0 &= \bigl(0, \llbracket |w_k - w'| \rrbracket_\mathcal{R}, \dots, \llbracket |w_k - w'|^p \rrbracket_\mathcal{R}\bigr), \\
    \vecb^0 &= \bigl(1, \llbracket |w_k - w'| \rrbracket_\mathcal{S}, \dots, \llbracket |w_k - w'|^p \rrbracket_\mathcal{S}\bigr), \\
    \vecb^1 &= \bigl(|q_k - w'|^{p}, \binom{p}{1}|q_k - w'|^{p-1}, \dots, 1\bigr).
    \end{align*}
    Here, $\veca^0$ is known to the receiver, and $\vecb^0,\vecb^1$ are known to the sender. The binomial computation thus reduces to computing the inner product $\langle \veca^0 + \vecb^0, \vecb^1 \rangle$. This can be implemented using an inner product protocol of length $p+1$, which can in turn be realized via $p+1$ instances of Oblivious Linear Evaluation (OLE) protocol.

    To eliminate the $O((\log \delta)^d)$ expansion, we introduce a new primitive called Equality Conditional Inner-Product Share (ECIPS), analogous to ECSS. In ECIPS, the sender inputs $A = \{ \veca_{k,v}=(\veca_{k,v}^0, a_{k,v}^2) \}_{k \in [d], v \in [l]}$ and the receiver inputs $B = \{ \vecb_{k,v} = (\vecb_{k,v}^0, \vecb_{k,v}^1, b_{k,v}^2) \}_{k \in [d], v \in [l]}$. The superscript $2$ components are used for equality comparison, while the superscript $0,1$ components are used for the inner product. For each dimension $k$, the functionality identifies an index $v_k^* \in [l]$ such that $a_{k,v_k^*}^2 = b_{k,v_k^*}^2$; if none exists, a random $v_k^*$ is chosen. It then computes the inner product sum $\mathsf{IPSum} := \sum_{k \in [d]} \langle \veca_{k, v_k^*}^0 + \vecb_{k, v_k^*}^0, \vecb_{k, v_k^*}^1\rangle$ and outputs secret shares of $\mathsf{IPSum}$ to the two parties.

    Similar to our $L_\infty$ distance protocol, we first let the participants obtain the conditional additive sharing of $|w_k-q_k|^t$ for $t\in[p]$ via an OPPRF. Denote this sharing by $(r_{k,t},\,r_{k,v,t}')$, that is, if the prefix $q_{k,v}^*$ of $q_k$ equals some prefix of the interval $[w_k-\delta,w_k]$\footnote{For convenience, we assume $q_k\leq w_k$. The case $q_k> w_k$ is handled analogously by a second OPPRF.}, then $r_{k,t}+r_{k,v,t}'=|w_k-w'|^t$, otherwise the shares are random. 
    In addition, the parties also generate a conditional zero sharing, denoted $(r_{k,p+1},r_{k,v,p+1}')$, for later equality testing.
    With these shares, the sender defines $\vecb^0_{k,v}:=(1,r'_{k,v,1},\dots,r'_{k,v,p}),\vecb^1_{k,v}:=(\binom{p}{0}|q_k-\mathsf{UpBound}(q_{k,v}^*)|^p,\dots,\binom{p}{p-1}|q_k-\mathsf{UpBound}(q_{k,v}^*)|^1,1),b_{k,v}^2:=r'_{k,v,p+1}$, and the receiver defines $\veca^0_{k,v}:=(0,r_{k,1},\dots,r_{k,p}),a_{k,v}^2:=r_{k,p+1}$. The two parties then invoke the ECIPS functionality on these inputs, receiving outputs $\rand_\mathcal{S}$ and $\rand_\mathcal{R}$. By the functionality of ECIPS, if $\forall k\in[d], |q_k-w_k|\leq \delta$, then $\rand_\mathcal{S} + \rand_\mathcal{R} = \sum_{k\in[d]}|w_k-q_k|^p =\dist_p(\vecw,\vecq)^p$. Now, the parties can thus complete $L_p$ fuzzy matching by checking whether $\rand_\mathcal{S} \in [-\rand_\mathcal{R},\, -\rand_\mathcal{R} + \delta^p]$ via the fuzzy matching for interval functionality \cite{DBLP:conf/uss/ChakrabortiFR23,DBLP:conf/asiacrypt/GaoQLLW24}. 
    
    The ECIPS construction follows a pattern similar to ECSS, with overhead $O(dp\log\delta)$. Further details are provided in Section~\ref{subsec:ecips}.


    \item \textbf{Fuzzy PSI for Low Dimensional Spaces.} We illustrate the construction using FPSI for $L_\infty$ distance. The approach for $L_{p\in[1,\infty)}$ distances follows a similar pattern. Fuzzy matching corresponds to the special case of FPSI where each party holds a single element. When both parties have multiple elements, spatial hashing techniques~\cite{DBLP:conf/crypto/GarimellaRS22,DBLP:conf/eurocrypt/BaarsenP24,DBLP:conf/crypto/GarimellaGM24,AC/BP25,DBLP:conf/ccs/DZL25} are commonly employed in low-dimensional settings to reduce the number of pairwise comparisons. This method requires the points of the parties to be at least $2\delta$ or $4\delta$ apart for successful OKVS/OPPRF encoding. The core idea is to partition the space into a grid of cells with side length $2\delta$. For each of its elements, the sender assigns as an identifier (ID) the coordinates of the cell containing that element. The receiver, for each of its elements, considers all cells that intersect the $\delta$-radius ball centered at the element, yielding $2^d$ IDs per element. This encoding guarantees that if $\dist(\mathbf{w}_i, \mathbf{q}_j) \leq \delta$, the ID sets of $\mathbf{w}_i$ and $\mathbf{q}_j$ intersect.

    Once the ID sets are computed, the receiver encodes its IDs into an OKVS. Concretely, for each element $\mathbf{w}_i$, each of its $2^d$ associated IDs $\{id_{\vecw_i,t}\}_{t\in[2^d]}$, the receiver encodes $\{(id_{\vecw_i,t} || (w_{i,k}+v),\; \mathsf{Enc}_{pk}(0)\}_{i\in[n],t\in[2^d],v\in[-\delta,\delta]}$. The sender, for each of its elements $\mathbf{q}_i$, decodes using the query $id_{\vecq_i} || q_{i,k}$. This ensures that only when the IDs match does the sender obtain a $0$-ciphertext (or, when using OPPRF, shares of $0$). When the prefix optimization is applied, the receiver concatenates each ID with the corresponding prefix before encoding, and the sender similarly appends prefixes to its decoding queries. Because each dimension of each sender element requires $l = O(\log \delta)$ decoding attempts, the sender must examine all $O((\log \delta)^d)$ possible prefix combinations.


    As previously noted, eliminating the $O((\log\delta)^d)$ factor requires the parties to run an ECS protocol, which thereby introduces an interaction phase after OPPRF. In contrast, prior traversal-based methods avoided such interaction by having the sender locally enumerate all possible summation items. This interaction, however, raises a new challenge: the receiver must correctly match its encoded shares with the corresponding decoding results from the sender to form valid inputs for the ECS.

    To overcome this challenge, we employ Simple/Cuckoo hashing~\cite{DBLP:conf/uss/Pinkas0Z14}. Specifically, the sender inserts its elements into a Cuckoo hash table using each element's ID as the hash key. In parallel, the receiver inserts its elements into a Simple hash table, encoding shares of $0$ in each bin. Crucially, under the apart assumption, this ensures that in each bin of the receiver’s table, at most one element lies within distance $\delta$ of sender's element. Thus, ECS can be performed bin-by-bin, resolving the alignment problem introduced by departing from purely local traversal.

    Nevertheless, the introduction of Cuckoo hashing also reveals the sender's hashing positions, which leaks information about its entire set. We mitigate this leakage by incorporating a Permute+Share (PS) protocol~\cite{DBLP:conf/asiacrypt/ChaseGP20}. The sender locally chooses a random permutation and uses the PS protocol to jointly permute the order of bins with the receiver, thereby hiding the original cuckoo hashing order. We further propose an optimization based on our newly introduced shared OPRF (sOPRF) functionality. Using sOPRF, the parties can realize a shared OPPRF where the receiver’s PRF shares directly serve as the corresponding evaluation shares for the sender. In this setting, the parties only need to run ECS for non-empty bins, and the PS step can be omitted entirely. Additional details are provided in Sections~\ref{subsubsec:fpsi-apart-inf-soprf} and~\ref{subsubsec:fpsi-apart-p-soprf}.

    \item \textbf{Fuzzy PSI for High Dimensional Spaces.} Under the apart assumption, spatial hashing incurs a $2^d$ blow-up, as the receiver must generate $2^d$ IDs per element. Gao et al.~\cite{DBLP:conf/asiacrypt/GaoQLLW24} introduced a distributed ID generation (dIDG) protocol with expansion rate $1$ under the stronger separate assumption, generating only one ID per element for both parties. 
    
    Roughly speaking, their idea assigns to each element $\vecq$ an ID defined as ${id}_{\vecq} := F_Q(\vecq) + F_W(\vecq)$, where $F_Q(\cdot)$ and $F_W(\cdot)$ are functions constructed such that for any close pair $\dist(\vecq, \vecw) \leq \delta$, we have $F_{Q}(\vecq)=F_Q(\vecw)$ and $F_W(\vecq)=F_W(\vecw)$. Critically, the function $F_Q(\cdot)$ (resp. $F_W(\cdot)$) is keyed by the sender's set $Q$ (resp. receiver's set $W$). Since each party holds only its own key, the protocol requires the receiver to obliviously evaluate $F_Q(\cdot)$ on its own set $W$, and symmetrically, the sender to obliviously evaluate $F_W(\cdot)$ on $Q$. Taking the receiver side as an example, $F_Q$ is constructed as follows. For each dimension $k$, a random value $r_k$ is assigned to each projection interval $[q_k-\delta, q_k+\delta]$, and overlapping intervals in the same dimension are merged and assigned the same $r_k$. Then $F_Q(\mathbf{w}) = \sum_{k=1}^d r_k$, where $r_k$ corresponds to the interval containing $w_k$. This construction, however, introduces collisions: distinct $\mathbf{w}_1 \neq \mathbf{w}_2$ may yield $F_Q(\mathbf{w}_1) = F_Q(\mathbf{w}_2)$, revealing proximity information. To eliminate this leakage, Gao et al.~\cite{DBLP:conf/asiacrypt/GaoQLLW24} incorporate a DH key exchange, defining the final ID as $id_\mathbf{w} := (F_Q(\mathbf{w}) + F_W(\mathbf{w}))^{sk_s \cdot sk_r}$, where $sk_s$ and $sk_r$ are secret DH keys held by the sender and receiver. The receiver thus obtains $id_\mathbf{w}$ without learning $F_Q(\mathbf{w})$ directly, thereby concealing collisions.

    The construction of Gao et al. \cite{DBLP:conf/asiacrypt/GaoQLLW24} incurs a cost of $O(d\delta(m+n))$. Dang et al. \cite{DBLP:conf/ccs/DZL25} adopt a similar approach but incorporate a prefix technique, reducing complexity to $O(d\log\delta(m+n))$. However, we find their construction remains insecure. Beyond the earlier mentioned insecurity due to mask reuse, we identify a further security flaw in their distributed ID generation. Gao et al. \cite{DBLP:conf/asiacrypt/GaoQLLW24} define the final ID as ${id}_\vecw = (F_Q(\mathbf{w}) + F_W(\mathbf{w}))^{sk_s \cdot sk_r}$, which involves symmetric operations from both parties. In contrast, Dang et al. \cite{DBLP:conf/ccs/DZL25} effectively set ${id}_\vecw := F_Q(\mathbf{w})$, omitting both the receiver’s $F_W$ evaluation and the DH step. This simplification reintroduces the collision leakage that the DH exchange was designed to prevent. A full analysis is given in Appendix~\ref{app:flaw}. Additionally, we note that both protocols in \cite{DBLP:conf/asiacrypt/GaoQLLW24,DBLP:conf/ccs/DZL25} still rely on relatively inefficient AHE.


To address these issues, we design a dIDG protocol based on OT and symmetric-key primitives. The core idea is to compute all intermediate values in secret‑shared form. Both Gao et al. \cite{DBLP:conf/asiacrypt/GaoQLLW24} and Dang et al. \cite{DBLP:conf/ccs/DZL25} use an OKVS to encode AHE ciphertexts of the random values \(r_k\). Our first step is also to replace AHE with ASS. Correspondingly, we replace OKVS with an OPPRF. To incorporate the prefix technique, the receiver must be able to identify which prefix decodes correctly. Fortunately, our previously introduced ECSS functionality exactly fulfills this requirement: we first run an OPPRF, followed by an ECSS protocol, so that the parties obtain additive shares of \(F_Q(\vecw)\). The receiver then locally adds its own \(F_W(\vecw)\) to obtain shares of \(F_Q(\vecw)+F_W(\vecw)\). The next step is to derive the final IDs. In prior work, a DH key exchange is used to compute \(id_{\vecw} = (F_Q(\vecw)+F_W(\vecw))^{sk_s \cdot sk_r}\), which requires expensive group exponentiation. We abstract this operation as the evaluation of a weak pseudorandom function \(F_{sk}(x)=x^{sk}\), so that the final ID can be expressed as
\[
id_{\vecw} := F_{sk_s}\!\bigl(F_{sk_r}(F_Q(\vecw)+F_W(\vecw))\bigr).
\]
To avoid exponentiation, we introduce a new primitive called shared oblivious PRF (sOPRF), a variant of OPRF in which both the input and the output are additively secret‑shared. By invoking two sequential sOPRF instances on the shares of \(F_Q(\vecw)+F_W(\vecw)\), the parties finally obtain shares of \(id_{\vecw}\), which can then be opened to the receiver.

Once the dIDG protocol is executed, the parties proceed with the same operations as in our low‑dimensional FPSI protocol, but now using the IDs produced by dIDG instead of spatial‑hashing‑based IDs. This substitution removes the \(2^d\) factor inherent in spatial hashing. The complete protocol is described in Section~\ref{sec:fpsi-sep}.

The remaining question is how to construct an efficient sOPRF. We find that the recent shared‑output OPRF of Alamati et al. \cite{DBLP:conf/crypto/AlamatiPRR24}, built from OT and symmetric‑key operations, offers a promising starting point. We revisit their protocol and introduce a simple yet crucial twist to additionally support shared input, yielding our sOPRF construction. See Section~\ref{sec:soprf} for the full description.




\end{trivlist}

\section{Preliminaries}\label{sec:preliminaries}
\subsection{Notation}\label{subsec:notation}
We use $\kappa$ and $\lambda$ to denote the computational and statistical security parameters, respectively.
We use $[n]$ to denote the set $\{1,2,\dots,n\}$. 
For a bit string $b$ we let $b_i$ denote the $i$th bit. 
We use the abbreviation PPT to denote probabilistic polynomial-time. 
We say that a function $f$ is negligible in $\kappa$ if it vanishes faster than the inverse of any polynomial in $\kappa$, 
and write it as $f(\kappa) = \mathsf{negl}(\kappa)$. $a := b$ denotes that $a$ is assigned by $b$. The operator $||$ denotes concatenation. We use $Q = \{\vecq_1,\dots,\vecq_m\}\subseteq \mathbb{U}^{d}$ to denote a set of $m$ $d$-dimensional elements. 
In this work, we consider $L_\infty$ and $L_{p\in[1,\infty)}$ distance metrics, i.e., $\dist_\infty(\vecq,\vecw) = \mathsf{max}_{k\in[d]}|q_k-w_k|$ and $\dist_p(\vecq,\vecw) = (\sum_{k\in[d]}|q_k-w_k|^p)^{\frac{1}{p}}$, where $\vecq=(q_1,\dots,q_d),\vecw=(w_1,\dots,w_d)$. 
For a key-value set $A = \{(x_i,y_i)\}_{i\in[n]}$, we use $A[x_i]$ to denote $y_i$. 



\subsection{Security Model}\label{subsec:security-model}
Similar to most previous protocols for FPSI, we use the standard notion of security in the presence of semi-honest adversaries \cite{DBLP:books/cu/Goldreich2004,DBLP:books/sp/17/Lindell17}.

Let $\mathcal{F}$ be a functionality between a sender $\mathcal{S}$ with input $Y$ and a receiver $\mathcal{R}$ with input $X$. Let $\Pi$ be a two-party protocol for computing $\mathcal{F}$.

\noindent\textbf{Semi-honest Security}. Let $\mathsf{view}_{\mathcal{P}}^{\Pi}(X,Y)$
be the views of party $\mathcal{P}$ ($\mathcal{P}\in \{\mathcal{S},\mathcal{R}\}$) in the protocol, which consists of $\mathcal{P}$'s input, randomness tape, and received messages during the protocol. Let $\mathsf{output}(X,Y)$ be the output of both parties in the protocol. 
\begin{definition}
A protocol $\Pi$ is said to securely compute functionality $\mathcal{F}$ against semi-honest $\mathcal{P}$
if for every PPT adversary $\mathcal{A}$ that corrupting $\mathcal{P}$, there exists a PPT simulator 
$\mathsf{Sim}_\mathcal{P}$ such that for all inputs $X$ and $Y$, 

\begin{center}
$\{\mathsf{view}_{\mathcal{P}}^{\Pi}(X,Y),\mathsf{output}(X,Y)\} \approx_c \{\mathsf{Sim}_{\mathcal{P}}(In(\mathcal{P}),\mathcal{F}(X,Y)),\mathcal{F}(X,Y)\}$
\end{center}
where $In(\mathcal{P})$ denotes the input of $\mathcal{P}$.
\end{definition}

\subsection{Permute+ Share}\label{subsec:ps}
The Permute + Share (PS) functionality \cite{DBLP:conf/eurocrypt/MohasselS13,DBLP:conf/asiacrypt/ChaseGP20} takes as input a vector from the sender and a permutation from the receiver. It then provides both parties with additive shares of the permuted vector. The formal definition is given in Figure \ref{fig:fps}.

\begin{figure}[!hbth]
\begin{framed}
\begin{minipage}[center]{\textwidth}
\begin{trivlist}
\item \textbf{Parameters:} Sender $\mathcal{S}$, Receiver $\mathcal{R}$, a finite field $\mathbb{F}$, vector length $n$.
    
\item \textbf{Functionality:}
\begin{itemize}
    \item Wait for input $\mathbf{x} = (x_1,\dots, x_n) \in \mathbb{F}^n$ from the sender $\mathcal{S}$.
    \item Wait for input $\pi$ from the receiver $\mathcal{R}$.
    \item Pick random $a_i\leftarrow \mathbb{F}$ and compute $b_i = x_{\pi(i)}-a_i$ for $i\in [n]$.
    \item Give $\veca = (a_1,\dots, a_n)$ to the sender $\mathcal{S}$ and give $\vecb = (b_1,\dots, b_n)$ to the receiver $\mathcal{R}$. 
\end{itemize}
\end{trivlist}
\end{minipage}
\end{framed}
\caption{Permute + Share Functionality $\Funcps$}
\label{fig:fps}
\end{figure}

\subsection{(Secret-Shared) Private Equality Test.}
\label{subsec:peqt}

In the private equality test (PEQT) functionality, a sender and a receiver respectively provide inputs $a, b \in\{0,1\}^l$, and the receiver learns a bit indicating whether $a = b$. We also define a secret-shared variant, ssPEQT, which outputs an additive boolean share of the equality result to each party. The formal definitions are provided in Figure \ref{fig:fsspeqt}.

\begin{figure}[t]
\begin{framed}
\begin{minipage}[center]{\textwidth}
\begin{trivlist}
\item \textbf{Parameters:} Sender $\mathcal{S}$, Receiver $\mathcal{R}$, item length $l$.
    
\item \textbf{Functionality:}
\begin{itemize}
    \item Wait for input $a\in \{0,1\}^l$ from $\mathcal{S}$.
    \item Wait for input $b\in \{0,1\}^l$ from $\mathcal{R}$.
    \item Define $e=1$ if $a=b$ and $e=0$ otherwise. 
    \begin{itemize}
        \item For PEQT: Output $e$ to the receiver $\mathcal{R}$.
         \item For ssPEQT: Select random $e_0 \leftarrow \{0,1\}$ and compute $e_1 := e_0 \oplus e$. Output $e_0$ to the sender $\mathcal{S}$ and output $e_1$ to the receiver $\mathcal{R}$.
    \end{itemize}
   
\end{itemize}
\end{trivlist}
\end{minipage}
\end{framed}
\caption{(Secret-Shared) Private Equality Test Functionality $\mathcal{F}_{\mathsf{PEQT}}/\mathcal{F}_{\mathsf{ssPEQT}}$.}
\label{fig:fsspeqt}
\end{figure}

\subsection{Secure AND and OR.} 
\label{subsec:andor}
The AND (resp. OR) functionality takes as input $n$-bit shares $\{b_i^0\}_{i\in[n]}$ from party $P_0$ and $\{b_i^1\}_{i\in[n]}$ from party $P_1$. It then provides each party with a share of the result $\bigwedge_{i=1}^n (b_i^0\oplus b_i^1)$ (resp. $\bigvee_{i=1}^n (b_i^0\oplus b_i^1)$). The formal definition is provided in Figure \ref{fig:fand}.


\begin{figure}[t]
\begin{framed}
\begin{minipage}[center]{\textwidth}
\begin{trivlist}
\item \textbf{Parameters:} Parties $P_0$ and $P_1$, input length $n$.
    
\item \textbf{Functionality:}
\begin{itemize}
    \item Wait for input $\{b_i^0\}_{i\in[n]}$ from the $P_0$.
    \item Wait for input $\{b_i^1\}_{i\in[n]}$ from the $P_1$.
    \item Select random $e_0 \leftarrow \{0,1\}$ and \begin{itemize}
        \item For AND: Compute $e_1 := e_0 \oplus (\bigwedge_{i=1}^n (b_i^0\oplus b_i^1))$. 
        \item For OR: Compute $e_1 := e_0 \oplus (\bigvee_{i=1}^n (b_i^0\oplus b_i^1))$. 
    \end{itemize}
    \item Output $e_0$ to $P_0$ and output $e_1$ to $P_1$.
\end{itemize}
\end{trivlist}
\end{minipage}
\end{framed}
\caption{Secure AND/OR Functionality $\mathcal{F}_\mathsf{AND}/\mathcal{F}_\mathsf{OR}$}
\label{fig:fand}
\end{figure}

\subsection{Oblivious Transfer}\label{subsec:ot}
Oblivious Transfer (OT) \cite{DBLP:journals/iacr/Rabin05} is a fundamental cryptographic primitive widely used in secure multiparty computation. While the $1$-out-of-$2$ variant is the most common, we present the formal definition for the more general $1$-out-of-$l$ OT functionality in Figure \ref{fig:fot}.

\begin{figure}[!hbth]
\begin{framed}
\begin{minipage}[center]{\textwidth}
\begin{trivlist}
\item \textbf{Parameters:} Sender $\mathcal{S}$, Receiver $\mathcal{R}$, message length $\kappa$ 
    
\item \textbf{Functionality:}
\begin{itemize}
    \item Wait for input $b\in [l]$ from the receiver $\mathcal{R}$.
    \item Wait for input $(x_1,x_2,\dots,x_l)$ from the sender $\mathcal{S}$.
    \item Give $x_b$ to the receiver $\mathcal{R}$.
\end{itemize}
\end{trivlist}
\end{minipage}
\end{framed}
\caption{$1$-out-of-$l$ Oblivious Transfer Functionality $\FuncOT$}
\label{fig:fot}
\end{figure}

\subsection{Inner Product}
\label{subsec:ip}
The inner product (IP) functionality takes a length-$n$ vector from both a sender and a receiver, and outputs their inner product to the receiver. This functionality can be realized via $n$ instances of Oblivious Linear Evaluation (OLE). Its formal definition is given in Figure \ref{fig:fip}.

\begin{figure}[!hbth]
\begin{framed}
\begin{minipage}[center]{\textwidth}
\begin{trivlist}
\item \textbf{Parameters:} Sender $\mathcal{S}$, Receiver $\mathcal{R}$, a finite field $\mathbb{F}$, vector length $n$.
    
\item \textbf{Functionality:}
\begin{itemize}
    \item Wait for input $\veca = (a_1,\dots,a_n)\in \mathbb{F}^n$ from the sender $\mathcal{S}$.
    \item Wait for input $\vecb = (b_1,\dots,b_n)\in \mathbb{F}^n$ from the receiver $\mathcal{R}$.
    \item Compute $c=\langle\veca,\vecb\rangle=\sum_{i\in[n]}a_i\cdot b_i$.
    \item Give $c$ to the receiver $\mathcal{R}$. 
\end{itemize}
\end{trivlist}
\end{minipage}
\end{framed}
\caption{Inner Product Functionality $\mathcal{F}_\mathsf{IP}$}
\label{fig:fip}
\end{figure}

\subsection{Oblivious programmable PRF}
\label{subsec:opprf}
Programmable PRF (PPRF) \cite{DBLP:conf/ccs/KolesnikovMPRT17} consists of two algorithms $(\mathsf{Hint},F)$ and resembles a standard PRF, except it outputs programmed values on a specified set of input keys. The $\mathsf{Hint}$ algorithm takes a PRF key $k$, a set of programmed keys $K$, and corresponding values $V$, and outputs a hint $\hint$ that encodes these key-value pairs.

An Oblivious PPRF (OPPRF), based on a PPRF scheme $(\mathsf{Hint},F)$, is a two-party protocol between a sender and a receiver. The sender inputs $(K,V)$, the receiver inputs a query set $X = \{x_i\}_{i\in [t]}$, and the functionality samples a random key $k$, computes $\hint \leftarrow \mathsf{Hint}(k,K,V)$ and $f_i = F(k,\hint,x_i)$, then gives $k,\hint$ to the sender and $\hint,\{f_i\}_{i\in[t]}$ to the receiver. We give the formal definition of OPPRF functionality in Figure \ref{fig:fopprf}.


\begin{figure}[!hbtp]
\begin{framed}
\begin{minipage}[center]{\textwidth}
\begin{trivlist}
\item \textbf{Parameters.} Sender $\mathcal{S}$. Receiver $\mathcal{R}$. Program size $n$. Query size $t$. PPRF scheme $(\mathsf{Hint},F)$.
\item \textbf{Functionality:} 
\begin{itemize}
\item Wait for input $(K,V)$ from $\mathcal{S}$, where $K$ is distinct key set and $V$ is value set, $|K|= |V| =n$. 
\item Wait for input $X = \{x_1,\dots,x_t\} \subseteq \{0,1\}^l$ from $\mathcal{R}$,
\item Sample uniformly random key $k$.
\item Compute $\hint\leftarrow \mathsf{Hint}(k,K,V), f_i:= F(\hint,k,x_i),i\in [t]$.
\item Give $k,\hint$ to the sender $\mathcal{S}$.
\item Give $\hint,  \{f_i\}_{i\in[t]}$ to the receiver $\mathcal{R}$.
\end{itemize}
\end{trivlist}
\end{minipage}
\end{framed}
\caption{OPPRF Functionality $\Funcopprf$}
\label{fig:fopprf}
\end{figure}

\subsection{Oblivious Key-Value Stores}\label{subsec:okvs}
The Oblivious Key-Value Store (OKVS) \cite{DBLP:conf/eurocrypt/PinkasRTY20,DBLP:conf/crypto/GarimellaPRTY21,285493} is a data structure that maps a set of keys to corresponding values such that, when values are random, the structure's distribution reveals no information about the keys. The formal definition is as follows:

\begin{definition}[Oblivious Key-Value Store]
An OKVS is parameterized by a set $\mathcal{K}$ of keys, a set $\mathcal{V}$ of values, and consists of two algorithms:
\begin{itemize}
\item $\mathsf{Encode}(\{(x_1,y_1),\dots,(x_n,y_n)\})$: on input key-value pairs $\{(x_i, y_i)\}_{i\in [n]} \subseteq
    \mathcal{K}\times \mathcal{V}$, outputs an object $E$ (or, probability $2^{-\lambda}$, an error $\perp$).

\item $\mathsf{Decode}(E, x):$ on input $E$ and a key $x$, outputs a value $y \in \mathcal{V}$. 
\end{itemize}
\end{definition}

\begin{trivlist}
\item\textbf{Correctness.} For all $A\subseteq \mathcal{K}\times \mathcal{V}$ with distinct keys:
\begin{center}
$(x,y)\in A$ and $\perp\ne E \leftarrow \mathsf{Encode}(A)\Longrightarrow \mathsf{Decode}(E,x) = y$
\end{center}
    
\item\textbf{Obliviousness.} For all distinct $\{x_1^0,\dots,x_n^0\}$ and $\{x_1^1,\dots,x^1_n\}$, if $\mathsf{Encode}$ does not output $\perp$ for $\{x_1^0,\dots,x_n^0\}$ or $\{x_1^1,\dots,x_n^1\}$, then the following distributions are computationally indistinguishable: 
\begin{center}
    $\{E|y_i\leftarrow \mathcal{V},i\in [n], \mathsf{Encode}((x_1^0,y_1),\dots,(x_n^0,y_n))\}$ \\
    $\{E|y_i\leftarrow \mathcal{V},i\in [n], \mathsf{Encode}((x_1^1,y_1),\dots,(x_n^1,y_n))\}$
\end{center}

\item\textbf{$t$-Randomness \cite{cryptoeprint:2025/885}.} For any $A= \{(x_1,y_1), \dots, (x_n,y_n)\}$ and for all distinct $\{x_i^*\}_{i\in[t]}$ satisfying $x_i^* \notin \{x_1,\dots,x_n\},i\in[t]$, the distribution of $\{\mathsf{Decode}(E,x_i^*)\}_{i\in[t]}$ is statistically indistinguishable to that of uniform distribution over $\mathcal{V}^t$, where $E\leftarrow \mathsf{Encode}(A)$. 
\end{trivlist}

\subsection{Prefix Trie}\label{subsec:prefix}
The Prefix Trie technique \cite{DBLP:conf/uss/ChakrabortiFR23,AC/BP25} reduces the problem of testing integer interval membership to a set intersection problem between ${O}(\log \delta)$-sized sets. The key insight is that any length-$O(\delta)$ interval can be represented using $O(\log \delta)$ distinct prefixes. This approach uses two algorithms:
\begin{itemize}
    \item $\mathsf{PrefixTrie}(q-\delta,q+\delta)\rightarrow \{{q}_1^*,\dots,{q}_l^*\}$
    \item $\mathsf{PrefixPath}(w,\delta)\rightarrow \{{w}_1^*,\dots,{w}_{l'}^*\}$
\end{itemize}
where $l= O(\log \delta), l'= O(\log \delta)$. 

The $\mathsf{PrefixTrie}$ algorithm takes the two endpoints of an input interval and outputs a prefix set that represents this interval. The $\mathsf{PrefixPath}$ algorithm takes a number and an interval radius as inputs, and returns a set of prefixes associated with that number. Their correctness guarantees that for any integers $w,q\in \mathbb{Z}$, $w\in [q-\delta,q+\delta]$ if and only if there exists a unique $i\in [l']$ such that ${w}^*_i\in \{{q}^*_1,\dots,{q}^*_l\}$. The original construction of \cite{DBLP:conf/uss/ChakrabortiFR23} required $O(\delta)$ time complexity, and this was later improved by \cite{DBLP:conf/eurocrypt/BaarsenP24} to $O(\log \delta)$.

A prefix of the form $q^* = q_l q_{l-1} \dots q_t \in \{0,1\}^{l-t+1}$ denotes the interval whose left endpoint is $q_l q_{l-1} \dots q_t 0 \dots 0$ and whose right endpoint is $q_l q_{l-1} \dots q_t 1 \dots 1$. We define the right and left endpoints of the interval respectively as $\mathsf{Bound}(0,q^*) \;:=\; q_l q_{l-1} \dots q_t 1 \dots 1$, $
\mathsf{Bound}(1,q^*) \;:=\; q_l q_{l-1} \dots q_t 0 \dots 0$.

\subsection{(Secret-Shared) Fuzzy Matching for Interval}
\label{subsec:ifmap}
The fuzzy matching for interval functionality (IFmat) \cite{DBLP:conf/uss/ChakrabortiFR23,DBLP:conf/asiacrypt/GaoQLLW24} enables a sender with an interval and a receiver with a number to determine whether the number falls within that interval. We also define a secret-shared variant, referred to as ssIFmat, where the output is delivered to both parties in the form of additive shares. A formal definition is provided in Figure \ref{fig:fifmat}.

\begin{figure}[!hbth]
\begin{framed}
\begin{minipage}[center]{\textwidth}
\begin{trivlist}
\item \textbf{Parameters:} Sender $\mathcal{S}$, Receiver $\mathcal{R}$, threshold $\delta$.

\item \textbf{Functionality:}
\begin{itemize}
\item Wait for input $x\in \mathbb{Z}$ from the sender $\mathcal{S}$.
\item Wait for input $y\in \mathbb{Z}$ from the receiver $\mathcal{R}$.
\item Define $b=1$ if $y\in [x-\delta,x+\delta]$, otherwise $b=0$.
\begin{itemize}
    \item For IFmat: Give $b$ to the receiver $\mathcal{R}$.
    \item For ssIFmat: Pick random $b_0\leftarrow\{0,1\}$ and compute $b_1:=b\oplus b_0$. Give $b_0$ to the sender $\mathcal{S}$ and give $b_1$ to the receiver $\mathcal{R}$.
\end{itemize}

\end{itemize}
\end{trivlist}
\end{minipage}
\end{framed}
\caption{(Secret-Shared) Fuzzy Matching for Interval Functionality $\Funcifmat/\Funcssifmat$}
\label{fig:fifmat}
\end{figure}

The design of the existing IFmat protocol~\cite{DBLP:conf/uss/ChakrabortiFR23} proceeds as follows: both parties first compute their respective prefix sets, and then invoke a PSI‑Cardinality (PSI‑CA) protocol to obtain the output of IFmat. To obtain a secret‑shared output, we follow a similar high‑level approach, but with a key difference: we replace the PSI‑CA step with a circuit‑PSI protocol, which allows the parties to learn secret‑shares indicating whether each element belongs to the intersection. Because prefix encoding guarantees that the intersection contains at most one element, we can simplify existing circuit‑PSI protocols~\cite{DBLP:conf/eurocrypt/Pinkas0TY19,DBLP:conf/eurocrypt/RindalS21} by omitting the hashing phase. The parties can then simply XOR the shares of the corresponding indicator bits. A formal description of our ssIFmat protocol is given in Figure~\ref{fig:pissifmat}.

\begin{figure}[!hbth]
\begin{framed}
\begin{minipage}[center]{\textwidth}
\begin{trivlist}
\item \textbf{Parameters:} 
\begin{itemize}
\item Two parties: Sender $\mathcal{S}$, Receiver $\mathcal{R}$, threshold $\delta$.
\item Ideal $\Funcopprf$ and $\Funcsspeqt$ primitives specified in Figure \ref{fig:fopprf} and \ref{fig:fsspeqt}, respectively. 
\end{itemize}

\item Input of $\mathcal{S}$: $x\in \mathbb{Z}$

\item Input of $\mathcal{R}$: $y\in \mathbb{Z}$

\item \textbf{Protocol:}

\begin{enumerate}

\item $\mathcal{S}$ computes $X=\{\tilde{x}_1,\dots,\tilde{x}_{l}\}:=\mathsf{PrefixTrie}(x-\delta,x+\delta)$.
\item $\mathcal{R}$ computes $Y=\{\tilde{y}_1,\dots,\tilde{y}_{l'}\}:=\mathsf{PrefixPath}(y,\delta)$.

\item $\mathcal{S}$ selects random $r\leftarrow \{0,1\}^l$.
\item $\mathcal{S}$ and $\mathcal{R}$ invoke the OPPRF functionality $\Funcopprf$. $\mathcal{S}$ acts as the sender in OPPRF with input $\{(\tilde{x}_i,r)\}_{i\in[l]}$ , and learns $k,\hint$. $\mathcal{R}$ acts as the receiver in OPPRF with input $Y$ and receives $\{r'_j\}_{j\in[l']}$. 

\item The parties invoke $l'$ instances of $\Funcsspeqt$, where in the $j$-th instance $\mathcal{R}$ inputs $r_j'$ and $\mathcal{S}$ inputs $r$. As a result, $\mathcal{S}$ and $\mathcal{R}$ receives $b_j^0, b_j^1 \in \{0,1\}$ respectively.
\item $\mathcal{S}$ outputs $\oplus_{j\in[l']}b_j^0$ and $\mathcal{R}$ outputs $\oplus_{j\in[l']}b_j^1$.

\end{enumerate}
\end{trivlist}
\end{minipage}
\end{framed}
\caption{Secret-Shared Fuzzy Matching for Interval Protocol} 
\label{fig:pissifmat}
\end{figure}

\subsection{Cuckoo Hashing and Simple Hashing}
\label{subsec:cuckoo}
The hash-to-bin technique using Cuckoo/Simple hashing was introduced by Pinkas et al. \cite{DBLP:conf/uss/Pinkas0Z14,DBLP:conf/uss/Pinkas0SZ15} for PSI. The sender uses hash functions $h_1,\dots,h_\alpha$ to map $m$ items into $m_c = (1+\epsilon)m$ bins via Cuckoo hashing \cite{DBLP:journals/jal/PaghR04}, ensuring at most one item per bin, while the receiver assigns each item $w\in W$ to bins $h_1(w),\dots,h_\alpha(w)$ via simple hashing. Following \cite{DBLP:journals/tissec/PinkasSZ18}, we use three hash functions with zero stash size, achieving failure probability $2^{-\lambda}$.

We generalize this scheme by decoupling bin index computation from stored items. Let $Y^*\leftarrow\mathsf{Cuckoo}_{h_1,h_2,h_3}^{m_c}(\{(x_i,y_i)\}_{i\in[m]})$ denote building a Cuckoo table on $X=\{x_i\}_{i\in [m]}$ while storing $Y=\{y_i\}_{i\in[m]}$ in bins, with similar notation for $\mathsf{Simple}$. When $X=Y$, this reduces to traditional simple/cuckoo hashing, abbreviated as $\mathsf{Cuckoo}_{h_1,h_2,h_3}^{m_c}(X)$ and $\mathsf{Simple}_{h_1,h_2,h_3}^{m_c}(X)$.


\subsection{Spatial Hashing}\label{subsec:sphash}

Spatial hashing \cite{DBLP:conf/crypto/GarimellaRS22,DBLP:conf/eurocrypt/BaarsenP24} partitions the space into grid cells, limiting point matching to a bounded number of neighboring cells. This technique plays a role in FPSI analogous to Cuckoo/Simple hashing in standard PSI. However, since the number of relevant cells grows exponentially with dimension, the approach is primarily suitable for low-dimensional spaces.

Formally, for vectors $\vecq = (q_1,\dots,q_d)$ and $\vecw = (w_1,\dots,w_d)$, define:
\begin{itemize}
    \item $\mathsf{cell}_\delta(\vecq) := (\lfloor q_1/2\delta\rfloor,\dots,\lfloor q_d/2\delta\rfloor)$
    \item $\mathsf{neighbor}_\delta(\vecw):=\{(\lfloor w_1/2\delta\rceil-b_1,\dots,\lfloor w_d/2\delta\rceil-b_d)|i\in[d],b_i\in \{0,1\}\}$
\end{itemize}

Baarsen and Pu \cite{DBLP:conf/eurocrypt/BaarsenP24} established the following lemmas.

\begin{lemma}[Maximal Distance in a Cell]\label{lemma:1}
    Given two points $\vecq,\vecw\in \mathbb{U}^d$ located in the same cell with side length $2\delta$, then the distance between them is $\mathsf{dist}_p(\vecw, \vecq) < 2\delta d^{\frac{1}{p}}$ where $p \in [1,\infty]$. Specifically, if $p = \infty, \mathsf{dist}_\infty(\vecw, \vecq) < 2\delta$.
\end{lemma}

\begin{lemma}[Unique Center]\label{lemma:2}
   Suppose there are multiple $L_p$ balls $(p \in [1,\infty])$ with radius $\delta$ lying in a $d$-dimensional space which is tiled by cells with side length $2\delta$. If these balls’ centers are at least $2\delta d^{\frac{1}{p}}$ apart from each other, then for each cell, there is at most one center lying in this cell. Specifically, if $p =\infty $, then this holds for disjoint balls, since $2\delta d^{\frac{1}{p}}$ degrades to $2\delta$ in this case.
\end{lemma}

\begin{lemma}[Unique Ball]\label{lemma:3}
   Suppose there are multiple $L_p$ balls $(p \in [1,\infty])$ with radius $\delta$ lying in a $d$-dimensional space which is tiled by cells with side length $2\delta$. If these balls’ centers are at least $2\delta (d^{\frac{1}{p}}+1)$ apart from each other, then there exists at most one ball intersecting with the same cell. Specifically, if $p =\infty $, then this holds for $L_\infty$ balls with $4\delta$-apart centers.
\end{lemma}


\subsection{Fuzzy Mapping and Distributed ID Generation}
\label{subsec:fmap}
Fuzzy mapping \cite{DBLP:conf/asiacrypt/GaoQLLW24,cryptoeprint:2025/885} generates a set of IDs for each element, ensuring that if two elements are within a distance $\delta$, their corresponding ID sets have a non-empty intersection. We give the formal definition in \cite{cryptoeprint:2025/885} below.

\begin{definition}[Fuzzy Mapping]
    A $(\rs,\rr)$-Fuzzy Mapping scheme for distance metric $\dist$ consists a tuple of algorithms $\mathsf{Fmap} = (\setup, \ids,\idr)$:
    \begin{itemize}
\item $\setup(Q,W)$: on input two sets $Q = \{\vecq_j\}_{j\in[m]}\subseteq \mathbb{U}^{d}, W = \{\vecw_i\}_{i\in[n]}\subseteq \mathbb{U}^{d}$, outputs a state $\sta$.
    
\item $\ids(\vecq,\sta)$: on input an item $\vecq\in \mathbb{U}^d$ and a state $\sta$, outputs a size-$\rs$ ID set $\id_\vecq = \{id^{\mathcal{S}}_1,\dots,id^{\mathcal{S}}_{\rs}\}$.

\item $\idr(\vecw,\sta)$: on input an item $\vecw\in \mathbb{U}^d$ and a state $\sta$, outputs a size-$\rr$ ID set $\id_\vecw = \{id^{\mathcal{R}}_1,\dots,id^{\mathcal{R}}_{\rr}\}$.
\end{itemize}
$\rs$ and $\rr$ are the expansion rates of the sender and receiver, respectively. The expansion rate of the Fuzzy mapping scheme is $\mathsf{r}:=\mathsf{max}(\rs,\rr)$.
\end{definition}

\begin{trivlist}
\item\textbf{Correctness.} For any $\vecq\in Q$ and $\vecw\in W$, if $\dist(\vecq,\vecw)\leq \delta$, then $\id_\vecq\cap \id_\vecw \neq \emptyset$, where $\sta\leftarrow \setup(Q,W), \id_\vecq\leftarrow \ids(\vecq,\sta),$ $\id_\vecw\leftarrow \idr(\vecw,\sta)$.
    
\item\textbf{Distinctiveness.}  $\forall i,i'\in [n],i\ne i'$, the probability of $\id_{\vecw_i}\cap \id_{\vecw_{i'}} \ne \emptyset$ is negligible, where $\sta\leftarrow \setup(Q,W),\id_{\vecw_i}\leftarrow \idr(\sta,\vecw_{i}),\id_{\vecw_{i'}}\leftarrow \idr(\sta,\vecw_{i'}) $.

\item\textbf{Randomness.}  For any $Q = \{\vecq_j\}_{j\in[m]}\subseteq \mathbb{U}^{d}, W = \{\vecw_i\}_{i\in[n]}\subseteq \mathbb{U}^{d}$, the distribution of $\{\id_\vecq\}_{\vecq\in Q}$ and $\{\id_\vecw\}_{\vecw\in W}$ is statistically indistinguishable to that of uniform distribution, where $\sta\leftarrow \setup(Q,W), \id_\vecq\leftarrow \ids(\vecq,\sta),\id_\vecw\leftarrow \idr(\vecw,\sta),\vecq\in Q, \vecw\in W$.

\item \textbf{Silence.} The output of algorithm $\setup$ is $\perp$. In other words, the algorithms $\ids(\vecq,\sta) $ and $\idr(\vecw,\sta)$ do not require a common input and work independently.

\end{trivlist}


A fuzzy mapping scheme must guarantee correctness and distinctiveness. Randomness and silence are optional, with a key constraint: a non-silent scheme must be random. 

Fuzzy mapping serves to reduce the number of required element comparisons in FPSI. Rather than comparing distances for all $m \cdot n$ element pairs, only those sharing an ID need to be compared. 

Spatial hashing is one instantiation of fuzzy mapping, satisfying the properties of correctness, distinctiveness, and silence, but it incurs an expansion rate of $2^d$. To achieve a constant expansion rate of $1$, Gao et al. \cite{DBLP:conf/asiacrypt/GaoQLLW24} introduced an interactive method for ID generation, termed distributed ID generation (dIDG). We formally define this functionality in Figure \ref{fig:fdidg}.

\begin{figure}[!hbth]
\begin{framed}
\begin{minipage}[center]{\textwidth}
\begin{trivlist}
\item \textbf{Parameters:} Sender $\mathcal{S}$, Receiver $\mathcal{R}$, set size $m,n$, a $(\rs,\rr)$-fuzzy mapping scheme $\mathsf{Fmap}=(\setup,\ids,\idr)$.
    
\item \textbf{Functionality:}
\begin{itemize}
    \item Wait for input $Q = \{\vecq_j\}_{j\in[m]}\subseteq \mathbb{U}^{d}$ from the sender $\mathcal{S}$.
    \item Wait for input $W = \{\vecw_i\}_{i\in[n]}\subseteq\mathbb{U}^{d}$ from the receiver $\mathcal{R}$.
    \item Compute $\sta\leftarrow \setup(Q,W)$. Define $\id_Q := \{\ids(\vecq_j,\sta)\}_{j\in [m]}$ and $\id_W := \{\idr(\vecw_i,\sta)\}_{i\in [n]}$. 
    \item Give $\id_Q$ to the sender $\mathcal{S}$ and give $\id_W$ to the receiver $\mathcal{R}$.
\end{itemize}
\end{trivlist}
\end{minipage}
\end{framed}
\caption{Distributed ID Generation Functionality $\Funcdidg$}
\label{fig:fdidg}
\end{figure}

\subsection{Fuzzy Private Set Intersection and Its Variants}
\label{subsec:fpsi}
In this section, we define the functionality of FPSI and its variants, including FPSI, FPSI-cardinality (FPSI-CA), labeled FPSI (LFPSI), and FPSI with sender privacy (FPSI-SP). The formal definition of these functionalities is given in Figure \ref{fig:ffpsi}.

\begin{figure}[!hbth]
\begin{framed}
\begin{minipage}[center]{\textwidth}
\begin{trivlist}
\item \textbf{Parameters:} Sender $\mathcal{S}$, Receiver $\mathcal{R}$, set sizes $m$ and $n$

\item \textbf{Functionality:}
\begin{itemize}
\item Wait for input $Q = \{\vecq_j\}_{j\in[m]}\subseteq \mathbb{U}^{d}$ from the sender $\mathcal{S}$.\begin{itemize}
    \item For LFPSI, wait another input $\mathsf{Label}_Q = \{\mathsf{label}_j\}\subseteq \{0,1\}^{l}$ from $\mathcal{S}$.
\end{itemize}
\item Wait for input $W = \{\vecw_i\}_{i\in[n]}\subseteq\mathbb{U}^{d}$ from the receiver $\mathcal{R}$.
 \begin{itemize}
    \item For FPSI: Give output $I:=\{\vecq_j|\exists i\in[n], \dist(\vecq_j,\vecw_i)\leq \delta\}$ to the receiver $\mathcal{R}$.
    \item For FPSI-SP: Give output $I:=\{\vecw_i|\exists j\in[m], \dist(\vecq_j,\vecw_i)\leq \delta\}$ to the receiver $\mathcal{R}$.
    \item For FPSI-CA: Give output $I:=\sum_{i\in[n],j\in [m]}\mathsf{1}(\dist(\vecq_j,\vecw_i)\leq \delta)$ to the receiver $\mathcal{R}$.
    \item For LFPSI: Give output $I:=\{\mathsf{label}_j|\exists i\in[n], \dist(\vecq_j,\vecw_i)\leq \delta\}$ to the receiver $\mathcal{R}$.
\end{itemize}
\end{itemize}
\end{trivlist}
\end{minipage}
\end{framed}
\caption{Fuzzy PSI and Its Variants Functionalities $\Funcfpsi,\Funcfpsisp,\Funcfpsica,\Funclfpsi$}
\label{fig:ffpsi}
\end{figure}

\section{Equality Conditional Sum/Inner-Product Share}
\label{sec:ecss}

In this section, we introduce the equality conditional sum share (ECSS) and equality conditional inner-product share (ECIPS), two new primitives that serve as the core building blocks for our FPSI protocols supporting the $L_\infty$ and $L_{p\in[1,\infty)}$ distance metrics, respectively.

\subsection{Equality Conditional Sum Share}
\label{subsec:ecss}

We define equality conditional secret sharing (ECSS) and its variant, equality conditional sum (ECS). As the name suggests, the difference is that ECS sends the summation result directly to the receiver, while ECSS outputs a secret sharing between the parties.

In ECS, the sender and receiver input sets of two-dimensional vectors: $A = \{ \veca_{k,v} = (a_{k,v}^0, a_{k,v}^1) \}_{k \in [d], v \in [l]}$ and $B = \{ \vecb_{k,v} = (b_{k,v}^0, b_{k,v}^1) \}_{k \in [d], v \in [l]}$. The superscript $1$ is used for comparison and superscript $0$ is used for summation.

For each dimension $k$, the functionality finds $v_k^* \in [l]$ such that $a_{k,v_k^*}^1 = b_{k,v_k^*}^1$. If multiple exist, $v_k^*$ is set as their XOR\footnote{In our FPSI, at most one such $v_k^*$ exists per dimension. We define the functionality in this general form because our ECS construction inherently satisfies this property.}; otherwise $v_k^*$ is chosen randomly. It then computes $\mathsf{Sum} := \sum_{k \in [d]} (a_{k, v_k^*}^0 + b_{k, v_k^*}^0)$ and sends $\mathsf{Sum}$ to the receiver. For ECSS, the parties learn secret shares of $\mathsf{Sum}$.

The ideal functionality for ECS(S) is given in Figure~\ref{fig:fecss}.

\begin{figure}[!hbth]
\begin{framed}
\begin{minipage}[center]{\textwidth}
\begin{trivlist}
\item \textbf{Parameters:} Sender $\mathcal{S}$, Receiver $\mathcal{R}$, a finite field $\mathbb{F}$, the number of summation terms $d$, the size of each comparison set $l$
    
\item \textbf{Functionality:}
\begin{itemize}
    \item Wait for input $A = \{\veca_{k,v} = (a_{k,v}^0,a_{k,v}^1)\}_{k\in[d],v\in[l]}\subseteq \mathbb{F}^{2dl}$ from the sender $\mathcal{S}$.
    \item Wait for input $B = \{\vecb_{k,v} = (b_{k,v}^0,b_{k,v}^1)\}_{k\in[d],v\in[l]}\subseteq \mathbb{F}^{2dl}$ from the receiver $\mathcal{R}$.
    \item For $k\in[d]$:\begin{itemize}
        \item  Define $E_k:=\{v\in[l]:a_{k,v}^1 = b_{k,v}^1\}$
        \item  If $E_k = \emptyset$: $v_k^*\leftarrow [l]$
        \item  Else, $v_k^*:= \oplus_{i\in E_k} i$
    \end{itemize}
    \item Compute $\mathsf{Sum}:= \sum_{k\in[d]} (a_{k,v_k^*}^0+b_{k,v_k^*}^0)$
    \begin{itemize}
        \item For $\Funcecs$: Give $\mathsf{Sum}$ to $\mathcal{R}$.
        \item For $\Funcecss$: Sample $o_s\leftarrow \mathbb{F}$ and compute $o_r:= \mathsf{Sum}-o_s$. Give $o_s$ to $\mathcal{S}$ and give $o_r$ to $\mathcal{R}$.
    \end{itemize}
\end{itemize}
\end{trivlist}
\end{minipage}
\end{framed}
\caption{Equality Conditional Sum (Share) Functionality $\mathcal{F}_{\mathsf{ECS(S)}}$}
\label{fig:fecss}
\end{figure}

We present our ECS(S) protocol in Figure \ref{fig:piecss}, which requires calling a sharing conversion algorithm $\mathsf{ShareTrans}$ to convert the $l$-bit vector\footnote{We assume $l=2^\eta$ for some integer $\eta$.} sharing into the sharing in $[l]$, representing the XOR at position $1$ of this vector.\footnote{For example, for $l=4$, the algorithm $\mathsf{ShareTrans}$ will convert the sharing of $0011$ into the sharing of $1\oplus 2$.} The $\mathsf{ShareTrans}$ algorithm is given in Figure \ref{alg:sharetrans}.

\begin{figure}[!hbth]
\begin{framed}
\begin{minipage}[center]{\textwidth}
$\mathsf{ShareTrans}(e_1e_2\dots e_l\in \{0,1\}^l)$:
\begin{enumerate}
\item Define $S_\alpha:=\{x_\eta\dots x_{\alpha+1}1x_{\alpha-1}\dots x_1: x_1,\dots,x_{\alpha-1},x_{\alpha+1},\dots,x_\eta\in \{0,1\}\},\alpha\in [\eta] $.
\item Compute $\tilde{e}_\alpha:= \oplus_{i\in S_\alpha}e_i, \alpha\in [\eta]$.
\item Output $\tilde{e}:=\tilde{e}_\eta\dots \tilde{e}_1\in [l]$.
\end{enumerate}
\end{minipage}
\end{framed}
\caption{Algorithm $\mathsf{ShareTrans}$} 
\label{alg:sharetrans}
\end{figure}

We argue the correctness of the $\mathsf{ShareTrans}$. 
If $e_\beta$ is a share of $1$, and the $\alpha$-th bit of $\beta$ is $1$. Then $\beta \in S_\alpha$, so $e_\beta$ is XORed into $\tilde{e}_\alpha$. This means the $\alpha$-th bit of $\tilde{e}$ accumulates the effect of a share of $1$. Consequently, $\tilde{e}$ can be viewed as the share representation of the XOR over all $\beta$ for which $e_\beta$ is a share of $1$. For the case where all $e_\beta$ are shares of $0$, it follows that $\tilde{e}$ will also be a share of $0$.

As outlined in the introduction, we construct the ECS protocol as follows. First, the parties invoke PS to permute the index $v$. They then run ssPEQT, $\mathsf{ShareTrans}$, and 2-choice-bit OT to give the receiver the permuted $v_k^*$. Finally, the parties use the mask-then-unmask trick to obtain the $\mathsf{Sum}$ for the receiver. We note that the OR in step 4 is to handle cases where multiple $v_k^*$ satisfy $a_{k,v_k^*}^1 = b_{k,v_k^*}^1$. In our FPSI construction, at most one such $v_k^*$ exists, so the parties can locally XOR their results without invoking the OR functionality.

\begin{figure}[!hbth]
\begin{framed}
\begin{minipage}[center]{\textwidth}
\begin{trivlist}
\item \textbf{Parameters:} 
\begin{itemize}
\item Two parties: sender $\mathcal{S}$ and receiver $\mathcal{R}$, a finite field $\mathbb{F}$, the number of summation terms $d$, the size of each comparison set $l$.
\item Ideal $\Funcps$, $\Funcsspeqt$, $\Funcor$ and $\FuncOT$ primitives specified in Figure \ref{fig:fps}, Figure \ref{fig:fsspeqt}, Figure \ref{fig:fand} and Figure \ref{fig:fot}, respectively. 
\end{itemize}

\item Input of $\mathcal{S}$: $A = \{\veca_{k,v} = (a_{k,v}^0,a_{k,v}^1)\}_{k\in[d],v\in[l]}\subseteq \mathbb{F}^{2dl}$

\item Input of $\mathcal{R}$: $B = \{\vecb_{k,v} = (b_{k,v}^0,b_{k,v}^1)\}_{k\in[d],v\in[l]}\subseteq \mathbb{F}^{2dl}$

\item \textbf{Protocol:}

\begin{enumerate}
\item For $k\in[d]$: $\mathcal{S}$ picks a random permutation $\pi_k$ over $[l]$. Then, the parties invoke the PS functionality $\Funcps$. $\mathcal{S}$ acts as receiver with input $\pi_k$ and $\mathcal{R}$ acts as
sender with input $\{\vecb_{k,v}\}_{v\in[l]}$. As a result, $\mathcal{S}$ receives $\{\vecb^\mathcal{S}_{k,v} = (b^{\mathcal{S},0}_{k,v},b^{\mathcal{S},1}_{k,v})\}_{v\in[l]}$ and $\mathcal{R}$ receives $\{\vecb^\mathcal{R}_{k,v}=(b^{\mathcal{R},0}_{k,v},b^{\mathcal{R},1}_{k,v})\}_{v\in[l]}$, where $\vecb^\mathcal{S}_{k,v} + \vecb^\mathcal{R}_{k,v} = \vecb_{k,\pi_k(v)}$.

\item $\mathcal{S}$ defines $\tilde{\veca}_{k,v}= (\tilde{a}^{0}_{k,v},\tilde{a}^{1}_{k,v}) := (a^0_{k,\pi_k(v)}+b^{\mathcal{S},0}_{k,v},a^1_{k,\pi_k(v)}-b^{\mathcal{S},1}_{k,v}), k\in[d],v\in[l]$.

\item For $k\in[d],v\in[l]$: $\mathcal{S}$ and $\mathcal{R}$ invoke the ssPEQT functionality $\Funcsspeqt$. $\mathcal{S}$ inputs $\tilde{a}_{k,v}^1$ and $\mathcal{R}$ inputs $b^{\mathcal{R},1}_{k,v}$. As a result,  $\mathcal{S}$ and $\mathcal{R}$ receives $e_{k,v}^\mathcal{S}, e_{k,v}^{\mathcal{R}} \in \{0,1\}$ respectively.

\item For $k\in[d]$: $\mathcal{S}$ and $\mathcal{R}$ invoke the secure OR functionality $\Funcor$. $\mathcal{S}$ inputs $\{e_{k,v}^{\mathcal{S}}\}_{v\in[l]}$ and $\mathcal{R}$ inputs $\{e_{k,v}^{\mathcal{R}}\}_{v\in[l]}$. As a result,  $\mathcal{S}$ and $\mathcal{R}$ receives $\tilde{e}_{k,0}^\mathcal{S}, \tilde{e}_{k,0}^{\mathcal{R}} \in \{0,1\}$ respectively.

\item For $k\in[d]$: $\mathcal{S}$ computes $\tilde{e}_k^{\mathcal{S}}: = \mathsf{ShareTrans}(e^{\mathcal{S}}_{k,1}\dots e^{\mathcal{S}}_{k,l})$. $\mathcal{R}$ computes $\tilde{e}_k^{\mathcal{R}}: = \mathsf{ShareTrans}(e^{\mathcal{R}}_{k,1}\dots e^{\mathcal{R}}_{k,l})$.

\item For $k\in[d]$: $\mathcal{S}$ picks random $r_k\leftarrow [l]$. Then, it defines $m_{k,0}:=r_k, m_{k,1} = \tilde{e}^{\mathcal{S}}_k$ if $\tilde{e}^{\mathcal{S}}_{k,0}=0$, otherwise it defines $m_{k,0}:=\tilde{e}^{\mathcal{S}}_k, m_{k,1} = r_k$. 

\item  For $k\in[d]$: $\mathcal{S}$ and $\mathcal{R}$ invoke the $1$-out-of-$2$ OT functionality $\FuncOT$. $\mathcal{S}$ inputs $(m_{k,0},m_{k,1})$ and $\mathcal{R}$ inputs $\tilde{e}_{k,0}^{\mathcal{R}}$. As a result,  $\mathcal{S}$ learns nothing and $\mathcal{R}$ learns $m_k^* = m_{k,\tilde{e}_{k,0}^{\mathcal{R}}}$. Then, $\mathcal{R}$ computes $\tilde{v}_k:=m_k^*\oplus \tilde{e}^\mathcal{R}_k$.

\item \begin{itemize}
    \item For ECSS: $\mathcal{S}$ picks random $\mathsf{mask}_k\leftarrow \mathbb{F}$, $k\in[d]$.
    \item For ECS: $\mathcal{S}$ picks random $\mathsf{mask}_k\leftarrow \mathbb{F}$, $k\in[d-1]$. Then, it computes $\mathsf{mask}_d:=-\sum_{k\in[d-1]}\mathsf{mask}_k$.
\end{itemize}

\item  For $k\in[d]$: $\mathcal{S}$ and $\mathcal{R}$ invoke the $1$-out-of-$l$ OT functionality $\FuncOT$. $\mathcal{S}$ inputs $\{\tilde{a}_{k,v}^0+\mathsf{mask}_k\}_{v\in[l]}$ and $\mathcal{R}$ inputs $\tilde{v}_k$. As a result, $\mathcal{S}$ learns nothing and $\mathcal{R}$ learns $\gamma_k=\tilde{a}_{k,\tilde{v}_k}^0+\mathsf{mask}_k$. 

\item \begin{itemize}
    \item For ECSS: $\mathcal{S}$ outputs $o_s:=-\sum_{k\in[d]}\mathsf{mask}_k$ and $\mathcal{R}$ outputs $o_r:=\sum_{k\in[d]}(b^{\mathcal{R},0}_{k,\tilde{v}_k}+\gamma_k)$.
    \item For ECS: $\mathcal{R}$ outputs $\mathsf{Sum}:=\sum_{k\in[d]}(b^{\mathcal{R},0}_{k,\tilde{v}_k}+ \gamma_k)$.
\end{itemize}

\end{enumerate}
\end{trivlist}
\end{minipage}
\end{framed}
\caption{Equality Conditional Sum (Share) Protocol $\Pi_{\mathsf{ECS(S)}}$} 
\label{fig:piecss}
\end{figure}

we argue the correctness of ECS described in Figure \ref{fig:piecss}. The correctness for ECSS is similar.
\begin{trivlist}

\item \textbf{Correctness.} Let $E_k$ and $v_k^*$ be defined as in Figure~\ref{fig:fecss}. For $k \in [d], v \in [l]$, the correctness of PS and ssPEQT ensures that:
\[
e_{k,v}^\mathcal{S} \oplus e_{k,v}^{\mathcal{R}} = 1 \Longleftrightarrow a^1_{k,\pi_k(v)} = b^1_{k,\pi_k(v)}.
\]
By the correctness of the secure OR functionality, we have:
\[
\tilde{e}^\mathcal{S}_{k,0} \oplus \tilde{e}^\mathcal{R}_{k,0} = 1 \Longleftrightarrow \exists \tilde{v}_k \in [l] \text{ such that } a^1_{k,\pi_k(\tilde{v}_k)} = b^1_{k,\pi_k(\tilde{v}_k)}.
\]
From the correctness of $\mathsf{ShareTrans}$, $\tilde{e}^{\mathcal{S}}_k$ and $\tilde{e}^{\mathcal{R}}_k$ form shares of $\tilde{v}_k = \bigoplus_{i \in \pi_k^{-1}(E_k)} i = \pi_k^{-1}(v_k^*)$ if $\tilde{e}^\mathcal{S}_{k,0} \oplus \tilde{e}^\mathcal{R}_{k,0} = 1$; otherwise, they form shares of $0$. The OT in Step 7 ensures $\mathcal{R}$ learns $\tilde{v}_k = \pi_k^{-1}(v_k^*)$ if the condition holds, or a random value otherwise. Therefore, the final sum is computed correctly:
\begin{align*}
\mathsf{Sum} &= \sum_{k \in [d]} \left( b^{\mathcal{R},0}_{k,\tilde{v}_k} + \gamma_k \right) \\
&= \sum_{k \in [d]} \left( b^{\mathcal{R},0}_{k,\tilde{v}_k} + \tilde{a}_{k,\tilde{v}_k}^0 + \mathsf{mask}_k \right) \\
&= \sum_{k \in [d]} \left( b^{\mathcal{R},0}_{k,\tilde{v}_k} + a^0_{k,\pi_k(\tilde{v}_k)} + b^{\mathcal{S},0}_{k,\tilde{v}_k}  \right) \\
&= \sum_{k \in [d]} \left( a^0_{k,v_k^*} + b^0_{k,v_k^*} \right)
\end{align*}

\end{trivlist}

We prove the security of the protocol for the case of ECS. The security proof of ECSS is similar.
\begin{theorem}\label{thm:ecss}
    The protocol in Figure \ref{fig:piecss} securely computes $\Funcecs$ against semi-honest adversaries in the $(\Funcps,\Funcsspeqt,\Funcor,\FuncOT)$-hybrid model.
\end{theorem}

Below we give details of the proof of Theorem \ref{thm:ecss}.

\begin{proof}
We exhibit simulators $\mathsf{Sim}_{\mathcal{S}}$ and $\mathsf{Sim}_{\mathcal{R}}$ 
for simulating corrupt $\mathcal{S}$ and $\mathcal{R}$ respectively, 
and argue the indistinguishability of the produced transcript from the real execution.
\begin{trivlist}
\item \underline{Corrupt Sender:}  $\mathsf{Sim}_\mathcal{S}(A = \{\veca_{k,v} = (a_{k,v}^0,a_{k,v}^1)\}_{k\in[d],v\in[l]})$ 
    simulates the view of corrupt semi-honest sender. It executes as follows:
\begin{enumerate}
\item In step 1, $\mathsf{Sim}_\mathcal{S}$ selects random permutations $\pi_k$ and $\vecb_{k,v}^\mathcal{S}\leftarrow \mathbb{F}^2$ for $k\in[d],v\in[l]$. Then, it invokes PS receiver's simulator $\mathsf{Sim}_\mathsf{PS}^\mathcal{R}(\pi_k,\{\vecb_{k,v}^\mathcal{S}\}_{v\in[l]})$ and appends the output to the view.

\item In step 2, $\mathsf{Sim}_\mathcal{S}$ executes like an honest sender and learns $\tilde{\veca}_{k,v},k\in[d],v\in[l]$.

\item In step 3, $\mathsf{Sim}_\mathcal{S}$ selects random $e^{\mathcal{S}}_{k,v}\leftarrow \{0,1\}, k\in[d],v\in[l]$. Then, it invokes the ssPEQT simulator $\mathsf{Sim}_\mathsf{ssPEQT}(\tilde{a}_{k,v}^1,e_{k,v}^\mathcal{S})$ and appends the output to the view.

\item In step 4, $\mathsf{Sim}_\mathcal{S}$ selects random $\tilde{e}^{\mathcal{S}}_{k,0}\leftarrow \{0,1\}, k\in[d]$. Then, $\mathsf{Sim}_\mathcal{S}$ invokes the OR simulator $\mathsf{Sim}_\mathsf{OR}(\{e^\mathcal{S}_{k,v}\}_{v\in[l]},\tilde{e}_{k,0}^\mathcal{S})$ and appends the output to the view.

\item In step 5-6, $\mathsf{Sim}_\mathcal{S}$ executes like an honest sender and learns $\tilde{e}^\mathcal{S}_k,m_{k,0},m_{k,1},k\in[d]$.

\item In step 7, $\mathsf{Sim}_{\mathcal{S}}$ invokes the OT sender's simulator $\mathsf{Sim}_\mathsf{OT}^\mathcal{S}((m_{k,0},m_{k,1}),\perp)$ and appends the output to the view.

\item In step 8, $\mathsf{Sim}_\mathcal{S}$ executes like an honest sender and learns $\mathsf{mask}_k,k\in[d]$.

\item In step 9, $\mathsf{Sim}_{\mathcal{S}}$ invokes the OT sender's simulator $\mathsf{Sim}_\mathsf{OT}^\mathcal{S}(\{\tilde{a}^0_{k,v}+\mathsf{mask}_k\}_{v\in[l]},\perp)$ and appends the output to the view.

\end{enumerate}
Now we argue that the view output by $\mathsf{Sim}_{\mathcal{S}}$ is indistinguishable from the real one. In the ECS protocol, the sender has no output, and all intermediate messages are secret shares. Consequently, the security of the underlying components guarantees that the simulator's output is computationally indistinguishable from the view in the real protocol.

\item \underline{Corrupt Receiver:} $\mathsf{Sim}_{\mathcal{R}}(B = \{\vecb_{k,v} = (b_{k,v}^0,b_{k,v}^1)\}_{k\in[d],v\in[l]},\mathsf{Sum})$ 
    simulates the view of corrupt semi-honest receiver. It executes as follows:
\begin{enumerate}

\item In step 1, $\mathsf{Sim}_\mathcal{R}$ selects random $\vecb_{k,v}^\mathcal{R}\leftarrow \mathbb{F}^2$ for $k\in[d],v\in[l]$. Then, it invokes PS sender's simulator $\mathsf{Sim}_\mathsf{PS}^\mathcal{S}(\{\vecb_{k,v}\}_{v\in[l]},\{\vecb_{k,v}^\mathcal{R}\}_{v\in[l]})$ and appends the output to the view.

\item In step 3, $\mathsf{Sim}_\mathcal{R}$ selects random $e^{\mathcal{R}}_{k,v}\leftarrow \{0,1\}, k\in[d],v\in[l]$. Then, it invokes the ssPEQT simulator $\mathsf{Sim}_\mathsf{ssPEQT}(b_{k,v}^{\mathcal{R},1},e_{k,v}^\mathcal{R})$ and appends the output to the view.

\item In step 4, $\mathsf{Sim}_\mathcal{R}$ selects random $\tilde{e}^{\mathcal{R}}_{k,0}\leftarrow \{0,1\}, k\in[d]$. Then, $\mathsf{Sim}_\mathcal{R}$ invokes the OR simulator $\mathsf{Sim}_\mathsf{OR}(\{e^\mathcal{R}_{k,v}\}_{v\in[l]},\tilde{e}_{k,0}^\mathcal{R})$ and appends the output to the view.

\item In step 5, $\mathsf{Sim}_\mathcal{R}$ executes like an honest receiver and learns $\tilde{e}^\mathcal{R}_k,k\in[d]$.

\item In step 7, $\mathsf{Sim}_{\mathcal{R}}$ picks random $m_k^*\leftarrow [l],k\in[d]$. Then, it invokes the OT receiver's simulator $\mathsf{Sim}_\mathsf{OT}^\mathcal{R}(\tilde{e}^\mathcal{R}_{k,0},m_k^*)$ and appends the output to the view. $\mathsf{Sim}_{\mathcal{R}}$ also computes $\tilde{v}_k:= m_k^*\oplus \tilde{e}_k^{\mathcal{R}},k\in[d]$.

\item In step 9, $\mathsf{Sim}_{\mathcal{R}}$ picks random $\gamma_k\leftarrow \mathbb{F}$ for $k\in[d-1]$ and computes $\gamma_d:=\mathsf{Sum}-\sum_{k\in[d]}b_{k,\tilde{v}_k}^{\mathcal{R},0}-\sum_{k\in[d-1]}\gamma_k$. Then, it invokes the OT receiver's simulator $\mathsf{Sim}_\mathsf{OT}^\mathcal{R}(\tilde{v}_k,\gamma_k)$ and appends the output to the view.

\end{enumerate}
\end{trivlist}
Now we argue that the view output by $\mathsf{Sim}_{\mathcal{R}}$ is indistinguishable from the real one. 

We formally prove this by a standard hybrid argument method. 
We define four hybrid transcripts $T_0, T_1, T_2,T_3$ where $T_0$ is real view of $\mathcal{R}$, 
and $T_3$ is the output of $\mathsf{Sim}_{\mathcal{R}}$.

\begin{itemize}
\item $\text{Hybrid}_0$. The first hybrid is the real interaction described in Figure \ref{fig:piecss}. 
    Here, an honest $\mathcal{R}$ uses input $B$, honestly interacts with the corrupt $\mathcal{S}$. 
    Let $T_0$ denote the real view of $\mathcal{R}$.

\item $\text{Hybrid}_1$.  Let $T_1$ be the same as $T_0$, except that the PS, ssPEQT, and OR execution are replaced by simulator $\mathsf{Sim}^\mathcal{S}_{\mathsf{PS}}, \mathsf{Sim}^\mathcal{R}_{\mathsf{ssPEQT}},\mathsf{Sim}^\mathcal{R}_{\mathsf{OR}}$. 
Since the receiver only learns random shares, the security of PS, ssPEQT, and OR functionality
guarantee this view is indistinguishable from $T_0$.

\item $\text{Hybrid}_2$.  Let $T_2$ be the same as $T_1$, except that the 1-out-of-2 OT execution are replaced by simulator $\mathsf{Sim}_\mathsf{OT}^\mathcal{R}(\tilde{e}^\mathcal{R}_{k,0},m_k^*)$. If $\forall \tilde{v}_k\in [l]$, $\tilde{a}_{k,v}^1\ne b_{k,v}^{\mathcal{R},1}$, we know that $\mathcal{R}$ will learn a random $m_k^*$. The security of OT functionality guarantees this view is indistinguishable from $T_1$. If $\exists \tilde{v}_k\in [l]$, $\tilde{a}_{k,\tilde{v}_k}^1 =  b_{k,\tilde{v}_k}^{\mathcal{R},1}$, we know that $\mathcal{R}$ will learn this $\tilde{v}_k$. Since the sender has permuted the positions of $a_{k,v}^1 = b_{k,v}^1$. As a result, $\tilde{v}_k$ is also a random index in the view of $\mathcal{R}$. The security of OT functionality guarantees this view is indistinguishable from $T_1$. 

\item $\text{Hybrid}_3$.  Let $T_3$ be the same as $T_2$, except that the 1-out-of-$l$ OT execution are replaced by simulator $\mathsf{Sim}_\mathsf{OT}^\mathcal{R}(\tilde{v}_k,\gamma_k)$. The randomness of $\mathsf{mask}_k$ guarantees each $\gamma_k$ is also random, conditioned on $\sum_{k\in[d]}(b_{k,\tilde{v}_k}^{\mathcal{R},0}+\gamma_k)=\mathsf{Sum}$. As a result, the security of OT functionality guarantees this view is indistinguishable from $T_2$. This hybrid is exactly the view output by the simulator.

\end{itemize}
\end{proof}

\begin{trivlist}
    \item \textbf{Complexity.} When using state-of-the-art PS \cite{DBLP:conf/pkc/PecenyRRS25} and ssPEQT \cite{DBLP:journals/popets/ChandranGS22,DBLP:conf/uss/lkz+25}, our ECS(S) protocol in Figure \ref{fig:piecss} has computational complexity $O(dl)$ and communication complexity $O(dl)$.
\end{trivlist}

\subsection{Equality Conditional Inner-Product Share}
\label{subsec:ecips}

In this section, we formally define the equality conditional inner-product share (ECIPS). 
It is worth noting that our definition slightly diverges from the naming convention, as the first vector in the inner product is secret-shared between the parties. This adjustment is made to align with the requirements of our FPSI protocol. 
The ideal functionality for ECIPS is given in Figure \ref{fig:fecips}.

\begin{figure}[!hbth]
\begin{framed}
\begin{minipage}[center]{\textwidth}
\begin{trivlist}
\item \textbf{Parameters:} Sender $\mathcal{S}$, Receiver $\mathcal{R}$, a finite field $\mathbb{F}$, the number of inner product terms $d$, the size of each comparison set $l$, the vector length $p$
    
\item \textbf{Functionality:}
\begin{itemize}
    \item Wait for input $A = \{\veca_{k,v} = (\veca_{k,v}^0,a_{k,v}^2)\}_{k\in[d],v\in[l]}\subseteq \mathbb{F}^{(p+1)dl}$ from the sender $\mathcal{S}$, where $\veca_{k,v}^0 = (a_{k,v,1}^0,\dots,a_{k,v,p}^0)$.
    \item Wait for input $B = \{\vecb_{k,v} = (\vecb_{k,v}^0,\vecb_{k,v}^1,b_{k,v}^2)\}_{k\in[d],v\in[l]}\subseteq \mathbb{F}^{(2p+1)dl}$ from the receiver $\mathcal{R}$, where $\vecb_{k,v}^0 = (b_{k,v,1}^0,\dots,b_{k,v,p}^0),\vecb_{k,v}^1 = (b_{k,v,1}^1,\dots,b_{k,v,p}^1)$.
    \item For $k\in[d]$:\begin{itemize}
        \item  Define $E_k:=\{v\in[l]:a_{k,v}^2 = b_{k,v}^2\}$
        \item  If $E_k = \emptyset$: $v_k^*\leftarrow [l]$
        \item  Else, $v_k^*:= \oplus_{i\in E_k} i$
    \end{itemize}
    \item Compute $\mathsf{IPSum}:= \sum_{k\in[d]} \langle \veca_{k,v_k^*}^0+\vecb_{k,v_k^*}^0, \vecb_{k,v_k^*}^1 \rangle$
        \item 
        Sample $o_s\leftarrow \mathbb{F}$ and compute $o_r:= \mathsf{IPSum}-o_s$. Give $o_s$ to $\mathcal{S}$ and give $o_r$ to $\mathcal{R}$.
\end{itemize}
\end{trivlist}
\end{minipage}
\end{framed}
\caption{Equality Conditional Inner-Product Share Functionality $\mathcal{F}_{\mathsf{ECIPS}}$}
\label{fig:fecips}
\end{figure}

We present our ECIPS protocol in Figure \ref{fig:piecips}. 
Our ECIPS construction follows a structure similar to that of ECSS, with a key procedural difference: after the receiver obtains $v_k^*$, the parties invoke an inner product protocol to compute the equality-conditional inner product, rather than locally summing the values as in ECSS.

\begin{figure}[!hbth]
\begin{framed}
\begin{minipage}[center]{\textwidth}
\begin{trivlist}
\item \textbf{Parameters:} 
\begin{itemize}
\item Two parties: sender $\mathcal{S}$ and receiver $\mathcal{R}$, a finite field $\mathbb{F}$, the number of summation terms $d$, the size of each comparison set $l$, vector length $p$.
\item Ideal $\Funcps$, $\Funcsspeqt$, $\Funcor$, $\FuncOT$ and $\Funcip$ primitives specified in Figure \ref{fig:fps}, Figure \ref{fig:fsspeqt}, Figure \ref{fig:fand}, Figure \ref{fig:fot} and Figure \ref{fig:fip}, respectively. 
\end{itemize}

\item Input of $\mathcal{S}$: $A = \{\veca_{k,v} = (\veca_{k,v}^0,a_{k,v}^2)\}_{k\in[d],v\in[l]}\subseteq \mathbb{F}^{(p+1)dl}$

\item Input of $\mathcal{R}$: $B = \{\vecb_{k,v} = (\vecb_{k,v}^0,\vecb_{k,v}^1,b_{k,v}^2)\}_{k\in[d],v\in[l]}\subseteq \mathbb{F}^{(2p+1)dl}$

\item \textbf{Protocol:}

\begin{enumerate}
\item For $k\in[d]$: $\mathcal{S}$ picks a random permutation $\pi_k$ over $[l]$. Then, the parties invoke the PS functionality $\Funcps$. $\mathcal{S}$ acts as receiver with input $\pi_k$ and $\mathcal{R}$ acts as
sender with input $\{\vecb_{k,v}\}_{v\in[l]}$. As a result, $\mathcal{S}$ receives $\{\vecb^\mathcal{S}_{k,v} = (\vecb^{\mathcal{S},0}_{k,v},\vecb^{\mathcal{S},1}_{k,v},b^{\mathcal{S},2}_{k,v})\}_{v\in[l]}$ and $\mathcal{R}$ receives $\{\vecb^\mathcal{R}_{k,v}=(\vecb^{\mathcal{R},0}_{k,v},\vecb^{\mathcal{R},1}_{k,v},b^{\mathcal{R},2}_{k,v})\}_{v\in[l]}$, where $\vecb^\mathcal{S}_{k,v} + \vecb^\mathcal{R}_{k,v} = \vecb_{k,\pi_k(v)}$.

\item $\mathcal{S}$ defines $\tilde{\veca}_{k,v}^0:= \vecb^{\mathcal{S},0}_{k,v}+\veca^0_{k,\pi_k(v)}, \tilde{a}_{k,v}^2:= a^2_{k,\pi_k(v)}-b^{\mathcal{S},2}_{k,v}, k\in[d],v\in[l]$.

\item For $k\in[d],v\in[l]$: $\mathcal{S}$ and $\mathcal{R}$ invoke the ssPEQT functionality $\Funcsspeqt$. $\mathcal{S}$ inputs $\tilde{a}_{k,v}^2$ and $\mathcal{R}$ inputs $b^{\mathcal{R},2}_{k,v}$. As a result,  $\mathcal{S}$ and $\mathcal{R}$ receives $e_{k,v}^\mathcal{S}, e_{k,v}^{\mathcal{R}} \in \{0,1\}$ respectively.

\item For $k\in[d]$: $\mathcal{S}$ and $\mathcal{R}$ invoke the secure OR functionality $\Funcor$. $\mathcal{S}$ inputs $\{e_{k,v}^{\mathcal{S}}\}_{v\in[l]}$ and $\mathcal{R}$ inputs $\{e_{k,v}^{\mathcal{R}}\}_{v\in[l]}$. As a result,  $\mathcal{S}$ and $\mathcal{R}$ receives $\tilde{e}_{k,0}^\mathcal{S}, \tilde{e}_{k,0}^{\mathcal{R}} \in \{0,1\}$ respectively.

\item For $k\in[d]$: $\mathcal{S}$ computes $\tilde{e}_k^{\mathcal{S}}: = \mathsf{ShareTrans}(e^{\mathcal{S}}_{k,1}\dots e^{\mathcal{S}}_{k,l})$. $\mathcal{R}$ computes $\tilde{e}_k^{\mathcal{R}}: = \mathsf{ShareTrans}(e^{\mathcal{R}}_{k,1}\dots e^{\mathcal{R}}_{k,l})$.

\item For $k\in[d]$: $\mathcal{S}$ picks random $r_k\leftarrow [l]$. Then, it defines $m_{k,0}:=r_k, m_{k,1} = \tilde{e}^{\mathcal{S}}_k$ if $\tilde{e}^{\mathcal{S}}_{k,0}=0$, otherwise it defines $m_{k,0}:=\tilde{e}^{\mathcal{S}}_k, m_{k,1} = r_k$. 

\item  For $k\in[d]$: $\mathcal{S}$ and $\mathcal{R}$ invoke the $1$-out-of-$2$ OT functionality $\FuncOT$. $\mathcal{S}$ inputs $(m_{k,0},m_{k,1})$ and $\mathcal{R}$ inputs $\tilde{e}_{k,0}^{\mathcal{R}}$. As a result,  $\mathcal{S}$ learns nothing and $\mathcal{R}$ learns $m_k^* = m_{k,\tilde{e}_{k,0}^{\mathcal{R}}}$. Then, $\mathcal{R}$ computes $\tilde{v}_k:=m_k^*\oplus \tilde{e}^\mathcal{R}_k$.

\item $\mathcal{S}$ picks random $\mathsf{\mathbf{mask}}_k^0 ,\mathsf{\mathbf{mask}}_k^1 \leftarrow \mathbb{F}^{p},\mathsf{mask}_k^2 \leftarrow \mathbb{F}$, $k\in[d]$.

\item  For $k\in[d]$: $\mathcal{S}$ and $\mathcal{R}$ invoke the $1$-out-of-$l$ OT functionality $\FuncOT$. $\mathcal{S}$ inputs $\{(\tilde{\veca}_{k,v}^0+\mathsf{\mathbf{mask}}_k^0,\vecb_{k,v}^{\mathcal{S},1}+\mathsf{\mathbf{mask}}_k^1,\langle \tilde{\veca}_{k,v}^0, \vecb_{k,v}^{\mathcal{S},1} \rangle+\mathsf{mask}_k^2)\}_{v\in[l]}$ and $\mathcal{R}$ inputs $\tilde{v}_k$. As a result, $\mathcal{S}$ learns nothing and $\mathcal{R}$ learns $(\mathbf{x}_k,\mathbf{y}_k,z_k) = (\tilde{\veca}_{k,\tilde{v}_k}^0+\mathsf{\mathbf{mask}}_k^0,\vecb_{k,\tilde{v}_k}^{\mathcal{S},1}+\mathsf{\mathbf{mask}}_k^1,\langle \tilde{\veca}_{k,\tilde{v}_k}^0, \vecb_{k,\tilde{v}_k}^{\mathcal{S},1} \rangle+\mathsf{mask}_k^2)$.

\item For $k\in[d]$: $\mathcal{S}$ and $\mathcal{R}$ invoke the IP functionality $\Funcip$. $\mathcal{S}$ inputs $(\mathbf{mask}_k^0,\mathbf{mask}_k^1)\in\mathbb{F}^{2p}$ and $\mathcal{R}$ inputs $(\vecb^{\mathcal{R},1}_{k,\tilde{v}_k},\vecb^{\mathcal{R},0}_{k,\tilde{v}_k})\in \mathbb{F}^{2p}$. As a result, $\mathcal{R}$ receives $h_k = \langle \mathbf{mask}_k^0||\mathbf{mask}_k^1, \vecb^{\mathcal{R},1}_{k,\tilde{v}_k}||\vecb^{\mathcal{R},0}_{k,\tilde{v}_k}  \rangle$.

\item  $\mathcal{R}$ outputs $o_r:= \sum_{k\in[d]} (\langle \vecb^{\mathcal{R},0}_{k,\tilde{v}_k},\vecb^{\mathcal{R},1}_{k,\tilde{v}_k}  \rangle+\langle \mathbf{x}_k,\vecb^{\mathcal{R},1}_{k,\tilde{v}_k} \rangle +\langle \mathbf{y}_k,\vecb^{\mathcal{R},0}_{k,\tilde{v}_k} \rangle +z_k-h_k) $ and $\mathcal{S}$ outputs $o_s:=-\sum_{k\in[d]}\mathsf{mask}_k^2$.
  
\end{enumerate}
\end{trivlist}
\end{minipage}
\end{framed}
\caption{Equality Conditional Inner-Product Share Protocol $\Pi_{\mathsf{ECIPS}}$} 
\label{fig:piecips}
\end{figure}

we argue the correctness of ECIPS described in Figure \ref{fig:piecips}. 
\begin{trivlist}

\item \textbf{Correctness.} 
Similar to ECSS protocol, we have $\tilde{v}_k = \pi_k^{-1}(v_k^*)$. Therefore, the final sum is computed correctly:
\begin{align*}
o_s+o_r &=  \sum_{k\in[d]} (\langle \vecb^{\mathcal{R},0}_{k,\tilde{v}_k},\vecb^{\mathcal{R},1}_{k,\tilde{v}_k}  \rangle+\langle \mathbf{x}_k,\vecb^{\mathcal{R},1}_{k,\tilde{v}_k} \rangle +\langle \mathbf{y}_k,\vecb^{\mathcal{R},0}_{k,\tilde{v}_k} \rangle +z_k-h_k-\mathsf{mask}_k^2)\\
&= \sum_{k\in[d]} (\langle \vecb^{\mathcal{R},0}_{k,\tilde{v}_k},\vecb^{\mathcal{R},1}_{k,\tilde{v}_k}  \rangle+\langle \tilde{\veca}_{k,\tilde{v}_k}^0+\mathsf{\mathbf{mask}}_k^0,\vecb^{\mathcal{R},1}_{k,\tilde{v}_k} \rangle +\langle \vecb_{k,\tilde{v}_k}^{\mathcal{S},1}+\mathsf{\mathbf{mask}}_k^1,\vecb^{\mathcal{R},0}_{k,\tilde{v}_k} \rangle 
\\ &+\langle \tilde{\veca}_{k,\tilde{v}_k}^0, \vecb_{k,\tilde{v}_k}^{\mathcal{S},1} \rangle+\mathsf{mask}_k^2-\langle \mathbf{mask}_k^0||\mathbf{mask}_k^1, \vecb^{\mathcal{R},1}_{k,\tilde{v}_k}||\vecb^{\mathcal{R},0}_{k,\tilde{v}_k}  \rangle-\mathsf{mask}_k^2)\\
&= \sum_{k\in[d]} (\langle \vecb^{\mathcal{R},0}_{k,\tilde{v}_k},\vecb^{\mathcal{R},1}_{k,\tilde{v}_k}  \rangle+\langle \tilde{\veca}_{k,\tilde{v}_k}^0,\vecb^{\mathcal{R},1}_{k,\tilde{v}_k} \rangle +\langle \vecb_{k,\tilde{v}_k}^{\mathcal{S},1},\vecb^{\mathcal{R},0}_{k,\tilde{v}_k} \rangle +\langle \tilde{\veca}_{k,\tilde{v}_k}^0, \vecb_{k,\tilde{v}_k}^{\mathcal{S},1} \rangle)\\
&= \sum_{k \in [d]} (\langle \vecb^{\mathcal{R},0}_{k,\tilde{v}_k} + \tilde{\veca}_{k,\tilde{v}_k}^0,\vecb^{\mathcal{R},1}_{k,\tilde{v}_k} + \vecb_{k,\tilde{v}_k}^{\mathcal{S},1} \rangle) \\
&= \sum_{k \in [d]} (\langle \vecb^{0}_{k,\pi_k(\tilde{v}_k)}  + \veca_{k,\pi_k(\tilde{v}_k)}^0,\vecb^{1}_{k,\pi_k(\tilde{v}_k)} \rangle) \\
&= \sum_{k \in [d]} (\langle \vecb^{0}_{k,v_k^*}  + \veca_{k,v_k^*}^0,\vecb^{1}_{k,v_k^*} \rangle)
\end{align*}

\end{trivlist}

We prove the security of the protocol described in Figure \ref{fig:piecips}.
\begin{theorem}\label{thm:ecips}
    The protocol in Figure \ref{fig:piecips} securely computes $\Funcecips$ against semi-honest adversaries in the $(\Funcps,\Funcsspeqt,\Funcor,\FuncOT,\Funcip)$-hybrid model.
\end{theorem}

Below we give details of the proof of Theorem \ref{thm:ecips}.

\begin{proof}
We exhibit simulators $\mathsf{Sim}_{\mathcal{S}}$ and $\mathsf{Sim}_{\mathcal{R}}$ 
for simulating corrupt $\mathcal{S}$ and $\mathcal{R}$ respectively, 
and argue the indistinguishability of the produced transcript from the real execution.
\begin{trivlist}
\item \underline{Corrupt Sender:}  $\mathsf{Sim}_\mathcal{S}(A = \{\veca_{k,v} = (\veca_{k,v}^0,a_{k,v}^2)\}_{k\in[d],v\in[l]},o_s)$ 
    simulates the view of corrupt semi-honest sender. It executes as follows:
\begin{enumerate}
\item In step 1, $\mathsf{Sim}_\mathcal{S}$ selects random permutations $\pi_k$ and $\vecb_{k,v}^\mathcal{S}\leftarrow \mathbb{F}^{2p+1}$ for $k\in[d],v\in[l]$. Then, it invokes PS receiver's simulator $\mathsf{Sim}_\mathsf{PS}^\mathcal{R}(\pi_k,\{\vecb_{k,v}^\mathcal{S}\}_{v\in[l]})$ and appends the output to the view.

\item In step 2, $\mathsf{Sim}_\mathcal{S}$ executes like an honest sender and learns $\tilde{\veca}_{k,v}^0,{a}_{k,v}^2,k\in[d],v\in[l]$.

\item In step 3, $\mathsf{Sim}_\mathcal{S}$ selects random $e^{\mathcal{S}}_{k,v}\leftarrow \{0,1\}, k\in[d],v\in[l]$. Then, it invokes the ssPEQT simulator $\mathsf{Sim}_\mathsf{ssPEQT}(\tilde{a}_{k,v}^2,e_{k,v}^\mathcal{S})$ and appends the output to the view.

\item In step 4, $\mathsf{Sim}_\mathcal{S}$ selects random $\tilde{e}^{\mathcal{S}}_{k,0}\leftarrow \{0,1\}, k\in[d]$. Then, $\mathsf{Sim}_\mathcal{S}$ invokes the OR simulator $\mathsf{Sim}_\mathsf{OR}(\{e^\mathcal{S}_{k,v}\}_{v\in[l]},\tilde{e}_{k,0}^\mathcal{S})$ and appends the output to the view.

\item In step 5-6, $\mathsf{Sim}_\mathcal{S}$ executes like an honest sender and learns $\tilde{e}^\mathcal{S}_k,m_{k,0},m_{k,1},k\in[d]$.

\item In step 7, $\mathsf{Sim}_{\mathcal{S}}$ invokes the OT sender's simulator $\mathsf{Sim}_\mathsf{OT}^\mathcal{S}((m_{k,0},m_{k,1}),\perp)$ and appends the output to the view.

\item In step 8, $\mathsf{Sim}_\mathcal{S}$ picks random $\mathbf{mask}_k^0,\mathbf{mask}_k^1\leftarrow\mathbb{F}^{p},k\in[d]$. $\mathsf{Sim}_\mathcal{S}$ also picks random $\mathsf{mask}_k^2\leftarrow\mathbb{F}$ for $k\in[d-1]$ and computes $\mathsf{mask}_d^2:= o_s-\sum_{k\in[d-1]}\mathsf{mask}_k^{2}$.

\item In step 9, $\mathsf{Sim}_{\mathcal{S}}$ invokes the OT sender's simulator $\mathsf{Sim}_\mathsf{OT}^\mathcal{S}(\{(\tilde{\veca}_{k,v}^0+\mathsf{\mathbf{mask}}_k^0,\vecb_{k,v}^{\mathcal{S},1}+\mathsf{\mathbf{mask}}_k^1,\langle \tilde{\veca}_{k,v}^0, \vecb_{k,v}^{\mathcal{S},1} \rangle+\mathsf{mask}_k^2)\}_{v\in[l]},\perp)$ and appends the output to the view.

\item In step 10, $\mathsf{Sim}_{\mathcal{S}}$ invokes the IP sender's simulator $\mathsf{Sim}_\mathsf{IP}^\mathcal{S}((\mathbf{mask}_k^0,\mathbf{mask}_k^1),\perp)$ and appends the output to the view.

\end{enumerate}
Now we argue that the view output by $\mathsf{Sim}_{\mathcal{S}}$ is indistinguishable from the real one. 
Note that the output of the sender $\mathcal{S}$ in ECIPS is a random share, i.e., $o_s$. In a real protocol, the sender randomly selects $\mathsf{mask}_k^2,k\in[d]$ and computes $o_s=\sum_{k\in[d]}\mathsf{mask}_k^2$. In the simulation, we have $\mathsf{mask}_k^2,k\in[d-1]$ and $o_s$ are random, $\mathsf{mask}_d^2 =o_s-\sum_{k\in[d-1]}\mathsf{mask}_k^d$. Therefore, these two distributions are identical. From the security of the underlying components, we have that the simulator's output is computationally indistinguishable from the view in the real protocol.

\item \underline{Corrupt Receiver:} $\mathsf{Sim}_{\mathcal{R}}(B = \{\vecb_{k,v} = (\vecb_{k,v}^0,\vecb_{k,v}^1,b_{k,v}^2)\}_{k\in[d],v\in[l]},o_r)$ 
    simulates the view of corrupt semi-honest receiver. It executes as follows:
\begin{enumerate}

\item In step 1, $\mathsf{Sim}_\mathcal{R}$ selects random $\vecb_{k,v}^\mathcal{R}\leftarrow \mathbb{F}^{2p+1}$ for $k\in[d],v\in[l]$. Then, it invokes PS sender's simulator $\mathsf{Sim}_\mathsf{PS}^\mathcal{S}(\{\vecb_{k,v}\}_{v\in[l]},\{\vecb_{k,v}^\mathcal{R}\}_{v\in[l]})$ and appends the output to the view.

\item In step 3, $\mathsf{Sim}_\mathcal{R}$ selects random $e^{\mathcal{R}}_{k,v}\leftarrow \{0,1\}, k\in[d],v\in[l]$. Then, it invokes the ssPEQT simulator $\mathsf{Sim}_\mathsf{ssPEQT}(b_{k,v}^{\mathcal{R},2},e_{k,v}^\mathcal{R})$ and appends the output to the view.

\item In step 4, $\mathsf{Sim}_\mathcal{R}$ selects random $\tilde{e}^{\mathcal{R}}_{k,0}\leftarrow \{0,1\}, k\in[d]$. Then, $\mathsf{Sim}_\mathcal{R}$ invokes the OR simulator $\mathsf{Sim}_\mathsf{OR}(\{e^\mathcal{R}_{k,v}\}_{v\in[l]},\tilde{e}_{k,0}^\mathcal{R})$ and appends the output to the view.

\item In step 5, $\mathsf{Sim}_\mathcal{R}$ executes like an honest receiver and learns $\tilde{e}^\mathcal{R}_k,k\in[d]$.

\item In step 7, $\mathsf{Sim}_{\mathcal{R}}$ picks random $m_k^*\leftarrow [l],k\in[d]$. Then, it invokes the OT receiver's simulator $\mathsf{Sim}_\mathsf{OT}^\mathcal{R}(\tilde{e}^\mathcal{R}_{k,0},m_k^*)$ and appends the output to the view. $\mathsf{Sim}_{\mathcal{R}}$ also computes $\tilde{v}_k:= m_k^*\oplus \tilde{e}_k^{\mathcal{R}},k\in[d]$.

\item In step 9, $\mathsf{Sim}_{\mathcal{R}}$ picks random $(\mathbf{x}_k,\mathbf{y}_k,z_k)\leftarrow \mathbb{F}^{2p+1}$. Then, it invokes the OT receiver's simulator $\mathsf{Sim}_\mathsf{OT}^\mathcal{R}(\tilde{v}_k,(\mathbf{x}_k,\mathbf{y}_k,z_k))$ and appends the output to the view.

\item In step 10, $\mathsf{Sim}_{\mathcal{R}}$ picks random $h_k\leftarrow \mathbb{F}$ for $k\in[d-1]$ and computes $h_d:=o_r-\sum_{k\in[d]}(\langle \vecb^{\mathcal{R},0}_{k,\tilde{v}_k},\vecb^{\mathcal{R},1}_{k,\tilde{v}_k}  \rangle+ \langle \mathbf{x}_k,\vecb^{\mathcal{R},1}_{k,\tilde{v}_k} \rangle+\langle \mathbf{y}_k,\vecb^{\mathcal{R},0}_{k,\tilde{v}_k} \rangle+ z_k)-\sum_{k\in[d-1]} h_k$. Then, $\mathsf{Sim}_{\mathcal{R}}$ invokes the IP receiver's simulator $\mathsf{Sim}_\mathsf{IP}^\mathcal{R}((\vecb^{\mathcal{R},1}_{k,\tilde{v}_k},\vecb^{\mathcal{R},0}_{k,\tilde{v}_k}),h_k)$ and appends the output to the view.

\end{enumerate}
\end{trivlist}
Now we argue that the view output by $\mathsf{Sim}_{\mathcal{R}}$ is indistinguishable from the real one. 

We formally prove this by a standard hybrid argument method. 
We define five hybrid transcripts $T_0, T_1, T_2,T_3,T_4$ where $T_0$ is real view of $\mathcal{R}$, 
and $T_4$ is the output of $\mathsf{Sim}_{\mathcal{R}}$.

\begin{itemize}
\item $\text{Hybrid}_0$. The first hybrid is the real interaction described in Figure \ref{fig:piecips}. 
    Here, an honest $\mathcal{R}$ uses input $B$, honestly interacts with the corrupt $\mathcal{S}$. 
    Let $T_0$ denote the real view of $\mathcal{R}$.

\item $\text{Hybrid}_1$.  Let $T_1$ be the same as $T_0$, except that the PS, ssPEQT, and OR execution are replaced by simulator $\mathsf{Sim}^\mathcal{S}_{\mathsf{PS}}, \mathsf{Sim}^\mathcal{R}_{\mathsf{ssPEQT}},\mathsf{Sim}^\mathcal{R}_{\mathsf{OR}}$. 
Since the receiver only learns random shares, the security of PS, ssPEQT, and OR functionality
guarantee this view is indistinguishable from $T_0$.

\item $\text{Hybrid}_2$.  Let $T_2$ be the same as $T_1$, except that the 1-out-of-2 OT execution are replaced by simulator $\mathsf{Sim}_\mathsf{OT}^\mathcal{R}(\tilde{e}^\mathcal{R}_{k,0},m_k^*)$. If $\forall \tilde{v}_k\in [l]$, $\tilde{a}_{k,v}^2\ne b_{k,v}^{\mathcal{R},2}$, we know that $\mathcal{R}$ will learn a random $m_k^*$. The security of OT functionality guarantees this view is indistinguishable from $T_1$. If $\exists \tilde{v}_k\in [l]$, $\tilde{a}_{k,\tilde{v}_k}^2 =  b_{k,\tilde{v}_k}^{\mathcal{R},2}$, we know that $\mathcal{R}$ will learn this $\tilde{v}_k$. Since the sender has permuted the positions of $a_{k,v}^2 = b_{k,v}^2$. As a result, $\tilde{v}_k$ is also a random index in the view of $\mathcal{R}$. The security of OT functionality guarantees this view is indistinguishable from $T_1$. 

\item $\text{Hybrid}_3$.  Let $T_3$ be the same as $T_2$, except that the 1-out-of-$l$ OT execution are replaced by simulator $\mathsf{Sim}_\mathsf{OT}^\mathcal{R}(\tilde{v}_k,(\mathbf{x}_k,\mathbf{y}_k,z_k))$. The randomness of $\mathbf{mask}_k^0,\mathbf{mask}_k^1,\mathsf{mask}_k^2$ guarantees $(\mathbf{x}_k,\mathbf{y}_k,z_k)$ is also random. The security of OT functionality guarantees this view is indistinguishable from $T_2$.

\item $\text{Hybrid}_4$.  Let $T_3$ be the same as $T_2$, except that the IP execution are replaced by simulator $\mathsf{Sim}_\mathsf{IP}^\mathcal{R}((\vecb^{\mathcal{R},1}_{k,\tilde{v}_k},\vecb^{\mathcal{R},0}_{k,\tilde{v}_k}),h_k)$. Note that the output of the receiver $\mathcal{R}$ in ECIPS is a random share, i.e., $o_r$. In a real protocol, this share is defined as $o_r:= \sum_{k\in[d]} (\langle \vecb^{\mathcal{R},0}_{k,\tilde{v}_k},\vecb^{\mathcal{R},1}_{k,\tilde{v}_k}  \rangle+\langle \mathbf{x}_k,\vecb^{\mathcal{R},1}_{k,\tilde{v}_k} \rangle +\langle \mathbf{y}_k,\vecb^{\mathcal{R},0}_{k,\tilde{v}_k} \rangle +z_k-h_k)$, where $\langle \mathbf{x}_k,\vecb^{\mathcal{R},1}_{k,\tilde{v}_k} \rangle, \langle \mathbf{y}_k,\vecb^{\mathcal{R},0}_{k,\tilde{v}_k} \rangle, z_k$ and $h_k$ are random, conditioned on these values sum to $o_r$. In the simulation, we have $\{\langle \mathbf{x}_k,\vecb^{\mathcal{R},1}_{k,\tilde{v}_k} \rangle, \langle \mathbf{y}_k,\vecb^{\mathcal{R},0}_{k,\tilde{v}_k} \rangle, z_k\}_k\in[d], \{h_k\}_{k\in[d-1]}$ and $o_r$ are random, $h_d =o_r-\sum_{k\in[d]}(\langle \vecb^{\mathcal{R},0}_{k,\tilde{v}_k},\vecb^{\mathcal{R},1}_{k,\tilde{v}_k}  \rangle+ \langle \mathbf{x}_k,\vecb^{\mathcal{R},1}_{k,\tilde{v}_k} \rangle+\langle \mathbf{y}_k,\vecb^{\mathcal{R},0}_{k,\tilde{v}_k} \rangle+ z_k)-\sum_{k\in[d-1]} h_k$. Therefore, these two distributions are identical. From the security of IP, we have that the simulator's output is computationally indistinguishable from the view in the real protocol. This hybrid is exactly the view output by the simulator.

\end{itemize}
\end{proof}

\begin{trivlist}
    \item \textbf{Complexity.} When using state-of-the-art PS \cite{DBLP:conf/pkc/PecenyRRS25} and ssPEQT \cite{DBLP:journals/popets/ChandranGS22,DBLP:conf/uss/lkz+25}, our ECIPS protocol in Figure \ref{fig:piecss} has computation complexity $O(dpl)$ and communication complexity $O(dpl)$.
\end{trivlist}

\section{Shared Oblivious PRF}
\label{sec:soprf}
This section introduces a variant of OPRF called shared OPRF (sOPRF). Unlike standard OPRF, sOPRF accepts queries in an additively secret-shared form and returns additive secret shares of the corresponding PRF values to the parties. Consequently, the PRF's domain and range must be additive groups.

The ideal functionality for sOPRF is depicted in Figure~\ref{fig:fsoprf}.

\begin{figure}[!hbt]
\begin{framed}
\begin{minipage}[c]{\textwidth}
\begin{trivlist}
\item \textbf{Parameters:} Sender $\mathcal{S}$, Receiver $\mathcal{R}$, a PRF $F$, query size $n$.

\item \textbf{Functionality:}
\begin{itemize}
    \item Wait for input $k,\{x^0_1,\dots,x^0_n\}$ from sender $\mathcal{S}$.
    \item Wait for input $\{x^1_1,\dots,x^1_n\}$ from receiver $\mathcal{R}$.
    \item Compute $x_i := x_i^0 + x_i^1$ for $i \in [n]$.
    \item Pick random $f^0_i$ from the range of the PRF and compute $f^1_i := F_k(x_i) - f^0_i$ for $i \in [n]$.
    \item Give $\{f^0_i\}_{i\in [n]}$ to sender $\mathcal{S}$ and $\{f^1_i\}_{i\in [n]}$ to receiver $\mathcal{R}$.
\end{itemize}
\end{trivlist}
\end{minipage}
\end{framed}
\caption{Shared OPRF functionality $\Funcsoprf$.}
\label{fig:fsoprf}
\end{figure}

Recently, Alamati et al.~\cite{DBLP:conf/crypto/AlamatiPRR24} introduced a weak\textsuperscript{\,1} PRF construction based on the alternating-moduli paradigm~\cite{DBLP:conf/tcc/BonehIPSW18}. They also presented several oblivious PRF (OPRF) protocols with shared output that rely solely on OT and symmetric-key operations, achieving high efficiency. We observe that their protocol can be extended to support both shared input and output---resulting in a shared OPRF (sOPRF).

\footnotetext[1]{Any weak PRF $F_k(\cdot)$ can be transformed into a standard PRF by composing it with a random oracle $H$ as $F_k(H(\cdot))$.}

Alamati et al.~\cite{DBLP:conf/crypto/AlamatiPRR24} proposed the following weak PRF construction based on the alternating-moduli paradigm~\cite{DBLP:conf/tcc/BonehIPSW18}:
\begin{definition}[\cite{DBLP:conf/crypto/AlamatiPRR24}]
    Let $n,m,t \in \mathbb{N}$, and let $p, q$ be distinct primes. The $(\mathbb{F}_p,\mathbb{F}_q)$-wPRF is defined as
    \[
        F_k(x) := \mathbf{B} \cdot_q \bigl( \mathbf{A} \cdot_p [k \odot_p x] \bigr),
    \]
    where $x,k \in \mathbb{F}_p^n$ and $\mathbf{A} \in \mathbb{F}_p^{m \times n}, \mathbf{B} \in \mathbb{F}_q^{t \times m}$ are uniformly distributed. Here $\cdot_p$ and $\odot_p$ denote multiplication and component-wise multiplication modulo $p$, respectively.
\end{definition}

Alamati et al. \cite{DBLP:conf/crypto/AlamatiPRR24} gave two instantiations, denoted $(\mathbb{F}_3,\mathbb{F}_2)$-wPRF and $(\mathbb{F}_2,\mathbb{F}_3)$-wPRF, by setting $(p,q)=(3,2)$ and $(p,q)=(2,3)$, respectively. For $(\mathbb{F}_3,\mathbb{F}_2)$-wPRF, they showed that security is preserved even when the key $k \in \mathbb{F}_2^n$, which leads to simplifications in the protocol.

We recall their OPRF with shared output construction for $(\mathbb{F}_3,\mathbb{F}_2)$-wPRF. The protocol invokes an OT-based $\mathbb{F}_3 \rightarrow \mathbb{F}_2$ modulus conversion functionality $\mathcal{F}_{\mathsf{ot}\text{-}3\rightarrow2\mathsf{conv}}$, defined in Figure~\ref{fig:fot32}. The construction for $(\mathbb{F}_2,\mathbb{F}_3)$-wPRF is analogous and is omitted. The formal description appears in Figure~\ref{fig:pioprfso1}.

\begin{figure}[!hbt]
\begin{framed}
\begin{minipage}[c]{\textwidth}
\begin{trivlist}
\item \textbf{Parameters:}
\begin{itemize}
    \item Two parties: $P_0$ and $P_1$; public matrices $\mathbf{A} \in \mathbb{F}_3^{m \times n}$, $\mathbf{B} \in \mathbb{F}_2^{t \times m}$.
    \item Ideal functionalities $\FuncOT$ and $\mathcal{F}_{\mathsf{ot}\text{-}3\rightarrow2\mathsf{conv}}$ as specified in Figures~\ref{fig:fot} and~\ref{fig:fot32}.
\end{itemize}

\item \textbf{Input:} $P_0$ holds $k \in \mathbb{F}_2^n$; $P_1$ holds $x \in \mathbb{F}_3^n$.

\item \textbf{Protocol:}
\begin{enumerate}
    \item The parties execute $n$ random OTs: $P_0$ acts as receiver with choice bits $k_i$; $P_1$ receives two random strings $\sigma_{i,0}, \sigma_{i,1} \in \{0,1\}^\kappa$, and $P_0$ receives $\sigma_{i,k_i}$.
    \item Let $G_{i,0}, G_{i,1}$ be stateful PRGs with output in $\mathbb{F}_3$, held by $P_1$ and seeded with $\sigma_{i,0}, \sigma_{i,1}$, respectively.
    \item Let $G_{i}'$ be a stateful PRG with output in $\mathbb{F}_3$, held by $P_0$ and seeded with $\sigma_{i,k_i}$.
    \item $P_1$ computes:
        \begin{enumerate}
            \item $h_{i,0} \leftarrow G_{i,0}$ for $i \in [n]$; set $h_0 := (h_{1,0},\dots,h_{n,0})$.
            \item $h_{i,1} \leftarrow G_{i,1}$ for $i \in [n]$; set $h_1 := (h_{1,1},\dots,h_{n,1})$.
            \item $f := x -_3 h_0 -_3 h_1$.
            \item $w_1 := \mathbf{A} \cdot_3 h_0$.
        \end{enumerate}
    \item $P_1$ sends $f \in \mathbb{F}_3^n$ to $P_0$.
    \item $P_0$ computes:
        \begin{enumerate}
            \item $t_i \leftarrow G_{i}'$ for $i \in [n]$; set $t := (t_1,\dots,t_n)$.
            \item $w_0 := \mathbf{A} \cdot_3 \bigl( (k \odot_3 f) +_3 t \bigr)$.
        \end{enumerate}
    \item $P_0$ and $P_1$ invoke $\mathcal{F}_{\mathsf{ot}\text{-}3\rightarrow2\mathsf{conv}}$ with inputs $w_0,w_1$; let $v_i$ be the output of $P_i$ ($i \in \{0,1\}$).
    \item $P_i$ outputs $\mathbf{B} v_i$ ($i \in \{0,1\}$).
\end{enumerate}
\end{trivlist}
\end{minipage}
\end{framed}
\caption{OPRF with shared output protocol for $(\mathbb{F}_3,\mathbb{F}_2)$-wPRF~\cite{DBLP:conf/crypto/AlamatiPRR24}.}
\label{fig:pioprfso1}
\end{figure}

\begin{figure}[!hbt]
\begin{framed}
\begin{minipage}[c]{\textwidth}
\begin{trivlist}
\item \textbf{Parameters:} Two parties $P_0$, $P_1$; input length $m$.

\item \textbf{Functionality:}
\begin{itemize}
    \item Wait for input $w_0 \in \mathbb{F}_3^m$ from $P_0$.
    \item Wait for input $w_1 \in \mathbb{F}_3^m$ from $P_1$.
    \item Compute $v := (w_0 +_3 w_1) \bmod 2$.
    \item Pick random $v_0 \leftarrow \mathbb{F}_2^m$ and compute $v_1 := v \oplus v_0$.
    \item Give $v_0$ to $P_0$ and $v_1$ to $P_1$.
\end{itemize}
\end{trivlist}
\end{minipage}
\end{framed}
\caption{OT-based $\mathbb{F}_3 \rightarrow \mathbb{F}_2$ modulus conversion functionality $\mathcal{F}_{\mathsf{ot}\text{-}3\rightarrow2\mathsf{conv}}$~\cite{DBLP:conf/crypto/AlamatiPRR24}.}
\label{fig:fot32}
\end{figure}

We now adapt the protocol in Figure~\ref{fig:pioprfso1} into a shared OPRF protocol that also supports shared input. The key observation is that $P_1$'s input $x$ is only used in the computation of $f := x - h_0 - h_1$. Because $P_0$ knows only one of $\{h_0, h_1\}$, this step effectively one-time-pad encrypts $x$ using $h_0 + h_1$, leaking no information about $x$.

To extend the protocol to shared inputs, where $x = x_0 + x_1$ with $P_0$ holding $x_0$ and $P_1$ holding $x_1$, we modify the computation as follows:
\begin{enumerate}
    \item $P_1$ computes $f_1 := x_1 - h_0 - h_1$ (which, analogously, leaks nothing about $x_1$) and sends $f_1$ to $P_0$.
    \item $P_0$, upon receiving $f_1$, computes $f := f_1 + x_0$ and continues the original protocol from Figure~\ref{fig:pioprfso1}.
\end{enumerate}

Security follows directly from the one-time-pad encryption, while correctness is verified by $f = f_1 + x_0 = (x_1 - h_0 - h_1) + x_0 = x - h_0 - h_1$.
Thus, we obtain a secure sOPRF protocol supporting both input and output sharing.
\section{Fuzzy PSI in Low Dimension}
\label{sec:fpsi-apart}

For low-dimensional space, we use spatial hashing \cite{DBLP:conf/crypto/GarimellaRS22,DBLP:conf/eurocrypt/BaarsenP24} to avoid quadratic costs, adding an $O(2^d)$ factor to the communication and receiver's computation. 
We present two FPSI constructions for $L_\infty$ and $L_{p\in[1,\infty)}$ distances, leveraging OPPRF \cite{DBLP:conf/ccs/KolesnikovMPRT17} and sOPRF, respectively.

\subsection{Fuzzy PSI for $L_\infty$ Distance}
\label{subsec:fpsi-apart-inf}

\subsubsection{Construction from OPPRF.}
\label{subsubsec:fpsi-apart-inf-opprf}
In this section, we give our construction of the FPSI (and its variants) for $L_\infty$ distance from OPPRF. 
As outlined in the introduction, our protocol proceeds in two main stages. First, we combine OPPRF, prefix trie techniques, and our newly proposed ECS protocol to achieve single-element fuzzy matching. We then extend this to handle multiple elements by employing spatial hashing and cuckoo hashing techniques to align the inputs across both parties.
The formal description is given in Figure \ref{fig:pifpsi-inf-apart-opprf}.


\begin{figure}[!ht]
\begin{framed}
\begin{minipage}[center]{\textwidth}
\begin{trivlist}
\item \textbf{Parameters:} 
\begin{itemize}
\item Two parties: sender $\mathcal{S}$ and receiver $\mathcal{R}$, set size $m$ and $n$, dimension $d$, threshold $\delta$.
\item Ideal $\Funcopprf$,$\Funcecs$, $\Funcps$, $\Funcsspeqt$, and $\FuncOT$ primitives specified in Figure \ref{fig:fopprf}, Figure \ref{fig:fecss}, Figure \ref{fig:fps}, Figure \ref{fig:fsspeqt}, and Figure \ref{fig:fot} respectively. 
\end{itemize}

\item Input of $\mathcal{S}$: $Q=\{\vecq_1,\dots,\vecq_{m}\}\subseteq \mathbb{U}^d$. 

\item Input of $\mathcal{R}$: $W=\{\vecw_1,\dots,\vecw_{n}\}\subseteq \mathbb{U}^d$.

\item \textbf{Protocol:}

\begin{enumerate}

\item The sender $\mathcal{S}$ computes the Cuckoo hash table $Q^*:= \mathsf{Cuckoo}_{h_1,h_2,h_3}^{m_c}(\{(\mathsf{cell}_\delta(\vecq_j),\vecq_j)\}_{j\in [m]})$ and fills empty bins with the dummy item. Let $m_c=(1+\epsilon)m$ denote the length of the Cuckoo hash table. $\mathcal{S}$ denotes the item in $u$-th bin as $Q^*[u]$ for $u\in [m_c]$. Let $\tau: [m]\rightarrow [m_c]$ denote a random injective function such that $\tau(1),\dots, \tau(m)$ are the non-dummy item bins of $Q^*$. 

\item $\mathcal{S}$ computes the prefix $\{q_{\tau(j),k,v}^*\}_{v\in [l]}:=\mathsf{PrefixPath}(Q^*[\tau(j)]_k,\delta)$ for $j\in[m],k\in[d]$.

\item $\mathcal{R}$ computes the prefix $\{w_{i,k,v}^*\}_{v\in [l']}:=\mathsf{PrefixTrie}(w_{i,k}-\delta,w_{i,k}+\delta)$ for $i\in[n],k\in[d]$. $\mathcal{R}$ also computes $\{\mathcal{C}_{i,t}\}_{t\in [2^d]}:= \mathsf{neighbor}_\delta(\vecw_i), i\in[n]$. Then, $\mathcal{R}$ picks random $r_{u,k}^\mathcal{R}\leftarrow \mathbb{F}$ and computes $\randr_u:=\sum_{k\in[d]}r_{u,k}^\mathcal{R}$.

\item  $\mathcal{R}$ defines $\mathsf{List}:=\{(\mathcal{C}_{i,t}||k||w_{i,k,v}^*||\alpha,r^\mathcal{R}_{h_\alpha(\mathcal{C}_{i,t}),k})\}_{i\in [n],k\in [d],t\in [2^d],v\in [l'],\alpha\in[3]})$. Then, $\mathcal{S}$ and $\mathcal{R}$ invoke the OPPRF functionality $\Funcopprf$. The receiver acts as the sender in OPPRF with input $\mathsf{List}$ and learns $\hint, sk_\mathcal{R}$. The sender $\mathcal{S}$ acts as the receiver in OPPRF with input $\{\mathsf{cell}_\delta (Q^*[\tau(j)])||k||q^*_{\tau(j),k,v}||\alpha_{\tau(j)}\}_{j\in [m],k\in [d],v\in[l]}$, where $\alpha_{\tau(j)}\in[3]$ is the index of hash function used to insert $\mathsf{cell}_\delta (Q^*[\tau(j)])$, i.e., $h_{\alpha_{\tau(j)}}(\mathsf{cell}_\delta (Q^*[\tau(j)])) = \tau(j)$, and learns $\hint, \{r_{\tau(j),k,v}^\mathcal{S}\}_{j\in [m],k\in [d],v\in[l]}$. 

\item For $u\in [m_c]\setminus \{\tau(j)\}_{j\in [m]}$, $\mathcal{S}$ picks random $r_{u,k,v}^\mathcal{S}\leftarrow \mathbb{F}$, $k\in[d],v\in[l]$.

\item For $u\in[m_c]$, $\mathcal{S}$ and $\mathcal{R}$ invoke the ECS functionality $\Funcecs$. $\mathcal{R}$ acts as the sender in ECS with input $\{\veca_{u,k,v}=(0,r_{u,k}^\mathcal{R})\}_{k\in[d],v\in[l]}$ and learns nothing. $\mathcal{S}$ acts as the receiver in ECS with input $\{\vecb_{u,k,v}=(r_{u,k,v}^\mathcal{S},r_{u,k,v}^\mathcal{S})\}_{k\in[d],v\in[l]}$ and learns $\rands_u$.

\item For $u\in[m_c]$, $\mathcal{S}$ and $\mathcal{R}$ invoke the ssPEQT functionality $\Funcsspeqt$. $\mathcal{S}$ and $\mathcal{R}$ take inputs $\rands_u$ and $\randr_u$, and learns $\tilde{b}^\mathcal{S}_u$ and $\tilde{b}^\mathcal{R}_u$, respectively.

\item $\mathcal{S}$ picks a random permutation $\pi$ over $[m_c]$. Then, $\mathcal{S}$ and $\mathcal{R}$ invoke the PS functionality $\Funcps$. $\mathcal{R}$ acts as the sender in PS with input $\{\tilde{b}^\mathcal{R}_u\}_{u\in[m_c]}$ and learns $\{\bar{b}^\mathcal{R}_u\}_{u\in[m_c]}$. $\mathcal{S}$ acts as the receiver in PS with input $\pi$ and learns  $\{\bar{b}^\mathcal{S}_u\}_{u\in[m_c]}$.

\item $\mathcal{S}$ computes $b_u^\mathcal{S}:=\bar{b}_u^\mathcal{S}\oplus \tilde{b}_{\pi(u)}^\mathcal{S}$ for $u\in [m_c]$. $\mathcal{R}$ initializes set $I:=\{\}$.

\item $\mathcal{S}$ and $\mathcal{R}$ invoke a batch of $m_c$ OT instances $\FuncOT$. In $u$-th OT, $\mathcal{S}$ inputs $(\perp,Q^*[\pi(u)])$ if $b^\mathcal{S}_{u}=0$ and $(Q^*[\pi(u)],\perp)$ if $b^\mathcal{S}_{u} = 1$. $\mathcal{R}$ acts as receiver with input $\bar{b}^{\mathcal{R}}_{u}$ and receives $\mathbf{z}_{u}$. $\mathcal{R}$ sets $I :=I \cup \{\mathbf{z}_{u}\}$ if $\mathbf{z}_{u}\ne \perp$. Finally, $\mathcal{R}$ outputs $I$.

\end{enumerate}
\end{trivlist}
\end{minipage}
\end{framed}
\caption{Fuzzy PSI Protocol $\Protocolfpsi$ for $L_\infty$ in Low Dimension Space from OPPRF} \label{fig:pifpsi-inf-apart-opprf}
\end{figure}

We argue the correctness of FPSI for $L_{\infty}$ in Figure \ref{fig:pifpsi-inf-apart-opprf}.
\begin{trivlist}
    \item \textbf{Correctness.} For $u\in[m_c]$, if $\exists \vecw_i\in W$, s.t. $\dist(Q^*[\pi(u)],\vecw_i)\leq \delta$, then the correctness of spatial hashing ensures $\exists t\in[2^d]$, s.t. $ \mathsf{cell}_\delta(Q^*[\pi(u)]) = \mathcal{C}_{i,t}$, thus
    $\forall k\in[d], \mathsf{cell}_\delta(Q^*[\pi(u)])||k||\alpha_{\pi(u)} = \mathcal{C}_{i,t}||k||\alpha_{\pi(u)}$ and $|Q^*[\pi(u)]_k-w_{i,k}|\leq \delta$. By the property of prefix and the correctness of OPPRF, $\exists !v\in[l]$, s.t. $r^\mathcal{S}_{\pi(u),k,v}=r^\mathcal{R}_{\pi(u),k}$. The correctness of ECS guarantees $\rands_{\pi(u)} = \randr_{\pi(u)}$. Then, the correctness of ssPEQT and PS ensures $\tilde{b}_{\pi(u)}^\mathcal{S}\oplus \tilde{b}_{\pi(u)}^\mathcal{R}=1$ and $b_u^{\mathcal{S}}\oplus \bar{b}_u^{\mathcal{R}}=1$. Finally, the correctness of OT ensures the receiver learns ${Q}^*[\pi(u)]$ if $b_u^{\mathcal{S}}\oplus \bar{b}_u^{\mathcal{R}}=1$. 
    If $\forall i\in [n], \dist({Q}^*[\pi(u)],\vecw_i)> \delta$, we have $\exists k\in[d]$, s.t. $\forall v\in[l], q_{\pi(u),k,v}^*\notin \{w_{i,k,v'}\}_{v'\in [l']}$. Therefore, the pseudorandomness of PRF in OPPRF guarantees $r^\mathcal{S}_{\pi(u),k,v}$ is pseudorandom. By setting $\mathsf{len}= \log |\mathbb{F}|= \lambda +\log mn +\log l$, a union bound shows the probability of $\exists i\in [n], u\in[m_c], v\in [l], r^\mathcal{S}_{\pi(u),k,v}=r^\mathcal{R}_{\pi(u),k} $ is negligible $2^{-\lambda}$. By the correctness of ECS, $\rands_{\pi(u)}$ is also pseudorandom. As a result, $\rands_{\pi(u)}\ne \randr_{\pi(u)}$ with overwhelming probability and $\mathcal{R}$ learns $\perp$ in the final OT.

\end{trivlist}

We prove the security of the protocol described in Figure \ref{fig:pifpsi-inf-apart-opprf}.
\begin{theorem}\label{thm:fpsi-inf-apart-opprf}
    The protocol in Figure \ref{fig:pifpsi-inf-apart-opprf} securely computes $\Funcfpsi$ for $L_{\infty}$ against semi-honest adversaries in the ($\Funcopprf$,$\Funcecs$, $\Funcps$, $\Funcsspeqt$,$\FuncOT$)-hybrid model.
\end{theorem}

Below we give details of the proof of Theorem \ref{thm:fpsi-inf-apart-opprf}.

\begin{proof}
We exhibit simulators $\mathsf{Sim}_{\mathcal{S}}$ and $\mathsf{Sim}_{\mathcal{R}}$ 
for simulating corrupt $\mathcal{S}$ and $\mathcal{R}$ respectively, 
and argue the indistinguishability of the produced transcript from the real execution.
\begin{trivlist}
\item \underline{Corrupt Sender:}  $\mathsf{Sim}_\mathcal{S}(Q=\{\vecq_j\}_{j\in [m]})$ 
    simulates the view of corrupt semi-honest sender. It executes as follows:
    
\begin{enumerate}
\item In step 1-2, $\mathsf{Sim}_\mathcal{S}$ executes like an honest sender and learns $Q^*,\tau,\{q^*_{\tau(j),k,v}\}_{j\in[m],k\in[d],v\in[l]}$.

\item In step 4, $\mathsf{Sim}_\mathcal{S}$ selects random $\hint, r^{\mathcal{S}}_{\tau(j),k,v},j\in [m],k\in[d],v\in[l]$. Then, it invokes the OPPRF's simulator $\mathsf{Sim}_\mathsf{OPPRF}^\mathcal{R}(\{\mathsf{cell}_\delta (Q^*[\tau(j)])||k||q^*_{\tau(j),k,v}||\alpha_{\tau(j)}\}_{j\in [m],k\in [d],v\in[l]},(\hint, \{r_{\tau(j),k,v}^\mathcal{S}\}_{j\in[m],k\in[d],v\in[l]}))$ and appends the output to the view.

\item In step 5, $\mathsf{Sim}_\mathcal{S}$ selects random $r_{u,k,v}^\mathcal{S}\leftarrow \mathbb{F},u\in [m_c]\setminus \{\tau(j)\}_{j\in[m]},k\in[d],v\in[l]$.


\item In step 6, $\mathsf{Sim}_{\mathcal{S}}$ selects random $\rands_u\leftarrow \mathbb{F},u\in[m_c]$. Then, it invokes the ECS receiver's simulator $\mathsf{Sim}_\mathsf{ECS}^\mathcal{R}(\{\vecb_{u,k,v}=(r^\mathcal{S}_{u,k,v},r_{u,k,v}^\mathcal{S})\}_{k\in[d],v\in[l]},\rands_{u})$ and appends the output to the view.

\item In step 7, $\mathsf{Sim}_\mathcal{S}$ selects random $\tilde{b}^{\mathcal{S}}_{u}\leftarrow \{0,1\}, u\in[m_c]$. Then, it invokes the ssPEQT simulator $\mathsf{Sim}_\mathsf{ssPEQT}(\rands_u,\tilde{b}_{u}^{\mathcal{S}})$ and appends the output to the view.

\item In step 8, $\mathsf{Sim}_\mathcal{S}$ selects a random permutation $\pi$ over $[m_c]$, random bits $\bar{b}_u^{\mathcal{S}}\leftarrow \{0,1\},u\in[m_c]$. Then, it invokes PS receiver's simulator $\mathsf{Sim}_\mathsf{PS}^\mathcal{R}(\pi,\{\bar{b}_u^{\mathcal{S}}\}_{u\in [m_c]})$ and appends the output to the view. $\mathsf{Sim}_\mathcal{S}$ also computes $b_u^\mathcal{S}=\bar{b}_u^\mathcal{S}\oplus \tilde{b}_{\pi(u)}^\mathcal{S},u\in[m_c]$.

\item In step 10, for $u\in [m_c]$, $\mathsf{Sim}_{\mathcal{S}}$ invokes OT sender's simulator $\mathsf{Sim}_\mathsf{OT}^\mathcal{S}(\perp,Q^*[\pi(u)])$ if $b_u^\mathcal{S}=0$ and $\mathsf{Sim}_\mathsf{OT}^\mathcal{S}(Q^*[\pi(u)],\perp)$ if $b_u^\mathcal{S}=1$. Then, it appends the output to the view.
\end{enumerate}

Now we argue that the view output by $\mathsf{Sim}_{\mathcal{S}}$ is indistinguishable from the real one. We formally prove this by a standard hybrid argument method. 
We define three hybrid transcripts $T_0, T_1, T_2$ where $T_0$ is real view of $\mathcal{S}$, 
and $T_2$ is the output of $\mathsf{Sim}_{\mathcal{S}}$.

\begin{itemize}
\item $\text{Hybrid}_0$. The first hybrid is the real interaction described in Figure \ref{fig:pifpsi-inf-apart-opprf}. 
    Here, an honest $\mathcal{R}$ uses input $W$, honestly interacts with the corrupt $\mathcal{S}$. 
    Let $T_0$ denote the real view of $\mathcal{S}$.

\item $\text{Hybrid}_1$.  Let $T_1$ be the same as $T_0$, except that all OPPRF values $r_{\tau(j),k,v}^\mathcal{S}$ are replaced by randomly selected values. Since the set $Q$ satisfies the apart assumption, that is, $\mathsf{cell}_\delta(Q^*[\tau(j)])$ are distinct, resulting in $r_{\tau(j),k,v}^\mathcal{S}$ are also distinct. If $\exists i\in[n],v'\in[l']$, s.t. $q_{\tau(j),k,v}^*=w_{i,k,v'}^*$, we have $r_{\tau(j),k,v}^\mathcal{S} = r_{\tau(j),k}^\mathcal{R}$, which is truly random. Otherwise, if $\forall i\in[n],v'\in[l'], q_{\tau(j),k,v}^*\neq w_{i,k,v'}^*$, from the pseudorandomness of OPPRF, $r_{\tau(j),k,v}^\mathcal{S}$ is a pseudorandom value. As a result, by the pseudorandomness of PRF in OPPRF, this hybrid is computationally indistinguishable from $T_0$.

\item $\text{Hybrid}_2$.  Let $T_2$ be the same as $T_1$, except that the OPPRF, ECS, ssPEQT, PS and OT execution are replaced by simulator $\mathsf{Sim}^\mathcal{R}_{\mathsf{OPPRF}},\mathsf{Sim}^\mathcal{S}_{\mathsf{ECS}},\mathsf{Sim}^\mathcal{R}_{\mathsf{PS}}, \mathsf{Sim}^\mathcal{S}_{\mathsf{ssPEQT}},\mathsf{Sim}^\mathcal{S}_{\mathsf{OT}}$. 
The security of OPPRF, ECS, ssPEQT, PS and OT functionality
guarantee this view is computationally indistinguishable from $T_1$. This hybrid is exactly the view output by the simulator.

\end{itemize}

\item \underline{Corrupt Receiver:} $\mathsf{Sim}_{\mathcal{R}}(W,I)$ 
    simulates the view of corrupt semi-honest receiver. It executes as follows:
\begin{enumerate}

\item In step 3, $\mathsf{Sim}_\mathcal{R}$ executes like an honest receiver and learns $\{w^*_{i,k,v}\}_{i\in[n],k\in[d],v\in[l']},\{\mathcal{C}_{i,t}\}_{i\in[n],t\in[2^d]},$
$\{r_{u,k}^\mathcal{R}\}_{u\in[m_c],k\in[d]},\{\randr_u\}_{u\in[m_c]}$.


\item In step 4, $\mathsf{Sim}_\mathcal{R}$ defines $\mathsf{List}$ as an honest receiver and picks random $\hint,sk_\mathcal{R}$. Then, it invokes the OPPRF sender's simulator $\mathsf{Sim}_\mathsf{OPPRF}^\mathcal{S}(\mathsf{List},(\hint,sk_\mathcal{R}))$ and appends the output to the view.

\item In step 6, $\mathsf{Sim}_{\mathcal{R}}$ invokes the ECS sender's simulator $\mathsf{Sim}_\mathsf{ECS}^\mathcal{S}(\{\veca_{u,k,v}=(0,r_{u,k}^\mathcal{R})\}_{k\in[d],v\in[l]},\randr_{u})$ and appends the output to the view.

\item In step 7, $\mathsf{Sim}_\mathcal{R}$ selects random $\tilde{b}^{\mathcal{R}}_{u}\leftarrow \{0,1\}, u\in[m_c]$. Then, it invokes the ssPEQT simulator $\mathsf{Sim}_\mathsf{ssPEQT}(\randr_u,\tilde{b}_{u}^{\mathcal{R}})$ and appends the output to the view.

\item In step 8, $\mathsf{Sim}_\mathcal{R}$ selects random values $\bar{b}_u^{\mathcal{R}}\leftarrow \{0,1\},u\in[m_c]$. Then, it invokes PS sender's simulator $\mathsf{Sim}_\mathsf{PS}^\mathcal{S}(\{\tilde{b}_u^\mathcal{R}\}_{u\in [m_c]},\{\bar{b}_u^{\mathcal{R}}\}_{u\in [m_c]})$ and appends the output to the view. 

\item In step 9, $\mathsf{Sim}_\mathcal{R}$ uses $\perp$ to pad $I$ to $m_c$ elements and permutes these elements randomly. 
    Let $I=\{\mathbf{z}_1,\dots,\mathbf{z}_{m_c}\}$. Then, $\mathsf{Sim}_\mathcal{R}$ invokes OT receiver's simulator 
    $\mathsf{Sim}_\mathsf{OT}^\mathcal{R}(\bar{b}_u^\mathcal{R},\mathbf{z}_u)$ and appends the output to the view.

\end{enumerate}
\end{trivlist}
Now we argue that the view output by $\mathsf{Sim}_{\mathcal{R}}$ is indistinguishable from the real one. In the simulation, the way $\mathcal{R}$ obtains the elements in $I$ is identical to the real execution since the elements in $I$ are randomly permuted. By the underlying simulators' indistinguishability, the simulated view is computationally indistinguishable from the real. 
\end{proof}

\begin{trivlist}
\item \textbf{FPSI variants.} The FPSI protocol described in Figure~\ref{fig:pifpsi-inf-apart-opprf} can be easily adapted to support several FPSI variants with only minor adjustments. For LFPSI, the sender simply replaces $Q^*[\pi(u)]$ with the corresponding label in the final OT step. For FPSI-CA, the sender sends $b_u^\mathcal{S}$ to the receiver in Step~9, allowing the receiver to compute $b_u := b_u^\mathcal{S} \oplus \bar{b}_u^{\mathcal{R}}$ and output the Hamming weight of $\vecb = (b_1, \dots, b_{m_c})$. For FPSI-SP, the roles are effectively reversed: the receiver sends $\bar{b}_u^{\mathcal{R}}$ to the sender, who then outputs the set $\{ Q^*[\pi(u)] : b_u^\mathcal{S} \oplus \bar{b}_u^{\mathcal{R}} = 1,u\in[m_c] \}$. In this case, the sender functionally acts as the receiver in the FPSI-SP definition.
\item \textbf{Trade-off from the execution order of PS and PEQT.} In our protocol shown in Figure~\ref{fig:pifpsi-inf-apart-opprf}, the parties currently use ssPEQT in step 7 to test whether $\rands_u$ and $\randr_u$ are equal, then apply PS to the indicator bits in step 8. These two steps can be swapped—the parties may first permute $\randr_u$ and then test equality between the permuted result using standard PEQT. This reordering improves computational efficiency by replacing ssPEQT with PEQT, but at the expense of communication, since PS would then operate on $O(\lambda)$-bit strings rather than single bits.
\item \textbf{Optimization from AND.} Since testing the $L_\infty$ distance can be decomposed into independent interval tests for each dimension, we can simplify the protocol as follows. Instead of invoking the ECS protocol, the parties perform ssPEQT on $r_{u,k,v}^\mathcal{S}$ and $r_{u,k}^\mathcal{R}$ for each dimension, then XOR the results across all prefixes per dimension. The resulting outputs are combined across dimensions using an AND protocol to produce a secret sharing of the fuzzy matching indication bit. This approach is both simpler and more efficient than the ECS-based construction. The formal description is provided in Figure~\ref{fig:pifpsi-inf-apart-opprf-and}.
\end{trivlist}

\begin{figure}[!ht]
\begin{framed}
\begin{minipage}[center]{\textwidth}
\begin{trivlist}
\item \textbf{Parameters:} 
\begin{itemize}
\item Two parties: sender $\mathcal{S}$ and receiver $\mathcal{R}$, set size $m$ and $n$, dimension $d$, threshold $\delta$.
\item Ideal $\Funcopprf$,$\Funcand$, $\Funcps$, $\Funcsspeqt$, and $\FuncOT$ primitives specified in Figure \ref{fig:fopprf}, Figure \ref{fig:fand}, Figure \ref{fig:fps}, Figure \ref{fig:fsspeqt}, and Figure \ref{fig:fot} respectively. 
\end{itemize}

\item Input of $\mathcal{S}$: $Q=\{\vecq_1,\dots,\vecq_{m}\}\subseteq \mathbb{U}^d$. 

\item Input of $\mathcal{R}$: $W=\{\vecw_1,\dots,\vecw_{n}\}\subseteq \mathbb{U}^d$.

\item \textbf{Protocol:}

\begin{enumerate}
    \item Steps 1-5 are identical to the protocol described in Figure \ref{fig:pifpsi-inf-apart-opprf}.
\end{enumerate}

\begin{enumerate}[start = 6]

\item For $u\in[m_c],k\in[d],v\in[l]$, $\mathcal{S}$ and $\mathcal{R}$ invoke the ssPEQT functionality $\Funcsspeqt$. $\mathcal{S}$ and $\mathcal{R}$ take inputs $r^{\mathcal{S}}_{u,k,v}$ and $r^{\mathcal{R}}_{u,k}$, and learns $b^\mathcal{S}_{u,k,v}$ and $b^\mathcal{R}_{u,k,v}$, respectively. Then, $\mathcal{S}$ computes $b_{u,k}^\mathcal{S}:=\oplus_{v\in[l]}b^\mathcal{S}_{u,k,v}$ and $\mathcal{R}$ computes $b_{u,k}^\mathcal{R}:=\oplus_{v\in[l]}b^\mathcal{R}_{u,k,v}$.

\item For $u\in[m_c]$, $\mathcal{S}$ and $\mathcal{R}$ invoke the AND functionality $\Funcand$. $\mathcal{S}$ and $\mathcal{R}$ take inputs $\{b_{u,k}^\mathcal{S}\}_{k\in[d]}$ and $\{b_{u,k}^\mathcal{R}\}_{k\in[d]}$, and learns $\tilde{b}^\mathcal{S}_u$ and $\tilde{b}^\mathcal{R}_u$, respectively.

\item Steps 8-10 are identical to the protocol described in Figure \ref{fig:pifpsi-inf-apart-opprf}.

\end{enumerate}
\end{trivlist}
\end{minipage}
\end{framed}
\caption{Optimized Fuzzy PSI Protocol $\Protocolfpsi$ for $L_\infty$ in Low Dimension Space from OPPRF} \label{fig:pifpsi-inf-apart-opprf-and}
\end{figure}

We argue the correctness of FPSI for $L_{\infty}$ in Figure \ref{fig:pifpsi-inf-apart-opprf-and}.
\begin{trivlist}
    \item \textbf{Correctness.} For $u\in[m_c]$, if $\exists \vecw_i\in W$, s.t. $\dist(Q^*[\pi(u)],\vecw_i)\leq \delta$, then the correctness of spatial hashing ensures $\exists t\in[2^d]$, s.t. $ \mathsf{cell}_\delta(Q^*[\pi(u)]) = \mathcal{C}_{i,t}$, thus
    $\forall k\in[d], \mathsf{cell}_\delta(Q^*[\pi(u)])||k||\alpha_{\pi(u)} = \mathcal{C}_{i,t}||k||\alpha_{\pi(u)}$ and $|Q^*[\pi(u)]_k-w_{i,k}|\leq \delta$. By the property of prefix and the correctness of OPPRF, $\exists !v\in[l]$, s.t. $r^\mathcal{S}_{\pi(u),k,v}=r^\mathcal{R}_{\pi(u),k}$. Therefore, $\forall k\in [d], \exists! v\in[l]$, s.t. $b_{\pi(u),k,v}^\mathcal{S}\oplus b_{\pi(u),k,v}^\mathcal{R}=1$ and $b_{\pi(u),k}^\mathcal{S} \oplus b_{\pi(u),k}^\mathcal{R}=1$. By the correctness of AND, $\tilde{b}^{\mathcal{S}}_{\pi(u)}\oplus \tilde{b}^{\mathcal{R}}_{\pi(u)}=1$. Same as Figure \ref{fig:pifpsi-inf-apart-opprf}, the receiver learns ${Q}^*[\pi(u)]$ if $b_u^{\mathcal{S}}\oplus \bar{b}_u^{\mathcal{R}}=1$. If $\forall i\in [n], \dist({Q}^*[\pi(u)],\vecw_i)> \delta$, we have $\exists k\in[d]$, s.t. $\forall v\in[l], q_{\pi(u),k,v}^*\notin \{w_{i,k,v'}\}_{v'\in [l']}$. Therefore, the pseudorandomness of PRF in OPPRF guarantees $r^\mathcal{S}_{\pi(u),k,v}$ is pseudorandom. By setting $\mathsf{len}= \log |\mathbb{F}|= \lambda +\log mn +\log l$, a union bound shows the probability of $\exists i\in [n], u\in[m_c], v\in [l], r^\mathcal{S}_{\pi(u),k,v}=r^\mathcal{R}_{\pi(u),k} $ is negligible $2^{-\lambda}$. By the correctness of ssPEQT and AND, $\forall v\in[l], b_{\pi(u),k,v}^\mathcal{S}\oplus b_{\pi(u),k,v}^\mathcal{R}=0$ and $\tilde{b}_{\pi(u)}^\mathcal{S}\oplus \tilde{b}_{\pi(u)}^\mathcal{R}=0$. As a result, $\mathcal{R}$ learns $\perp$ in the final OT.

\end{trivlist}

We prove the security of the protocol described in Figure \ref{fig:pifpsi-inf-apart-opprf-and}.
\begin{theorem}\label{thm:fpsi-inf-apart-opprf-and}
    The protocol in Figure \ref{fig:pifpsi-inf-apart-opprf-and} securely computes $\Funcfpsi$ for $L_{\infty}$ against semi-honest adversaries in the ($\Funcopprf$,$\Funcand$, $\Funcsspeqt$,$\FuncOT$)-hybrid model.
\end{theorem}

Below we give details of the proof of Theorem \ref{thm:fpsi-inf-apart-opprf-and}.

\begin{proof}
We exhibit simulators $\mathsf{Sim}_{\mathcal{S}}$ and $\mathsf{Sim}_{\mathcal{R}}$ 
for simulating corrupt $\mathcal{S}$ and $\mathcal{R}$ respectively, 
and argue the indistinguishability of the produced transcript from the real execution.
\begin{trivlist}
\item \underline{Corrupt Sender:}  $\mathsf{Sim}_\mathcal{S}(Q=\{\vecq_j\}_{j\in [m]})$ 
    simulates the view of corrupt semi-honest sender. It executes as follows:
    
\begin{enumerate}
\item In step 1-2, $\mathsf{Sim}_\mathcal{S}$ executes like an honest sender and learns $Q^*,\tau,\{q^*_{\tau(j),k,v}\}_{j\in[m],k\in[d],v\in[l]}$.

\item In step 4, $\mathsf{Sim}_\mathcal{S}$ selects random $\hint, r^{\mathcal{S}}_{\tau(j),k,v},j\in [m],k\in[d],v\in[l]$. Then, it invokes OPPRF's simulator $\mathsf{Sim}_\mathsf{OPPRF}^\mathcal{R}(\{\mathsf{cell}_\delta (Q^*[\tau(j)])||k||q^*_{\tau(j),k,v}||\alpha_{\tau(j)}\}_{j\in [m],k\in [d],v\in[l]},(\hint, \{r_{\tau(j),k,v}^\mathcal{S}\}_{j\in[m],k\in[d],v\in[l]}))$ and appends the output to the view.

\item In step 5, $\mathsf{Sim}_\mathcal{S}$ selects random $r_{u,k,v}^\mathcal{S}\leftarrow \mathbb{F},u\in [m_c]\setminus \{\tau(j)\}_{j\in[m]},k\in[d],v\in[l]$.


\item In step 6, $\mathsf{Sim}_\mathcal{S}$ selects random $b^{\mathcal{S}}_{u,k,v}\leftarrow \{0,1\}, u\in[m_c],k\in[d],v\in[l]$. Then, it invokes the ssPEQT simulator $\mathsf{Sim}_\mathsf{ssPEQT}(r^\mathcal{S}_{u,k,v},b^{\mathcal{S}}_{u,k,v})$ and appends the output to the view. $\mathsf{Sim}_\mathcal{S}$ also computes $b_{u,k}^\mathcal{S}:=\oplus_{v\in[l]}b^\mathcal{S}_{u,k,v},u\in[m_c],k\in[d]$.

\item In step 7, $\mathsf{Sim}_\mathcal{S}$ selects random $\tilde{b}^{\mathcal{S}}_{u}\leftarrow \{0,1\}, u\in[m_c]$. Then, it invokes the AND simulator $\mathsf{Sim}_\mathsf{AND}(\{b^{\mathcal{S}}_{u,k}\}_{k\in[d]},\tilde{b}_u^\mathcal{S})$ and appends the output to the view.

\item In step 8, $\mathsf{Sim}_\mathcal{S}$ selects a random permutation $\pi$ over $[m_c]$, random bits $\bar{b}_u^{\mathcal{S}}\leftarrow \{0,1\},u\in[m_c]$. Then, it invokes PS receiver's simulator $\mathsf{Sim}_\mathsf{PS}^\mathcal{R}(\pi,\{\bar{b}_u^{\mathcal{S}}\}_{u\in [m_c]})$ and appends the output to the view. $\mathsf{Sim}_\mathcal{S}$ also computes $b_u^\mathcal{S}=\bar{b}_u^\mathcal{S}\oplus \tilde{b}_{\pi(u)}^\mathcal{S},u\in[m_c]$.

\item In step 10, for $u\in [m_c]$, $\mathsf{Sim}_{\mathcal{S}}$ invokes OT sender's simulator $\mathsf{Sim}_\mathsf{OT}^\mathcal{S}(\perp,Q^*[\pi(u)])$ if $b_u^\mathcal{S}=0$ and $\mathsf{Sim}_\mathsf{OT}^\mathcal{S}(Q^*[\pi(u)],\perp)$ if $b_u^\mathcal{S}=1$. Then, it appends the output to the view.
\end{enumerate}

Now we argue that the view output by $\mathsf{Sim}_{\mathcal{S}}$ is indistinguishable from the real one. We formally prove this by a standard hybrid argument method. 
We define three hybrid transcripts $T_0, T_1, T_2$ where $T_0$ is real view of $\mathcal{S}$, 
and $T_2$ is the output of $\mathsf{Sim}_{\mathcal{S}}$.

\begin{itemize}
\item $\text{Hybrid}_0$. The first hybrid is the real interaction described in Figure \ref{fig:pifpsi-inf-apart-opprf-and}. 
    Here, an honest $\mathcal{R}$ uses input $W$, honestly interacts with the corrupt $\mathcal{S}$. 
    Let $T_0$ denote the real view of $\mathcal{S}$.

\item $\text{Hybrid}_1$.  Let $T_1$ be the same as $T_0$, except that all OPPRF values $r_{\tau(j),k,v}^\mathcal{S}$ are replaced by randomly selected values. Since the set $Q$ satisfies the apart assumption, that is, $\mathsf{cell}_\delta(Q^*[\tau(j)])$ are distinct, resulting in $r_{\tau(j),k,v}^\mathcal{S}$ are also distinct. If $\exists i\in[n],v'\in[l']$, s.t. $q_{\tau(j),k,v}^*=w_{i,k,v'}^*$, we have $r_{\tau(j),k,v}^\mathcal{S} = r_{\tau(j),k}^\mathcal{R}$, which is truly random. Otherwise, if $\forall i\in[n],v'\in[l'], q_{\tau(j),k,v}^*\neq w_{i,k,v'}^*$, from the pseudorandomness of OPPRF, $r_{\tau(j),k,v}^\mathcal{S}$ is a pseudorandom value. As a result, by the pseudorandomness of PRF in OPPRF, this hybrid is computationally indistinguishable from $T_0$.

\item $\text{Hybrid}_2$.  Let $T_2$ be the same as $T_1$, except that the OPPRF, ssPEQT, AND, PS and OT execution are replaced by simulator $\mathsf{Sim}^\mathcal{R}_{\mathsf{OPPRF}}, \mathsf{Sim}^\mathcal{S}_{\mathsf{ssPEQT}},\mathsf{Sim}^\mathcal{S}_{\mathsf{AND}},\mathsf{Sim}^\mathcal{R}_{\mathsf{PS}},\mathsf{Sim}^\mathcal{S}_{\mathsf{OT}}$. 
The security of OPPRF, ssPEQT, AND, PS and OT functionality
guarantee this view is computationally indistinguishable from $T_1$. This hybrid is exactly the view output by the simulator.

\end{itemize}

\item \underline{Corrupt Receiver:} $\mathsf{Sim}_{\mathcal{R}}(W,I)$ 
    simulates the view of corrupt semi-honest receiver. It executes as follows:
\begin{enumerate}

\item In step 3, $\mathsf{Sim}_\mathcal{R}$ executes like an honest receiver and learns $\{w^*_{i,k,v}\}_{i\in[n],k\in[d],v\in[l']},\{\mathcal{C}_{i,t}\}_{i\in[n],t\in[2^d]},$
$\{r_{u,k}^\mathcal{R}\}_{u\in[m_c],k\in[d]},\{\randr_u\}_{u\in[m_c]}$.


\item In step 4, $\mathsf{Sim}_\mathcal{R}$ defines $\mathsf{List}$ as an honest receiver and picks random $\hint,sk_\mathcal{R}$. Then, it invokes the OPPRF sender's simulator $\mathsf{Sim}_\mathsf{OPPRF}^\mathcal{S}(\mathsf{List},(\hint,sk_\mathcal{R}))$ and appends the output to the view.

\item In step 6, $\mathsf{Sim}_\mathcal{R}$ selects random $b^{\mathcal{R}}_{u,k,v}\leftarrow \{0,1\}, u\in[m_c],k\in[d],v\in[l]$. Then, it invokes the ssPEQT simulator $\mathsf{Sim}_\mathsf{ssPEQT}(r^\mathcal{R}_{u,k,v},b^{\mathcal{R}}_{u,k,v})$ and appends the output to the view. $\mathsf{Sim}_\mathcal{R}$ also computes $b_{u,k}^\mathcal{R}:=\oplus_{v\in[l]}b^\mathcal{R}_{u,k,v},u\in[m_c],k\in[d]$.

\item In step 7, $\mathsf{Sim}_\mathcal{R}$ selects random $\tilde{b}^{\mathcal{R}}_{u}\leftarrow \{0,1\}, u\in[m_c]$. Then, it invokes the AND simulator $\mathsf{Sim}_\mathsf{AND}(\{b^{\mathcal{R}}_{u,k}\}_{k\in[d]},\tilde{b}_u^\mathcal{R})$ and appends the output to the view.

\item In step 8, $\mathsf{Sim}_\mathcal{R}$ selects random values $\bar{b}_u^{\mathcal{R}}\leftarrow \{0,1\},u\in[m_c]$. Then, it invokes PS sender's simulator $\mathsf{Sim}_\mathsf{PS}^\mathcal{S}(\{\tilde{b}_u^\mathcal{R}\}_{u\in [m_c]},\{\bar{b}_u^{\mathcal{R}}\}_{u\in [m_c]})$ and appends the output to the view. 

\item In step 9, $\mathsf{Sim}_\mathcal{R}$ uses $\perp$ to pad $I$ to $m_c$ elements and permutes these elements randomly. 
    Let $I=\{\mathbf{z}_1,\dots,\mathbf{z}_{m_c}\}$. Then, $\mathsf{Sim}_\mathcal{R}$ invokes OT receiver's simulator 
    $\mathsf{Sim}_\mathsf{OT}^\mathcal{R}(\bar{b}_u^\mathcal{R},\mathbf{z}_u)$ and appends the output to the view.

\end{enumerate}
\end{trivlist}
Now we argue that the view output by $\mathsf{Sim}_{\mathcal{R}}$ is indistinguishable from the real one. In the simulation, the way $\mathcal{R}$ obtains the elements in $I$ is identical to the real execution since the elements in $I$ are randomly permuted. By the underlying simulators' indistinguishability, the simulated view is computationally indistinguishable from the real. 
\end{proof}

\subsubsection{Construction from sOPRF.}
\label{subsubsec:fpsi-apart-inf-soprf}

In this section, we give our construction of the FPSI (and its variants) for $L_\infty$ distance from sOPRF. As discussed in introduction, sOPRF naturally provides the receiver with secret shares of the sender’s evaluation only for non‑empty bins, eliminating the need for dummy values in empty bins. The parties then run ECSS solely on those non‑empty bins. Since empty bins are never involved, the sender’s Cuckoo‑hashing arrangement remains hidden, and no global PS step is required. The formal protocol is described in Figure~\ref{fig:pifpsi-inf-apart-soprf}.

\begin{figure}[!ht]
\begin{framed}
\begin{minipage}[center]{\textwidth}
\begin{trivlist}
\item \textbf{Parameters:} 
\begin{itemize}
\item Two parties: sender $\mathcal{S}$ and receiver $\mathcal{R}$, set size $m$ and $n$, dimension $d$, threshold $\delta$.
\item Ideal $\Funcsoprf$,$\Funcecss$, $\Funcpeqt$, and $\FuncOT$ primitives specified in Figure \ref{fig:fopprf}, Figure \ref{fig:fecss}, Figure \ref{fig:fps}, Figure \ref{fig:fsspeqt}, and Figure \ref{fig:fot} respectively. 
\end{itemize}

\item Input of $\mathcal{S}$: $Q=\{\vecq_1,\dots,\vecq_{m}\}\subseteq \mathbb{U}^d$. 

\item Input of $\mathcal{R}$: $W=\{\vecw_1,\dots,\vecw_{n}\}\subseteq \mathbb{U}^d$.

\item \textbf{Protocol:}

\begin{enumerate}

\item The sender $\mathcal{S}$ computes the Cuckoo hash table $Q^*:= \mathsf{Cuckoo}_{h_1,h_2,h_3}^{m_c}(\{(\mathsf{cell}_\delta(\vecq_j),\vecq_j)\}_{j\in [m]})$ and fills empty bins with the dummy item. Let $m_c=(1+\epsilon)m$ denote the length of the Cuckoo hash table. $\mathcal{S}$ denotes the item in $u$-th bin as $Q^*[u]$ for $u\in [m_c]$. Let $\tau: [m]\rightarrow [m_c]$ denote a random injective function such that $\tau(1),\dots, \tau(m)$ are the non-dummy item bins of $Q^*$. 

\item $\mathcal{S}$ computes the prefix $\{q_{\tau(j),k,v}^*\}_{v\in [l]}:=\mathsf{PrefixPath}(Q^*[\tau(j)]_k,\delta)$ for $j\in[m],k\in[d]$.

\item $\mathcal{R}$ computes the prefix $\{w_{i,k,v}^*\}_{v\in [l']}:=\mathsf{PrefixTrie}(w_{i,k}-\delta,w_{i,k}+\delta)$ for $i\in[n],k\in[d]$. $\mathcal{R}$ also computes $\{\mathcal{C}_{i,t}\}_{t\in [2^d]}:= \mathsf{neighbor}_\delta(\vecw_i), i\in[n]$. 

\item  $\mathcal{R}$ picks a random PRF key $sk_\mathcal{R}$. Then, $\mathcal{S}$ and $\mathcal{R}$ invoke the sOPRF functionality $\Funcsoprf$. The receiver acts as the sender in sOPRF with input $sk_\mathcal{R}$ and learns $\{f_{j,k,v}^1\}_{j\in[m],k\in[d],v\in[l]}$. The sender $\mathcal{S}$ acts as the receiver in sOPRF with input $\{\mathsf{cell}_\delta (Q^*[\tau(j)])||k||q^*_{\tau(j),k,v}||\alpha_{\tau(j)}\}_{j\in [m],k\in [d],v\in[l]}$, where $\alpha_{\tau(j)}\in[3]$ is the index of hash function used to insert $\mathsf{cell}_\delta (Q^*[\tau(j)])$, i.e., $h_{\alpha_{\tau(j)}}(\mathsf{cell}_\delta (Q^*[\tau(j)])) = \tau(j)$, and learns $\{f_{j,k,v}^0\}_{j\in[m],k\in[d],v\in[l]}$.

\item  $\mathcal{R}$ defines $\mathsf{List}:=\{(\mathcal{C}_{i,t}||k||w_{i,k,v}^*||\alpha, f_{sk_\mathcal{R}}(\mathcal{C}_{i,t}||k||w_{i,k,v}^*||\alpha) )\}_{i\in [n],k\in [d],t\in [2^d],v\in [l'],\alpha\in[3]})$. Then, $\mathcal{R}$ computes $E_\mathcal{R}:=\mathsf{Encode}(\mathsf{List})$ and sends $E_\mathcal{R}$ to $\mathcal{S}$. 

\item $\mathcal{S}$ computes $r_{j,k,v}^\mathcal{S}:=\mathsf{Decode}(E_\mathcal{R}, \mathsf{cell}_\delta (Q^*[\tau(j)])||k||q^*_{\tau(j),k,v}||\alpha_{\tau(j)})-f_{j,k,v}^0, j\in[m],k\in[d],v\in[l]$.

\item For $j\in[m]$, $\mathcal{S}$ and $\mathcal{R}$ invoke the ECSS functionality $\Funcecss$. $\mathcal{R}$ acts as the sender in ECSS with input $\{\veca_{j,k,v}=(-f^1_{j,k,v},f^1_{j,k,v})\}_{k\in[d],v\in[l]}$ and learns $\randr_j$. $\mathcal{S}$ acts as the receiver in ECSS with input $\{\vecb_{j,k,v}=(r_{j,k,v}^\mathcal{S},r_{j,k,v}^\mathcal{S})\}_{k\in[d],v\in[l]}$ and learns $\rands_j$.

\item For $j\in[m]$, $\mathcal{S}$ and $\mathcal{R}$ invoke the PEQT functionality $\Funcpeqt$. $\mathcal{S}$ inputs $-\rands_j$ and learns nothing. $\mathcal{R}$ inputs $\randr_j$ and learns $b_j$.

\item $\mathcal{R}$ initializes set $I:=\{\}$. Then, $\mathcal{S}$ and $\mathcal{R}$ invoke a batch of $m$ OT instances $\FuncOT$. In $j$-th OT, $\mathcal{S}$ inputs $(\perp,Q^*[\tau(j)])$. $\mathcal{R}$ acts as receiver with input $b_j$ and receives $\mathbf{z}_{j}$. $\mathcal{R}$ sets $I :=I \cup \{\mathbf{z}_{j}\}$ if $\mathbf{z}_{j}\ne \perp$. Finally, $\mathcal{R}$ outputs $I$.

\end{enumerate}
\end{trivlist}
\end{minipage}
\end{framed}
\caption{Fuzzy PSI Protocol $\Protocolfpsi$ for $L_\infty$ in Low Dimension Space from sOPRF} \label{fig:pifpsi-inf-apart-soprf}
\end{figure}

We argue the correctness of FPSI for $L_{\infty}$ in Figure \ref{fig:pifpsi-inf-apart-soprf}.
\begin{trivlist}
    \item \textbf{Correctness.} For $j\in[m]$, if $\exists \vecw_i\in W$, s.t. $\dist(Q^*[\tau(j)],\vecw_i)\leq \delta$, then the correctness of spatial hashing ensures $\exists t\in[2^d]$, s.t. $ \mathsf{cell}_\delta(Q^*[\tau(j)]) = \mathcal{C}_{i,t}$, thus
    $\forall k\in[d], \mathsf{cell}_\delta(Q^*[\tau(j)])||k||\alpha_{\tau(j)} = \mathcal{C}_{i,t}||k||\alpha_{\tau(j)}$ and $|Q^*[\tau(j)]_k-w_{i,k}|\leq \delta$. By the property of prefix and the correctness of sOPRF and OKVS, $\exists !v\in[l]$, s.t. $r^\mathcal{S}_{j,k,v}=f^1_{j,k,v}$. The correctness of ECSS guarantees $\rands_{j} + \randr_{j}=r^\mathcal{S}_{j,k,v}-f^1_{j,k,v}=0$. Then, the correctness of PEQT ensures $b_j=1$. Finally, the correctness of OT ensures the receiver learns ${Q}^*[\tau(j)]$. 
    If $\forall i\in [n], \dist({Q}^*[\tau(j)],\vecw_i)> \delta$, we have $\exists k\in[d]$, s.t. $\forall v\in[l], q_{\tau(j),k,v}^*\notin \{w_{i,k,v'}\}_{v'\in [l']}$. Therefore, the randomness of OKVS guarantees $r^\mathcal{S}_{j,k,v}$ is random. By setting $\mathsf{len}= \log |\mathbb{F}|= \lambda +\log mn +\log l$, a union bound shows the probability of $\exists i\in [n], u\in[m_c], v\in [l], r^\mathcal{S}_{j,k,v}-f^1_{j,k,v}=0 $ is negligible $2^{-\lambda}$. By the correctness of ECS, $\rands_{j}$ is also pseudorandom. As a result, $\rands_{j}\ne \randr_{j}$ with overwhelming probability. $\mathcal{R}$ receives $b_j=0$ and learns $\perp$ in the final OT.

\end{trivlist}

We prove the security of the protocol described in Figure \ref{fig:pifpsi-inf-apart-soprf}.
\begin{theorem}\label{thm:fpsi-inf-apart-soprf}
    The protocol in Figure \ref{fig:pifpsi-inf-apart-soprf} securely computes $\Funcfpsi$ for $L_{\infty}$ against semi-honest adversaries in the ($\Funcsoprf$,$\Funcecss$, $\Funcpeqt$,$\FuncOT$)-hybrid model.
\end{theorem}

Below we give details of the proof of Theorem \ref{thm:fpsi-inf-apart-soprf}.

\begin{proof}
We exhibit simulators $\mathsf{Sim}_{\mathcal{S}}$ and $\mathsf{Sim}_{\mathcal{R}}$ 
for simulating corrupt $\mathcal{S}$ and $\mathcal{R}$ respectively, 
and argue the indistinguishability of the produced transcript from the real execution.
\begin{trivlist}
\item \underline{Corrupt Sender:}  $\mathsf{Sim}_\mathcal{S}(Q=\{\vecq_j\}_{j\in [m]})$ 
    simulates the view of corrupt semi-honest sender. It executes as follows:
    
\begin{enumerate}
\item In step 1-2, $\mathsf{Sim}_\mathcal{S}$ executes like an honest sender and learns $Q^*,\tau,\{q^*_{\tau(j),k,v}\}_{j\in[m],k\in[d],v\in[l]}$.

\item In step 4, $\mathsf{Sim}_\mathcal{S}$ selects random $ f^0_{j,k,v},j\in [m],k\in[d],v\in[l]$. Then, it invokes the sOPRF receiver's simulator $\mathsf{Sim}_\mathsf{sOPRF}^\mathcal{R}(\{\mathsf{cell}_\delta (Q^*[\tau(j)])||k||q^*_{\tau(j),k,v}||\alpha_{\tau(j)}\}_{j\in [m],k\in [d],v\in[l]}, \{f_{j,k,v}^0\}_{j\in[m],k\in[d],v\in[l]})$ and appends the output to the view.

\item In step 5, $\mathsf{Sim}_\mathcal{S}$ selects $3nd2^dl'$ random key-value pairs $\mathsf{List}$ and computes $E_\mathcal{R}=\mathsf{Encode}(\mathsf{List})$. Then, it appends $E_\mathcal{R}$ to the view.

\item In step 6, $\mathsf{Sim}_\mathcal{S}$ executes like an honest sender and learns $r_{j,k,v}^\mathcal{S},j\in[m],k\in[d],v\in[l]$.

\item In step 7, $\mathsf{Sim}_{\mathcal{S}}$ selects random $\rands_j\leftarrow \mathbb{F},j\in[m]$. Then, it invokes the ECSS receiver's simulator $\mathsf{Sim}_\mathsf{ECSS}^\mathcal{R}(\{\vecb_{j,k,v}=(r^\mathcal{S}_{j,k,v},r_{j,k,v}^\mathcal{S})\}_{k\in[d],v\in[l]},\rands_{j})$ and appends the output to the view.

\item In step 8, $\mathsf{Sim}_\mathcal{S}$ invokes the PEQT sender's simulator $\mathsf{Sim}^\mathcal{S}_\mathsf{PEQT}(-\rands_j,\perp)$ and appends the output to the view.

\item In step 9, for $j\in [m]$, $\mathsf{Sim}_{\mathcal{S}}$ invokes OT sender's simulator $\mathsf{Sim}_\mathsf{OT}^\mathcal{S}(\perp,Q^*[\tau(j)])$. Then, it appends the output to the view.
\end{enumerate}

Now we argue that the view output by $\mathsf{Sim}_{\mathcal{S}}$ is indistinguishable from the real one. We formally prove this by a standard hybrid argument method. 
We define four hybrid transcripts $T_0, T_1, T_2,T_3$ where $T_0$ is real view of $\mathcal{S}$, 
and $T_3$ is the output of $\mathsf{Sim}_{\mathcal{S}}$.

\begin{itemize}
\item $\text{Hybrid}_0$. The first hybrid is the real interaction described in Figure \ref{fig:pifpsi-inf-apart-soprf}. 
    Here, an honest $\mathcal{R}$ uses input $W$, honestly interacts with the corrupt $\mathcal{S}$. 
    Let $T_0$ denote the real view of $\mathcal{S}$.

\item $\text{Hybrid}_1$.  Let $T_1$ be the same as $T_0$, except that all PRF values are replaced by randomly selected values, i.e., $f^0_{j,k,v}+f^1_{j,k,v}$ and $f_{sk_\mathcal{R}}(\mathcal{C}_{i,t}||k||w_{i,k,v}^*||\alpha)$ are replaced by truly random values. By the pseudorandomness of PRF in sOPRF, this hybrid is computationally indistinguishable from $T_0$.

\item $\text{Hybrid}_2$.  Let $T_2$ be the same as $T_1$, except that the OKVS $E_\mathcal{R}$ is generated from randomly selected key-value pairs. Since $f_{sk_\mathcal{R}}(\mathcal{C}_{i,t}||k||w_{i,k,v}^*||\alpha)$ has been replaced by truly random values, the obliviousness of OKVS scheme
guarantees this view is statistically indistinguishable from $T_2$. 

\item $\text{Hybrid}_3$.  Let $T_3$ be the same as $T_2$, except that the sOPRF, ECSS, PEQT and OT execution are replaced by simulator $\mathsf{Sim}^\mathcal{R}_{\mathsf{sOPRF}},\mathsf{Sim}^\mathcal{S}_{\mathsf{ECSS}},\mathsf{Sim}^\mathcal{S}_{\mathsf{PEQT}},\mathsf{Sim}^\mathcal{S}_{\mathsf{OT}}$. 
The security of sOPRF, ECSS, PEQT and OT functionality
guarantee this view is computationally indistinguishable from $T_2$. This hybrid is exactly the view output by the simulator.

\end{itemize}

\item \underline{Corrupt Receiver:} $\mathsf{Sim}_{\mathcal{R}}(W,I)$ 
    simulates the view of corrupt semi-honest receiver. It executes as follows:
\begin{enumerate}

\item In step 3, $\mathsf{Sim}_\mathcal{R}$ executes like an honest receiver and learns $\{w^*_{i,k,v}\}_{i\in[n],k\in[d],v\in[l']},\{\mathcal{C}_{i,t}\}_{i\in[n],t\in[2^d]}$.


\item In step 4, $\mathsf{Sim}_\mathcal{R}$ selects random $sk_{\mathcal{R}}$ and $f_{j,k,v}^1,j\in[m],k\in[d],v\in[l]$. Then, it invokes the sOPRF sender's simulator $\mathsf{Sim}_\mathsf{sOPRF}^\mathcal{S}(sk_\mathcal{R}, \{f_{j,k,v}^1\}_{j\in[m],k\in[d],v\in[l]})$ and appends the output to the view.

\item In step 5, $\mathsf{Sim}_\mathcal{R}$ executes like an honest receiver and learns $E_\mathcal{R}$.

\item In step 7, $\mathsf{Sim}_{\mathcal{R}}$ selects random $\randr_j\leftarrow \mathbb{F},j\in[m]$. Then, it invokes the ECSS sender's simulator $\mathsf{Sim}_\mathsf{ECSS}^\mathcal{S}(\{\veca_{j,k,v}=(-f^1_{j,k,v},f_{j,k,v}^1)\}_{k\in[d],v\in[l]},\randr_{j})$ and appends the output to the view.

\item In step 8-9, $\mathsf{Sim}_\mathcal{R}$ uses $\perp$ to pad $I$ to $m$ elements and permutes these elements randomly. 
    Let $I=\{\mathbf{z}_1,\dots,\mathbf{z}_{m}\}$. For $j\in [m]$, $\mathsf{Sim}_\mathcal{R}$ sets $b_j = 1$ if and only if $\mathbf{z}_j\notin W$, otherwise, $b_j=0$. Then, $\mathsf{Sim}_\mathcal{R}$ invokes the PEQT receiver's simulator $\mathsf{Sim}_\mathsf{PEQT}^\mathcal{R}(\randr_j,b_j)$ and appends the output to the view. $\mathsf{Sim}_\mathcal{R}$ also invokes OT receiver's simulator 
    $\mathsf{Sim}_\mathsf{OT}^\mathcal{R}(b_j,\mathbf{z}_j)$ and appends the output to the view.

\end{enumerate}
\end{trivlist}
Now we argue that the view output by $\mathsf{Sim}_{\mathcal{R}}$ is indistinguishable from the real one. In the real execution, the sender selects a random injection $\tau$ to shuffle its set elements. In the simulation, the way $\mathcal{R}$ obtains the elements in $I$ is identical to the real execution since the elements in $I$ are randomly permuted. By the underlying simulators' indistinguishability, the simulated view is computationally indistinguishable from the real. 
\end{proof}

\begin{trivlist}
\item \textbf{FPSI variants.} Similar to the OPPRF-based construction, our sOPRF-based FPSI protocol (Figure \ref{fig:pifpsi-inf-apart-soprf}) can also be adapted to other FPSI variants with minimal modifications. For LFPSI, the sender replaces $Q^*[\tau(u)]$ with the corresponding label in the final OT step. For FPSI-CA, the receiver directly outputs the Hamming weight of $\vec{b} = (b_1, \dots, b_{m})$. For FPSI-SP, the roles are effectively reversed: the sender acts as the receiver in the PEQT step (step 8) to learn each $b_j$, and then outputs the set ${ Q^*[\tau(j)] : b_j = 1, j \in [m] }$. In this case, the sender functionally acts as the receiver in the FPSI-SP definition.
\item \textbf{Optimization from AND.} Similar to the OPPRF-based construction, our sOPRF-based FPSI protocol (Figure \ref{fig:pifpsi-inf-apart-soprf}) can also be optimized using AND functionality. The formal description is given in Figure \ref{fig:pifpsi-inf-apart-soprf-and}.
\end{trivlist}

\begin{figure}[!ht]
\begin{framed}
\begin{minipage}[center]{\textwidth}
\begin{trivlist}
\item \textbf{Parameters:} 
\begin{itemize}
\item Two parties: sender $\mathcal{S}$ and receiver $\mathcal{R}$, set size $m$ and $n$, dimension $d$, threshold $\delta$.
\item Ideal $\Funcopprf$,$\Funcand$, $\Funcsspeqt$, and $\FuncOT$ primitives specified in Figure \ref{fig:fopprf}, Figure \ref{fig:fand}, Figure \ref{fig:fsspeqt}, and Figure \ref{fig:fot} respectively. 
\end{itemize}

\item Input of $\mathcal{S}$: $Q=\{\vecq_1,\dots,\vecq_{m}\}\subseteq \mathbb{U}^d$. 

\item Input of $\mathcal{R}$: $W=\{\vecw_1,\dots,\vecw_{n}\}\subseteq \mathbb{U}^d$.

\item \textbf{Protocol:}

\begin{enumerate}
    \item Steps 1-6 are identical to the protocol described in Figure \ref{fig:pifpsi-inf-apart-soprf}.
\end{enumerate}

\begin{enumerate}[start = 7]

\item For $j\in[m],k\in[d],v\in[l]$, $\mathcal{S}$ and $\mathcal{R}$ invoke the ssPEQT functionality $\Funcsspeqt$. $\mathcal{S}$ and $\mathcal{R}$ take inputs $r^{\mathcal{S}}_{j,k,v}$ and $f^1_{j,k,v}$, and learns $b^\mathcal{S}_{j,k,v}$ and $b^\mathcal{R}_{j,k,v}$, respectively. Then, $\mathcal{S}$ computes $b_{j,k}^\mathcal{S}:=\oplus_{v\in[l]}b^\mathcal{S}_{j,k,v}$ and $\mathcal{R}$ computes $b_{j,k}^\mathcal{R}:=\oplus_{v\in[l]}b^\mathcal{R}_{j,k,v}$.

\item For $j\in[m]$, $\mathcal{S}$ and $\mathcal{R}$ invoke the AND functionality $\Funcand$. $\mathcal{S}$ and $\mathcal{R}$ take inputs $\{b_{j,k}^\mathcal{S}\}_{k\in[d]}$ and $\{b_{j,k}^\mathcal{R}\}_{k\in[d]}$, and learns $\tilde{b}^\mathcal{S}_j$ and $\tilde{b}^\mathcal{R}_j$, respectively.

\item $\mathcal{R}$ initializes set $I:=\{\}$.Then, $\mathcal{S}$ and $\mathcal{R}$ invoke a batch of $m$ OT instances $\FuncOT$. In $j$-th OT, $\mathcal{S}$ inputs $(\perp,Q^*[\tau(j)])$ if $\tilde{b}^\mathcal{S}_{j}=0$ and $(Q^*[\tau(j)],\perp)$ if $\tilde{b}^\mathcal{S}_{j} = 1$. $\mathcal{R}$ acts as receiver with input $\tilde{b}^{\mathcal{R}}_{j}$ and receives $\mathbf{z}_{j}$. $\mathcal{R}$ sets $I :=I \cup \{\mathbf{z}_{j}\}$ if $\mathbf{z}_{j}\ne \perp$. Finally, $\mathcal{R}$ outputs $I$.

\end{enumerate}
\end{trivlist}
\end{minipage}
\end{framed}
\caption{Optimized Fuzzy PSI Protocol $\Protocolfpsi$ for $L_\infty$ in Low Dimension Space from sOPRF} \label{fig:pifpsi-inf-apart-soprf-and}
\end{figure}

We argue the correctness of FPSI for $L_{\infty}$ in Figure \ref{fig:pifpsi-inf-apart-soprf-and}.
\begin{trivlist}
    \item \textbf{Correctness.} For $j\in[m]$, if $\exists \vecw_i\in W$, s.t. $\dist(Q^*[\tau(j)],\vecw_i)\leq \delta$, then the correctness of spatial hashing ensures $\exists t\in[2^d]$, s.t. $ \mathsf{cell}_\delta(Q^*[\tau(j)]) = \mathcal{C}_{i,t}$, thus
    $\forall k\in[d], \mathsf{cell}_\delta(Q^*[\tau(j)])||k||\alpha_{\tau(j)} = \mathcal{C}_{i,t}||k||\alpha_{\tau(j)}$ and $|Q^*[\tau(j)]_k-w_{i,k}|\leq \delta$. By the property of prefix and the correctness of sOPRF and OKVS, $\exists !v\in[l]$, s.t. $r^\mathcal{S}_{j,k,v}=f^1_{j,k,v}$. 
    Therefore, $\forall k\in [d], \exists! v\in[l]$, s.t. $b_{j,k,v}^\mathcal{S}\oplus b_{j,k,v}^\mathcal{R}=1$ and $b_{j,k}^\mathcal{S} \oplus b_{j,k}^\mathcal{R}=1$. By the correctness of AND, $\tilde{b}^{\mathcal{S}}_{j}\oplus \tilde{b}^{\mathcal{R}}_{j}=1$. Same as Figure \ref{fig:pifpsi-inf-apart-opprf}, the receiver learns ${Q}^*[\tau(j)]$ if $b_j^{\mathcal{S}}\oplus \bar{b}_j^{\mathcal{R}}=1$. 
    If $\forall i\in [n], \dist({Q}^*[\tau(j)],\vecw_i)> \delta$, we have $\exists k\in[d]$, s.t. $\forall v\in[l], q_{\tau(j),k,v}^*\notin \{w_{i,k,v'}\}_{v'\in [l']}$. Therefore, the randomness of OKVS guarantees $r^\mathcal{S}_{j,k,v}$ is random. By setting $\mathsf{len}= \log |\mathbb{F}|= \lambda +\log mn +\log l$, a union bound shows the probability of $\exists i\in [n], u\in[m_c], v\in [l], r^\mathcal{S}_{j,k,v}-f^1_{j,k,v}=0 $ is negligible $2^{-\lambda}$. By the correctness of ssPEQT and AND, $\forall v\in[l], b_{j,k,v}^\mathcal{S}\oplus b_{j,k,v}^\mathcal{R}=0$ and $\tilde{b}_{j}^\mathcal{S}\oplus \tilde{b}_{j}^\mathcal{R}=0$. As a result, $\mathcal{R}$ learns $\perp$ in the final OT.
\end{trivlist}

We prove the security of the protocol described in Figure \ref{fig:pifpsi-inf-apart-soprf-and}.
\begin{theorem}\label{thm:fpsi-inf-apart-soprf-and}
    The protocol in Figure \ref{fig:pifpsi-inf-apart-soprf-and} securely computes $\Funcfpsi$ for $L_{\infty}$ against semi-honest adversaries in the ($\Funcsoprf$,$\Funcand$, $\Funcsspeqt$,$\FuncOT$)-hybrid model.
\end{theorem}

Below we give details of the proof of Theorem \ref{thm:fpsi-inf-apart-soprf-and}.

\begin{proof}
We exhibit simulators $\mathsf{Sim}_{\mathcal{S}}$ and $\mathsf{Sim}_{\mathcal{R}}$ 
for simulating corrupt $\mathcal{S}$ and $\mathcal{R}$ respectively, 
and argue the indistinguishability of the produced transcript from the real execution.
\begin{trivlist}
\item \underline{Corrupt Sender:}  $\mathsf{Sim}_\mathcal{S}(Q=\{\vecq_j\}_{j\in [m]})$ 
    simulates the view of corrupt semi-honest sender. It executes as follows:
    
\begin{enumerate}
\item In step 1-2, $\mathsf{Sim}_\mathcal{S}$ executes like an honest sender and learns $Q^*,\tau,\{q^*_{\tau(j),k,v}\}_{j\in[m],k\in[d],v\in[l]}$.

\item In step 4, $\mathsf{Sim}_\mathcal{S}$ selects random $ f^0_{j,k,v},j\in [m],k\in[d],v\in[l]$. Then, it invokes the sOPRF receiver's simulator $\mathsf{Sim}_\mathsf{sOPRF}^\mathcal{R}(\{\mathsf{cell}_\delta (Q^*[\tau(j)])||k||q^*_{\tau(j),k,v}||\alpha_{\tau(j)}\}_{j\in [m],k\in [d],v\in[l]}, \{f_{j,k,v}^0\}_{j\in[m],k\in[d],v\in[l]})$ and appends the output to the view.

\item In step 5, $\mathsf{Sim}_\mathcal{S}$ selects $3nd2^dl'$ random key-value pairs $\mathsf{List}$ and computes $E_\mathcal{R}=\mathsf{Encode}(\mathsf{List})$. Then, it appends $E_\mathcal{R}$ to the view.

\item In step 6, $\mathsf{Sim}_\mathcal{S}$ executes like an honest sender and learns $r_{j,k,v}^\mathcal{S},j\in[m],k\in[d],v\in[l]$.

\item In step 7, $\mathsf{Sim}_\mathcal{S}$ selects random $b^{\mathcal{S}}_{j,k,v}\leftarrow \{0,1\}, j\in[m],k\in[d],v\in[l]$. Then, it invokes the ssPEQT simulator $\mathsf{Sim}_\mathsf{ssPEQT}(r^\mathcal{S}_{j,k,v},b^{\mathcal{S}}_{j,k,v})$ and appends the output to the view. $\mathsf{Sim}_\mathcal{S}$ also computes $b_{j,k}^\mathcal{S}:=\oplus_{v\in[l]}b^\mathcal{S}_{j,k,v},j\in[m],k\in[d]$.

\item In step 8, $\mathsf{Sim}_\mathcal{S}$ selects random $\tilde{b}^{\mathcal{S}}_{j}\leftarrow \{0,1\}, j\in[m]$. Then, it invokes the AND simulator $\mathsf{Sim}_\mathsf{AND}(\{b^{\mathcal{S}}_{j,k}\}_{k\in[d]},\tilde{b}_j^\mathcal{S})$ and appends the output to the view.

\item In step 9, for $j\in [m]$, $\mathsf{Sim}_{\mathcal{S}}$ invokes OT sender's simulator $\mathsf{Sim}_\mathsf{OT}^\mathcal{S}(\perp,Q^*[\tau(j)])$ if $\tilde{b}_j^\mathcal{S}=0$ and $\mathsf{Sim}_\mathsf{OT}^\mathcal{S}(Q^*[\tau(j)],\perp)$ if $\tilde{b}_j^\mathcal{S}=1$. Then, it appends the output to the view.

\end{enumerate}

Now we argue that the view output by $\mathsf{Sim}_{\mathcal{S}}$ is indistinguishable from the real one. We formally prove this by a standard hybrid argument method. 
We define four hybrid transcripts $T_0, T_1, T_2,T_3$ where $T_0$ is real view of $\mathcal{S}$, 
and $T_3$ is the output of $\mathsf{Sim}_{\mathcal{S}}$.

\begin{itemize}
\item $\text{Hybrid}_0$. The first hybrid is the real interaction described in Figure \ref{fig:pifpsi-inf-apart-soprf-and}. 
    Here, an honest $\mathcal{R}$ uses input $W$, honestly interacts with the corrupt $\mathcal{S}$. 
    Let $T_0$ denote the real view of $\mathcal{S}$.

\item $\text{Hybrid}_1$.  Let $T_1$ be the same as $T_0$, except that all PRF values are replaced by randomly selected values, i.e., $f^0_{j,k,v}+f^1_{j,k,v}$ and $f_{sk_\mathcal{R}}(\mathcal{C}_{i,t}||k||w_{i,k,v}^*||\alpha)$ are replaced by truly random values. By the pseudorandomness of PRF in sOPRF, this hybrid is computationally indistinguishable from $T_0$.

\item $\text{Hybrid}_2$.  Let $T_2$ be the same as $T_1$, except that the OKVS $E_\mathcal{R}$ is generated from randomly selected key-value pairs. Since $f_{sk_\mathcal{R}}(\mathcal{C}_{i,t}||k||w_{i,k,v}^*||\alpha)$ has been replaced by truly random values, the obliviousness of OKVS scheme
guarantees this view is statistically indistinguishable from $T_2$. 

\item $\text{Hybrid}_3$.  Let $T_3$ be the same as $T_2$, except that the sOPRF, ssPEQT, AND and OT execution are replaced by simulator $\mathsf{Sim}^\mathcal{R}_{\mathsf{sOPRF}},\mathsf{Sim}^\mathcal{S}_{\mathsf{ssPEQT}},\mathsf{Sim}^\mathcal{S}_{\mathsf{AND}},\mathsf{Sim}^\mathcal{S}_{\mathsf{OT}}$. 
The security of sOPRF, ssPEQT, AND and OT functionality
guarantee this view is computationally indistinguishable from $T_2$. This hybrid is exactly the view output by the simulator.

\end{itemize}

\item \underline{Corrupt Receiver:} $\mathsf{Sim}_{\mathcal{R}}(W,I)$ 
    simulates the view of corrupt semi-honest receiver. It executes as follows:
\begin{enumerate}

\item In step 3, $\mathsf{Sim}_\mathcal{R}$ executes like an honest receiver and learns $\{w^*_{i,k,v}\}_{i\in[n],k\in[d],v\in[l']},\{\mathcal{C}_{i,t}\}_{i\in[n],t\in[2^d]}$.

\item In step 4, $\mathsf{Sim}_\mathcal{R}$ selects random $sk_{\mathcal{R}}$ and $f_{j,k,v}^1,j\in[m],k\in[d],v\in[l]$. Then, it invokes the sOPRF sender's simulator $\mathsf{Sim}_\mathsf{sOPRF}^\mathcal{S}(sk_\mathcal{R}, \{f_{j,k,v}^1\}_{j\in[m],k\in[d],v\in[l]})$ and appends the output to the view.

\item In step 5, $\mathsf{Sim}_\mathcal{R}$ executes like an honest receiver and learns $E_\mathcal{R}$.

\item In step 7, $\mathsf{Sim}_\mathcal{R}$ selects random $b^{\mathcal{R}}_{j,k,v}\leftarrow \{0,1\}, j\in[m],k\in[d],v\in[l]$. Then, it invokes the ssPEQT simulator $\mathsf{Sim}_\mathsf{ssPEQT}(f^1_{j,k,v},b^{\mathcal{R}}_{j,k,v})$ and appends the output to the view. $\mathsf{Sim}_\mathcal{R}$ also computes $b_{j,k}^\mathcal{R}:=\oplus_{v\in[l]}b^\mathcal{R}_{j,k,v},j\in[m],k\in[d]$.

\item In step 8, $\mathsf{Sim}_\mathcal{R}$ selects random $\tilde{b}^{\mathcal{R}}_{j}\leftarrow \{0,1\}, j\in[m]$. Then, it invokes the AND simulator $\mathsf{Sim}_\mathsf{AND}(\{b^{\mathcal{R}}_{j,k}\}_{k\in[d]},\tilde{b}_j^\mathcal{R})$ and appends the output to the view.

\item In step 9, $\mathsf{Sim}_\mathcal{R}$ uses $\perp$ to pad $I$ to $m$ elements and permutes these elements randomly. 
    Let $I=\{\mathbf{z}_1,\dots,\mathbf{z}_{m}\}$. Then, $\mathsf{Sim}_\mathcal{R}$ invokes OT receiver's simulator 
    $\mathsf{Sim}_\mathsf{OT}^\mathcal{R}(\tilde{b}_j^\mathcal{R},\mathbf{z}_j)$ and appends the output to the view.

\end{enumerate}
\end{trivlist}
Now we argue that the view output by $\mathsf{Sim}_{\mathcal{R}}$ is indistinguishable from the real one. In the real execution, the sender selects a random injection $\tau$ to shuffle its set elements. In the simulation, the way $\mathcal{R}$ obtains the elements in $I$ is identical to the real execution since the elements in $I$ are randomly permuted. By the underlying simulators' indistinguishability, the simulated view is computationally indistinguishable from the real. 
\end{proof}

\subsection{Fuzzy PSI for $L_{p\in[1,\infty)}$ Distance}
\label{subsec:fpsi-apart-p}

\subsubsection{Construction from OPPRF}
\label{subsubsec:fpsi-apart-p-opprf}
In this section, we give our construction of the FPSI (and its variants) for $L_{p\in[1,\infty)}$ distance from OPPRF. 

The formal description is given in Figure \ref{fig:pifpsi-p-apart-opprf}.

\begin{figure}[!ht]
\begin{framed}
\begin{minipage}[center]{\textwidth}
\begin{trivlist}
\item \textbf{Parameters:} 
\begin{itemize}
\item Two parties: sender $\mathcal{S}$ and receiver $\mathcal{R}$, set size $m$ and $n$, dimension $d$, threshold $\delta$.
\item Ideal $\Funcopprf$,$\Funcecips$, $\Funcps$, $\Funcifmat$, and $\FuncOT$ primitives specified in Figure \ref{fig:fopprf}, Figure \ref{fig:fecips}, Figure \ref{fig:fps}, Figure \ref{fig:fifmat}, and Figure \ref{fig:fot} respectively. 
\end{itemize}

\item Input of $\mathcal{S}$: $Q=\{\vecq_1,\dots,\vecq_{m}\}\subseteq \mathbb{U}^d$. 

\item Input of $\mathcal{R}$: $W=\{\vecw_1,\dots,\vecw_{n}\}\subseteq \mathbb{U}^d$.

\item \textbf{Protocol:}

\begin{enumerate}

\item The sender $\mathcal{S}$ computes the Cuckoo hash table $Q^*:= \mathsf{Cuckoo}_{h_1,h_2,h_3}^{m_c}(\{(\mathsf{cell}_\delta(\vecq_j),\vecq_j)\}_{j\in [m]})$ and fills empty bins with the dummy item. Let $m_c=(1+\epsilon)m$ denote the length of the Cuckoo hash table. $\mathcal{S}$ denotes the item in $u$-th bin as $Q^*[u]$ for $u\in [m_c]$. Let $\tau: [m]\rightarrow [m_c]$ denote a random injective function such that $\tau(1),\dots, \tau(m)$ are the non-dummy item bins of $Q^*$. 

\item $\mathcal{S}$ computes the prefix $\{q_{\tau(j),k,v}^*\}_{v\in [l]}:=\mathsf{PrefixPath}(Q^*[\tau(j)]_k,\delta/2)$ for $j\in[m],k\in[d]$.

\item $\mathcal{R}$ computes the prefix $\{w_{i,0,k,v}^*\}_{v\in [l']}:=\mathsf{PrefixTrie}(w_{i,k}-\delta,w_{i,k})$ and $\{w_{i,1,k,v}^*\}_{v\in [l']}:=\mathsf{PrefixTrie}(w_{i,k}+1,w_{i,k}+\delta)$ for $i\in[n],k\in[d]$. $\mathcal{R}$ also computes $\{\mathcal{C}_{i,t}\}_{t\in [2^d]}:= \mathsf{neighbor}_\delta(\vecw_i), i\in[n]$. Then, $\mathcal{R}$ picks random $r_{u,k,1}^\mathcal{R},\dots,r_{u,k,p+1}^\mathcal{R}\leftarrow \mathbb{F}$. 

\item  $\mathcal{R}$ defines $\mathsf{List}:=\{(\mathcal{C}_{i,t}||\sigma||k||w_{i,\sigma,k,v}^*||\alpha,r^\mathcal{R}_{h_\alpha(\mathcal{C}_{i,t}),k,1}+|w_{i,k}-\mathsf{Bound}(\sigma,w^*_{i,\sigma,k,v})|^1 ||\dots|| r^\mathcal{R}_{h_\alpha(\mathcal{C}_{i,t}),k,p}+|w_{i,k}-\mathsf{Bound}(\sigma,w^*_{i,\sigma,k,v})|^p||r^\mathcal{R}_{h_\alpha(\mathcal{C}_{i,t}),k,p+1})\}_{i\in [n],k\in [d],\sigma\in\{0,1\},t\in [2^d],v\in [l'],\alpha\in[3]})$. Then, $\mathcal{S}$ and $\mathcal{R}$ invoke the OPPRF functionality $\Funcopprf$. The receiver acts as the sender in OPPRF with input $\mathsf{List}$ and learns $\hint, sk_\mathcal{R}$. The sender $\mathcal{S}$ acts as the receiver in OPPRF with input $\{\mathsf{cell}_\delta (Q^*[\tau(j)])||\sigma||k||q^*_{\tau(j),k,v}||\alpha_{\tau(j)}\}_{j\in [m],\sigma\in\{0,1\},k\in [d],v\in[l]}$, where $\alpha_{\tau(j)}\in[3]$ is the index of hash function used to insert $\mathsf{cell}_\delta (Q^*[\tau(j)])$, i.e., $h_{\alpha_{\tau(j)}}(\mathsf{cell}_\delta (Q^*[\tau(j)])) = \tau(j)$, and learns $\hint, \{\mathbf{r}_{\tau(j),\sigma,k,v}^\mathcal{S}\}_{j\in [m],\sigma\in\{0,1\},k\in [d],v\in[l]}$. 
Parse $\mathbf{r}_{\tau(j),\sigma,k,v}^\mathcal{S}=r_{\tau(j),\sigma,k,v,1}^\mathcal{S}||\dots||r_{\tau(j),\sigma,k,v,p+1}^\mathcal{S}$.

\item $\mathcal{S}$ defines $\vecb_{\tau(j),\sigma,k,v}^0:=(1,r_{\tau(j),\sigma,k,v,1}^\mathcal{S},\dots,r_{\tau(j),\sigma,k,v,p}^\mathcal{S})\in \mathbb{F}^{p+1}, \vecb_{\tau(j),\sigma,k,v}^1:=(|Q^*[\tau(j)]_k-\mathsf{Bound}(\sigma,q^*_{\tau(j),k,v})|^p,\binom{p}{1}|Q^*[\tau(j)]_k-\mathsf{Bound}(\sigma,q^*_{\tau(j),k,v})|^{p-1},\dots,\binom{p}{p})\in \mathbb{F}^{p+1}, b_{\tau(j),\sigma,k,v}^2:= r_{\tau(j),\sigma,k,v,p+1}^\mathcal{S}$, $j\in[m],\sigma\in\{0,1\},k\in[d],v\in[l]$. 

\item For $u\in [m_c]\setminus \{\tau(j)\}_{j\in[m]}$, $\mathcal{S}$ picks random $\vecb_{u,\sigma,k,v}^0,\vecb_{u,\sigma,k,v}^1\leftarrow \mathbb{F}^{p+1}, b_{u,\sigma,k,v}^2\leftarrow \mathbb{F}, \sigma\in \{0,1\},k\in[d],v\in[l]$.

\item $\mathcal{R}$ defines $\veca_{u,\sigma,k,v}^0:=(0,-r_{u,k,1}^\mathcal{R},\dots,-r_{u,k,p}^\mathcal{R})\in \mathbb{F}^{p+1}, a_{u,\sigma,k,v}^2:= r_{u,k,p+1}^\mathcal{R}$, $u\in[m_c],\sigma\in\{0,1\},k\in[d],v\in[l]$. 

\item For $u\in[m_c]$, $\mathcal{S}$ and $\mathcal{R}$ invoke the ECIPS functionality $\Funcecips$. $\mathcal{R}$ acts as the sender in ECIPS with input $\{(\veca_{u,\sigma,k,v}^0,a_{u,\sigma,k,v}^2)\}_{\sigma\in \{0,1\},k\in[d],v\in[l]}$ and learns $\randr_u$. $\mathcal{S}$ acts as the receiver in ECIPS with input $\{(\vecb_{u,\sigma,k,v}^0,\vecb_{u,\sigma,k,v}^1,b_{u,\sigma,k,v}^2)\}_{\sigma\in\{0,1\},k\in[d],v\in[l]}$ and learns $\rands_u$.

\item For $u\in[m_c]$, $\mathcal{S}$ and $\mathcal{R}$ invoke the ssIFmat functionality $\Funcssifmat$. $\mathcal{S}$ and $\mathcal{R}$ take inputs $\rands_u$ and $\randr_u$, and learns $\tilde{b}^\mathcal{S}_u$ and $\tilde{b}^\mathcal{R}_u$, respectively.

\item $\mathcal{S}$ picks a random permutation $\pi$ over $[m_c]$. Then, $\mathcal{S}$ and $\mathcal{R}$ invoke the PS functionality $\Funcps$. $\mathcal{R}$ acts as the sender in PS with input $\{\tilde{b}^\mathcal{R}_u\}_{u\in[m_c]}$ and learns $\{\bar{b}^\mathcal{R}_u\}_{u\in[m_c]}$. $\mathcal{S}$ acts as the receiver in PS with input $\pi$ and learns  $\{\bar{b}^\mathcal{S}_u\}_{u\in[m_c]}$.

\item $\mathcal{S}$ computes $b_u^\mathcal{S}:=\bar{b}_u^\mathcal{S}\oplus \tilde{b}_{\pi(u)}^\mathcal{S}$ for $u\in [m_c]$. $\mathcal{R}$ initializes set $I:=\{\}$.

\item $\mathcal{S}$ and $\mathcal{R}$ invoke a batch of $m_c$ OT instances $\FuncOT$. In $u$-th OT, $\mathcal{S}$ inputs $(\perp,Q^*[\pi(u)])$ if $b^\mathcal{S}_{u}=0$ and $(Q^*[\pi(u)],\perp)$ if $b^\mathcal{S}_{u} = 1$. $\mathcal{R}$ acts as receiver with input $\bar{b}^{\mathcal{R}}_{u}$ and receives $\mathbf{z}_{u}$. $\mathcal{R}$ sets $I :=I \cup \{\mathbf{z}_{u}\}$ if $\mathbf{z}_{u}\ne \perp$. Finally, $\mathcal{R}$ outputs $I$.

\end{enumerate}
\end{trivlist}
\end{minipage}
\end{framed}
\caption{Fuzzy PSI Protocol $\Protocolfpsi$ for $L_{p\in[1,\infty)}$ in Low Dimension Space from OPPRF} \label{fig:pifpsi-p-apart-opprf}
\end{figure}

We argue the correctness of FPSI for $L_{p\in[1,\infty)}$ in Figure \ref{fig:pifpsi-p-apart-opprf}.
\begin{trivlist}
    \item \textbf{Correctness.} For $u\in[m_c]$, if $\exists \vecw_i\in W$, s.t. $\dist(Q^*[\pi(u)],\vecw_i)\leq \delta$, then the correctness of spatial hashing ensures $\exists t\in[2^d]$, s.t. $ \mathsf{cell}_\delta(Q^*[\pi(u)]) = \mathcal{C}_{i,t}$, thus
    $\forall k\in[d], \mathsf{cell}_\delta(Q^*[\pi(u)])||\sigma||k||\alpha_{\pi(u)} = \mathcal{C}_{i,t}||\sigma||k||\alpha_{\pi(u)}$ and $|Q^*[\pi(u)]_k-w_{i,k}|\leq \delta$.  
    By the property of prefix, $\exists !v\in[l],v'\in [l'],\sigma\in\{0,1\}$, s.t. $q_{\pi(u),k,v}^*=w_{i,\sigma,k,v'}^*$. By the correctness of OPPRF, we have $\forall \gamma\in[p], r^{\mathcal{S}}_{\pi(u),\sigma,k,v,\gamma}=r^{\mathcal{R}}_{\pi(u),k,\gamma}+|w_{i,k} -\mathsf{Bound}(\sigma,w_{i,\sigma,k,v'})|^\gamma$ and $r^{\mathcal{S}}_{\pi(u),\sigma,k,v,p+1}=r^{\mathcal{R}}_{\pi(u),k,p+1}$.
    Then, the correctness of ECIPS guarantees 
    \begin{align*}
        \rands_{\pi(u)}+\randr_{\pi(u)} &= \sum_{k\in[d]}\langle \veca^0_{\pi(u),\sigma,k,v}+ \vecb^0_{\pi(u),\sigma,k,v}, \vecb^1_{\pi(u),\sigma,k,v} \rangle \\
        &= \sum_{k\in[d]}\sum_{\gamma\in[0,p]} \binom{p}{\gamma}|w_{i,k} - \mathsf{Bound}(\sigma,w_{i,\sigma,k,v'}^*)|^\gamma \cdot |Q^*[\pi(u)]_k-\mathsf{Bound}(\sigma,q_{\pi(u),k,v}^*)|^{p-\gamma}\\
        &=\sum_{k\in[d]} |w_{i,k}-Q^*[\pi(u)]_k|^p\\
        &=\mathsf{dist}(\vecw_{i},Q^*[\pi(u)])^p
    \end{align*}
    By the correctness of ssIFmat, $\tilde{b}^\mathcal{S}_{\pi(u)}\oplus \tilde{b}_{\pi(u)}^\mathcal{R}=1$ iff $\rands_{\pi(u)}\in [-\randr_{\pi(u)} ,-\randr_{\pi(u)}+\delta^p]$, which means $\mathsf{dist}(\vecw_{i},Q^*[\pi(u)])^p\in [0,\delta^p]$. Then, the correctness of ssPEQT and PS ensures $\tilde{b}_{\pi(u)}^\mathcal{S}\oplus \tilde{b}_{\pi(u)}^\mathcal{R}=1$ and $b_u^{\mathcal{S}}\oplus \bar{b}_u^{\mathcal{R}}=1$. Finally, the correctness of OT ensures the receiver learns ${Q}^*[\pi(u)]$ if $b_u^{\mathcal{S}}\oplus \bar{b}_u^{\mathcal{R}}=1$. 
    If $\forall i\in [n], \dist({Q}^*[\pi(u)],\vecw_i)> \delta$, we have $\exists k\in[d]$, s.t. $\forall v\in[l], q_{\pi(u),k,v}^*\notin \{w_{i,k,v'}\}_{v'\in [l']}$. Therefore, the pseudorandomness of PRF in OPPRF guarantees $r^\mathcal{S}_{\pi(u),k,v}$ is pseudorandom, resulting in $\vecb_{\pi(u),\sigma,k,v}^0$ being also pseudorandom (except the first component). The correctness of ECIPS guarantees $\rands_{\pi(u)}+\randr_{\pi(u)}$ is also pseudorandom.
    By setting $\mathsf{len}= \log |\mathbb{F}|= \lambda +\log mn +\log l+p\log \delta$, a union bound shows the probability of $\exists i\in [n], u\in[m_c], v\in [l], \rands_{\pi(u)}+\randr_{\pi(u)}\in [0,\delta^p]$ is negligible $2^{-\lambda}$. As a result, $b_u^{\mathcal{S}}\oplus \bar{b}_u^{\mathcal{R}}=0$ with overwhelming probability. $\mathcal{R}$ learns $\perp$ in the final OT.
\end{trivlist}

We prove the security of the protocol described in Figure \ref{fig:pifpsi-p-apart-opprf}.
\begin{theorem}\label{thm:fpsi-p-apart-opprf}
    The protocol in Figure \ref{fig:pifpsi-p-apart-opprf} securely computes $\Funcfpsi$ for $L_{p\in[1,\infty)}$ against semi-honest adversaries in the ($\Funcopprf$,$\Funcecips$, $\Funcps$, $\Funcsspeqt$,$\FuncOT$)-hybrid model.
\end{theorem}

Below we give details of the proof of Theorem \ref{thm:fpsi-p-apart-opprf}.

\begin{proof}
We exhibit simulators $\mathsf{Sim}_{\mathcal{S}}$ and $\mathsf{Sim}_{\mathcal{R}}$ 
for simulating corrupt $\mathcal{S}$ and $\mathcal{R}$ respectively, 
and argue the indistinguishability of the produced transcript from the real execution.
\begin{trivlist}
\item \underline{Corrupt Sender:}  $\mathsf{Sim}_\mathcal{S}(Q=\{\vecq_j\}_{j\in [m]})$ 
    simulates the view of corrupt semi-honest sender. It executes as follows:
    
\begin{enumerate}
\item In step 1-2, $\mathsf{Sim}_\mathcal{S}$ executes like an honest sender and learns $Q^*,\tau,\{q^*_{\tau(j),k,v}\}_{j\in[m],k\in[d],v\in[l]}$.

\item In step 4, $\mathsf{Sim}_\mathcal{S}$ selects random $\hint, \mathbf{r}^{\mathcal{S}}_{\tau(j),\sigma,k,v},j\in [m],\sigma\in \{0,1\},k\in[d],v\in[l]$. Then, it invokes the OPPRF receiver's simulator $\mathsf{Sim}_\mathsf{OPPRF}^\mathcal{R}(\{\mathsf{cell}_\delta (Q^*[\tau(j)])||\sigma||k||q^*_{\tau(j),k,v}||\alpha_{\tau(j)}\}_{j\in [m],\sigma\in\{0,1\}, k\in [d],v\in[l]},$
$(\hint, \{\mathbf{r}_{\tau(j),\sigma,k,v}^\mathcal{S}\}_{j\in[m],\sigma\in\{0,1\},k\in[d],v\in[l]}))$ and appends the output to the view.

\item In step 5-6, $\mathsf{Sim}_\mathcal{S}$ executes like an honest sender and learns $\{\vecb^0_{u,\sigma,k,v},\vecb^1_{u,\sigma,k,v},b^2_{u,\sigma,k,v}\}_{u\in[m_c],\sigma\in \{0,1\},k\in[d],v\in[l]}$.

\item In step 8, $\mathsf{Sim}_{\mathcal{S}}$ selects random $\rands_u\leftarrow \mathbb{F},u\in[m_c]$. Then, it invokes the ECIPS receiver's simulator $\mathsf{Sim}_\mathsf{ECIPS}^\mathcal{R}(\{\vecb^0_{u,\sigma,k,v},\vecb^1_{u,\sigma,k,v},b^2_{u,\sigma,k,v}\}_{\sigma\in \{0,1\},k\in[d],v\in[l]},\rands_{u})$ and appends the output to the view.

\item In step 9, $\mathsf{Sim}_\mathcal{S}$ selects random $\tilde{b}^{\mathcal{S}}_{u}\leftarrow \{0,1\}, u\in[m_c]$. Then, it invokes the ssIFmat simulator $\mathsf{Sim}_\mathsf{ssIFmat}(\rands_u,\tilde{b}_{u}^{\mathcal{S}})$ and appends the output to the view.

\item In step 10, $\mathsf{Sim}_\mathcal{S}$ selects a random permutation $\pi$ over $[m_c]$, random bits $\bar{b}_u^{\mathcal{S}}\leftarrow \{0,1\},u\in[m_c]$. Then, it invokes PS receiver's simulator $\mathsf{Sim}_\mathsf{PS}^\mathcal{R}(\pi,\{\bar{b}_u^{\mathcal{S}}\}_{u\in [m_c]})$ and appends the output to the view. $\mathsf{Sim}_\mathcal{S}$ also computes $b_u^\mathcal{S}=\bar{b}_u^\mathcal{S}\oplus \tilde{b}_{\pi(u)}^\mathcal{S},u\in[m_c]$.

\item In step 12, for $u\in [m_c]$, $\mathsf{Sim}_{\mathcal{S}}$ invokes OT sender's simulator $\mathsf{Sim}_\mathsf{OT}^\mathcal{S}(\perp,Q^*[\pi(u)])$ if $b_u^\mathcal{S}=0$ and $\mathsf{Sim}_\mathsf{OT}^\mathcal{S}(Q^*[\pi(u)],\perp)$ if $b_u^\mathcal{S}=1$. Then, it appends the output to the view.
\end{enumerate}

Now we argue that the view output by $\mathsf{Sim}_{\mathcal{S}}$ is indistinguishable from the real one. We formally prove this by a standard hybrid argument method. 
We define three hybrid transcripts $T_0, T_1, T_2$ where $T_0$ is real view of $\mathcal{S}$, 
and $T_2$ is the output of $\mathsf{Sim}_{\mathcal{S}}$.

\begin{itemize}
\item $\text{Hybrid}_0$. The first hybrid is the real interaction described in Figure \ref{fig:pifpsi-p-apart-opprf}. 
    Here, an honest $\mathcal{R}$ uses input $W$, honestly interacts with the corrupt $\mathcal{S}$. 
    Let $T_0$ denote the real view of $\mathcal{S}$.

\item $\text{Hybrid}_1$.  Let $T_1$ be the same as $T_0$, except that all OPPRF values $\mathbf{r}_{\tau(j),\sigma,k,v}^\mathcal{S}$ are replaced by randomly selected values. Since the set $Q$ satisfies the apart assumption, that is, $\mathsf{cell}_\delta(Q^*[\tau(j)])$ are distinct, resulting in $\mathbf{r}_{\tau(j),\sigma,k,v}^\mathcal{S}$ are also distinct. If $\exists i\in[n],\sigma\in \{0,1\},v'\in[l']$, s.t. $q_{\tau(j),k,v}^*=w_{i,\sigma,k,v'}^*$, we have ${r}_{\tau(j),\sigma,k,v,1}^\mathcal{S} = {r}_{\tau(j),k,1}^\mathcal{R}+|w_{i,k}-\mathsf{Bound}(\sigma,w^*_{i,\sigma,k,v})|^1,\dots,{r}_{\tau(j),\sigma,k,v,p+1}^\mathcal{S} = {r}_{\tau(j),k,p+1}^\mathcal{R} $, which are truly random. Otherwise, if $\forall i\in[n],\sigma\in \{0,1\},v'\in[l'], q_{\tau(j),k,v}^*\neq w_{i,\sigma,k,v'}^*$, from the pseudorandomness of OPPRF, $\mathbf{r}_{\tau(j),\sigma,k,v}^\mathcal{S}$ is a pseudorandom value. As a result, by the pseudorandomness of PRF in OPPRF, this hybrid is computationally indistinguishable from $T_0$.

\item $\text{Hybrid}_2$.  Let $T_2$ be the same as $T_1$, except that the OPPRF, ECIPS, ssIFmat, PS and OT execution are replaced by simulator $\mathsf{Sim}^\mathcal{R}_{\mathsf{OPPRF}},\mathsf{Sim}^\mathcal{S}_{\mathsf{ECIPS}},\mathsf{Sim}^\mathcal{R}_{\mathsf{PS}}, \mathsf{Sim}^\mathcal{S}_{\mathsf{ssIFmat}},\mathsf{Sim}^\mathcal{S}_{\mathsf{OT}}$. 
The security of OPPRF, ECIPS, ssIFmat, PS and OT functionality
guarantee this view is computationally indistinguishable from $T_1$. This hybrid is exactly the view output by the simulator.

\end{itemize}

\item \underline{Corrupt Receiver:} $\mathsf{Sim}_{\mathcal{R}}(W,I)$ 
    simulates the view of corrupt semi-honest receiver. It executes as follows:
\begin{enumerate}

\item In step 3, the simulator $\mathsf{Sim}_\mathcal{R}$ executes like an honest receiver and learns $\{w^*_{i,\sigma,k,v}\}_{i\in[n],\sigma\in \{0,1\},k\in[d],v\in[l']},$
$\{\mathcal{C}_{i,t}\}_{i\in[n],t\in[2^d]},\{r_{u,k,1}^\mathcal{R},\dots,r_{u,k,p+1}^\mathcal{R}\}_{u\in[m_c],k\in[d]}$.


\item In step 4, $\mathsf{Sim}_\mathcal{R}$ defines $\mathsf{List}$ as an honest receiver and picks random $\hint,sk_\mathcal{R}$. Then, it invokes the OPPRF sender's simulator $\mathsf{Sim}_\mathsf{OPPRF}^\mathcal{S}(\mathsf{List},(\hint,sk_\mathcal{R}))$ and appends the output to the view.

\item In step 7, $\mathsf{Sim}_\mathcal{R}$ executes like an honest receiver and learns $\{\veca^0_{u,\sigma,k,v},a^2_{u,\sigma,k,v}\}_{u\in[m_c],\sigma\in \{0,1\},k\in[d],v\in[l]}$.

\item In step 8, $\mathsf{Sim}_{\mathcal{R}}$ picks random $\randr_u,u\in[m_c]$. Then, it invokes the ECIPS sender's simulator $\mathsf{Sim}_\mathsf{ECIPS}^\mathcal{S}(\{\veca^0_{u,\sigma,k,v},a^2_{u,\sigma,k,v}\}_{\sigma\in \{0,1\},k\in[d],v\in[l]},\randr_{u})$ and appends the output to the view.

\item In step 9, $\mathsf{Sim}_\mathcal{R}$ selects random $\tilde{b}^{\mathcal{R}}_{u}\leftarrow \{0,1\}, u\in[m_c]$. Then, it invokes the ssIFmat simulator $\mathsf{Sim}_\mathsf{ssIFmat}(\randr_u,\tilde{b}_{u}^{\mathcal{R}})$ and appends the output to the view.

\item In step 10, $\mathsf{Sim}_\mathcal{R}$ selects random values $\bar{b}_u^{\mathcal{R}}\leftarrow \{0,1\},u\in[m_c]$. Then, it invokes PS sender's simulator $\mathsf{Sim}_\mathsf{PS}^\mathcal{S}(\{\tilde{b}_u^\mathcal{R}\}_{u\in [m_c]},\{\bar{b}_u^{\mathcal{R}}\}_{u\in [m_c]})$ and appends the output to the view. 

\item In step 12, $\mathsf{Sim}_\mathcal{R}$ uses $\perp$ to pad $I$ to $m_c$ elements and permutes these elements randomly. 
    Let $I=\{\mathbf{z}_1,\dots,\mathbf{z}_{m_c}\}$. Then, $\mathsf{Sim}_\mathcal{R}$ invokes OT receiver's simulator 
    $\mathsf{Sim}_\mathsf{OT}^\mathcal{R}(\bar{b}_u^\mathcal{R},\mathbf{z}_u)$ and appends the output to the view.

\end{enumerate}
\end{trivlist}
Now we argue that the view output by $\mathsf{Sim}_{\mathcal{R}}$ is indistinguishable from the real one. In the simulation, the way $\mathcal{R}$ obtains the elements in $I$ is identical to the real execution since the elements in $I$ are randomly permuted. By the underlying simulators' indistinguishability, the simulated view is computationally indistinguishable from the real. 
\end{proof}

\begin{trivlist}
\item \textbf{FPSI variants.} The protocol presented in Figure \ref{fig:pifpsi-p-apart-opprf} can be similarly extended to support other FPSI variants with straightforward modifications. These adaptations follow exactly the same adjustments as described for the $L_\infty$ protocol in Figure \ref{fig:pifpsi-inf-apart-opprf}, and we omit them here.
\item \textbf{Trade-off from the execution order of PS and IFmat.} Similar to the $L_\infty$ protocol in Figure \ref{fig:pifpsi-inf-apart-opprf}, we can also achieve a trade-off between computation and communication by adjusting the order of (ss)IFmat and PS protocols. 
\end{trivlist}

\subsubsection{Construction from sOPRF}
\label{subsubsec:fpsi-apart-p-soprf}

In this section, we give our construction of the FPSI (and its variants) for $L_{p\in[1,\infty)}$ distance from sOPRF. Its idea is similar to the protocol for $L_\infty$ distance, which only requires ECIPS execution for non-empty bins. 
The formal description is given in Figure \ref{fig:pifpsi-p-apart-soprf}.

\begin{figure}[!ht]
\begin{framed}
\begin{minipage}[center]{\textwidth}
\begin{trivlist}
\item \textbf{Parameters:} 
\begin{itemize}
\item Two parties: sender $\mathcal{S}$ and receiver $\mathcal{R}$, set size $m$ and $n$, dimension $d$, threshold $\delta$.
\item Ideal $\Funcsoprf$,$\Funcecips$, $\Funcifmat$, and $\FuncOT$ primitives specified in Figure \ref{fig:fsoprf}, Figure \ref{fig:fecips}, Figure \ref{fig:fifmat}, and Figure \ref{fig:fot} respectively. 
\end{itemize}

\item Input of $\mathcal{S}$: $Q=\{\vecq_1,\dots,\vecq_{m}\}\subseteq \mathbb{U}^d$. 

\item Input of $\mathcal{R}$: $W=\{\vecw_1,\dots,\vecw_{n}\}\subseteq \mathbb{U}^d$.

\item \textbf{Protocol:}

\begin{enumerate}

\item The sender $\mathcal{S}$ computes the Cuckoo hash table $Q^*:= \mathsf{Cuckoo}_{h_1,h_2,h_3}^{m_c}(\{(\mathsf{cell}_\delta(\vecq_j),\vecq_j)\}_{j\in [m]})$ and fills empty bins with the dummy item. Let $m_c=(1+\epsilon)m$ denote the length of the Cuckoo hash table. $\mathcal{S}$ denotes the item in $u$-th bin as $Q^*[u]$ for $u\in [m_c]$. Let $\tau: [m]\rightarrow [m_c]$ denote a random injective function such that $\tau(1),\dots, \tau(m)$ are the non-dummy item bins of $Q^*$. 

\item $\mathcal{S}$ computes the prefix $\{q_{\tau(j),k,v}^*\}_{v\in [l]}:=\mathsf{PrefixPath}(Q^*[\tau(j)]_k,\delta/2)$ for $j\in[m],k\in[d]$.

\item $\mathcal{R}$ computes the prefix $\{w_{i,0,k,v}^*\}_{v\in [l']}:=\mathsf{PrefixTrie}(w_{i,k}-\delta,w_{i,k})$ and $\{w_{i,1,k,v}^*\}_{v\in [l']}:=\mathsf{PrefixTrie}(w_{i,k}+1,w_{i,k}+\delta)$ for $i\in[n],k\in[d]$. $\mathcal{R}$ also computes $\{\mathcal{C}_{i,t}\}_{t\in [2^d]}:= \mathsf{neighbor}_\delta(\vecw_i), i\in[n]$. 

\item  $\mathcal{R}$ picks a random PRF key $sk_\mathcal{R}$. Then, $\mathcal{S}$ and $\mathcal{R}$ invoke the sOPRF functionality $\Funcsoprf$. The receiver acts as the sender in sOPRF with input $sk_\mathcal{R}$ and learns $\{\mathsf{f}_{j,\sigma,k,v}^1\}_{j\in[m],\sigma\in\{0,1\},k\in[d],v\in[l]}$. The sender $\mathcal{S}$ acts as the receiver in sOPRF with input $\{\mathsf{cell}_\delta (Q^*[\tau(j)])||\sigma||k||q^*_{\tau(j),k,v}||\alpha_{\tau(j)}\}_{j\in [m],\sigma\in\{0,1\},k\in [d],v\in[l]}$, where $\alpha_{\tau(j)}\in[3]$ is the index of hash function used to insert $\mathsf{cell}_\delta (Q^*[\tau(j)])$, i.e., $h_{\alpha_{\tau(j)}}(\mathsf{cell}_\delta (Q^*[\tau(j)])) = \tau(j)$, and learns $\{\mathbf{f}_{j,\sigma,k,v}^0\}_{j\in[m],\sigma\in\{0,1\},k\in[d],v\in[l]}$. Parse $\mathbf{f}_{j,\sigma,k,v}^0=f_{j,\sigma,k,v,1}^0||\dots||f_{j,\sigma,k,v,p+1}^0,\mathbf{f}_{j,\sigma,k,v}^1=f_{j,\sigma,k,v,1}^1||\dots||f_{j,\sigma,k,v,p+1}^1$.

\item  $\mathcal{R}$ computes $\mathbf{f}_{i,t,\sigma,k,v,\alpha}:= f_{sk_\mathcal{R}}(\mathcal{C}_{i,t}||\sigma||k||w_{i,\sigma,k,v}^*||\alpha) = {f}_{i,t,\sigma,k,v,\alpha,1}||\dots||{f}_{i,t,\sigma,k,v,\alpha,p+1}\in \mathbb{F}^{p+1}, i\in [n],t\in [2^d],\sigma\in\{0,1\},k\in [d],v\in [l'],\alpha\in[3]$. Then, it defines $\mathsf{List}:=\{(\mathcal{C}_{i,t}||\sigma||k||w_{i,\sigma,k,v}^*||\alpha,|w_{i,k}-\mathsf{Bound}(\sigma,w^*_{i,\sigma,k,v})|^1 ||\dots||\ |w_{i,k}-\mathsf{Bound}(\sigma,w^*_{i,\sigma,k,v})|^p||0 + \mathbf{f}_{i,t,\sigma,k,v,\alpha}  )\}_{i\in [n],k\in [d],\sigma\in\{0,1\},t\in [2^d],v\in [l'],\alpha\in[3]})$.

\item $\mathcal{R}$ computes $E_\mathcal{R}:=\mathsf{Encode}(\mathsf{List})$ and sends $E_\mathcal{R}$ to $\mathcal{S}$. 

\item $\mathcal{S}$ computes $\mathbf{r}_{j,\sigma,k,v}^\mathcal{S}:=\mathsf{Decode}(E_\mathcal{R}, \mathsf{cell}_\delta (Q^*[\tau(j)])||\sigma||k||q^*_{\tau(j),k,v}||\alpha_{\tau(j)})-\mathbf{f}_{j,\sigma,k,v}^0, j\in[m],\sigma\in \{0,1\},k\in[d],v\in[l]$. Parse $\mathbf{r}_{j,\sigma,k,v}^\mathcal{S}=r_{j,\sigma,k,v,1}^\mathcal{S}||\dots||r_{j,\sigma,k,v,p+1}^\mathcal{S}$.

\item $\mathcal{S}$ defines $\vecb_{j,\sigma,k,v}^0:=(1,r_{j,\sigma,k,v,1}^\mathcal{S},\dots,r_{j,\sigma,k,v,p}^\mathcal{S})\in \mathbb{F}^{p+1}, \vecb_{j,\sigma,k,v}^1:=(|Q^*[\tau(j)]-\mathsf{Bound}(\sigma,q^*_{\tau(j),k,v})|^p,\binom{p}{1}|Q^*[\tau(j)]-\mathsf{Bound}(\sigma,q^*_{\tau(j),k,v})|^{p-1},\dots,\binom{p}{p})\in \mathbb{F}^{p+1}, b_{\tau(j),\sigma,k,v}^2:= r_{j,\sigma,k,v,p+1}^\mathcal{S}$, $j\in[m],\sigma\in\{0,1\},k\in[d],v\in[l]$. 

\item $\mathcal{R}$ defines $\veca_{j,\sigma,k,v}^0:=(0,-f_{j,\sigma,k,v,1}^1,\dots,-f_{j,\sigma,k,v,p}^1)\in \mathbb{F}^{p+1}, a_{j,\sigma,k,v}^2:= f_{j,\sigma,k,v,p+1}^1$, $j\in[m],\sigma\in\{0,1\},k\in[d],v\in[l]$. 

\item For $j\in[m]$, $\mathcal{S}$ and $\mathcal{R}$ invoke the ECIPS functionality $\Funcecips$. $\mathcal{R}$ acts as the sender in ECIPS with input $\{(\veca_{j,\sigma,k,v}^0,a_{j,\sigma,k,v}^2)\}_{\sigma\in \{0,1\},k\in[d],v\in[l]}$ and learns $\randr_j$. $\mathcal{S}$ acts as the receiver in ECIPS with input $\{(\vecb_{j,\sigma,k,v}^0,\vecb_{j,\sigma,k,v}^1,b_{j,\sigma,k,v}^2)\}_{\sigma\in\{0,1\},k\in[d],v\in[l]}$ and learns $\rands_j$.

\item For $j\in[m]$, $\mathcal{S}$ and $\mathcal{R}$ invoke the IFmat functionality $\Funcifmat$. $\mathcal{S}$ inputs $\rands_j$ and learns nothing. $\mathcal{R}$ inputs $\randr_j$ and learns $b_j$.

\item $\mathcal{R}$ initializes set $I:=\{\}$. Then, $\mathcal{S}$ and $\mathcal{R}$ invoke a batch of $m$ OT instances $\FuncOT$. In $j$-th OT, $\mathcal{S}$ inputs $(\perp,Q^*[\tau(j)])$. $\mathcal{R}$ acts as receiver with input $b_j$ and receives $\mathbf{z}_{j}$. $\mathcal{R}$ sets $I :=I \cup \{\mathbf{z}_{j}\}$ if $\mathbf{z}_{j}\ne \perp$. Finally, $\mathcal{R}$ outputs $I$.

\end{enumerate}
\end{trivlist}
\end{minipage}
\end{framed}
\caption{Fuzzy PSI Protocol $\Protocolfpsi$ for $L_{p\in[1,\infty)}$ in Low Dimension Space from sOPRF} \label{fig:pifpsi-p-apart-soprf}
\end{figure}

\begin{trivlist}
\item \textbf{FPSI variants.} The protocol presented in Figure \ref{fig:pifpsi-p-apart-soprf} can be similarly extended to support other FPSI variants with straightforward modifications. These adaptations follow exactly the same adjustments as described for the $L_\infty$ protocol in Figure \ref{fig:pifpsi-inf-apart-soprf}, and we omit them here.
\end{trivlist}

We argue the correctness of FPSI for $L_{p\in[1,\infty)}$ in Figure \ref{fig:pifpsi-p-apart-soprf}.
\begin{trivlist}
    \item \textbf{Correctness.} For $j\in[m]$, if $\exists \vecw_i\in W$, s.t. $\dist(Q^*[\tau(j)],\vecw_i)\leq \delta$, then the correctness of spatial hashing ensures $\exists t\in[2^d]$, s.t. $ \mathsf{cell}_\delta(Q^*[\tau(j)]) = \mathcal{C}_{i,t}$, thus
    $\forall k\in[d], \mathsf{cell}_\delta(Q^*[\tau(j)])||\sigma||k||\alpha_{\tau(j)} = \mathcal{C}_{i,t}||\sigma||k||\alpha_{\pi(u)}$ and $|Q^*[\tau(j)]_k-w_{i,k}|\leq \delta$.  
    By the property of prefix, $\exists !v\in[l],v'\in [l'],\sigma\in\{0,1\}$, s.t. $q_{\pi(u),k,v}^*=w_{i,\sigma,k,v'}^*$. By the correctness of sOPRF and OKVS, we have 
    $\forall \gamma\in[p], r^{\mathcal{S}}_{j,\sigma,k,v,\gamma}=f^{1}_{j,\sigma,k,v,\gamma}+|w_{i,k} -\mathsf{Bound}(\sigma,w_{i,\sigma,k,v'})|^\gamma$ and $r^{\mathcal{S}}_{j,\sigma,k,v,p+1}=f^{1}_{j,\sigma,k,v,p+1}$.
    Then, the correctness of ECIPS guarantees 
    \begin{align*}
        \rands_{j}+\randr_{j} &= \sum_{k\in[d]}\langle \veca^0_{j,\sigma,k,v}+ \vecb^0_{j,\sigma,k,v}, \vecb^1_{j,\sigma,k,v} \rangle \\
        &= \sum_{k\in[d]}\sum_{\gamma\in[0,p]} \binom{p}{\gamma}|w_{i,k} - \mathsf{Bound}(\sigma,w_{i,\sigma,k,v'}^*)|^\gamma \cdot |Q^*[\tau(j)]_k-\mathsf{Bound}(\sigma,q_{\tau(j),k,v}^*)|^{p-\gamma}\\
        &=\sum_{k\in[d]} |w_{i,k}-Q^*[\tau(j)]_k|^p\\
        &=\mathsf{dist}(\vecw_{i},Q^*[\tau(j)])^p
    \end{align*}
    By the correctness of IFmat, $b_j=1$ iff $\rands_{j}\in [-\randr_{j} ,-\randr_{j}+\delta^p]$, which means $\mathsf{dist}(\vecw_{i},Q^*[\tau(j)])^p\in [0,\delta^p]$. 
    Finally, the correctness of OT ensures the receiver learns ${Q}^*[\tau(j)]$. 
    If $\forall i\in [n], \dist({Q}^*[\tau(j)],\vecw_i)> \delta$, we have $\exists k\in[d]$, s.t. $\forall v\in[l], q_{\tau(j),k,v}^*\notin \{w_{i,k,v'}\}_{v'\in [l']}$. Therefore, the randomness of OKVS guarantees $\mathbf{r}^\mathcal{S}_{j,\sigma,k,v}$ is random, resulting in $\vecb_{j,\sigma,k,v}^0$ is also pseudorandom (except the first component). The correctness of ECIPS guarantees $\rands_{j}+\randr_{j}$ is also random.
    By setting $\mathsf{len}= \log |\mathbb{F}|= \lambda +\log mn +\log l+p\log \delta$, a union bound shows the probability of $\exists i\in [n], j\in[m], v\in [l], \rands_{j}+\randr_{j}\in [0,\delta^p]$ is negligible $2^{-\lambda}$. As a result, $\mathcal{R}$ receives $b_j=0$ in IFmat and learns $\perp$ in the final OT.
\end{trivlist}

We prove the security of the protocol described in Figure \ref{fig:pifpsi-p-apart-soprf}.
\begin{theorem}\label{thm:fpsi-p-apart-soprf}
    The protocol in Figure \ref{fig:pifpsi-p-apart-soprf} securely computes $\Funcfpsi$ for $L_{p\in[1,\infty)}$ against semi-honest adversaries in the ($\Funcsoprf$, $\Funcecips$, $\Funcifmat$, $\FuncOT$)-hybrid model.
\end{theorem}

Below we give details of the proof of Theorem \ref{thm:fpsi-p-apart-soprf}.

\begin{proof}
We exhibit simulators $\mathsf{Sim}_{\mathcal{S}}$ and $\mathsf{Sim}_{\mathcal{R}}$ 
for simulating corrupt $\mathcal{S}$ and $\mathcal{R}$ respectively, 
and argue the indistinguishability of the produced transcript from the real execution.
\begin{trivlist}
\item \underline{Corrupt Sender:}  $\mathsf{Sim}_\mathcal{S}(Q=\{\vecq_j\}_{j\in [m]})$ 
    simulates the view of corrupt semi-honest sender. It executes as follows:
    
\begin{enumerate}
\item In step 1-2, $\mathsf{Sim}_\mathcal{S}$ executes like an honest sender and learns $Q^*,\tau,\{q^*_{\tau(j),k,v}\}_{j\in[m],k\in[d],v\in[l]}$.

\item In step 4, $\mathsf{Sim}_\mathcal{S}$ selects random $ \mathbf{f}^0_{j,\sigma,k,v},j\in [m],\sigma\in \{0,1\},k\in[d],v\in[l]$. Then, it invokes the sOPRF's simulator $\mathsf{Sim}_\mathsf{sOPRF}^\mathcal{R}(\{\mathsf{cell}_\delta (Q^*[\tau(j)])||\sigma||k||q^*_{\tau(j),k,v}||\alpha_{\tau(j)}\}_{j\in [m],k\in [d],v\in[l]}, \{\mathbf{f}_{j,\sigma,k,v}^0\}_{j\in[m],k\in[d],v\in[l]})$ and appends the output to the view.

\item In step 6, $\mathsf{Sim}_\mathcal{S}$ selects $3nd2^dl'$ random key-value pairs $\mathsf{List}$ and computes $E_\mathcal{R}=\mathsf{Encode}(\mathsf{List})$. Then, it appends $E_\mathcal{R}$ to the view.

\item In step 7, $\mathsf{Sim}_\mathcal{S}$ executes like an honest sender and learns $\mathbf{r}_{j,\sigma,k,v}^\mathcal{S}\in \mathbb{F}^{p+1},j\in[m],\sigma\in \{0,1\},k\in[d],v\in[l]$.

\item In step 8, $\mathsf{Sim}_\mathcal{S}$ executes like an honest sender and learns $\{\vecb^0_{j,\sigma,k,v},\vecb^1_{j,\sigma,k,v},b^2_{j,\sigma,k,v}\}_{j\in[m],\sigma\in \{0,1\},k\in[d],v\in[l]}$.

\item In step 10, $\mathsf{Sim}_{\mathcal{S}}$ selects random $\rands_j\leftarrow \mathbb{F},j\in[m]$. Then, it invokes the ECIPS receiver's simulator $\mathsf{Sim}_\mathsf{ECIPS}^\mathcal{R}(\{\vecb^0_{j,\sigma,k,v},\vecb^1_{j,\sigma,k,v},b^2_{j,\sigma,k,v}\}_{\sigma\in \{0,1\},k\in[d],v\in[l]},\rands_{j})$ and appends the output to the view.

\item In step 11, $\mathsf{Sim}_\mathcal{S}$ invokes the IFmat sender's simulator $\mathsf{Sim}_\mathsf{IFmat}^\mathcal{S}(\rands_j,\perp)$ and appends the output to the view.

\item In step 12, for $j\in [m]$, $\mathsf{Sim}_{\mathcal{S}}$ invokes OT sender's simulator $\mathsf{Sim}_\mathsf{OT}^\mathcal{S}(\perp,Q^*[\tau(j)])$. Then, it appends the output to the view.
\end{enumerate}

Now we argue that the view output by $\mathsf{Sim}_{\mathcal{S}}$ is indistinguishable from the real one. We formally prove this by a standard hybrid argument method. 
We define four hybrid transcripts $T_0, T_1, T_2,T_3$ where $T_0$ is real view of $\mathcal{S}$, 
and $T_3$ is the output of $\mathsf{Sim}_{\mathcal{S}}$.

\begin{itemize}
\item $\text{Hybrid}_0$. The first hybrid is the real interaction described in Figure \ref{fig:pifpsi-p-apart-soprf}. 
    Here, an honest $\mathcal{R}$ uses input $W$, honestly interacts with the corrupt $\mathcal{S}$. 
    Let $T_0$ denote the real view of $\mathcal{S}$.

\item $\text{Hybrid}_1$.  Let $T_1$ be the same as $T_0$, except that all PRF values are replaced by randomly selected values, i.e., $\mathbf{f}^0_{j,\sigma,k,v}+\mathbf{f}^1_{j,\sigma,k,v}$ and $f_{sk_\mathcal{R}}(\mathcal{C}_{i,t}||\sigma||k||w_{i,\sigma,k,v}^*||\alpha)$ are replaced by truly random values. By the pseudorandomness of PRF in sOPRF, this hybrid is computationally indistinguishable from $T_0$.

\item $\text{Hybrid}_2$.  Let $T_2$ be the same as $T_1$, except that the OKVS $E_\mathcal{R}$ is generated from randomly selected key-value pairs. Since $f_{sk_\mathcal{R}}(\mathcal{C}_{i,t}||\sigma||k||w_{i,\sigma,k,v}^*||\alpha)$ has been replaced by truly random values, the obliviousness of OKVS scheme
guarantees this view is statistically indistinguishable from $T_2$. 

\item $\text{Hybrid}_3$.  Let $T_3$ be the same as $T_2$, except that the sOPRF, ECIPS, IFmat and OT execution are replaced by simulator $\mathsf{Sim}^\mathcal{R}_{\mathsf{sOPRF}},\mathsf{Sim}^\mathcal{S}_{\mathsf{ECIPS}},\mathsf{Sim}^\mathcal{S}_{\mathsf{IFmat}},\mathsf{Sim}^\mathcal{S}_{\mathsf{OT}}$. 
The security of sOPRF, ECIPS, IFmat and OT functionality
guarantee this view is computationally indistinguishable from $T_2$. This hybrid is exactly the view output by the simulator.

\end{itemize}

\item \underline{Corrupt Receiver:} $\mathsf{Sim}_{\mathcal{R}}(W,I)$ 
    simulates the view of corrupt semi-honest receiver. It executes as follows:
\begin{enumerate}

\item In step 3, $\mathsf{Sim}_\mathcal{R}$ executes like an honest receiver and learns $\{w^*_{i,\sigma,k,v}\}_{i\in[n],\sigma\in \{0,1\},k\in[d],v\in[l']},\{\mathcal{C}_{i,t}\}_{i\in[n],t\in[2^d]}$.

\item In step 4, $\mathsf{Sim}_\mathcal{R}$ selects random $sk_{\mathcal{R}}$ and $\mathbf{f}_{j,\sigma,k,v}^1,j\in[m],\sigma\in \{0,1\},k\in[d],v\in[l]$. Then, it invokes the sOPRF sender's simulator $\mathsf{Sim}_\mathsf{sOPRF}^\mathcal{S}(sk_\mathcal{R}, \{\mathbf{f}_{j,\sigma,k,v}^1\}_{j\in[m],\sigma\in \{0,1\},k\in[d],v\in[l]})$ and appends the output to the view.

\item In step 5-6, $\mathsf{Sim}_\mathcal{R}$ executes like an honest receiver and learns $E_\mathcal{R}$.

\item In step 9, $\mathsf{Sim}_\mathcal{R}$ executes like an honest receiver and learns $\{\veca^0_{j,\sigma,k,v},a^2_{j,\sigma,k,v}\}_{j\in[m],\sigma\in \{0,1\},k\in[d],v\in[l]}$.

\item In step 10, $\mathsf{Sim}_{\mathcal{R}}$ picks random $\randr_j,j\in[m]$. Then, it invokes the ECIPS sender's simulator $\mathsf{Sim}_\mathsf{ECIPS}^\mathcal{S}(\{\veca^0_{j,\sigma,k,v},a^2_{j,\sigma,k,v}\}_{\sigma\in \{0,1\},k\in[d],v\in[l]},\randr_{j})$ and appends the output to the view.

\item In step 11-12, $\mathsf{Sim}_\mathcal{R}$ uses $\perp$ to pad $I$ to $m$ elements and permutes these elements randomly. 
    Let $I=\{\mathbf{z}_1,\dots,\mathbf{z}_{m}\}$. For $j\in [m]$, $\mathsf{Sim}_\mathcal{R}$ sets $b_j = 1$ if and only if $\mathbf{z}_j\notin W$, otherwise, $b_j=0$. Then, $\mathsf{Sim}_\mathcal{R}$ invokes the IFmat receiver's simulator $\mathsf{Sim}_\mathsf{IFmat}^\mathcal{R}(\randr_j,b_j)$ and appends the output to the view. $\mathsf{Sim}_\mathcal{R}$ also invokes OT receiver's simulator 
    $\mathsf{Sim}_\mathsf{OT}^\mathcal{R}(b_j,\mathbf{z}_j)$ and appends the output to the view.

\end{enumerate}
\end{trivlist}
Now we argue that the view output by $\mathsf{Sim}_{\mathcal{R}}$ is indistinguishable from the real one. In the real execution, the sender selects a random injection $\tau$ to shuffle its set elements. In the simulation, the way $\mathcal{R}$ obtains the elements in $I$ is identical to the real execution since the elements in $I$ are randomly permuted. By the underlying simulators' indistinguishability, the simulated view is computationally indistinguishable from the real. 
\end{proof}
\section{Fuzzy PSI in High Dimension}
\label{sec:fpsi-sep}

For high-dimensional space, we use distributed ID generation (dIDG) \cite{DBLP:conf/asiacrypt/GaoQLLW24,cryptoeprint:2025/885,DBLP:conf/ccs/DZL25} to avoid $O(2^d)$ costs. Similar to our FPSI protocols in low dimension, we present two FPSI constructions for $L_\infty$ and $L_p$ distances, leveraging OPPRF \cite{DBLP:conf/ccs/KolesnikovMPRT17} and sOPRF \cite{cryptoeprint:2025/885}, respectively.

\subsection{Distributed ID Generation}
\label{subsec:didg}

In this section, we present the construction of our dIDG protocol, which achieves only $O(d\log\delta(m+n))$ complexity. 
The formal description of our construction is given in Figure \ref{fig:pididg}.

\begin{figure}[!hbth]
\begin{framed}
\begin{minipage}[center]{\textwidth}
\begin{trivlist}
\item \textbf{Parameters:} 
\begin{itemize}
\item Two parties: sender $\mathcal{S}$ and receiver $\mathcal{R}$, set size $m,n$, item universe $\mathbb{U}^d$, dimension $d$, distance threshold $\delta$, an OKVS scheme $(\mathsf{Encode},\mathsf{Decode})$.
\item Ideal $\Funcsoprf$ and $\Funcecss$ primitives specified in Figure \ref{fig:fsoprf} and Figure \ref{fig:fecss} respectively. 
\end{itemize}



\item \textbf{Protocol:}

\begin{enumerate}


\item $\mathcal{S}$ computes the prefix $\{q_{j,k,v}^*\}_{v\in [l]}:=\mathsf{PrefixPath}(q_{j,k},\delta)$ for $j\in[m],k\in[d]$. 
\item $\mathcal{R}$ computes the prefix $\{w_{i,k,v}^*\}_{v\in [l']}:=\mathsf{PrefixTrie}(w_{i,k}-\delta,w_{i,k}+\delta)$ for $i\in[n],k\in[d]$.

\item $\mathcal{S}$ and $\mathcal{R}$ invoke the sOPRF functionality $\Funcsoprf$. The receiver $\mathcal{R}$ acts as the receiver in sOPRF with input $\{k||w_{i,k,v}^*\}_{i\in [n],k\in[d],v\in[l']}$, and learns $\{\mathbf{f}_{\mathcal{R},i,k,v}^1= {f}_{\mathcal{R},i,k,v,0}^1||{f}_{\mathcal{R},i,k,v,1}^1\}_{i\in [n],k\in[d],v\in[l']} $. The sender $\mathcal{S}$ selects a random PRF key $sk_{\mathcal{S}}$ and set its input sharing as $0$, and learns $\{\mathbf{f}_{\mathcal{R},i,k,v}^0 = {f}_{\mathcal{R},i,k,v,0}^0||{f}_{\mathcal{R},i,k,v,1}^0\}_{i\in [n],k\in[d],v\in[l']}$. We have $\mathbf{f}_{\mathcal{R},i,k,v}^0 +\mathbf{f}_{\mathcal{R},i,k,v}^1= F_{sk_{\mathcal{S}}}(k||w_{i,k,v}^*)$.


\item $\mathcal{S}$ picks random $\rands_{j,k}\leftarrow \mathbb{F}$ and defines $A_{j,k}:=\{(k||q_{j,k,v}^*, \rands_{j,k})\}_{v\in[l]}$ for $j\in[m],k\in [d]$. For $k\in[d],j_1\in[m]$: if $\exists j_2\in[j_1],v_1,v_2\in [l]$, s.t. $q_{j_1,k,v_1}^* = q_{j_2,k,v_2}^*$, then update $A_{j_2,k}:=\{(k||q_{j_2,k,v}^*, \rands_{j_1,k})\}_{v\in[l]}$. Let $A:= \cup_{j\in[m],k\in[d]}A_{j,k}$ and pad $A$ with dummy random key-value pairs to size of $dml$. Then, $\mathcal{S}$ computes $\pids_j:= \sum_{k\in [d]}A[k||q_{j,k,1}^*], j\in [m]$.

\item $\mathcal{S}$ defines $A':=\{(x,y||0+F_{sk_{\mathcal{S}}}(x)):(x,y)\in A\}$ and computes $E_\mathcal{S}:=\mathsf{Encode}(A')$. Then, $\mathcal{S}$ sends $E_\mathcal{S}$ to $\mathcal{R}$.

\item $\mathcal{R}$ computes $\mathbf{r}_{\mathcal{R},i,k,v}:=(\mathsf{Decode}(E_\mathcal{S},k||w^*_{i,k,v}) - \mathbf{f}_{\mathcal{R},i,k,v}^1)= {r}_{\mathcal{R},i,k,v,0}||{r}_{\mathcal{R},i,k,v,1} ,i\in [n],k\in[d],v\in[l']$. 

\item Symmetrically, $\mathcal{S}$ and $\mathcal{R}$ invoke the sOPRF functionality $\Funcsoprf$. The sender $\mathcal{S}$ acts as the receiver in sOPRF with input $\{k||q_{j,k,v}^*\}_{j\in[m],k\in[d],v\in[l]}$, and learns $\{\mathbf{f}_{\mathcal{S},j,k,v}^0={f}_{\mathcal{S},j,k,v,0}^0||{f}_{\mathcal{S},j,k,v,1}^0\}_{j\in[m],k\in[d],v\in[l]}$. 
The receiver $\mathcal{R}$ selects a random PRF key $sk_{\mathcal{R}}$ and set its input sharing as $0$, and learns $\{\mathbf{f}_{\mathcal{S},j,k,v}^1={f}_{\mathcal{S},j,k,v,0}^1||{f}_{\mathcal{S},j,k,v,1}^1\}_{j\in[m],k\in[d],v\in[l]}$. We have $\mathbf{f}_{\mathcal{S},j,k,v}^0 +\mathbf{f}_{\mathcal{S},j,k,v}^1= F_{sk_{\mathcal{R}}}(k||q_{j,k,v}^*)$.

\item $\mathcal{R}$ picks random $\randr_{i,k}\leftarrow \mathbb{F}$ and defines $B_{i,k}:=\{(k||w_{i,k,v}^*, \randr_{i,k})\}_{v\in[l']}$ for $i\in[n],k\in [d]$. For $k\in[d],i_1\in[n]$: if $\exists i_2\in[i_1],v_1,v_2\in [l']$, s.t. $w_{i_1,k,v_1}^* = w_{i_2,k,v_2}^*$, then update $B_{i_2,k}:=\{(k||w_{i_2,k,v}^*, \randr_{i_1,k})\}_{v\in[l']}$. Let $B:= \cup_{i\in[n],k\in[d]}B_{i,k}$ and pad $B$ with dummy random key-value pairs to size of $dnl'$. Then, $\mathcal{R}$ computes $\pidr_i:= \sum_{k\in [d]}B[k||w_{i,k,1}^*], i\in [n]$.

\item $\mathcal{R}$ defines $B':=\{(x,y||0+F_{sk_{\mathcal{R}}}(x)):(x,y)\in B\}$ and computes $E_\mathcal{R}:=\mathsf{Encode}(B')$. Then, $\mathcal{R}$ sends $E_\mathcal{R}$ to $\mathcal{S}$.

\item $\mathcal{S}$ computes $\mathbf{r}_{\mathcal{S},j,k,v}:=(\mathsf{Decode}(E_\mathcal{R},k||q^*_{j,k,v}) - \mathbf{f}_{\mathcal{S},j,k,v}^0)= {r}_{\mathcal{S},j,k,v,0}||{r}_{\mathcal{S},j,k,v,1} ,j\in [m],k\in[d],v\in[l]$.

\item For $i\in[n]$, $\mathcal{S}$ and $\mathcal{R}$ invoke the ECSS functionality $\Funcecss$. $\mathcal{R}$ acts as the sender in ECSS with input $\{\mathbf{r}_{\mathcal{R},i,k,v}=({r}_{\mathcal{R},i,k,v,0},{r}_{\mathcal{R},i,k,v,1})\}_{k\in[d],v\in[l']}$ and learns $\sigma_{\mathcal{R},i}'$. $\mathcal{S}$ acts as the receiver in ECSS with input $\{ \bar{\mathbf{f}}_{\mathcal{R},i,k,v}^0 = (-{f}_{\mathcal{R},i,k,v,0}^0,{f}_{\mathcal{R},i,k,v,1}^0) \}_{k\in[d],v\in[l']}$ and learns $\sigma_{\mathcal{R},i}^0$.

\item For $j\in[m]$, $\mathcal{S}$ and $\mathcal{R}$ invoke the ECSS functionality $\Funcecss$. $\mathcal{S}$ acts as the sender in ECSS with input $\{\mathbf{r}_{\mathcal{S},j,k,v}=({r}_{\mathcal{S},j,k,v,0},{r}_{\mathcal{S},j,k,v,1})\}_{k\in[d],v\in[l]}$ and learns $\sigma_{\mathcal{S},j}'$. $\mathcal{R}$ acts as the receiver in ECSS with input $\{ \bar{\mathbf{f}}_{\mathcal{S},j,k,v}^1 = (-{f}_{\mathcal{S},j,k,v,0}^1,{f}_{\mathcal{S},j,k,v,1}^1) \}_{k\in[d],v\in[l]}$ and learns $\sigma_{\mathcal{S},j}^1$.

\item $\mathcal{S}$ computes $\sigma_{\mathcal{S},j}^0:=\sigma_{\mathcal{S},j}'+\pids_j, j\in[m]$. $\mathcal{R}$ computes $\sigma_{\mathcal{R},i}^1:=\sigma_{\mathcal{R},i}'+\pidr_i, i\in[n]$.

\item $\mathcal{S}$ picks a random PRF key $sk_\mathcal{S}'$. Then, $\mathcal{S}$ and $\mathcal{R}$ invoke the sOPRF functionality $\Funcsoprf$. 
The sender $\mathcal{S}$ acts as the sender in sOPRF with input $sk_{\mathcal{S}}', \{\sigma_{\mathcal{S},j}^0\}_{j\in [m]}\cup \{\sigma_{\mathcal{R},i}^0\}_{i\in[n]}$, and learns $\{\rho_{\mathcal{S},j}^0\}_{j\in [m]}\cup \{\rho_{\mathcal{R},i}^0\}_{i\in[n]}$. 
The receiver $\mathcal{R}$ acts as the receiver in sOPRF with input $\{\sigma_{\mathcal{S},j}^1\}_{j\in [m]}\cup \{\sigma_{\mathcal{R},i}^1\}_{i\in[n]}$, and learns $\{\rho_{\mathcal{S},j}^1\}_{j\in [m]}\cup \{\rho_{\mathcal{R},i}^1\}_{i\in[n]}$. 
We have $\rho_{\mathcal{S},j}^0 + \rho_{\mathcal{S},j}^1  = F_{sk_{\mathcal{S}}'}(\sigma_{\mathcal{S},j}^0+\sigma_{\mathcal{S},j}^1)$, $\rho_{\mathcal{R},i}^0 + \rho_{\mathcal{R},i}^1  = F_{sk_{\mathcal{S}}'}(\sigma_{\mathcal{R},i}^0+\sigma_{\mathcal{R},i}^1)$.

\item $\mathcal{R}$ picks a random PRF key $sk_\mathcal{R}'$. Then, $\mathcal{S}$ and $\mathcal{R}$ invoke the sOPRF functionality $\Funcsoprf$. 
The receiver $\mathcal{R}$ acts as the sender in sOPRF with input $sk_{\mathcal{R}}', \{\rho_{\mathcal{S},j}^1\}_{j\in [m]}\cup \{\rho_{\mathcal{R},i}^1\}_{i\in[n]}$, and learns $\{\theta_{\mathcal{S},j}^1\}_{j\in [m]}\cup \{\theta_{\mathcal{R},i}^1\}_{i\in[n]}$. 
The sender $\mathcal{S}$ acts as the receiver in sOPRF with input $\{\rho_{\mathcal{S},j}^0\}_{j\in [m]}\cup \{\rho_{\mathcal{R},i}^0\}_{i\in[n]}$, and learns $\{\theta_{\mathcal{S},j}^0\}_{j\in [m]}\cup \{\theta_{\mathcal{R},i}^0\}_{i\in[n]}$. 
We have $\theta_{\mathcal{S},j}^0 + \theta_{\mathcal{S},j}^1  = F_{sk_{\mathcal{R}}'}(\rho_{\mathcal{S},j}^0+\rho_{\mathcal{S},j}^1)$, $\theta_{\mathcal{R},i}^0 + \theta_{\mathcal{R},i}^1  = F_{sk_{\mathcal{R}}'}(\rho_{\mathcal{R},i}^0+\rho_{\mathcal{R},i}^1)$.

\item The sender $\mathcal{S}$ sends $\{\theta_{\mathcal{R},i}^0\}_{i\in[n]}$ to the receiver $\mathcal{R}$. The receiver $\mathcal{R}$ sends $\{\theta_{\mathcal{S},j}^1\}_{j\in[m]}$ to the sender $\mathcal{S}$.

\item The sender $\mathcal{S}$ defines $id_{\vecq_j}:=\theta_{\mathcal{S},j}^0 + \theta_{\mathcal{S},j}^1$ and outputs $\id_{\vecq_j}:= \{ id_{\vecq_j}\}, j\in [m]$.

\item The receiver $\mathcal{R}$ defines $id_{\vecw_i}:=\theta_{\mathcal{R},i}^0 + \theta_{\mathcal{R},i}^1$ and outputs $\id_{\vecw_i}:= \{id_{\vecw_i}\}, i\in [n]$.
\end{enumerate}
\end{trivlist}
\end{minipage}
\end{framed}
\caption{Distributed ID Generation Protocol $\Protocoldidg$} 
\label{fig:pididg}
\end{figure}

\begin{trivlist}
    \item \textbf{Correctness.} If $\exists \vecq_j\in Q, \vecw_i\in W$, s.t. $\dist(\vecq_j,\vecw_i)\leq \delta$, we have $\forall k\in[d], q_{j,k} \in [w_{i,k}-\delta,w_{i,k}+\delta]$. By the property of prefix, we have $\forall k\in[d], \exists ! v\in [l],v'\in[l'], q_{j,k,v}^*  = w_{i,k,v'}^*$
    Then, from the definition of $B$, we have $B[k||q_{j,k,v}^*]=B[k||w_{i,k,v'}^*] = B[k||w_{i,k,1}^*],k\in[d]$.
    By the correctness of sOPRF and OKVS, we have $r_{\mathcal{S},j,k,v,0} = B[k||w_{i,k,v'}^*]+f_{\mathcal{S},j,k,v,0}^1$ and $r_{\mathcal{S},j,k,v,1} =f_{\mathcal{S},j,k,v,1}^1$.
    By the correctness of ECSS, we have $\sigma_{\mathcal{S},j}'+\sigma_{\mathcal{S},j}^1= \sum_{k\in[d]} (r_{\mathcal{S},j,k,v,0} -f_{\mathcal{S},j,k,v,0}^1 )=\sum_{k\in[d]} B[k||w_{i,k,v'}^*]=\pidr_i$. By the correctness of final two sOPRF, we have $id_{\vecq_j} = F_{sk_{\mathcal{R}}'}(F_{sk_{\mathcal{S}}'}( \pidr_i+\pids_j ))$. Symmetrically, we also have $id_{\vecw_i} = F_{sk_{\mathcal{R}}'}(F_{sk_{\mathcal{S}}'}( \pidr_i+ \pids_j)) = id_{\vecq_j}$.
    \item \textbf{Distinctiveness.} For any $i \in [n]$, the $d$-separate set assumption guarantees the existence of a $k_i \in [d]$ such that the interval $[w_{i,k_i} - \delta, w_{i,k_i} + \delta]$ does not overlap with any other interval $[w_{i',k_i} - \delta, w_{i',k_i} + \delta]$ for $i' \neq i$. By the property of prefix, we have $\forall i,i'\in[n], \{w_{i,k_i,v}^*\}_{v\in [l']}\cap \{w_{i',k_{i'},v}^*\}_{v\in [l']}=\emptyset$. Consequently, a unique and uniform random value $\randr_{i,k_i}$ is added to $\pidr_{i}$, ensuring that $\pidr_{i}$ is independent of $\pidr_{i'}$ for all $i' \neq i$. By setting $|\mathbb{F}| \geq \lambda + \log n$, the union bound demonstrates that the probability of $\exists i' \neq i$ such that $\pidr_{i'} = \pidr_i$ is negligible $2^{-\lambda}$. Since $\id_{\vecw_i} = F_{sk_{\mathcal{R}}'}(F_{sk_{\mathcal{S}}'}( \pidr_i+\Delta_i ))$, where $\Delta_i$ is the remaining addend, it follows that the probability of $\exists i' \neq i$ such that $\id_{\vecw_{i'}} = \id_{\vecw_i}$ is also negligible $2^{-\lambda}$.
    \item \textbf{Randomness.} The randomness can be easily obtained by combining the distinctiveness and the pseudorandomness of PRFs.
\end{trivlist}

\begin{theorem}\label{thm:didg}
    The protocol in Figure \ref{fig:pididg} securely computes $\Funcdidg$ against semi-honest adversaries in the $(\Funcsoprf,\Funcecss)$-hybrid model.
\end{theorem}

Below we give details of the proof of Theorem \ref{thm:didg}.
\begin{proof}
Since the protocol is symmetric, we only exhibit the simulator $\mathsf{Sim}_{\mathcal{S}}$ for simulating corrupt $\mathcal{S}$ and argue the indistinguishability of the produced transcript from the real execution. 
\begin{trivlist}
\item \underline{Corrupt Sender:} $\mathsf{Sim}_\mathcal{S}(Q,\{\id_{q_j}\}_{j\in [m]})$ 
    simulates the view of corrupt semi-honest sender. It executes as follows:
\begin{enumerate}

\item In step 1, $\mathsf{Sim}_\mathcal{S}$ executes like an honest sender, and learns $\{q_{j,k,v}^*\}_{j\in[m],k\in[d],v\in[l]}$.

\item In step 3, $\mathsf{Sim}_\mathcal{S}$ generates a PRF key $sk_{\mathcal{S}}$ and selects random $\{\mathbf{f}_{\mathcal{R},i,k,v}^0\}_{i\in[n],k\in[d],v\in[l']}$. 
Then, it invokes sOPRF sender's simulator $\mathsf{Sim}_\mathsf{sOPRF}^\mathcal{S}((sk_{\mathcal{S}},0),\{\mathbf{f}_{\mathcal{R},i,k,v}^0\}_{i\in[n],k\in[d],v\in[l']})$ and appends the output to the view.

\item In step 4-6, $\mathsf{Sim}_\mathcal{S}$ executes like an honest sender, and learns $A, A', \rands_{j,k},\pids_{j}, j\in[m],k\in[d]$.

\item In step 7, $\mathsf{Sim}_\mathcal{S}$ selects random $\mathbf{f}_{\mathcal{S},j,k,v}^0\leftarrow \mathbb{F}^2,j\in [m],k\in[d],v\in[l]$. Then, it invokes the sOPRF receiver's simulator $\mathsf{Sim}_\mathsf{sOPRF}^\mathcal{R}(\{k||q_{j,k,v}^*\}_{j\in[m],k\in[d],v\in[l]},\{\mathbf{f}_{\mathcal{S},j,k,v}^0\}_{j\in[m],k\in[d],v\in[l]})$ and appends the output to the view.

\item In step 9, $\mathsf{Sim}_\mathcal{S}$ picks $dnl'$ random key-value pairs and computes the corresponding OKVS $E_\mathcal{R}$. Then it appends $E_{\mathcal{R}}$ to the view.

\item In step 10, $\mathsf{Sim}_\mathcal{S}$ honestly computes $\mathbf{r}_{\mathcal{S},j,k,v}^0={r}_{\mathcal{S},j,k,v,0}^0||{r}_{\mathcal{S},j,k,v,1}^0,j\in[m],k\in[d],v\in[l]$.

\item In step 11, $\mathsf{Sim}_{\mathcal{S}}$ selects random $\sigma_{\mathcal{R},i}^0\leftarrow \mathbb{F},i\in[n]$. Then, it invokes the ECSS receiver's simulator $\mathsf{Sim}_\mathsf{ECSS}^\mathcal{R}(\{\bar{\mathbf{f}}_{\mathcal{R},i,k,v}=(-{f}^0_{\mathcal{R},i,k,v,0},{f}^0_{\mathcal{R},i,k,v,1})\}_{k\in[d],v\in[l']},\sigma_{\mathcal{R},i}^0)$ and appends the output to the view.

\item In step 12, $\mathsf{Sim}_{\mathcal{S}}$ selects random $\sigma_{\mathcal{S},j}'\leftarrow \mathbb{F},i\in[n]$. Then, it invokes the ECSS sender's simulator $\mathsf{Sim}_\mathsf{ECSS}^\mathcal{S}(\{\mathbf{r}_{\mathcal{S},j,k,v}\}_{k\in[d],v\in[l]},\sigma_{\mathcal{S},j}')$ and appends the output to the view.

\item In step 13, $\mathsf{Sim}_\mathcal{S}$ honestly computes $\sigma_{\mathcal{S},j}^0:=\sigma_{\mathcal{S},j}'+\pids_j, j\in[m]$. 

\item In step 14, $\mathsf{Sim}_\mathcal{S}$ generates a PRF key $sk_{\mathcal{S}}'$ and selects random $\{\rho_{\mathcal{S},j}^0\}_{j\in[m]}\cup \{\rho_{\mathcal{R},i}^0\}_{i\in[n]}$. Then, it invokes sOPRF sender's simulator $\mathsf{Sim}_\mathsf{sOPRF}^\mathcal{S}(sk_{\mathcal{S}}',\{\sigma_{\mathcal{S},j}^0\}_{j\in[m]}\cup \{\sigma_{\mathcal{R},i}^0\}_{i\in[n]},\{\rho_{\mathcal{S},j}^0\}_{j\in[m]}\cup \{\rho_{\mathcal{R},i}^0\}_{i\in[n]})$ and appends the output to the view.

\item In step 15, $\mathsf{Sim}_\mathcal{S}$ selects random $\{\theta_{\mathcal{S},j}^0\}_{j\in[m]}\cup \{\theta_{\mathcal{R},i}^0\}_{i\in[n]}$. Then, it invokes sOPRF receiver's simulator $\mathsf{Sim}_\mathsf{sOPRF}^\mathcal{S}(\{\rho_{\mathcal{S},j}^0\}_{j\in[m]}\cup \{\rho_{\mathcal{R},i}^0\}_{i\in[n]},\{\theta_{\mathcal{S},j}^0\}_{j\in[m]}\cup \{\theta_{\mathcal{R},i}^0\}_{i\in[n]})$ and appends the output to the view.

\item In step 16, $\mathsf{Sim}_\mathcal{S}$ sets $\theta_{\mathcal{S},j}^1:=\id_{\vecq_j} - \theta_{\mathcal{S},j}^0$ and append $\{\theta_{\mathcal{S},j}^1\}$ to the view.
    
\end{enumerate}
Now we argue that the view output by $\mathsf{Sim}_{\mathcal{S}}$ is indistinguishable from the real one. 
We formally prove this by a standard hybrid argument method. 
We define five hybrid transcripts $T_0, T_1, T_2,T_3,T_4$ where $T_0$ is real view of $\mathcal{S}$, 
and $T_4$ is the output of $\mathsf{Sim}_{\mathcal{S}}$.

\begin{itemize}
\item $\text{Hybrid}_0$. The first hybrid is the real interaction described in Figure \ref{fig:pididg}. 
    Here, an honest $\mathcal{R}$ uses input $W$, honestly interacts with the corrupt $\mathcal{S}$. 
    Let $T_0$ denote the real view of $\mathcal{S}$.

\item $\text{Hybrid}_1$.  Let $T_1$ be the same as $T_0$, except that all PRF values $F_{sk_{\mathcal{R}}}(\cdot)$ are replaced by randomly selected values. 
This hybrid is computationally indistinguishable from $T_0$ by the pseudorandomness of the PRF.

\item $\text{Hybrid}_2$.  Let $T_2$ be the same as $T_1$, except that the OKVS $E_{\mathcal{R}}$ is computed from $dnl'$ random key-value pairs.
Note that in the previous hybrid, the key-value pairs used to compute OKVS are $\{(k||w_{i,k,v}^*,\randr_{i,k}||0+\mathbf{f}_{i,k,v})\}_{i\in[n],k\in[d],t\in[-\delta,\delta]}$, where $\mathbf{f}_{i,k,v}$ is truly random value. Note that $\mathcal{S}$ knows nothing about any $\mathbf{f}_{i,k,v}$ because it only learns secret shares of some of these values (i.e., those $\mathbf{f}_{i,k,v}$ satisfying $\exists i\in[n],k\in[d],v\in[l],v'\in[l']$, s.t. $q_{j,k,v}^*=w_{i,k,v'}^* $). By the obliviousness property of OKVS, $T_1$ and $T_2$ are statistically indistinguishable. 

\item $\text{Hybrid}_3$.  Let $T_3$ be the same as $T_2$, except that all PRF values $F_{sk_{\mathcal{R}}'}(\cdot)$ are replaced by randomly selected values. 
This hybrid is computationally indistinguishable from $T_2$ by the pseudorandomness of the PRF.

\item $\text{Hybrid}_4$.  Let $T_4$ be the same as $T_3$, except that the generation of $\theta_{\mathcal{S},j}^1$ is now the same as the simulation. In other words, instead of computing $id_{\vecq_j}$ as in step 17, where $\theta_{\mathcal{S},j}^1$ are uniform random, we instead compute $id_{\vecq_j}$ randomly and then program $\theta_{\mathcal{S},j}^1$ to contain the correct value. This change has no effect on sender's view distribution. $T_2$ and $T_3$ are identical.

\item $\text{Hybrid}_5$.  Let $T_5$ be the same as $T_4$, except that all ECSS and sOPRF executions are replaced by ECSS and sOPRF's simulator. Since the outputs of ECSS and sOPRF are both random secret shares, the security of ECSS and sOPRF functionality
guarantee this view is indistinguishable from $T_4$. This hybrid is exactly the view output by the simulator.

\end{itemize}
\end{trivlist}
\end{proof}

\subsection{Fuzzy PSI for $L_\infty$ Distance}
\label{subsec:fpsi-sep-inf}
In this section, we present our FPSI protocol (and its variants) for $L_\infty$ distance. The construction largely mirrors that of Section \ref{sec:fpsi-apart}, with the key modification being the replacement of the spatial hashing with the dIDG protocol (Figure \ref{fig:pididg}). Consequently, we detail only the OPPRF-based construction here, as other trade-offs and optimizations apply similarly and are thus omitted. The formal protocol is described in Figure \ref{fig:pifpsi-inf-sep-opprf}.

\begin{figure}[!ht]
\begin{framed}
\begin{minipage}[center]{\textwidth}
\begin{trivlist}
\item \textbf{Parameters:} 
\begin{itemize}
\item Two parties: sender $\mathcal{S}$ and receiver $\mathcal{R}$, set size $m$ and $n$, dimension $d$, threshold $\delta$.
\item Ideal $\Funcdidg$, $\Funcopprf$,$\Funcecs$, $\Funcps$, $\Funcsspeqt$, and $\FuncOT$ primitives specified in Figure \ref{fig:fdidg}, Figure \ref{fig:fopprf}, Figure \ref{fig:fecss}, Figure \ref{fig:fps}, Figure \ref{fig:fsspeqt}, and Figure \ref{fig:fot} respectively. 
\end{itemize}

\item Input of $\mathcal{S}$: $Q=\{\vecq_1,\dots,\vecq_{m}\}\subseteq \mathbb{U}^d$. 

\item Input of $\mathcal{R}$: $W=\{\vecw_1,\dots,\vecw_{n}\}\subseteq \mathbb{U}^d$.

\item \textbf{Protocol:}

\begin{enumerate}

\item $\mathcal{S}$ and $\mathcal{R}$ invoke the distributed ID generation functionality $\Funcdidg$ with input $Q$ and $W$, respectively. As a result, the sender $\mathcal{S}$ learns $\id_{\vecq_j} = \{id_{\vecq_j}\}$ for $j\in[m]$. The receiver $\mathcal{R}$ learns $\id_{\vecw_i} = \{id_{\vecw_i}\}$ for $i\in[n]$.

\item The sender $\mathcal{S}$ computes the Cuckoo hash table $Q^*:= \mathsf{Cuckoo}_{h_1,h_2,h_3}^{m_c}(\{(id_{\vecq_j},\vecq_j)\}_{j\in [m]})$ and fills empty bins with the dummy item. Let $m_c=(1+\epsilon)m$ denote the length of the Cuckoo hash table. $\mathcal{S}$ denotes the item in $u$-th bin as $Q^*[u]$ for $u\in [m_c]$. Let $\tau: [m]\rightarrow [m_c]$ denote a random injective function such that $\tau(1),\dots, \tau(m)$ are the non-dummy item bins of $Q^*$. 

\item $\mathcal{S}$ computes the prefix $\{q_{\tau(j),k,v}^*\}_{v\in [l]}:=\mathsf{PrefixPath}(Q^*[\tau(j)]_k,\delta)$ for $j\in[m],k\in[d]$.

\item $\mathcal{R}$ computes the prefix $\{w_{i,k,v}^*\}_{v\in [l']}:=\mathsf{PrefixTrie}(w_{i,k}-\delta,w_{i,k}+\delta)$ for $j\in[m],k\in[d]$. Then, $\mathcal{R}$ picks random $r_{u,k}^\mathcal{R}\leftarrow \mathbb{F}$ and computes $\randr_u:=\sum_{k\in[d]}r_{u,k}^\mathcal{R}$.

\item  $\mathcal{R}$ defines $\mathsf{List}:=\{(id_{\vecw_i}||k||w_{i,k,v}^*||\alpha,r_{h_\alpha(id_{\vecw_i}),k})\}_{i\in [n],k\in [d],v\in [l'],\alpha\in[3]})$. Then, $\mathcal{S}$ and $\mathcal{R}$ invoke the OPPRF functionality $\Funcopprf$. The receiver acts as the sender in OPPRF with input $\mathsf{List}$ and learns $\hint, sk_\mathcal{R}$. The sender $\mathcal{S}$ acts as the receiver in OPPRF with input $\{ id_{Q^*[\tau(j)]}||k||q^*_{\tau(j),k,v}||\alpha_{\tau(j)}\}_{j\in [m],k\in [d],v\in[l]}$, where $\alpha_{\tau(j)}\in[3]$ is the index of hash function used to insert $id_{Q^*[\tau(j)]}$, i.e., $h_{\alpha_{\tau(j)}}( id_{Q^*[\tau(j)]}) = \tau(j)$, and learns $\hint, \{r_{\tau(j),k,v}^\mathcal{S}\}_{j\in [m],k\in [d],v\in[l]}$. 

\item For $u\in [m_c]\setminus \{\tau(j)\}_{j\in [m]}$, $\mathcal{S}$ picks random $r_{u,k,v}^\mathcal{S}\leftarrow \mathbb{F}$, $k\in[d],v\in[l]$.

\item For $u\in[m_c]$, $\mathcal{S}$ and $\mathcal{R}$ invoke the ECS functionality $\Funcecs$. $\mathcal{R}$ acts as the sender in ECS with input $\{\veca_{u,k,v}=(0,r_{u,k}^\mathcal{R})\}_{k\in[d],v\in[l]}$ and learns nothing. $\mathcal{S}$ acts as the receiver in ECS with input $\{\vecb_{u,k,v}=(r_{u,k,v}^\mathcal{S},r_{u,k,v}^\mathcal{S})\}_{k\in[d],v\in[l]}$ and learns $\rands_u$.

\item For $u\in[m_c]$, $\mathcal{S}$ and $\mathcal{R}$ invoke the ssPEQT functionality $\Funcsspeqt$. $\mathcal{S}$ and $\mathcal{R}$ take inputs $\rands_u$ and $\randr_u$, and learns $\tilde{b}^\mathcal{S}_u$ and $\tilde{b}^\mathcal{R}_u$, respectively.

\item $\mathcal{S}$ picks a random permutation $\pi$ over $[m_c]$. Then, $\mathcal{S}$ and $\mathcal{R}$ invoke the PS functionality $\Funcps$. $\mathcal{R}$ acts as the sender in PS with input $\{\tilde{b}^\mathcal{R}_u\}_{u\in[m_c]}$ and learns $\{\bar{b}^\mathcal{R}_u\}_{u\in[m_c]}$. $\mathcal{S}$ acts as the receiver in PS with input $\pi$ and learns  $\{\bar{b}^\mathcal{S}_u\}_{u\in[m_c]}$.

\item $\mathcal{S}$ computes $b_u^\mathcal{S}:=\bar{b}_u^\mathcal{S}\oplus \tilde{b}_{\pi(u)}^\mathcal{S}$ for $u\in [m_c]$. $\mathcal{R}$ initializes set $I:=\{\}$.

\item $\mathcal{S}$ and $\mathcal{R}$ invoke a batch of $m_c$ OT instances $\FuncOT$. In $u$-th OT, $\mathcal{S}$ inputs $(\perp,Q^*[\pi(u)])$ if $b^\mathcal{S}_{u}=0$ and $(Q^*[\pi(u)],\perp)$ if $b^\mathcal{S}_{u} = 1$. $\mathcal{R}$ acts as receiver with input $\bar{b}^{\mathcal{R}}_{u}$ and receives $\mathbf{z}_{u}$. $\mathcal{R}$ sets $I :=I \cup \{\mathbf{z}_{u}\}$ if $\mathbf{z}_{u}\ne \perp$. Finally, $\mathcal{R}$ outputs $I$.

\end{enumerate}
\end{trivlist}
\end{minipage}
\end{framed}
\caption{Fuzzy PSI Protocol $\Protocolfpsi$ for $L_\infty$ in High Dimension Space from OPPRF} \label{fig:pifpsi-inf-sep-opprf}
\end{figure}

\subsection{Fuzzy PSI for $L_{p\in[1,\infty)}$ Distance}
\label{subsec:fpsi-sep-p}

In this section, we give our construction of the FPSI (and its variants) for $L_{p\in[1,\infty)}$ distance from OPPRF. Similar to protocols for $L_\infty$ distance, we only give the OPPRF-based construction here, as other trade-offs and optimizations apply similarly and are thus omitted. The formal description is given in Figure \ref{fig:pifpsi-p-sep-opprf}.

\begin{figure}[!ht]
\begin{framed}
\begin{minipage}[center]{\textwidth}
\begin{trivlist}
\item \textbf{Parameters:} 
\begin{itemize}
\item Two parties: sender $\mathcal{S}$ and receiver $\mathcal{R}$, set size $m$ and $n$, dimension $d$, threshold $\delta$.
\item Ideal $\Funcdidg$, $\Funcopprf$,$\Funcecips$, $\Funcps$, $\Funcifmat$, and $\FuncOT$ primitives specified in Figure \ref{fig:fdidg}, Figure \ref{fig:fopprf}, Figure \ref{fig:fecips}, Figure \ref{fig:fps}, Figure \ref{fig:fifmat}, and Figure \ref{fig:fot} respectively. 
\end{itemize}

\item Input of $\mathcal{S}$: $Q=\{\vecq_1,\dots,\vecq_{m}\}\subseteq \mathbb{U}^d$. 

\item Input of $\mathcal{R}$: $W=\{\vecw_1,\dots,\vecw_{n}\}\subseteq \mathbb{U}^d$.

\item \textbf{Protocol:}

\begin{enumerate}

\item $\mathcal{S}$ and $\mathcal{R}$ invoke the inconsistent distributed ID generation functionality $\Funcdidg$ with input $Q$ and $W$, respectively. As a result, the sender $\mathcal{S}$ learns $\id_{\vecq_j} = \{id_{\vecq_j}\}$ for $j\in[m]$. The receiver $\mathcal{R}$ learns $\id_{\vecw_i} = \{id_{\vecw_i}\}$ for $i\in[n]$.

\item The sender $\mathcal{S}$ computes the Cuckoo hash table $Q^*:= \mathsf{Cuckoo}_{h_1,h_2,h_3}^{m_c}(\{(id_{\vecq_j},\vecq_j)\}_{j\in [m]})$ and fills empty bins with the dummy item. Let $m_c=(1+\epsilon)m$ denote the length of the Cuckoo hash table. $\mathcal{S}$ denotes the item in $u$-th bin as $Q^*[u]$ for $u\in [m_c]$. Let $\tau: [m]\rightarrow [m_c]$ denote a random injective function such that $\tau(1),\dots, \tau(m)$ are the non-dummy item bins of $Q^*$. 

\item $\mathcal{S}$ computes the prefix $\{q_{\tau(j),k,v}^*\}_{v\in [l]}:=\mathsf{PrefixPath}(Q^*[\tau(j)]_k,\delta/2)$ for $j\in[m],k\in[d]$.

\item $\mathcal{R}$ computes the prefix $\{w_{i,0,k,v}^*\}_{v\in [l']}:=\mathsf{PrefixTrie}(w_{i,k}-\delta,w_{i,k})$ and $\{w_{i,1,k,v}^*\}_{v\in [l']}:=\mathsf{PrefixTrie}([w_{i,k}+1,w_{i,k}+\delta])$ for $j\in[m],k\in[d]$. Then, $\mathcal{R}$ picks random $r_{u,k,1}^\mathcal{R},\dots,r_{u,k,p+1}^\mathcal{R}\leftarrow \mathbb{F}$.

\item  $\mathcal{R}$ defines $\mathsf{List}:=\{(id_{\vecw_i}||\sigma||k||w_{i,\sigma,k,v}^*||\alpha,r_{h_\alpha(id_{\vecw_i}),k,1}+|w_{i,k}-\mathsf{Bound}(\sigma,w^*_{i,\sigma,k,v})|^1 ||\dots|| r_{h_\alpha(id_{\vecw_i}),k,p}+|w_{i,k}-\mathsf{Bound}(\sigma,w^*_{i,\sigma,k,v})|^p||r_{h_\alpha(id_{\vecw_i}),k,p+1})\}_{i\in [n],k\in [d],\sigma\in\{0,1\},v\in [l'],\alpha\in[3]})$. Then, $\mathcal{S}$ and $\mathcal{R}$ invoke the OPPRF functionality $\Funcopprf$. The receiver acts as the sender in OPPRF with input $\mathsf{List}$ and learns $\hint, sk_\mathcal{R}$. The sender $\mathcal{S}$ acts as the receiver in OPPRF with input $\{id_{Q^*[\tau(j)]}||\sigma||k||q^*_{\tau(j),k,v}||\alpha_{\tau(j)}\}_{j\in [m],\sigma\in\{0,1\},k\in [d],v\in[l]}$, where $\alpha_{\tau(j)}\in[3]$ is the index of hash function used to insert $id_{Q^*[\tau(j)]}$, i.e., $h_{\alpha_{\tau(j)}}(id_{Q^*[\tau(j)]}) = \tau(j)$, and learns $\hint, \{\mathbf{r}_{\tau(j),\sigma,k,v}^\mathcal{S}\}_{j\in [m],\sigma\in\{0,1\},k\in [d],v\in[l]}$. 
Parse $\mathbf{r}_{\tau(j),\sigma,k,v}^\mathcal{S}=r_{\tau(j),\sigma,k,v,1}^\mathcal{S}||\dots||r_{\tau(j),\sigma,k,v,p+1}^\mathcal{S}$.

\item $\mathcal{S}$ defines $\vecb_{\tau(j),\sigma,k,v}^0:=(1,r_{\tau(j),\sigma,k,v,1}^\mathcal{S},\dots,r_{\tau(j),\sigma,k,v,p}^\mathcal{S})\in \mathbb{F}^{p+1}, \vecb_{\tau(j),\sigma,k,v}^1:=(|Q^*[\tau(j)]_k-\mathsf{Bound}(\sigma,q^*_{\tau(j),k,v})|^p,\binom{p}{1}|Q^*[\tau(j)]_k-\mathsf{Bound}(\sigma,q^*_{\tau(j),k,v})|^{p-1},\dots,\binom{p}{p})\in \mathbb{F}^{p+1}, b_{\tau(j),\sigma,k,v}^2:= r_{\tau(j),\sigma,k,v,p+1}^\mathcal{S}$, $j\in[m],\sigma\in\{0,1\},k\in[d],v\in[l]$. 

\item For $u\in [m_c]\setminus \{\tau(j)\}_{j\in[m]}$, $\mathcal{S}$ picks random $\vecb_{u,\sigma,k,v}^0,\vecb_{u,\sigma,k,v}^1\leftarrow \mathbb{F}^{p+1}, b_{u,\sigma,k,v}^2\leftarrow \mathbb{F}, \sigma\in \{0,1\},k\in[d],v\in[l]$.

\item $\mathcal{R}$ defines $\veca_{u,\sigma,k,v}^0:=(0,-r_{u,k,1}^\mathcal{R},\dots,-r_{u,k,p}^\mathcal{R})\in \mathbb{F}^{p+1}, a_{u,\sigma,k,v}^2:= r_{u,k,p+1}^\mathcal{R}$, $u\in[m_c],\sigma\in\{0,1\},k\in[d],v\in[l]$. 

\item For $u\in[m_c]$, $\mathcal{S}$ and $\mathcal{R}$ invoke the ECIPS functionality $\Funcecips$. $\mathcal{R}$ acts as the sender in ECIPS with input $\{(\veca_{u,\sigma,k,v}^0,a_{u,\sigma,k,v}^2)\}_{\sigma\in \{0,1\},k\in[d],v\in[l]}$ and learns $\randr_u$. $\mathcal{S}$ acts as the receiver in ECIPS with input $\{(\vecb_{u,\sigma,k,v}^0,\vecb_{u,\sigma,k,v}^1,b_{u,\sigma,k,v}^2)\}_{\sigma\in\{0,1\},k\in[d],v\in[l]}$ and learns $\rands_u$.

\item For $u\in[m_c]$, $\mathcal{S}$ and $\mathcal{R}$ invoke the ssIFmat functionality $\Funcssifmat$. $\mathcal{S}$ and $\mathcal{R}$ take inputs $\rands_u$ and $\randr_u$, and learns $\tilde{b}^\mathcal{S}_u$ and $\tilde{b}^\mathcal{R}_u$, respectively.

\item $\mathcal{S}$ picks a random permutation $\pi$ over $[m_c]$. Then, $\mathcal{S}$ and $\mathcal{R}$ invoke the PS functionality $\Funcps$. $\mathcal{R}$ acts as the sender in PS with input $\{\tilde{b}^\mathcal{R}_u\}_{u\in[m_c]}$ and learns $\{\bar{b}^\mathcal{R}_u\}_{u\in[m_c]}$. $\mathcal{S}$ acts as the receiver in PS with input $\pi$ and learns  $\{\bar{b}^\mathcal{S}_u\}_{u\in[m_c]}$.

\item $\mathcal{S}$ computes $b_u^\mathcal{S}:=\bar{b}_u^\mathcal{S}\oplus \tilde{b}_{\pi(u)}^\mathcal{S}$ for $u\in [m_c]$. $\mathcal{R}$ initializes set $I:=\{\}$.

\item $\mathcal{S}$ and $\mathcal{R}$ invoke a batch of $m_c$ OT instances $\FuncOT$. In $u$-th OT, $\mathcal{S}$ inputs $(\perp,Q^*[\pi(u)])$ if $b^\mathcal{S}_{u}=0$ and $(Q^*[\pi(u)],\perp)$ if $b^\mathcal{S}_{u} = 1$. $\mathcal{R}$ acts as receiver with input $\bar{b}^{\mathcal{R}}_{u}$ and receives $\mathbf{z}_{u}$. $\mathcal{R}$ sets $I :=I \cup \{\mathbf{z}_{u}\}$ if $\mathbf{z}_{u}\ne \perp$. Finally, $\mathcal{R}$ outputs $I$.

\end{enumerate}
\end{trivlist}
\end{minipage}
\end{framed}
\caption{Fuzzy PSI Protocol $\Protocolfpsi$ for $L_{p\in[1,\infty)}$ in High Dimension Space from OPPRF} \label{fig:pifpsi-p-sep-opprf}
\end{figure}

\section{Implementation and Performance}
\label{sec:implementation-and-performance}



We conduct an experimental evaluation of our FPSI protocols and compare them against state-of-the-art FPSI schemes \cite{DBLP:conf/asiacrypt/GaoQLLW24,AC/BGMP25,DBLP:conf/ccs/DZL25}.
The work \cite{AC/BP25} is excluded from the comparison as their protocol was not implemented. Additionally, we do not include \cite{DBLP:conf/ccs/piske+25} due to concerns regarding potential issues in their implementation.
Our examination of the code provided in \cite{DBLP:conf/ccs/piske+25}\footnote{\url{https://github.com/asu-crypto/daOT-fuzzyPSI}} reveals anomalous performance trends with respect to the parameter $\delta$. Specifically, when $\delta$ is increased from $10$ to $10,000$, the communication cost remains unchanged, and the running time for large $\delta$ values is nearly identical to—or in some cases even shorter than—that for small $\delta$ values.\footnote{This is also reflected in their paper. For example, Table~2 in \cite{DBLP:conf/ccs/piske+25} reports for set size $2^{12}$ and dimension $d=10$ a runtime of $129.17$ seconds for $\delta=10$, compared to $125.02$ seconds for $\delta=30$.}
This contradicts the theoretical analysis, which indicates that the protocol’s cost should scale linearly with $\delta$, suggesting possible inaccuracies in their implementation.

We have implemented all our proposed FPSI protocols. Due to space limitations, we only compare FPSI protocols for $L_\infty$ distances in this section. We have included a comparison of FPSI protocols for $L_1$ and $L_2$ distances in Appendix \ref{app:l1l2}. For clarity, we adopt the following naming scheme: each protocol is denoted as $\Pi_{\mathsf{ID}}^{\mathsf{assumption}}$, where the superscript indicates the underlying assumption and the subscript abbreviates the main cryptographic components. Under the \textit{apart} assumption, the five concrete protocols are:

\begin{itemize}
    \item $\Pi_{\mathsf{OE}\text{-}\mathsf{1}}^{\mathsf{apart}}$: The FPSI construction in Figure \ref{fig:pifpsi-inf-apart-opprf}, it is based on OPPRF and ECS;
    \item $\Pi_{\mathsf{OE}\text{-}\mathsf{2}}^{\mathsf{apart}}$: Same as Figure \ref{fig:pifpsi-inf-apart-opprf} but use the trade-off we discussed in Section \ref{subsubsec:fpsi-apart-inf-opprf}, that is, it use PS + PEQT instead of ssPEQT + PS;
    \item $\Pi_{\mathsf{OA}}^{\mathsf{apart}}$: The FPSI construction in Figure \ref{fig:pifpsi-inf-apart-opprf-and}, it is based on OPPRF and AND;
    \item $\Pi_{\mathsf{SE}}^{\mathsf{apart}}$: The FPSI construction in Figure \ref{fig:pifpsi-inf-apart-soprf}, it is based on sOPRF and ECSS;
    \item $\Pi_{\mathsf{SA}}^{\mathsf{apart}}$: The FPSI construction in Figure \ref{fig:pifpsi-inf-apart-soprf-and}, it is based on sOPRF and AND.
\end{itemize}

The protocols under the \textit{separate} assumption follow the same naming pattern, only with the superscript changed to $\mathsf{sep}$.


\begin{table}[!htbp]
\renewcommand\arraystretch{1}
	\centering
 \resizebox{0.83\linewidth}{!}{
\begin{tabular}{|c|c|c|cccccccccc|}
\hline
 &
   &
   &
  \multicolumn{10}{c|}{Threshold $\delta$} \\ \cline{4-13} 
 &
   &
   &
  \multicolumn{2}{c|}{16} &
  \multicolumn{2}{c|}{32} &
  \multicolumn{2}{c|}{64} &
  \multicolumn{2}{c|}{128} &
  \multicolumn{2}{c|}{256} \\ \cline{4-13} 
\multirow{-3}{*}{\begin{tabular}[c]{@{}c@{}}Set Size\\ $m=n$\end{tabular}} &
  \multirow{-3}{*}{\begin{tabular}[c]{@{}c@{}}Dimension\\ $d$\end{tabular}} &
  \multirow{-3}{*}{Protocol} &
  \multicolumn{1}{c|}{Comm.} &
  \multicolumn{1}{c|}{Time} &
  \multicolumn{1}{c|}{Comm.} &
  \multicolumn{1}{c|}{Time} &
  \multicolumn{1}{c|}{Comm.} &
  \multicolumn{1}{c|}{Time} &
  \multicolumn{1}{c|}{Comm.} &
  \multicolumn{1}{c|}{Time} &
  \multicolumn{1}{c|}{Comm.} &
  Time \\ \hline
 &
   &
  \cite{AC/BGMP25} &
  \multicolumn{1}{c|}{2.882} &
  \multicolumn{1}{c|}{1.282} &
  \multicolumn{1}{c|}{3.349} &
  \multicolumn{1}{c|}{1.285} &
  \multicolumn{1}{c|}{3.521} &
  \multicolumn{1}{c|}{1.347} &
  \multicolumn{1}{c|}{4.115} &
  \multicolumn{1}{c|}{1.391} &
  \multicolumn{1}{c|}{4.286} &
  1.506 \\ \cline{3-13} 
 &
   &
  \cite{DBLP:conf/ccs/DZL25}-Low &
  \multicolumn{1}{c|}{7.930} &
  \multicolumn{1}{c|}{0.618} &
  \multicolumn{1}{c|}{9.346} &
  \multicolumn{1}{c|}{0.815} &
  \multicolumn{1}{c|}{10.72} &
  \multicolumn{1}{c|}{0.896} &
  \multicolumn{1}{c|}{11.90} &
  \multicolumn{1}{c|}{1.012} &
  \multicolumn{1}{c|}{13.46} &
  1.183 \\ \cline{3-13} 
 &
   &
  $\Pi_{\mathsf{OE}\text{-1}}^{\mathsf{apart}}$ &
  \multicolumn{1}{c|}{2.237} &
  \multicolumn{1}{c|}{0.439} &
  \multicolumn{1}{c|}{2.391} &
  \multicolumn{1}{c|}{0.439} &
  \multicolumn{1}{c|}{2.589} &
  \multicolumn{1}{c|}{0.442} &
  \multicolumn{1}{c|}{2.814} &
  \multicolumn{1}{c|}{0.464} &
  \multicolumn{1}{c|}{2.985} &
  0.499 \\ \cline{3-13} 
 &
   &
  $\Pi_{\mathsf{OE}\text{-2}}^{\mathsf{apart}}$ &
  \multicolumn{1}{c|}{2.673} &
  \multicolumn{1}{c|}{0.437} &
  \multicolumn{1}{c|}{2.832} &
  \multicolumn{1}{c|}{0.412} &
  \multicolumn{1}{c|}{3.024} &
  \multicolumn{1}{c|}{0.423} &
  \multicolumn{1}{c|}{3.224} &
  \multicolumn{1}{c|}{0.448} &
  \multicolumn{1}{c|}{3.415} &
  0.452 \\ \cline{3-13} 
 &
   &
  $\Pi_{\mathsf{OA}}^{\mathsf{apart}}$ &
  \multicolumn{1}{c|}{\cellcolor[HTML]{C7ECFF}1.894} &
  \multicolumn{1}{c|}{0.368} &
  \multicolumn{1}{c|}{\cellcolor[HTML]{C7ECFF}2.035} &
  \multicolumn{1}{c|}{0.375} &
  \multicolumn{1}{c|}{\cellcolor[HTML]{C7ECFF}2.224} &
  \multicolumn{1}{c|}{0.386} &
  \multicolumn{1}{c|}{\cellcolor[HTML]{C7ECFF}2.395} &
  \multicolumn{1}{c|}{0.395} &
  \multicolumn{1}{c|}{\cellcolor[HTML]{C7ECFF}2.558} &
  0.412 \\ \cline{3-13} 
 &
   &
  $\Pi_{\mathsf{SE}}^{\mathsf{apart}}$ &
  \multicolumn{1}{c|}{2.195} &
  \multicolumn{1}{c|}{\cellcolor[HTML]{C7ECFF}0.225} &
  \multicolumn{1}{c|}{2.391} &
  \multicolumn{1}{c|}{\cellcolor[HTML]{C7ECFF}0.234} &
  \multicolumn{1}{c|}{2.593} &
  \multicolumn{1}{c|}{\cellcolor[HTML]{C7ECFF}0.268} &
  \multicolumn{1}{c|}{2.853} &
  \multicolumn{1}{c|}{\cellcolor[HTML]{C7ECFF}0.307} &
  \multicolumn{1}{c|}{3.057} &
  \cellcolor[HTML]{C7ECFF}0.328 \\ \cline{3-13} 
 &
  \multirow{-7}{*}{2} &
  $\Pi_{\mathsf{SA}}^{\mathsf{apart}}$ &
  \multicolumn{1}{c|}{\cellcolor[HTML]{99BEFF}1.421} &
  \multicolumn{1}{c|}{\cellcolor[HTML]{99BEFF}0.185} &
  \multicolumn{1}{c|}{\cellcolor[HTML]{99BEFF}1.603} &
  \multicolumn{1}{c|}{\cellcolor[HTML]{99BEFF}0.203} &
  \multicolumn{1}{c|}{\cellcolor[HTML]{99BEFF}1.804} &
  \multicolumn{1}{c|}{\cellcolor[HTML]{99BEFF}0.238} &
  \multicolumn{1}{c|}{\cellcolor[HTML]{99BEFF}1.994} &
  \multicolumn{1}{c|}{\cellcolor[HTML]{99BEFF}0.239} &
  \multicolumn{1}{c|}{\cellcolor[HTML]{99BEFF}2.191} &
  \cellcolor[HTML]{99BEFF}0.252 \\ \cline{2-13} 
 &
   &
  \cite{AC/BGMP25} &
  \multicolumn{1}{c|}{5.573} &
  \multicolumn{1}{c|}{1.368} &
  \multicolumn{1}{c|}{9.274} &
  \multicolumn{1}{c|}{1.594} &
  \multicolumn{1}{c|}{9.550} &
  \multicolumn{1}{c|}{1.783} &
  \multicolumn{1}{c|}{15.50} &
  \multicolumn{1}{c|}{1.993} &
  \multicolumn{1}{c|}{15.75} &
  2.316 \\ \cline{3-13} 
 &
   &
  \cite{DBLP:conf/ccs/DZL25}-Low &
  \multicolumn{1}{c|}{40.94} &
  \multicolumn{1}{c|}{1.438} &
  \multicolumn{1}{c|}{{\color[HTML]{FE0000} 54.39}} &
  \multicolumn{1}{c|}{{\color[HTML]{FE0000} 1.542}} &
  \multicolumn{1}{c|}{{\color[HTML]{FE0000} 38.39}} &
  \multicolumn{1}{c|}{{\color[HTML]{FE0000} 1.486}} &
  \multicolumn{1}{c|}{48.45} &
  \multicolumn{1}{c|}{1.682} &
  \multicolumn{1}{c|}{61.85} &
  2.005 \\ \cline{3-13} 
 &
   &
  $\Pi_{\mathsf{OE}\text{-1}}^{\mathsf{apart}}$ &
  \multicolumn{1}{c|}{3.784} &
  \multicolumn{1}{c|}{0.501} &
  \multicolumn{1}{c|}{4.200} &
  \multicolumn{1}{c|}{0.508} &
  \multicolumn{1}{c|}{4.897} &
  \multicolumn{1}{c|}{0.655} &
  \multicolumn{1}{c|}{5.406} &
  \multicolumn{1}{c|}{0.638} &
  \multicolumn{1}{c|}{5.902} &
  0.708 \\ \cline{3-13} 
 &
   &
  $\Pi_{\mathsf{OE}\text{-2}}^{\mathsf{apart}}$ &
  \multicolumn{1}{c|}{4.176} &
  \multicolumn{1}{c|}{0.486} &
  \multicolumn{1}{c|}{4.666} &
  \multicolumn{1}{c|}{0.504} &
  \multicolumn{1}{c|}{5.333} &
  \multicolumn{1}{c|}{0.613} &
  \multicolumn{1}{c|}{5.875} &
  \multicolumn{1}{c|}{0.665} &
  \multicolumn{1}{c|}{6.348} &
  0.674 \\ \cline{3-13} 
 &
   &
  $\Pi_{\mathsf{OA}}^{\mathsf{apart}}$ &
  \multicolumn{1}{c|}{3.359} &
  \multicolumn{1}{c|}{0.433} &
  \multicolumn{1}{c|}{3.786} &
  \multicolumn{1}{c|}{0.449} &
  \multicolumn{1}{c|}{4.48} &
  \multicolumn{1}{c|}{0.557} &
  \multicolumn{1}{c|}{4.862} &
  \multicolumn{1}{c|}{0.591} &
  \multicolumn{1}{c|}{5.374} &
  0.607 \\ \cline{3-13} 
 &
   &
  $\Pi_{\mathsf{SE}}^{\mathsf{apart}}$ &
  \multicolumn{1}{c|}{\cellcolor[HTML]{C7ECFF}3.334} &
  \multicolumn{1}{c|}{\cellcolor[HTML]{C7ECFF}0.291} &
  \multicolumn{1}{c|}{\cellcolor[HTML]{C7ECFF}3.727} &
  \multicolumn{1}{c|}{\cellcolor[HTML]{C7ECFF}0.353} &
  \multicolumn{1}{c|}{\cellcolor[HTML]{C7ECFF}4.310} &
  \multicolumn{1}{c|}{\cellcolor[HTML]{C7ECFF}0.355} &
  \multicolumn{1}{c|}{\cellcolor[HTML]{C7ECFF}4.812} &
  \multicolumn{1}{c|}{\cellcolor[HTML]{C7ECFF}0.398} &
  \multicolumn{1}{c|}{\cellcolor[HTML]{C7ECFF}5.190} &
  \cellcolor[HTML]{C7ECFF}0.455 \\ \cline{3-13} 
 &
  \multirow{-7}{*}{3} &
  $\Pi_{\mathsf{SA}}^{\mathsf{apart}}$ &
  \multicolumn{1}{c|}{\cellcolor[HTML]{99BEFF}2.497} &
  \multicolumn{1}{c|}{\cellcolor[HTML]{99BEFF}0.257} &
  \multicolumn{1}{c|}{\cellcolor[HTML]{99BEFF}2.876} &
  \multicolumn{1}{c|}{\cellcolor[HTML]{99BEFF}0.291} &
  \multicolumn{1}{c|}{\cellcolor[HTML]{99BEFF}3.441} &
  \multicolumn{1}{c|}{\cellcolor[HTML]{99BEFF}0.308} &
  \multicolumn{1}{c|}{\cellcolor[HTML]{99BEFF}3.813} &
  \multicolumn{1}{c|}{\cellcolor[HTML]{99BEFF}0.349} &
  \multicolumn{1}{c|}{\cellcolor[HTML]{99BEFF}4.227} &
  \cellcolor[HTML]{99BEFF}0.389 \\ \cline{2-13} 
 &
   &
  \cite{AC/BGMP25} &
  \multicolumn{1}{c|}{19.02} &
  \multicolumn{1}{c|}{3.736} &
  \multicolumn{1}{c|}{52.07} &
  \multicolumn{1}{c|}{5.047} &
  \multicolumn{1}{c|}{52.34} &
  \multicolumn{1}{c|}{7.893} &
  \multicolumn{1}{c|}{121.7} &
  \multicolumn{1}{c|}{10.53} &
  \multicolumn{1}{c|}{122.1} &
  16.48 \\ \cline{3-13} 
 &
   &
  \cite{DBLP:conf/ccs/DZL25}-Low &
  \multicolumn{1}{c|}{108.2} &
  \multicolumn{1}{c|}{3.359} &
  \multicolumn{1}{c|}{{\color[HTML]{FE0000} 144.0}} &
  \multicolumn{1}{c|}{{\color[HTML]{FE0000} 3.725}} &
  \multicolumn{1}{c|}{{\color[HTML]{FE0000} 100.3}} &
  \multicolumn{1}{c|}{{\color[HTML]{FE0000} 3.074}} &
  \multicolumn{1}{c|}{127.2} &
  \multicolumn{1}{c|}{3.648} &
  \multicolumn{1}{c|}{162.9} &
  4.634 \\ \cline{3-13} 
 &
   &
  $\Pi_{\mathsf{OE}\text{-1}}^{\mathsf{apart}}$ &
  \multicolumn{1}{c|}{7.530} &
  \multicolumn{1}{c|}{0.707} &
  \multicolumn{1}{c|}{8.720} &
  \multicolumn{1}{c|}{0.737} &
  \multicolumn{1}{c|}{9.830} &
  \multicolumn{1}{c|}{0.803} &
  \multicolumn{1}{c|}{11.42} &
  \multicolumn{1}{c|}{0.899} &
  \multicolumn{1}{c|}{12.51} &
  0.981 \\ \cline{3-13} 
 &
   &
  $\Pi_{\mathsf{OE}\text{-2}}^{\mathsf{apart}}$ &
  \multicolumn{1}{c|}{7.949} &
  \multicolumn{1}{c|}{0.708} &
  \multicolumn{1}{c|}{9.11} &
  \multicolumn{1}{c|}{0.756} &
  \multicolumn{1}{c|}{10.26} &
  \multicolumn{1}{c|}{0.829} &
  \multicolumn{1}{c|}{11.86} &
  \multicolumn{1}{c|}{0.902} &
  \multicolumn{1}{c|}{13.02} &
  0.985 \\ \cline{3-13} 
 &
   &
  $\Pi_{\mathsf{OA}}^{\mathsf{apart}}$ &
  \multicolumn{1}{c|}{6.976} &
  \multicolumn{1}{c|}{0.642} &
  \multicolumn{1}{c|}{8.104} &
  \multicolumn{1}{c|}{0.687} &
  \multicolumn{1}{c|}{9.330} &
  \multicolumn{1}{c|}{0.743} &
  \multicolumn{1}{c|}{10.82} &
  \multicolumn{1}{c|}{0.844} &
  \multicolumn{1}{c|}{11.88} &
  0.916 \\ \cline{3-13} 
 &
   &
  $\Pi_{\mathsf{SE}}^{\mathsf{apart}}$ &
  \multicolumn{1}{c|}{\cellcolor[HTML]{C7ECFF}5.631} &
  \multicolumn{1}{c|}{\cellcolor[HTML]{C7ECFF}0.432} &
  \multicolumn{1}{c|}{\cellcolor[HTML]{C7ECFF}6.443} &
  \multicolumn{1}{c|}{\cellcolor[HTML]{C7ECFF}0.488} &
  \multicolumn{1}{c|}{\cellcolor[HTML]{C7ECFF}7.227} &
  \multicolumn{1}{c|}{\cellcolor[HTML]{C7ECFF}0.571} &
  \multicolumn{1}{c|}{\cellcolor[HTML]{C7ECFF}8.175} &
  \multicolumn{1}{c|}{\cellcolor[HTML]{C7ECFF}0.633} &
  \multicolumn{1}{c|}{\cellcolor[HTML]{C7ECFF}8.982} &
  \cellcolor[HTML]{C7ECFF}0.715 \\ \cline{3-13} 
\multirow{-21}{*}{$2^8$} &
  \multirow{-7}{*}{4} &
  $\Pi_{\mathsf{SA}}^{\mathsf{apart}}$ &
  \multicolumn{1}{c|}{\cellcolor[HTML]{99BEFF}4.704} &
  \multicolumn{1}{c|}{\cellcolor[HTML]{99BEFF}0.387} &
  \multicolumn{1}{c|}{\cellcolor[HTML]{99BEFF}5.557} &
  \multicolumn{1}{c|}{\cellcolor[HTML]{99BEFF}0.463} &
  \multicolumn{1}{c|}{\cellcolor[HTML]{99BEFF}6.329} &
  \multicolumn{1}{c|}{\cellcolor[HTML]{99BEFF}0.527} &
  \multicolumn{1}{c|}{\cellcolor[HTML]{99BEFF}7.110} &
  \multicolumn{1}{c|}{\cellcolor[HTML]{99BEFF}0.584} &
  \multicolumn{1}{c|}{\cellcolor[HTML]{99BEFF}7.917} &
  \cellcolor[HTML]{99BEFF}0.648 \\ \hline
 &
   &
  \cite{AC/BGMP25} &
  \multicolumn{1}{c|}{27.67} &
  \multicolumn{1}{c|}{2.252} &
  \multicolumn{1}{c|}{35.55} &
  \multicolumn{1}{c|}{2.667} &
  \multicolumn{1}{c|}{38.17} &
  \multicolumn{1}{c|}{3.122} &
  \multicolumn{1}{c|}{47.55} &
  \multicolumn{1}{c|}{4.354} &
  \multicolumn{1}{c|}{50.18} &
  6.172 \\ \cline{3-13} 
 &
   &
  \cite{DBLP:conf/ccs/DZL25}-Low &
  \multicolumn{1}{c|}{126.3} &
  \multicolumn{1}{c|}{12.35} &
  \multicolumn{1}{c|}{149.0} &
  \multicolumn{1}{c|}{14.55} &
  \multicolumn{1}{c|}{171.5} &
  \multicolumn{1}{c|}{16.86} &
  \multicolumn{1}{c|}{189.1} &
  \multicolumn{1}{c|}{19.32} &
  \multicolumn{1}{c|}{207.9} &
  21.97 \\ \cline{3-13} 
 &
   &
  $\Pi_{\mathsf{OE}\text{-1}}^{\mathsf{apart}}$ &
  \multicolumn{1}{c|}{19.50} &
  \multicolumn{1}{c|}{1.702} &
  \multicolumn{1}{c|}{22.69} &
  \multicolumn{1}{c|}{1.918} &
  \multicolumn{1}{c|}{25.86} &
  \multicolumn{1}{c|}{2.072} &
  \multicolumn{1}{c|}{29.62} &
  \multicolumn{1}{c|}{2.318} &
  \multicolumn{1}{c|}{32.89} &
  2.527 \\ \cline{3-13} 
 &
   &
  $\Pi_{\mathsf{OE}\text{-2}}^{\mathsf{apart}}$ &
  \multicolumn{1}{c|}{19.97} &
  \multicolumn{1}{c|}{1.637} &
  \multicolumn{1}{c|}{23.19} &
  \multicolumn{1}{c|}{1.837} &
  \multicolumn{1}{c|}{26.33} &
  \multicolumn{1}{c|}{\cellcolor[HTML]{C7ECFF}2.059} &
  \multicolumn{1}{c|}{30.05} &
  \multicolumn{1}{c|}{\cellcolor[HTML]{C7ECFF}2.304} &
  \multicolumn{1}{c|}{33.33} &
  \cellcolor[HTML]{C7ECFF}2.489 \\ \cline{3-13} 
 &
   &
  $\Pi_{\mathsf{OA}}^{\mathsf{apart}}$ &
  \multicolumn{1}{c|}{\cellcolor[HTML]{99BEFF}17.12} &
  \multicolumn{1}{c|}{\cellcolor[HTML]{99BEFF}1.567} &
  \multicolumn{1}{c|}{\cellcolor[HTML]{99BEFF}20.20} &
  \multicolumn{1}{c|}{\cellcolor[HTML]{99BEFF}1.786} &
  \multicolumn{1}{c|}{\cellcolor[HTML]{99BEFF}23.14} &
  \multicolumn{1}{c|}{\cellcolor[HTML]{99BEFF}1.965} &
  \multicolumn{1}{c|}{\cellcolor[HTML]{99BEFF}26.13} &
  \multicolumn{1}{c|}{\cellcolor[HTML]{99BEFF}2.169} &
  \multicolumn{1}{c|}{\cellcolor[HTML]{99BEFF}29.17} &
  \cellcolor[HTML]{99BEFF}2.383 \\ \cline{3-13} 
 &
   &
  $\Pi_{\mathsf{SE}}^{\mathsf{apart}}$ &
  \multicolumn{1}{c|}{21.49} &
  \multicolumn{1}{c|}{1.633} &
  \multicolumn{1}{c|}{24.89} &
  \multicolumn{1}{c|}{1.881} &
  \multicolumn{1}{c|}{28.29} &
  \multicolumn{1}{c|}{2.183} &
  \multicolumn{1}{c|}{32.23} &
  \multicolumn{1}{c|}{2.481} &
  \multicolumn{1}{c|}{35.68} &
  2.852 \\ \cline{3-13} 
 &
  \multirow{-7}{*}{2} &
  $\Pi_{\mathsf{SA}}^{\mathsf{apart}}$ &
  \multicolumn{1}{c|}{\cellcolor[HTML]{C7ECFF}18.67} &
  \multicolumn{1}{c|}{\cellcolor[HTML]{C7ECFF}1.572} &
  \multicolumn{1}{c|}{\cellcolor[HTML]{C7ECFF}21.91} &
  \multicolumn{1}{c|}{\cellcolor[HTML]{C7ECFF}1.808} &
  \multicolumn{1}{c|}{\cellcolor[HTML]{C7ECFF}25.10} &
  \multicolumn{1}{c|}{2.108} &
  \multicolumn{1}{c|}{\cellcolor[HTML]{C7ECFF}28.27} &
  \multicolumn{1}{c|}{2.377} &
  \multicolumn{1}{c|}{\cellcolor[HTML]{C7ECFF}31.60} &
  2.654 \\ \cline{2-13} 
 &
   &
  \cite{AC/BGMP25} &
  \multicolumn{1}{c|}{71.11} &
  \multicolumn{1}{c|}{5.155} &
  \multicolumn{1}{c|}{130.6} &
  \multicolumn{1}{c|}{7.451} &
  \multicolumn{1}{c|}{134.5} &
  \multicolumn{1}{c|}{9.959} &
  \multicolumn{1}{c|}{229.9} &
  \multicolumn{1}{c|}{13.76} &
  \multicolumn{1}{c|}{233.9} &
  18.52 \\ \cline{3-13} 
 &
   &
  \cite{DBLP:conf/ccs/DZL25}-Low &
  \multicolumn{1}{c|}{655.3} &
  \multicolumn{1}{c|}{21.22} &
  \multicolumn{1}{c|}{{\color[HTML]{FE0000} 869.9}} &
  \multicolumn{1}{c|}{{\color[HTML]{FE0000} 26.22}} &
  \multicolumn{1}{c|}{{\color[HTML]{FE0000} 614.0}} &
  \multicolumn{1}{c|}{{\color[HTML]{FE0000} 25.95}} &
  \multicolumn{1}{c|}{775.2} &
  \multicolumn{1}{c|}{29.69} &
  \multicolumn{1}{c|}{989.7} &
  34.29 \\ \cline{3-13} 
 &
   &
  $\Pi_{\mathsf{OE}\text{-1}}^{\mathsf{apart}}$ &
  \multicolumn{1}{c|}{44.79} &
  \multicolumn{1}{c|}{3.147} &
  \multicolumn{1}{c|}{52.57} &
  \multicolumn{1}{c|}{3.651} &
  \multicolumn{1}{c|}{60.35} &
  \multicolumn{1}{c|}{4.178} &
  \multicolumn{1}{c|}{69.33} &
  \multicolumn{1}{c|}{4.888} &
  \multicolumn{1}{c|}{77.42} &
  5.441 \\ \cline{3-13} 
 &
   &
  $\Pi_{\mathsf{OE}\text{-2}}^{\mathsf{apart}}$ &
  \multicolumn{1}{c|}{45.33} &
  \multicolumn{1}{c|}{\cellcolor[HTML]{C7ECFF}3.138} &
  \multicolumn{1}{c|}{53.09} &
  \multicolumn{1}{c|}{\cellcolor[HTML]{C7ECFF}3.623} &
  \multicolumn{1}{c|}{60.86} &
  \multicolumn{1}{c|}{\cellcolor[HTML]{C7ECFF}4.149} &
  \multicolumn{1}{c|}{69.84} &
  \multicolumn{1}{c|}{\cellcolor[HTML]{C7ECFF}4.817} &
  \multicolumn{1}{c|}{77.75} &
  \cellcolor[HTML]{C7ECFF}5.368 \\ \cline{3-13} 
 &
   &
  $\Pi_{\mathsf{OA}}^{\mathsf{apart}}$ &
  \multicolumn{1}{c|}{41.51} &
  \multicolumn{1}{c|}{\cellcolor[HTML]{99BEFF}3.043} &
  \multicolumn{1}{c|}{49.06} &
  \multicolumn{1}{c|}{\cellcolor[HTML]{99BEFF}3.562} &
  \multicolumn{1}{c|}{56.50} &
  \multicolumn{1}{c|}{\cellcolor[HTML]{99BEFF}4.083} &
  \multicolumn{1}{c|}{64.29} &
  \multicolumn{1}{c|}{\cellcolor[HTML]{99BEFF}4.649} &
  \multicolumn{1}{c|}{75.13} &
  \cellcolor[HTML]{99BEFF}5.196 \\ \cline{3-13} 
 &
   &
  $\Pi_{\mathsf{SE}}^{\mathsf{apart}}$ &
  \multicolumn{1}{c|}{\cellcolor[HTML]{C7ECFF}39.70} &
  \multicolumn{1}{c|}{3.316} &
  \multicolumn{1}{c|}{\cellcolor[HTML]{C7ECFF}46.25} &
  \multicolumn{1}{c|}{3.919} &
  \multicolumn{1}{c|}{\cellcolor[HTML]{C7ECFF}52.89} &
  \multicolumn{1}{c|}{4.578} &
  \multicolumn{1}{c|}{\cellcolor[HTML]{C7ECFF}60.58} &
  \multicolumn{1}{c|}{5.201} &
  \multicolumn{1}{c|}{\cellcolor[HTML]{C7ECFF}67.34} &
  5.985 \\ \cline{3-13} 
 &
  \multirow{-7}{*}{3} &
  $\Pi_{\mathsf{SA}}^{\mathsf{apart}}$ &
  \multicolumn{1}{c|}{\cellcolor[HTML]{99BEFF}35.99} &
  \multicolumn{1}{c|}{3.181} &
  \multicolumn{1}{c|}{\cellcolor[HTML]{99BEFF}42.29} &
  \multicolumn{1}{c|}{3.846} &
  \multicolumn{1}{c|}{\cellcolor[HTML]{99BEFF}48.71} &
  \multicolumn{1}{c|}{4.474} &
  \multicolumn{1}{c|}{\cellcolor[HTML]{99BEFF}55.15} &
  \multicolumn{1}{c|}{5.093} &
  \multicolumn{1}{c|}{\cellcolor[HTML]{99BEFF}61.56} &
  5.767 \\ \cline{2-13} 
 &
   &
  \cite{AC/BGMP25} &
  \multicolumn{1}{c|}{284.2} &
  \multicolumn{1}{c|}{33.59} &
  \multicolumn{1}{c|}{814.3} &
  \multicolumn{1}{c|}{61.69} &
  \multicolumn{1}{c|}{819.6} &
  \multicolumn{1}{c|}{106.5} &
  \multicolumn{1}{c|}{1932} &
  \multicolumn{1}{c|}{147.4} &
  \multicolumn{1}{c|}{1937} &
  230.8 \\ \cline{3-13} 
 &
   &
  \cite{DBLP:conf/ccs/DZL25}-Low &
  \multicolumn{1}{c|}{1732} &
  \multicolumn{1}{c|}{47.62} &
  \multicolumn{1}{c|}{{\color[HTML]{FE0000} 2304}} &
  \multicolumn{1}{c|}{{\color[HTML]{FE0000} 61.83}} &
  \multicolumn{1}{c|}{{\color[HTML]{FE0000} 1602}} &
  \multicolumn{1}{c|}{{\color[HTML]{FE0000} 52.04}} &
  \multicolumn{1}{c|}{2033} &
  \multicolumn{1}{c|}{62.05} &
  \multicolumn{1}{c|}{2604} &
  75.78 \\ \cline{3-13} 
 &
   &
  $\Pi_{\mathsf{OE}\text{-1}}^{\mathsf{apart}}$ &
  \multicolumn{1}{c|}{102.2} &
  \multicolumn{1}{c|}{6.814} &
  \multicolumn{1}{c|}{121.5} &
  \multicolumn{1}{c|}{8.081} &
  \multicolumn{1}{c|}{140.1} &
  \multicolumn{1}{c|}{9.411} &
  \multicolumn{1}{c|}{160.6} &
  \multicolumn{1}{c|}{10.79} &
  \multicolumn{1}{c|}{179.9} &
  12.35 \\ \cline{3-13} 
 &
   &
  $\Pi_{\mathsf{OE}\text{-2}}^{\mathsf{apart}}$ &
  \multicolumn{1}{c|}{103.0} &
  \multicolumn{1}{c|}{\cellcolor[HTML]{C7ECFF}6.767} &
  \multicolumn{1}{c|}{121.7} &
  \multicolumn{1}{c|}{\cellcolor[HTML]{C7ECFF}7.973} &
  \multicolumn{1}{c|}{140.7} &
  \multicolumn{1}{c|}{\cellcolor[HTML]{C7ECFF}9.374} &
  \multicolumn{1}{c|}{161.3} &
  \multicolumn{1}{c|}{\cellcolor[HTML]{C7ECFF}10.73} &
  \multicolumn{1}{c|}{180.4} &
  \cellcolor[HTML]{C7ECFF}12.19 \\ \cline{3-13} 
 &
   &
  $\Pi_{\mathsf{OA}}^{\mathsf{apart}}$ &
  \multicolumn{1}{c|}{97.95} &
  \multicolumn{1}{c|}{\cellcolor[HTML]{99BEFF}6.643} &
  \multicolumn{1}{c|}{116.7} &
  \multicolumn{1}{c|}{\cellcolor[HTML]{99BEFF}7.971} &
  \multicolumn{1}{c|}{135.1} &
  \multicolumn{1}{c|}{\cellcolor[HTML]{99BEFF}9.279} &
  \multicolumn{1}{c|}{154.3} &
  \multicolumn{1}{c|}{\cellcolor[HTML]{99BEFF}10.68} &
  \multicolumn{1}{c|}{172.8} &
  \cellcolor[HTML]{99BEFF}12.01 \\ \cline{3-13} 
 &
   &
  $\Pi_{\mathsf{SE}}^{\mathsf{apart}}$ &
  \multicolumn{1}{c|}{\cellcolor[HTML]{C7ECFF}74.07} &
  \multicolumn{1}{c|}{7.063} &
  \multicolumn{1}{c|}{\cellcolor[HTML]{C7ECFF}87.29} &
  \multicolumn{1}{c|}{8.518} &
  \multicolumn{1}{c|}{\cellcolor[HTML]{C7ECFF}100.3} &
  \multicolumn{1}{c|}{9.945} &
  \multicolumn{1}{c|}{\cellcolor[HTML]{C7ECFF}114.7} &
  \multicolumn{1}{c|}{11.51} &
  \multicolumn{1}{c|}{\cellcolor[HTML]{C7ECFF}127.8} &
  13.01 \\ \cline{3-13} 
\multirow{-21}{*}{$2^{12}$} &
  \multirow{-7}{*}{4} &
  $\Pi_{\mathsf{SA}}^{\mathsf{apart}}$ &
  \multicolumn{1}{c|}{\cellcolor[HTML]{99BEFF}69.30} &
  \multicolumn{1}{c|}{6.936} &
  \multicolumn{1}{c|}{\cellcolor[HTML]{99BEFF}82.28} &
  \multicolumn{1}{c|}{8.409} &
  \multicolumn{1}{c|}{\cellcolor[HTML]{99BEFF}94.78} &
  \multicolumn{1}{c|}{9.842} &
  \multicolumn{1}{c|}{\cellcolor[HTML]{99BEFF}107.9} &
  \multicolumn{1}{c|}{11.62} &
  \multicolumn{1}{c|}{\cellcolor[HTML]{99BEFF}120.5} &
  12.83 \\ \hline
 &
   &
  \cite{AC/BGMP25} &
  \multicolumn{1}{c|}{416.0} &
  \multicolumn{1}{c|}{\cellcolor[HTML]{99BEFF}18.86} &
  \multicolumn{1}{c|}{542.1} &
  \multicolumn{1}{c|}{\cellcolor[HTML]{99BEFF}24.34} &
  \multicolumn{1}{c|}{584.1} &
  \multicolumn{1}{c|}{33.64} &
  \multicolumn{1}{c|}{734.2} &
  \multicolumn{1}{c|}{56.36} &
  \multicolumn{1}{c|}{-} &
  - \\ \cline{3-13} 
 &
   &
  \cite{DBLP:conf/ccs/DZL25}-Low &
  \multicolumn{1}{c|}{1946} &
  \multicolumn{1}{c|}{197.5} &
  \multicolumn{1}{c|}{2262} &
  \multicolumn{1}{c|}{238.7} &
  \multicolumn{1}{c|}{2614} &
  \multicolumn{1}{c|}{272.3} &
  \multicolumn{1}{c|}{2953} &
  \multicolumn{1}{c|}{312.9} &
  \multicolumn{1}{c|}{3295} &
  349.9 \\ \cline{3-13} 
 &
   &
  $\Pi_{\mathsf{OE}\text{-1}}^{\mathsf{apart}}$ &
  \multicolumn{1}{c|}{\cellcolor[HTML]{C7ECFF}297.7} &
  \multicolumn{1}{c|}{24.28} &
  \multicolumn{1}{c|}{\cellcolor[HTML]{C7ECFF}348.3} &
  \multicolumn{1}{c|}{27.75} &
  \multicolumn{1}{c|}{\cellcolor[HTML]{C7ECFF}398.8} &
  \multicolumn{1}{c|}{32.39} &
  \multicolumn{1}{c|}{460.0} &
  \multicolumn{1}{c|}{37.23} &
  \multicolumn{1}{c|}{510.8} &
  41.79 \\ \cline{3-13} 
 &
   &
  $\Pi_{\mathsf{OE}\text{-2}}^{\mathsf{apart}}$ &
  \multicolumn{1}{c|}{298.3} &
  \multicolumn{1}{c|}{23.18} &
  \multicolumn{1}{c|}{348.9} &
  \multicolumn{1}{c|}{26.84} &
  \multicolumn{1}{c|}{399.4} &
  \multicolumn{1}{c|}{\cellcolor[HTML]{C7ECFF}31.54} &
  \multicolumn{1}{c|}{460.8} &
  \multicolumn{1}{c|}{\cellcolor[HTML]{C7ECFF}36.35} &
  \multicolumn{1}{c|}{511.6} &
  \cellcolor[HTML]{99BEFF}40.09 \\ \cline{3-13} 
 &
   &
  $\Pi_{\mathsf{OA}}^{\mathsf{apart}}$ &
  \multicolumn{1}{c|}{\cellcolor[HTML]{99BEFF}262.0} &
  \multicolumn{1}{c|}{\cellcolor[HTML]{C7ECFF}22.79} &
  \multicolumn{1}{c|}{\cellcolor[HTML]{99BEFF}310.1} &
  \multicolumn{1}{c|}{\cellcolor[HTML]{C7ECFF}26.49} &
  \multicolumn{1}{c|}{\cellcolor[HTML]{99BEFF}357.8} &
  \multicolumn{1}{c|}{\cellcolor[HTML]{99BEFF}31.06} &
  \multicolumn{1}{c|}{\cellcolor[HTML]{99BEFF}406.4} &
  \multicolumn{1}{c|}{\cellcolor[HTML]{99BEFF}35.45} &
  \multicolumn{1}{c|}{\cellcolor[HTML]{99BEFF}454.3} &
  \cellcolor[HTML]{C7ECFF}41.55 \\ \cline{3-13} 
 &
   &
  $\Pi_{\mathsf{SE}}^{\mathsf{apart}}$ &
  \multicolumn{1}{c|}{334.4} &
  \multicolumn{1}{c|}{29.32} &
  \multicolumn{1}{c|}{388.8} &
  \multicolumn{1}{c|}{34.18} &
  \multicolumn{1}{c|}{443.1} &
  \multicolumn{1}{c|}{39.78} &
  \multicolumn{1}{c|}{508.3} &
  \multicolumn{1}{c|}{46.29} &
  \multicolumn{1}{c|}{562.7} &
  51.22 \\ \cline{3-13} 
 &
  \multirow{-7}{*}{2} &
  $\Pi_{\mathsf{SA}}^{\mathsf{apart}}$ &
  \multicolumn{1}{c|}{299.2} &
  \multicolumn{1}{c|}{29.15} &
  \multicolumn{1}{c|}{351.0} &
  \multicolumn{1}{c|}{33.61} &
  \multicolumn{1}{c|}{403.0} &
  \multicolumn{1}{c|}{39.06} &
  \multicolumn{1}{c|}{\cellcolor[HTML]{C7ECFF}455.0} &
  \multicolumn{1}{c|}{45.42} &
  \multicolumn{1}{c|}{\cellcolor[HTML]{C7ECFF}506.8} &
  51.22 \\ \cline{2-13} 
 &
   &
  \cite{AC/BGMP25} &
  \multicolumn{1}{c|}{1111} &
  \multicolumn{1}{c|}{78.08} &
  \multicolumn{1}{c|}{2062} &
  \multicolumn{1}{c|}{113.9} &
  \multicolumn{1}{c|}{2125} &
  \multicolumn{1}{c|}{162.5} &
  \multicolumn{1}{c|}{-} &
  \multicolumn{1}{c|}{-} &
  \multicolumn{1}{c|}{-} &
  - \\ \cline{3-13} 
 &
   &
  \cite{DBLP:conf/ccs/DZL25}-Low &
  \multicolumn{1}{c|}{-} &
  \multicolumn{1}{c|}{-} &
  \multicolumn{1}{c|}{-} &
  \multicolumn{1}{c|}{-} &
  \multicolumn{1}{c|}{-} &
  \multicolumn{1}{c|}{-} &
  \multicolumn{1}{c|}{-} &
  \multicolumn{1}{c|}{-} &
  \multicolumn{1}{c|}{-} &
  - \\ \cline{3-13} 
 &
   &
  $\Pi_{\mathsf{OE}\text{-1}}^{\mathsf{apart}}$ &
  \multicolumn{1}{c|}{704.0} &
  \multicolumn{1}{c|}{57.36} &
  \multicolumn{1}{c|}{828.5} &
  \multicolumn{1}{c|}{69.94} &
  \multicolumn{1}{c|}{953.4} &
  \multicolumn{1}{c|}{79.64} &
  \multicolumn{1}{c|}{1098} &
  \multicolumn{1}{c|}{91.64} &
  \multicolumn{1}{c|}{1226} &
  103.1 \\ \cline{3-13} 
 &
   &
  $\Pi_{\mathsf{OE}\text{-2}}^{\mathsf{apart}}$ &
  \multicolumn{1}{c|}{704.7} &
  \multicolumn{1}{c|}{\cellcolor[HTML]{C7ECFF}56.49} &
  \multicolumn{1}{c|}{828.8} &
  \multicolumn{1}{c|}{\cellcolor[HTML]{C7ECFF}69.28} &
  \multicolumn{1}{c|}{953.5} &
  \multicolumn{1}{c|}{\cellcolor[HTML]{C7ECFF}78.97} &
  \multicolumn{1}{c|}{1098} &
  \multicolumn{1}{c|}{\cellcolor[HTML]{C7ECFF}90.69} &
  \multicolumn{1}{c|}{1226} &
  \cellcolor[HTML]{C7ECFF}102.6 \\ \cline{3-13} 
 &
   &
  $\Pi_{\mathsf{OA}}^{\mathsf{apart}}$ &
  \multicolumn{1}{c|}{652.9} &
  \multicolumn{1}{c|}{\cellcolor[HTML]{99BEFF}55.88} &
  \multicolumn{1}{c|}{773.1} &
  \multicolumn{1}{c|}{\cellcolor[HTML]{99BEFF}68.55} &
  \multicolumn{1}{c|}{894.2} &
  \multicolumn{1}{c|}{\cellcolor[HTML]{99BEFF}78.12} &
  \multicolumn{1}{c|}{1019} &
  \multicolumn{1}{c|}{\cellcolor[HTML]{99BEFF}89.72} &
  \multicolumn{1}{c|}{1144} &
  \cellcolor[HTML]{99BEFF}101.8 \\ \cline{3-13} 
 &
   &
  $\Pi_{\mathsf{SE}}^{\mathsf{apart}}$ &
  \multicolumn{1}{c|}{\cellcolor[HTML]{C7ECFF}629.1} &
  \multicolumn{1}{c|}{65.24} &
  \multicolumn{1}{c|}{\cellcolor[HTML]{C7ECFF}734.6} &
  \multicolumn{1}{c|}{80.08} &
  \multicolumn{1}{c|}{\cellcolor[HTML]{C7ECFF}840.8} &
  \multicolumn{1}{c|}{90.52} &
  \multicolumn{1}{c|}{\cellcolor[HTML]{C7ECFF}965.1} &
  \multicolumn{1}{c|}{104.2} &
  \multicolumn{1}{c|}{\cellcolor[HTML]{C7ECFF}1073} &
  119.2 \\ \cline{3-13} 
 &
  \multirow{-7}{*}{3} &
  $\Pi_{\mathsf{SA}}^{\mathsf{apart}}$ &
  \multicolumn{1}{c|}{\cellcolor[HTML]{99BEFF}578.1} &
  \multicolumn{1}{c|}{64.73} &
  \multicolumn{1}{c|}{\cellcolor[HTML]{99BEFF}680.3} &
  \multicolumn{1}{c|}{79.69} &
  \multicolumn{1}{c|}{\cellcolor[HTML]{99BEFF}782.6} &
  \multicolumn{1}{c|}{90.69} &
  \multicolumn{1}{c|}{\cellcolor[HTML]{99BEFF}887.1} &
  \multicolumn{1}{c|}{104.3} &
  \multicolumn{1}{c|}{\cellcolor[HTML]{99BEFF}991.1} &
  116.7 \\ \cline{2-13} 
 &
   &
  \cite{AC/BGMP25} &
  \multicolumn{1}{c|}{4517} &
  \multicolumn{1}{c|}{520.7} &
  \multicolumn{1}{c|}{-} &
  \multicolumn{1}{c|}{-} &
  \multicolumn{1}{c|}{-} &
  \multicolumn{1}{c|}{-} &
  \multicolumn{1}{c|}{-} &
  \multicolumn{1}{c|}{-} &
  \multicolumn{1}{c|}{-} &
  - \\ \cline{3-13} 
 &
   &
  \cite{DBLP:conf/ccs/DZL25}-Low &
  \multicolumn{1}{c|}{-} &
  \multicolumn{1}{c|}{-} &
  \multicolumn{1}{c|}{-} &
  \multicolumn{1}{c|}{-} &
  \multicolumn{1}{c|}{-} &
  \multicolumn{1}{c|}{-} &
  \multicolumn{1}{c|}{-} &
  \multicolumn{1}{c|}{-} &
  \multicolumn{1}{c|}{-} &
  - \\ \cline{3-13} 
 &
   &
  $\Pi_{\mathsf{OE}\text{-1}}^{\mathsf{apart}}$ &
  \multicolumn{1}{c|}{1628} &
  \multicolumn{1}{c|}{140.8} &
  \multicolumn{1}{c|}{1930} &
  \multicolumn{1}{c|}{169.8} &
  \multicolumn{1}{c|}{2237} &
  \multicolumn{1}{c|}{200.5} &
  \multicolumn{1}{c|}{2563} &
  \multicolumn{1}{c|}{234.4} &
  \multicolumn{1}{c|}{2867} &
  267.4 \\ \cline{3-13} 
 &
   &
  $\Pi_{\mathsf{OE}\text{-2}}^{\mathsf{apart}}$ &
  \multicolumn{1}{c|}{1629} &
  \multicolumn{1}{c|}{\cellcolor[HTML]{99BEFF}138.9} &
  \multicolumn{1}{c|}{1931} &
  \multicolumn{1}{c|}{\cellcolor[HTML]{C7ECFF}167.9} &
  \multicolumn{1}{c|}{2235} &
  \multicolumn{1}{c|}{\cellcolor[HTML]{99BEFF}199.3} &
  \multicolumn{1}{c|}{2563} &
  \multicolumn{1}{c|}{\cellcolor[HTML]{C7ECFF}233.5} &
  \multicolumn{1}{c|}{2870} &
  \cellcolor[HTML]{99BEFF}265.5 \\ \cline{3-13} 
 &
   &
  $\Pi_{\mathsf{OA}}^{\mathsf{apart}}$ &
  \multicolumn{1}{c|}{1560} &
  \multicolumn{1}{c|}{\cellcolor[HTML]{C7ECFF}139.5} &
  \multicolumn{1}{c|}{1858} &
  \multicolumn{1}{c|}{\cellcolor[HTML]{99BEFF}167.5} &
  \multicolumn{1}{c|}{2158} &
  \multicolumn{1}{c|}{\cellcolor[HTML]{C7ECFF}200.1} &
  \multicolumn{1}{c|}{2460} &
  \multicolumn{1}{c|}{\cellcolor[HTML]{99BEFF}231.1} &
  \multicolumn{1}{c|}{2760} &
  \cellcolor[HTML]{C7ECFF}265.9 \\ \cline{3-13} 
 &
   &
  $\Pi_{\mathsf{SE}}^{\mathsf{apart}}$ &
  \multicolumn{1}{c|}{\cellcolor[HTML]{C7ECFF}1182} &
  \multicolumn{1}{c|}{151.4} &
  \multicolumn{1}{c|}{\cellcolor[HTML]{C7ECFF}1391} &
  \multicolumn{1}{c|}{183.0} &
  \multicolumn{1}{c|}{\cellcolor[HTML]{C7ECFF}1602} &
  \multicolumn{1}{c|}{213.3} &
  \multicolumn{1}{c|}{\cellcolor[HTML]{C7ECFF}1835} &
  \multicolumn{1}{c|}{250.9} &
  \multicolumn{1}{c|}{\cellcolor[HTML]{C7ECFF}2045} &
  282.9 \\ \cline{3-13} 
\multirow{-21}{*}{$2^{16}$} &
  \multirow{-7}{*}{4} &
  $\Pi_{\mathsf{SA}}^{\mathsf{apart}}$ &
  \multicolumn{1}{c|}{\cellcolor[HTML]{99BEFF}1116} &
  \multicolumn{1}{c|}{150.6} &
  \multicolumn{1}{c|}{\cellcolor[HTML]{99BEFF}1320} &
  \multicolumn{1}{c|}{181.7} &
  \multicolumn{1}{c|}{\cellcolor[HTML]{99BEFF}1526} &
  \multicolumn{1}{c|}{213.7} &
  \multicolumn{1}{c|}{\cellcolor[HTML]{99BEFF}1732} &
  \multicolumn{1}{c|}{248.6} &
  \multicolumn{1}{c|}{\cellcolor[HTML]{99BEFF}1938} &
  284.2 \\ \hline
\end{tabular}
}
\caption{Communication cost (in MB) and running time (in seconds) comparing our protocols to \cite{AC/BGMP25,DBLP:conf/ccs/DZL25}. Cells with - denote trials that ran out of memory. The best result is highlighted in \textcolor[rgb]{0,0.3,1}{blue}, the second best in \textcolor[rgb]{0,0.7,1}{cyan}, and data in \textcolor[rgb]{1,0,0}{red} font indicates abnormal values.}
\label{tab:infty-low-lan}
\end{table}

\begin{table}[!htbp]
\renewcommand\arraystretch{1}
	\centering
 \resizebox{0.83\linewidth}{!}{
\begin{tabular}{|c|c|c|cccccccccc|}
\hline
 &
   &
   &
  \multicolumn{10}{c|}{Threshold $\delta$} \\ \cline{4-13} 
 &
   &
   &
  \multicolumn{2}{c|}{16} &
  \multicolumn{2}{c|}{32} &
  \multicolumn{2}{c|}{64} &
  \multicolumn{2}{c|}{128} &
  \multicolumn{2}{c|}{256} \\ \cline{4-13} 
\multirow{-3}{*}{\begin{tabular}[c]{@{}c@{}}Set Size\\ $m=n$\end{tabular}} &
  \multirow{-3}{*}{\begin{tabular}[c]{@{}c@{}}Dimension\\ $d$\end{tabular}} &
  \multirow{-3}{*}{Protocol} &
  \multicolumn{1}{c|}{Comm.} &
  \multicolumn{1}{c|}{Time} &
  \multicolumn{1}{c|}{Comm.} &
  \multicolumn{1}{c|}{Time} &
  \multicolumn{1}{c|}{Comm.} &
  \multicolumn{1}{c|}{Time} &
  \multicolumn{1}{c|}{Comm.} &
  \multicolumn{1}{c|}{Time} &
  \multicolumn{1}{c|}{Comm.} &
  Time \\ \hline
 &
   &
  \cite{DBLP:conf/asiacrypt/GaoQLLW24} &
  \multicolumn{1}{c|}{34.67} &
  \multicolumn{1}{c|}{5.615} &
  \multicolumn{1}{c|}{68.09} &
  \multicolumn{1}{c|}{10.29} &
  \multicolumn{1}{c|}{134.9} &
  \multicolumn{1}{c|}{19.39} &
  \multicolumn{1}{c|}{268.6} &
  \multicolumn{1}{c|}{38.43} &
  \multicolumn{1}{c|}{535.9} &
  75.63 \\ \cline{3-13} 
 &
   &
  \cite{DBLP:conf/ccs/DZL25}-High &
  \multicolumn{1}{c|}{65.14} &
  \multicolumn{1}{c|}{7.138} &
  \multicolumn{1}{c|}{70.16} &
  \multicolumn{1}{c|}{7.274} &
  \multicolumn{1}{c|}{75.26} &
  \multicolumn{1}{c|}{7.885} &
  \multicolumn{1}{c|}{79.40} &
  \multicolumn{1}{c|}{7.992} &
  \multicolumn{1}{c|}{121.0} &
  13.65 \\ \cline{3-13} 
 &
   &
  $\Pi_{\mathsf{OE}\text{-1}}^{\mathsf{sep}}$ &
  \multicolumn{1}{c|}{11.71} &
  \multicolumn{1}{c|}{1.519} &
  \multicolumn{1}{c|}{13.52} &
  \multicolumn{1}{c|}{1.728} &
  \multicolumn{1}{c|}{\cellcolor[HTML]{C7ECFF}15.09} &
  \multicolumn{1}{c|}{1.806} &
  \multicolumn{1}{c|}{\cellcolor[HTML]{C7ECFF}16.83} &
  \multicolumn{1}{c|}{1.948} &
  \multicolumn{1}{c|}{\cellcolor[HTML]{C7ECFF}18.39} &
  2.134 \\ \cline{3-13} 
 &
   &
  $\Pi_{\mathsf{OE}\text{-2}}^{\mathsf{sep}}$ &
  \multicolumn{1}{c|}{12.13} &
  \multicolumn{1}{c|}{1.497} &
  \multicolumn{1}{c|}{13.98} &
  \multicolumn{1}{c|}{1.673} &
  \multicolumn{1}{c|}{15.55} &
  \multicolumn{1}{c|}{1.787} &
  \multicolumn{1}{c|}{17.29} &
  \multicolumn{1}{c|}{1.979} &
  \multicolumn{1}{c|}{18.85} &
  2.147 \\ \cline{3-13} 
 &
   &
  $\Pi_{\mathsf{OA}}^{\mathsf{sep}}$ &
  \multicolumn{1}{c|}{\cellcolor[HTML]{99BEFF}11.10} &
  \multicolumn{1}{c|}{1.481} &
  \multicolumn{1}{c|}{\cellcolor[HTML]{99BEFF}12.89} &
  \multicolumn{1}{c|}{1.606} &
  \multicolumn{1}{c|}{\cellcolor[HTML]{99BEFF}14.43} &
  \multicolumn{1}{c|}{1.726} &
  \multicolumn{1}{c|}{\cellcolor[HTML]{99BEFF}15.94} &
  \multicolumn{1}{c|}{1.896} &
  \multicolumn{1}{c|}{\cellcolor[HTML]{99BEFF}17.45} &
  2.028 \\ \cline{3-13} 
 &
   &
  $\Pi_{\mathsf{SE}}^{\mathsf{sep}}$ &
  \multicolumn{1}{c|}{12.46} &
  \multicolumn{1}{c|}{\cellcolor[HTML]{C7ECFF}1.254} &
  \multicolumn{1}{c|}{14.53} &
  \multicolumn{1}{c|}{\cellcolor[HTML]{C7ECFF}1.374} &
  \multicolumn{1}{c|}{16.46} &
  \multicolumn{1}{c|}{\cellcolor[HTML]{C7ECFF}1.574} &
  \multicolumn{1}{c|}{18.46} &
  \multicolumn{1}{c|}{\cellcolor[HTML]{C7ECFF}1.756} &
  \multicolumn{1}{c|}{20.29} &
  \cellcolor[HTML]{C7ECFF}1.872 \\ \cline{3-13} 
 &
  \multirow{-7}{*}{6} &
  $\Pi_{\mathsf{SA}}^{\mathsf{sep}}$ &
  \multicolumn{1}{c|}{\cellcolor[HTML]{C7ECFF}11.41} &
  \multicolumn{1}{c|}{\cellcolor[HTML]{99BEFF}1.233} &
  \multicolumn{1}{c|}{\cellcolor[HTML]{C7ECFF}13.44} &
  \multicolumn{1}{c|}{\cellcolor[HTML]{99BEFF}1.344} &
  \multicolumn{1}{c|}{15.35} &
  \multicolumn{1}{c|}{\cellcolor[HTML]{99BEFF}1.519} &
  \multicolumn{1}{c|}{17.16} &
  \multicolumn{1}{c|}{\cellcolor[HTML]{99BEFF}1.671} &
  \multicolumn{1}{c|}{18.94} &
  \cellcolor[HTML]{99BEFF}1.807 \\ \cline{2-13} 
 &
   &
  \cite{DBLP:conf/asiacrypt/GaoQLLW24} &
  \multicolumn{1}{c|}{46.16} &
  \multicolumn{1}{c|}{7.593} &
  \multicolumn{1}{c|}{90.71} &
  \multicolumn{1}{c|}{13.75} &
  \multicolumn{1}{c|}{179.8} &
  \multicolumn{1}{c|}{26.52} &
  \multicolumn{1}{c|}{358.0} &
  \multicolumn{1}{c|}{51.05} &
  \multicolumn{1}{c|}{714.4} &
  105.2 \\ \cline{3-13} 
 &
   &
  \cite{DBLP:conf/ccs/DZL25}-High &
  \multicolumn{1}{c|}{86.86} &
  \multicolumn{1}{c|}{9.531} &
  \multicolumn{1}{c|}{93.55} &
  \multicolumn{1}{c|}{9.612} &
  \multicolumn{1}{c|}{100.3} &
  \multicolumn{1}{c|}{10.18} &
  \multicolumn{1}{c|}{105.9} &
  \multicolumn{1}{c|}{10.59} &
  \multicolumn{1}{c|}{161.3} &
  18.32 \\ \cline{3-13} 
 &
   &
  $\Pi_{\mathsf{OE}\text{-1}}^{\mathsf{sep}}$ &
  \multicolumn{1}{c|}{\cellcolor[HTML]{C7ECFF}15.07} &
  \multicolumn{1}{c|}{1.831} &
  \multicolumn{1}{c|}{\cellcolor[HTML]{C7ECFF}17.11} &
  \multicolumn{1}{c|}{2.042} &
  \multicolumn{1}{c|}{\cellcolor[HTML]{C7ECFF}19.15} &
  \multicolumn{1}{c|}{2.224} &
  \multicolumn{1}{c|}{\cellcolor[HTML]{C7ECFF}21.76} &
  \multicolumn{1}{c|}{2.387} &
  \multicolumn{1}{c|}{\cellcolor[HTML]{C7ECFF}23.96} &
  2.657 \\ \cline{3-13} 
 &
   &
  $\Pi_{\mathsf{OE}\text{-2}}^{\mathsf{sep}}$ &
  \multicolumn{1}{c|}{15.51} &
  \multicolumn{1}{c|}{1.807} &
  \multicolumn{1}{c|}{17.51} &
  \multicolumn{1}{c|}{2.027} &
  \multicolumn{1}{c|}{19.62} &
  \multicolumn{1}{c|}{2.212} &
  \multicolumn{1}{c|}{22.18} &
  \multicolumn{1}{c|}{2.437} &
  \multicolumn{1}{c|}{24.40} &
  2.679 \\ \cline{3-13} 
 &
   &
  $\Pi_{\mathsf{OA}}^{\mathsf{sep}}$ &
  \multicolumn{1}{c|}{\cellcolor[HTML]{99BEFF}14.32} &
  \multicolumn{1}{c|}{1.786} &
  \multicolumn{1}{c|}{\cellcolor[HTML]{99BEFF}16.32} &
  \multicolumn{1}{c|}{1.993} &
  \multicolumn{1}{c|}{\cellcolor[HTML]{99BEFF}18.33} &
  \multicolumn{1}{c|}{2.183} &
  \multicolumn{1}{c|}{\cellcolor[HTML]{99BEFF}20.68} &
  \multicolumn{1}{c|}{2.359} &
  \multicolumn{1}{c|}{\cellcolor[HTML]{99BEFF}22.86} &
  2.603 \\ \cline{3-13} 
 &
   &
  $\Pi_{\mathsf{SE}}^{\mathsf{sep}}$ &
  \multicolumn{1}{c|}{16.41} &
  \multicolumn{1}{c|}{\cellcolor[HTML]{C7ECFF}1.589} &
  \multicolumn{1}{c|}{18.80} &
  \multicolumn{1}{c|}{\cellcolor[HTML]{99BEFF}1.748} &
  \multicolumn{1}{c|}{21.22} &
  \multicolumn{1}{c|}{\cellcolor[HTML]{C7ECFF}1.997} &
  \multicolumn{1}{c|}{23.99} &
  \multicolumn{1}{c|}{\cellcolor[HTML]{C7ECFF}2.219} &
  \multicolumn{1}{c|}{26.61} &
  \cellcolor[HTML]{C7ECFF}2.548 \\ \cline{3-13} 
 &
  \multirow{-7}{*}{8} &
  $\Pi_{\mathsf{SA}}^{\mathsf{sep}}$ &
  \multicolumn{1}{c|}{15.22} &
  \multicolumn{1}{c|}{\cellcolor[HTML]{99BEFF}1.514} &
  \multicolumn{1}{c|}{17.58} &
  \multicolumn{1}{c|}{\cellcolor[HTML]{C7ECFF}1.749} &
  \multicolumn{1}{c|}{19.96} &
  \multicolumn{1}{c|}{\cellcolor[HTML]{99BEFF}1.973} &
  \multicolumn{1}{c|}{22.56} &
  \multicolumn{1}{c|}{\cellcolor[HTML]{99BEFF}2.198} &
  \multicolumn{1}{c|}{25.10} &
  \cellcolor[HTML]{99BEFF}2.507 \\ \cline{2-13} 
 &
   &
  \cite{DBLP:conf/asiacrypt/GaoQLLW24} &
  \multicolumn{1}{c|}{57.65} &
  \multicolumn{1}{c|}{9.496} &
  \multicolumn{1}{c|}{113.3} &
  \multicolumn{1}{c|}{17.12} &
  \multicolumn{1}{c|}{224.7} &
  \multicolumn{1}{c|}{32.93} &
  \multicolumn{1}{c|}{447.5} &
  \multicolumn{1}{c|}{64.26} &
  \multicolumn{1}{c|}{893.0} &
  130.4 \\ \cline{3-13} 
 &
   &
  \cite{DBLP:conf/ccs/DZL25}-High &
  \multicolumn{1}{c|}{108.5} &
  \multicolumn{1}{c|}{11.79} &
  \multicolumn{1}{c|}{116.9} &
  \multicolumn{1}{c|}{11.85} &
  \multicolumn{1}{c|}{125.3} &
  \multicolumn{1}{c|}{13.12} &
  \multicolumn{1}{c|}{132.3} &
  \multicolumn{1}{c|}{13.16} &
  \multicolumn{1}{c|}{201.6} &
  22.73 \\ \cline{3-13} 
 &
   &
  $\Pi_{\mathsf{OE}\text{-1}}^{\mathsf{sep}}$ &
  \multicolumn{1}{c|}{\cellcolor[HTML]{C7ECFF}17.99} &
  \multicolumn{1}{c|}{2.157} &
  \multicolumn{1}{c|}{\cellcolor[HTML]{C7ECFF}20.70} &
  \multicolumn{1}{c|}{2.448} &
  \multicolumn{1}{c|}{\cellcolor[HTML]{C7ECFF}23.63} &
  \multicolumn{1}{c|}{2.691} &
  \multicolumn{1}{c|}{\cellcolor[HTML]{C7ECFF}26.76} &
  \multicolumn{1}{c|}{2.974} &
  \multicolumn{1}{c|}{\cellcolor[HTML]{C7ECFF}30.01} &
  3.185 \\ \cline{3-13} 
 &
   &
  $\Pi_{\mathsf{OE}\text{-2}}^{\mathsf{sep}}$ &
  \multicolumn{1}{c|}{18.45} &
  \multicolumn{1}{c|}{2.146} &
  \multicolumn{1}{c|}{21.12} &
  \multicolumn{1}{c|}{2.399} &
  \multicolumn{1}{c|}{24.09} &
  \multicolumn{1}{c|}{2.653} &
  \multicolumn{1}{c|}{27.13} &
  \multicolumn{1}{c|}{2.947} &
  \multicolumn{1}{c|}{30.50} &
  3.175 \\ \cline{3-13} 
 &
   &
  $\Pi_{\mathsf{OA}}^{\mathsf{sep}}$ &
  \multicolumn{1}{c|}{\cellcolor[HTML]{99BEFF}17.17} &
  \multicolumn{1}{c|}{2.059} &
  \multicolumn{1}{c|}{\cellcolor[HTML]{99BEFF}19.79} &
  \multicolumn{1}{c|}{2.379} &
  \multicolumn{1}{c|}{\cellcolor[HTML]{99BEFF}22.64} &
  \multicolumn{1}{c|}{2.594} &
  \multicolumn{1}{c|}{\cellcolor[HTML]{99BEFF}25.45} &
  \multicolumn{1}{c|}{2.868} &
  \multicolumn{1}{c|}{\cellcolor[HTML]{99BEFF}28.84} &
  3.122 \\ \cline{3-13} 
 &
   &
  $\Pi_{\mathsf{SE}}^{\mathsf{sep}}$ &
  \multicolumn{1}{c|}{19.97} &
  \multicolumn{1}{c|}{\cellcolor[HTML]{C7ECFF}1.873} &
  \multicolumn{1}{c|}{23.03} &
  \multicolumn{1}{c|}{\cellcolor[HTML]{C7ECFF}2.205} &
  \multicolumn{1}{c|}{26.31} &
  \multicolumn{1}{c|}{\cellcolor[HTML]{C7ECFF}2.475} &
  \multicolumn{1}{c|}{29.72} &
  \multicolumn{1}{c|}{\cellcolor[HTML]{C7ECFF}2.738} &
  \multicolumn{1}{c|}{33.16} &
  \cellcolor[HTML]{C7ECFF}3.022 \\ \cline{3-13} 
\multirow{-21}{*}{$2^8$} &
  \multirow{-7}{*}{10} &
  $\Pi_{\mathsf{SA}}^{\mathsf{sep}}$ &
  \multicolumn{1}{c|}{18.66} &
  \multicolumn{1}{c|}{\cellcolor[HTML]{99BEFF}1.838} &
  \multicolumn{1}{c|}{21.70} &
  \multicolumn{1}{c|}{\cellcolor[HTML]{99BEFF}2.159} &
  \multicolumn{1}{c|}{24.95} &
  \multicolumn{1}{c|}{\cellcolor[HTML]{99BEFF}2.422} &
  \multicolumn{1}{c|}{28.10} &
  \multicolumn{1}{c|}{\cellcolor[HTML]{99BEFF}2.714} &
  \multicolumn{1}{c|}{31.41} &
  \cellcolor[HTML]{99BEFF}2.982 \\ \hline
 &
   &
  \cite{DBLP:conf/asiacrypt/GaoQLLW24} &
  \multicolumn{1}{c|}{554.7} &
  \multicolumn{1}{c|}{112.1} &
  \multicolumn{1}{c|}{1089} &
  \multicolumn{1}{c|}{208.1} &
  \multicolumn{1}{c|}{2159} &
  \multicolumn{1}{c|}{389.3} &
  \multicolumn{1}{c|}{4297} &
  \multicolumn{1}{c|}{754.1} &
  \multicolumn{1}{c|}{-} &
  - \\ \cline{3-13} 
 &
   &
  \cite{DBLP:conf/ccs/DZL25}-High &
  \multicolumn{1}{c|}{1043} &
  \multicolumn{1}{c|}{110.7} &
  \multicolumn{1}{c|}{1123} &
  \multicolumn{1}{c|}{111.8} &
  \multicolumn{1}{c|}{1204} &
  \multicolumn{1}{c|}{123.5} &
  \multicolumn{1}{c|}{1271} &
  \multicolumn{1}{c|}{125.9} &
  \multicolumn{1}{c|}{1936} &
  218.2 \\ \cline{3-13} 
 &
   &
  $\Pi_{\mathsf{OE}\text{-1}}^{\mathsf{sep}}$ &
  \multicolumn{1}{c|}{\cellcolor[HTML]{C7ECFF}159.3} &
  \multicolumn{1}{c|}{15.39} &
  \multicolumn{1}{c|}{\cellcolor[HTML]{C7ECFF}185.5} &
  \multicolumn{1}{c|}{17.82} &
  \multicolumn{1}{c|}{\cellcolor[HTML]{C7ECFF}212.6} &
  \multicolumn{1}{c|}{20.21} &
  \multicolumn{1}{c|}{\cellcolor[HTML]{C7ECFF}241.3} &
  \multicolumn{1}{c|}{22.53} &
  \multicolumn{1}{c|}{\cellcolor[HTML]{C7ECFF}268.0} &
  25.08 \\ \cline{3-13} 
 &
   &
  $\Pi_{\mathsf{OE}\text{-2}}^{\mathsf{sep}}$ &
  \multicolumn{1}{c|}{159.7} &
  \multicolumn{1}{c|}{\cellcolor[HTML]{C7ECFF}15.34} &
  \multicolumn{1}{c|}{186.2} &
  \multicolumn{1}{c|}{\cellcolor[HTML]{C7ECFF}17.65} &
  \multicolumn{1}{c|}{213.0} &
  \multicolumn{1}{c|}{\cellcolor[HTML]{C7ECFF}20.15} &
  \multicolumn{1}{c|}{241.8} &
  \multicolumn{1}{c|}{\cellcolor[HTML]{C7ECFF}22.47} &
  \multicolumn{1}{c|}{268.5} &
  \cellcolor[HTML]{C7ECFF}25.06 \\ \cline{3-13} 
 &
   &
  $\Pi_{\mathsf{OA}}^{\mathsf{sep}}$ &
  \multicolumn{1}{c|}{\cellcolor[HTML]{99BEFF}153.0} &
  \multicolumn{1}{c|}{\cellcolor[HTML]{99BEFF}15.03} &
  \multicolumn{1}{c|}{\cellcolor[HTML]{99BEFF}179.0} &
  \multicolumn{1}{c|}{\cellcolor[HTML]{99BEFF}17.59} &
  \multicolumn{1}{c|}{\cellcolor[HTML]{99BEFF}205.4} &
  \multicolumn{1}{c|}{\cellcolor[HTML]{99BEFF}20.13} &
  \multicolumn{1}{c|}{\cellcolor[HTML]{99BEFF}231.8} &
  \multicolumn{1}{c|}{\cellcolor[HTML]{99BEFF}22.31} &
  \multicolumn{1}{c|}{\cellcolor[HTML]{99BEFF}258.0} &
  \cellcolor[HTML]{99BEFF}24.74 \\ \cline{3-13} 
 &
   &
  $\Pi_{\mathsf{SE}}^{\mathsf{sep}}$ &
  \multicolumn{1}{c|}{177.8} &
  \multicolumn{1}{c|}{16.21} &
  \multicolumn{1}{c|}{207.4} &
  \multicolumn{1}{c|}{18.79} &
  \multicolumn{1}{c|}{237.2} &
  \multicolumn{1}{c|}{21.54} &
  \multicolumn{1}{c|}{269.0} &
  \multicolumn{1}{c|}{24.03} &
  \multicolumn{1}{c|}{298.6} &
  26.69 \\ \cline{3-13} 
 &
  \multirow{-7}{*}{6} &
  $\Pi_{\mathsf{SA}}^{\mathsf{sep}}$ &
  \multicolumn{1}{c|}{171.1} &
  \multicolumn{1}{c|}{16.08} &
  \multicolumn{1}{c|}{200.1} &
  \multicolumn{1}{c|}{18.68} &
  \multicolumn{1}{c|}{229.5} &
  \multicolumn{1}{c|}{21.39} &
  \multicolumn{1}{c|}{259.0} &
  \multicolumn{1}{c|}{23.82} &
  \multicolumn{1}{c|}{288.2} &
  26.51 \\ \cline{2-13} 
 &
   &
  \cite{DBLP:conf/asiacrypt/GaoQLLW24} &
  \multicolumn{1}{c|}{738.5} &
  \multicolumn{1}{c|}{153.5} &
  \multicolumn{1}{c|}{1451} &
  \multicolumn{1}{c|}{275.4} &
  \multicolumn{1}{c|}{2877} &
  \multicolumn{1}{c|}{522.1} &
  \multicolumn{1}{c|}{5728} &
  \multicolumn{1}{c|}{1014} &
  \multicolumn{1}{c|}{-} &
  - \\ \cline{3-13} 
 &
   &
  \cite{DBLP:conf/ccs/DZL25}-High &
  \multicolumn{1}{c|}{1390} &
  \multicolumn{1}{c|}{147.7} &
  \multicolumn{1}{c|}{1497} &
  \multicolumn{1}{c|}{152.4} &
  \multicolumn{1}{c|}{1605} &
  \multicolumn{1}{c|}{162.5} &
  \multicolumn{1}{c|}{1694} &
  \multicolumn{1}{c|}{165.1} &
  \multicolumn{1}{c|}{2581} &
  291.5 \\ \cline{3-13} 
 &
   &
  $\Pi_{\mathsf{OE}\text{-1}}^{\mathsf{sep}}$ &
  \multicolumn{1}{c|}{\cellcolor[HTML]{C7ECFF}210.6} &
  \multicolumn{1}{c|}{20.44} &
  \multicolumn{1}{c|}{\cellcolor[HTML]{C7ECFF}246.1} &
  \multicolumn{1}{c|}{23.56} &
  \multicolumn{1}{c|}{\cellcolor[HTML]{C7ECFF}281.8} &
  \multicolumn{1}{c|}{26.67} &
  \multicolumn{1}{c|}{\cellcolor[HTML]{C7ECFF}320.1} &
  \multicolumn{1}{c|}{30.08} &
  \multicolumn{1}{c|}{\cellcolor[HTML]{C7ECFF}355.9} &
  33.16 \\ \cline{3-13} 
 &
   &
  $\Pi_{\mathsf{OE}\text{-2}}^{\mathsf{sep}}$ &
  \multicolumn{1}{c|}{211.1} &
  \multicolumn{1}{c|}{\cellcolor[HTML]{C7ECFF}20.38} &
  \multicolumn{1}{c|}{246.4} &
  \multicolumn{1}{c|}{\cellcolor[HTML]{C7ECFF}23.44} &
  \multicolumn{1}{c|}{282.1} &
  \multicolumn{1}{c|}{\cellcolor[HTML]{C7ECFF}26.57} &
  \multicolumn{1}{c|}{320.5} &
  \multicolumn{1}{c|}{\cellcolor[HTML]{C7ECFF}29.97} &
  \multicolumn{1}{c|}{356.5} &
  \cellcolor[HTML]{C7ECFF}33.09 \\ \cline{3-13} 
 &
   &
  $\Pi_{\mathsf{OA}}^{\mathsf{sep}}$ &
  \multicolumn{1}{c|}{\cellcolor[HTML]{99BEFF}202.6} &
  \multicolumn{1}{c|}{\cellcolor[HTML]{99BEFF}20.21} &
  \multicolumn{1}{c|}{\cellcolor[HTML]{99BEFF}237.1} &
  \multicolumn{1}{c|}{\cellcolor[HTML]{99BEFF}23.23} &
  \multicolumn{1}{c|}{\cellcolor[HTML]{99BEFF}272.3} &
  \multicolumn{1}{c|}{\cellcolor[HTML]{99BEFF}26.49} &
  \multicolumn{1}{c|}{\cellcolor[HTML]{99BEFF}307.4} &
  \multicolumn{1}{c|}{\cellcolor[HTML]{99BEFF}29.84} &
  \multicolumn{1}{c|}{\cellcolor[HTML]{99BEFF}342.7} &
  \cellcolor[HTML]{99BEFF}32.99 \\ \cline{3-13} 
 &
   &
  $\Pi_{\mathsf{SE}}^{\mathsf{sep}}$ &
  \multicolumn{1}{c|}{235.7} &
  \multicolumn{1}{c|}{21.64} &
  \multicolumn{1}{c|}{274.9} &
  \multicolumn{1}{c|}{25.05} &
  \multicolumn{1}{c|}{314.6} &
  \multicolumn{1}{c|}{28.17} &
  \multicolumn{1}{c|}{357.2} &
  \multicolumn{1}{c|}{32.13} &
  \multicolumn{1}{c|}{396.9} &
  35.38 \\ \cline{3-13} 
 &
  \multirow{-7}{*}{8} &
  $\Pi_{\mathsf{SA}}^{\mathsf{sep}}$ &
  \multicolumn{1}{c|}{227.0} &
  \multicolumn{1}{c|}{21.39} &
  \multicolumn{1}{c|}{265.7} &
  \multicolumn{1}{c|}{24.78} &
  \multicolumn{1}{c|}{304.9} &
  \multicolumn{1}{c|}{28.35} &
  \multicolumn{1}{c|}{344.1} &
  \multicolumn{1}{c|}{31.89} &
  \multicolumn{1}{c|}{383.1} &
  35.33 \\ \cline{2-13} 
 &
   &
  \cite{DBLP:conf/asiacrypt/GaoQLLW24} &
  \multicolumn{1}{c|}{922.4} &
  \multicolumn{1}{c|}{191.2} &
  \multicolumn{1}{c|}{1813} &
  \multicolumn{1}{c|}{344.8} &
  \multicolumn{1}{c|}{3595} &
  \multicolumn{1}{c|}{649.1} &
  \multicolumn{1}{c|}{7159} &
  \multicolumn{1}{c|}{1265} &
  \multicolumn{1}{c|}{-} &
  - \\ \cline{3-13} 
 &
   &
  \cite{DBLP:conf/ccs/DZL25}-High &
  \multicolumn{1}{c|}{1736} &
  \multicolumn{1}{c|}{185.1} &
  \multicolumn{1}{c|}{1870} &
  \multicolumn{1}{c|}{188.9} &
  \multicolumn{1}{c|}{2006} &
  \multicolumn{1}{c|}{205.2} &
  \multicolumn{1}{c|}{2117} &
  \multicolumn{1}{c|}{207.9} &
  \multicolumn{1}{c|}{3226} &
  358.7 \\ \cline{3-13} 
 &
   &
  $\Pi_{\mathsf{OE}\text{-1}}^{\mathsf{sep}}$ &
  \multicolumn{1}{c|}{\cellcolor[HTML]{C7ECFF}262.1} &
  \multicolumn{1}{c|}{25.46} &
  \multicolumn{1}{c|}{\cellcolor[HTML]{C7ECFF}306.4} &
  \multicolumn{1}{c|}{29.38} &
  \multicolumn{1}{c|}{\cellcolor[HTML]{C7ECFF}350.9} &
  \multicolumn{1}{c|}{33.27} &
  \multicolumn{1}{c|}{\cellcolor[HTML]{C7ECFF}398.8} &
  \multicolumn{1}{c|}{37.42} &
  \multicolumn{1}{c|}{\cellcolor[HTML]{C7ECFF}443.6} &
  41.19 \\ \cline{3-13} 
 &
   &
  $\Pi_{\mathsf{OE}\text{-2}}^{\mathsf{sep}}$ &
  \multicolumn{1}{c|}{262.4} &
  \multicolumn{1}{c|}{\cellcolor[HTML]{C7ECFF}25.44} &
  \multicolumn{1}{c|}{306.7} &
  \multicolumn{1}{c|}{\cellcolor[HTML]{C7ECFF}29.28} &
  \multicolumn{1}{c|}{351.4} &
  \multicolumn{1}{c|}{\cellcolor[HTML]{C7ECFF}33.24} &
  \multicolumn{1}{c|}{399.3} &
  \multicolumn{1}{c|}{\cellcolor[HTML]{C7ECFF}37.12} &
  \multicolumn{1}{c|}{444.4} &
  \cellcolor[HTML]{C7ECFF}41.08 \\ \cline{3-13} 
 &
   &
  $\Pi_{\mathsf{OA}}^{\mathsf{sep}}$ &
  \multicolumn{1}{c|}{\cellcolor[HTML]{99BEFF}252.1} &
  \multicolumn{1}{c|}{\cellcolor[HTML]{99BEFF}25.26} &
  \multicolumn{1}{c|}{\cellcolor[HTML]{99BEFF}295.5} &
  \multicolumn{1}{c|}{\cellcolor[HTML]{99BEFF}28.89} &
  \multicolumn{1}{c|}{\cellcolor[HTML]{99BEFF}339.5} &
  \multicolumn{1}{c|}{\cellcolor[HTML]{99BEFF}32.89} &
  \multicolumn{1}{c|}{\cellcolor[HTML]{99BEFF}383.4} &
  \multicolumn{1}{c|}{\cellcolor[HTML]{99BEFF}37.06} &
  \multicolumn{1}{c|}{\cellcolor[HTML]{99BEFF}427.4} &
  \cellcolor[HTML]{99BEFF}41.02 \\ \cline{3-13} 
 &
   &
  $\Pi_{\mathsf{SE}}^{\mathsf{sep}}$ &
  \multicolumn{1}{c|}{293.2} &
  \multicolumn{1}{c|}{26.99} &
  \multicolumn{1}{c|}{342.8} &
  \multicolumn{1}{c|}{31.31} &
  \multicolumn{1}{c|}{392.4} &
  \multicolumn{1}{c|}{35.68} &
  \multicolumn{1}{c|}{445.0} &
  \multicolumn{1}{c|}{39.97} &
  \multicolumn{1}{c|}{495.0} &
  44.04 \\ \cline{3-13} 
\multirow{-21}{*}{$2^{12}$} &
  \multirow{-7}{*}{10} &
  $\Pi_{\mathsf{SA}}^{\mathsf{sep}}$ &
  \multicolumn{1}{c|}{282.6} &
  \multicolumn{1}{c|}{26.95} &
  \multicolumn{1}{c|}{331.4} &
  \multicolumn{1}{c|}{31.18} &
  \multicolumn{1}{c|}{380.1} &
  \multicolumn{1}{c|}{35.42} &
  \multicolumn{1}{c|}{429.1} &
  \multicolumn{1}{c|}{39.86} &
  \multicolumn{1}{c|}{478.5} &
  44.24 \\ \hline
 &
   &
  \cite{DBLP:conf/asiacrypt/GaoQLLW24} &
  \multicolumn{1}{c|}{-} &
  \multicolumn{1}{c|}{-} &
  \multicolumn{1}{c|}{-} &
  \multicolumn{1}{c|}{-} &
  \multicolumn{1}{c|}{-} &
  \multicolumn{1}{c|}{-} &
  \multicolumn{1}{c|}{-} &
  \multicolumn{1}{c|}{-} &
  \multicolumn{1}{c|}{-} &
  - \\ \cline{3-13} 
 &
   &
  \cite{DBLP:conf/ccs/DZL25}-High &
  \multicolumn{1}{c|}{-} &
  \multicolumn{1}{c|}{-} &
  \multicolumn{1}{c|}{-} &
  \multicolumn{1}{c|}{-} &
  \multicolumn{1}{c|}{-} &
  \multicolumn{1}{c|}{-} &
  \multicolumn{1}{c|}{-} &
  \multicolumn{1}{c|}{-} &
  \multicolumn{1}{c|}{-} &
  - \\ \cline{3-13} 
 &
   &
  $\Pi_{\mathsf{OE}\text{-1}}^{\mathsf{sep}}$ &
  \multicolumn{1}{c|}{\cellcolor[HTML]{C7ECFF}2540} &
  \multicolumn{1}{c|}{245.9} &
  \multicolumn{1}{c|}{\cellcolor[HTML]{C7ECFF}2967} &
  \multicolumn{1}{c|}{285.2} &
  \multicolumn{1}{c|}{\cellcolor[HTML]{C7ECFF}3397} &
  \multicolumn{1}{c|}{319.9} &
  \multicolumn{1}{c|}{\cellcolor[HTML]{C7ECFF}3860} &
  \multicolumn{1}{c|}{\cellcolor[HTML]{C7ECFF}364.6} &
  \multicolumn{1}{c|}{\cellcolor[HTML]{C7ECFF}4292} &
  \cellcolor[HTML]{C7ECFF}400.6 \\ \cline{3-13} 
 &
   &
  $\Pi_{\mathsf{OE}\text{-2}}^{\mathsf{sep}}$ &
  \multicolumn{1}{c|}{2541} &
  \multicolumn{1}{c|}{\cellcolor[HTML]{C7ECFF}241.7} &
  \multicolumn{1}{c|}{2968} &
  \multicolumn{1}{c|}{\cellcolor[HTML]{C7ECFF}284.2} &
  \multicolumn{1}{c|}{3398} &
  \multicolumn{1}{c|}{\cellcolor[HTML]{C7ECFF}319.7} &
  \multicolumn{1}{c|}{\cellcolor[HTML]{C7ECFF}3860} &
  \multicolumn{1}{c|}{\cellcolor[HTML]{99BEFF}364.5} &
  \multicolumn{1}{c|}{4293} &
  \cellcolor[HTML]{99BEFF}400.2 \\ \cline{3-13} 
 &
   &
  $\Pi_{\mathsf{OA}}^{\mathsf{sep}}$ &
  \multicolumn{1}{c|}{\cellcolor[HTML]{99BEFF}2442} &
  \multicolumn{1}{c|}{\cellcolor[HTML]{99BEFF}240.5} &
  \multicolumn{1}{c|}{\cellcolor[HTML]{99BEFF}2862} &
  \multicolumn{1}{c|}{\cellcolor[HTML]{99BEFF}282.6} &
  \multicolumn{1}{c|}{\cellcolor[HTML]{99BEFF}3283} &
  \multicolumn{1}{c|}{\cellcolor[HTML]{99BEFF}318.4} &
  \multicolumn{1}{c|}{\cellcolor[HTML]{99BEFF}3707} &
  \multicolumn{1}{c|}{365.7} &
  \multicolumn{1}{c|}{\cellcolor[HTML]{99BEFF}4131} &
  402.3 \\ \cline{3-13} 
 &
   &
  $\Pi_{\mathsf{SE}}^{\mathsf{sep}}$ &
  \multicolumn{1}{c|}{2848} &
  \multicolumn{1}{c|}{262.9} &
  \multicolumn{1}{c|}{3324} &
  \multicolumn{1}{c|}{307.3} &
  \multicolumn{1}{c|}{3803} &
  \multicolumn{1}{c|}{348.1} &
  \multicolumn{1}{c|}{4316} &
  \multicolumn{1}{c|}{397.2} &
  \multicolumn{1}{c|}{4797} &
  434.9 \\ \cline{3-13} 
 &
  \multirow{-7}{*}{6} &
  $\Pi_{\mathsf{SA}}^{\mathsf{sep}}$ &
  \multicolumn{1}{c|}{2750} &
  \multicolumn{1}{c|}{261.8} &
  \multicolumn{1}{c|}{3219} &
  \multicolumn{1}{c|}{307.1} &
  \multicolumn{1}{c|}{3689} &
  \multicolumn{1}{c|}{345.6} &
  \multicolumn{1}{c|}{4163} &
  \multicolumn{1}{c|}{393.2} &
  \multicolumn{1}{c|}{4638} &
  433.4 \\ \cline{2-13} 
 &
   &
  \cite{DBLP:conf/asiacrypt/GaoQLLW24} &
  \multicolumn{1}{c|}{-} &
  \multicolumn{1}{c|}{-} &
  \multicolumn{1}{c|}{-} &
  \multicolumn{1}{c|}{-} &
  \multicolumn{1}{c|}{-} &
  \multicolumn{1}{c|}{-} &
  \multicolumn{1}{c|}{-} &
  \multicolumn{1}{c|}{-} &
  \multicolumn{1}{c|}{-} &
  - \\ \cline{3-13} 
 &
   &
  \cite{DBLP:conf/ccs/DZL25}-High &
  \multicolumn{1}{c|}{-} &
  \multicolumn{1}{c|}{-} &
  \multicolumn{1}{c|}{-} &
  \multicolumn{1}{c|}{-} &
  \multicolumn{1}{c|}{-} &
  \multicolumn{1}{c|}{-} &
  \multicolumn{1}{c|}{-} &
  \multicolumn{1}{c|}{-} &
  \multicolumn{1}{c|}{-} &
  - \\ \cline{3-13} 
 &
   &
  $\Pi_{\mathsf{OE}\text{-1}}^{\mathsf{sep}}$ &
  \multicolumn{1}{c|}{\cellcolor[HTML]{C7ECFF}3367} &
  \multicolumn{1}{c|}{324.6} &
  \multicolumn{1}{c|}{\cellcolor[HTML]{C7ECFF}3937} &
  \multicolumn{1}{c|}{375.1} &
  \multicolumn{1}{c|}{\cellcolor[HTML]{C7ECFF}4511} &
  \multicolumn{1}{c|}{426.8} &
  \multicolumn{1}{c|}{\cellcolor[HTML]{C7ECFF}5128} &
  \multicolumn{1}{c|}{\cellcolor[HTML]{C7ECFF}480.6} &
  \multicolumn{1}{c|}{\cellcolor[HTML]{C7ECFF}5704} &
  \cellcolor[HTML]{C7ECFF}539.9 \\ \cline{3-13} 
 &
   &
  $\Pi_{\mathsf{OE}\text{-2}}^{\mathsf{sep}}$ &
  \multicolumn{1}{c|}{3369} &
  \multicolumn{1}{c|}{\cellcolor[HTML]{C7ECFF}322.4} &
  \multicolumn{1}{c|}{3938} &
  \multicolumn{1}{c|}{\cellcolor[HTML]{C7ECFF}374.8} &
  \multicolumn{1}{c|}{\cellcolor[HTML]{C7ECFF}4511} &
  \multicolumn{1}{c|}{\cellcolor[HTML]{C7ECFF}425.8} &
  \multicolumn{1}{c|}{\cellcolor[HTML]{C7ECFF}5128} &
  \multicolumn{1}{c|}{\cellcolor[HTML]{99BEFF}479.5} &
  \multicolumn{1}{c|}{5705} &
  \cellcolor[HTML]{99BEFF}538.7 \\ \cline{3-13} 
 &
   &
  $\Pi_{\mathsf{OA}}^{\mathsf{sep}}$ &
  \multicolumn{1}{c|}{\cellcolor[HTML]{99BEFF}3239} &
  \multicolumn{1}{c|}{\cellcolor[HTML]{99BEFF}321.5} &
  \multicolumn{1}{c|}{\cellcolor[HTML]{99BEFF}3798} &
  \multicolumn{1}{c|}{\cellcolor[HTML]{99BEFF}373.9} &
  \multicolumn{1}{c|}{\cellcolor[HTML]{99BEFF}4361} &
  \multicolumn{1}{c|}{\cellcolor[HTML]{99BEFF}424.8} &
  \multicolumn{1}{c|}{\cellcolor[HTML]{99BEFF}4926} &
  \multicolumn{1}{c|}{481.4} &
  \multicolumn{1}{c|}{\cellcolor[HTML]{99BEFF}5492} &
  542.7 \\ \cline{3-13} 
 &
   &
  $\Pi_{\mathsf{SE}}^{\mathsf{sep}}$ &
  \multicolumn{1}{c|}{3778} &
  \multicolumn{1}{c|}{353.6} &
  \multicolumn{1}{c|}{4414} &
  \multicolumn{1}{c|}{406.8} &
  \multicolumn{1}{c|}{5053} &
  \multicolumn{1}{c|}{463.2} &
  \multicolumn{1}{c|}{5737} &
  \multicolumn{1}{c|}{521.3} &
  \multicolumn{1}{c|}{6379} &
  583.8 \\ \cline{3-13} 
 &
  \multirow{-7}{*}{8} &
  $\Pi_{\mathsf{SA}}^{\mathsf{sep}}$ &
  \multicolumn{1}{c|}{3649} &
  \multicolumn{1}{c|}{352.7} &
  \multicolumn{1}{c|}{4275} &
  \multicolumn{1}{c|}{406.6} &
  \multicolumn{1}{c|}{4904} &
  \multicolumn{1}{c|}{462.5} &
  \multicolumn{1}{c|}{5535} &
  \multicolumn{1}{c|}{517.9} &
  \multicolumn{1}{c|}{6167} &
  581.2 \\ \cline{2-13} 
 &
   &
  \cite{DBLP:conf/asiacrypt/GaoQLLW24} &
  \multicolumn{1}{c|}{-} &
  \multicolumn{1}{c|}{-} &
  \multicolumn{1}{c|}{-} &
  \multicolumn{1}{c|}{-} &
  \multicolumn{1}{c|}{-} &
  \multicolumn{1}{c|}{-} &
  \multicolumn{1}{c|}{-} &
  \multicolumn{1}{c|}{-} &
  \multicolumn{1}{c|}{-} &
  - \\ \cline{3-13} 
 &
   &
  \cite{DBLP:conf/ccs/DZL25}-High &
  \multicolumn{1}{c|}{-} &
  \multicolumn{1}{c|}{-} &
  \multicolumn{1}{c|}{-} &
  \multicolumn{1}{c|}{-} &
  \multicolumn{1}{c|}{-} &
  \multicolumn{1}{c|}{-} &
  \multicolumn{1}{c|}{-} &
  \multicolumn{1}{c|}{-} &
  \multicolumn{1}{c|}{-} &
  - \\ \cline{3-13} 
 &
   &
  $\Pi_{\mathsf{OE}\text{-1}}^{\mathsf{sep}}$ &
  \multicolumn{1}{c|}{\cellcolor[HTML]{C7ECFF}4196} &
  \multicolumn{1}{c|}{404.4} &
  \multicolumn{1}{c|}{\cellcolor[HTML]{C7ECFF}4908} &
  \multicolumn{1}{c|}{466.8} &
  \multicolumn{1}{c|}{\cellcolor[HTML]{C7ECFF}5624} &
  \multicolumn{1}{c|}{539.8} &
  \multicolumn{1}{c|}{\cellcolor[HTML]{C7ECFF}6396} &
  \multicolumn{1}{c|}{605.2} &
  \multicolumn{1}{c|}{\cellcolor[HTML]{C7ECFF}7118} &
  \cellcolor[HTML]{C7ECFF}673.1 \\ \cline{3-13} 
 &
   &
  $\Pi_{\mathsf{OE}\text{-2}}^{\mathsf{sep}}$ &
  \multicolumn{1}{c|}{\cellcolor[HTML]{C7ECFF}4196} &
  \multicolumn{1}{c|}{\cellcolor[HTML]{C7ECFF}403.6} &
  \multicolumn{1}{c|}{\cellcolor[HTML]{C7ECFF}4908} &
  \multicolumn{1}{c|}{\cellcolor[HTML]{99BEFF}465.5} &
  \multicolumn{1}{c|}{5625} &
  \multicolumn{1}{c|}{\cellcolor[HTML]{99BEFF}539.3} &
  \multicolumn{1}{c|}{\cellcolor[HTML]{C7ECFF}6396} &
  \multicolumn{1}{c|}{\cellcolor[HTML]{99BEFF}600.3} &
  \multicolumn{1}{c|}{\cellcolor[HTML]{C7ECFF}7118} &
  \cellcolor[HTML]{99BEFF}672.3 \\ \cline{3-13} 
 &
   &
  $\Pi_{\mathsf{OA}}^{\mathsf{sep}}$ &
  \multicolumn{1}{c|}{\cellcolor[HTML]{99BEFF}4036} &
  \multicolumn{1}{c|}{\cellcolor[HTML]{99BEFF}403.2} &
  \multicolumn{1}{c|}{\cellcolor[HTML]{99BEFF}4734} &
  \multicolumn{1}{c|}{\cellcolor[HTML]{C7ECFF}466.2} &
  \multicolumn{1}{c|}{\cellcolor[HTML]{99BEFF}5439} &
  \multicolumn{1}{c|}{\cellcolor[HTML]{C7ECFF}539.6} &
  \multicolumn{1}{c|}{\cellcolor[HTML]{99BEFF}6144} &
  \multicolumn{1}{c|}{\cellcolor[HTML]{C7ECFF}604.3} &
  \multicolumn{1}{c|}{\cellcolor[HTML]{99BEFF}6853} &
  673.6 \\ \cline{3-13} 
 &
   &
  $\Pi_{\mathsf{SE}}^{\mathsf{sep}}$ &
  \multicolumn{1}{c|}{4709} &
  \multicolumn{1}{c|}{438.5} &
  \multicolumn{1}{c|}{5502} &
  \multicolumn{1}{c|}{511.7} &
  \multicolumn{1}{c|}{6303} &
  \multicolumn{1}{c|}{585.7} &
  \multicolumn{1}{c|}{7157} &
  \multicolumn{1}{c|}{648.8} &
  \multicolumn{1}{c|}{7960} &
  725.9 \\ \cline{3-13} 
\multirow{-21}{*}{$2^{16}$} &
  \multirow{-7}{*}{10} &
  $\Pi_{\mathsf{SA}}^{\mathsf{sep}}$ &
  \multicolumn{1}{c|}{4549} &
  \multicolumn{1}{c|}{436.7} &
  \multicolumn{1}{c|}{5331} &
  \multicolumn{1}{c|}{505.7} &
  \multicolumn{1}{c|}{6117} &
  \multicolumn{1}{c|}{582.8} &
  \multicolumn{1}{c|}{6906} &
  \multicolumn{1}{c|}{645.3} &
  \multicolumn{1}{c|}{7696} &
  726.7 \\ \hline
\end{tabular}
}
\caption{Communication cost (in MB) and running time (in seconds) comparing our protocols to \cite{DBLP:conf/asiacrypt/GaoQLLW24,DBLP:conf/ccs/DZL25}. Cells with - denote trials that ran out of memory. The best result is highlighted in \textcolor[rgb]{0,0.3,1}{blue}, the second best in \textcolor[rgb]{0,0.7,1}{cyan}.}
\label{tab:infty-high-lan}
\end{table}


\subsection{Implementation Details}

We conduct our experiments on a server with an Intel(R) Xeon(R) Platinum 8575C CPU (8 physical cores) and 64 GB RAM. All experiments are executed with a single thread. We set the computational security parameter $\kappa = 128$ and the statistical security parameter $\lambda = 40$. We simulate the network connection using the Linux command $\mathsf{tc}$. We benchmark the full protocol runtime in a low-latency LAN environment (10Gbps bandwidth with 0.02ms RTT latency). We also conduct experiments under a WAN setting (400 Mbps bandwidth, 80 ms RTT latency), which are provided in Appendix \ref{app:wan}.

Our protocols are written in C++, and we use the following libraries in our implementation. 
Our complete implementation will be freely available on GitHub.
\begin{itemize}
    \item Shared OPRF. We use OPRF with shared-output in \cite{DBLP:conf/crypto/AlamatiPRR24} to instantiate our sOPRF. 
    \item OPPRF, OKVS and ssPEQT. We use state-of-the-art OPPRF, OKVS and ssPEQT implementation in \cite{DBLP:conf/ccs/RaghuramanR22,volepsi}.
    \item Permute + Share. We use the implementation of \cite{280022,psu} to implement Permute + Share functionality.
    \item Fuzzy matching for interval. We use the implementation of previous FPSI work \cite{DBLP:conf/asiacrypt/GaoQLLW24,fuzzypsi} to implement fuzzy matching for interval functionality \cite{DBLP:conf/uss/ChakrabortiFR23}.
    \item OT. We use SoftSpokenOT \cite{DBLP:conf/crypto/Roy22} implemented in libOTe \cite{libOTe}, and set field bits to 5 to balance computation and communication.
    \item For computing hash functions and PRG calls, we employ the $\mathsf{cryptoTools}$ library \cite{cryptotools}, while for network communication, we utilize $\mathsf{Coproto}$ \cite{coproto}.
\end{itemize}


\subsection{Performance Evaluation}

Specifically, our evaluation uses the following parameters: set size $m=n \in\{2^{8}, 2^{12}, 2^{16}\}$, distance threshold $\delta \in \{16,32,64,128,256\}$\footnote{Since \cite{AC/BGMP25} only supports $\delta\in\{10,30,60,120,250\}$, we align their results with ours using the closest matching $\delta$ values.}, and dimension $d$ chosen as $ \{2,3,4 \} $ for the \textit{apart} assumption and $ \{6,8,10 \} $ for the \textit{separate} assumption. The results are shown in Table \ref{tab:infty-low-lan} and Table \ref{tab:infty-high-lan}, respectively. 

\begin{trivlist}
    \item \textbf{Comparison in Low Dimension.} 
   As shown in Table \ref{tab:infty-low-lan}, we observed that, for $d\in\{3,4\}$, \cite{DBLP:conf/ccs/DZL25}-Low's data shows non-monotonic scaling with $\delta$, with costs dropping unexpectedly at $\delta=64$ (marked in red), suggesting potential implementation issues. Nevertheless, our protocols exhibit significant performance improvements in almost all parameters. For instance, with $n=m=2^{12}$, $d=4$, and $\delta=256$, our $\Pi_{\mathsf{SA}}^{\mathsf{apart}}$ achieves up to $18\times$ speedup and $21.6\times$ lower communication overhead. While $\Pi_{\mathsf{SA}}^{\mathsf{apart}}$ performs best at smaller set sizes ($n=2^8$), $\Pi_{\mathsf{OA}}^{\mathsf{apart}}$ scales more favorably as the set size increases, owing to the higher efficiency of the OPRF protocol within OPPRF under large parameter settings.
 
    \item \textbf{Comparison in High Dimension.} As shown in Table \ref{tab:infty-high-lan}, our protocols maintain substantial advantages in high-dimensional settings. For example, with $m=n=2^8$, $d=4$, and $\delta=256$, our $\Pi_{\mathsf{SA}}^{\mathsf{apart}}$ achieves up to $43.7\times$ speedup and $28.4\times$ lower communication overhead. Notably, both \cite{DBLP:conf/asiacrypt/GaoQLLW24} and \cite{DBLP:conf/ccs/DZL25}-High encountered memory overflow at $m=n=2^{16}$, whereas our protocols execute successfully under the same configuration.

\end{trivlist}



\bibliographystyle{alpha}
\bibliography{FPSI}

\newpage


\appendix

\section{Security Flaws of \cite{DBLP:conf/ccs/DZL25}}
\label{app:flaw}

In this section, we identify security flaws in the protocols of \cite{DBLP:conf/ccs/DZL25}. Their constructions contain two distinct flaws: one in the fuzzy matching protocol and another in the distributed ID generation protocol. Since all of their FPSI protocols adopt the same fuzzy matching idea, these security flaws affect every FPSI variant presented in their work, including both low- and high-dimensional cases for $L_\infty$ and $L_{p\in[1,\infty)}$ distances, thereby rendering them all insecure.

\subsection{Security Flaws in Fuzzy Matching}
\label{app:flaw-fmat}

We analyze the security flaws in their fuzzy matching protocol for $L_\infty$ distance as an example. The $L_p$ distance variant suffers from the same vulnerabilities, which we omit here. We begin by reviewing their $L_\infty$ fuzzy matching protocol, presented in Figure \ref{fig:pifmat-dzl25}.

\begin{figure}[!hbth]
\begin{framed}
\begin{minipage}[center]{\textwidth}
\begin{trivlist}
\item \textbf{Parameters:} 
\begin{itemize}
\item Two parties: The sender $\mathcal{S}$ and the receiver $\mathcal{R}$, universe $\mathbb{U}^d$, distance threshold $\delta$.
\item An OKVS scheme $(\mathsf{Encode},\mathsf{Decode})$.
\item A prefix trie scheme $(\mathsf{PrefixTrie}, \mathsf{PrefixPath})$.
\item An AHE scheme $(\mathsf{KeyGen},\mathsf{Enc},\mathsf{Dec},\boxplus)$ with plaintext space $\mathcal{P}$ and ciphertext space $\mathcal{C}$.
\item Two universal hash functions $H_{\gamma_1}: \{0,1\}^*\rightarrow \{0,1\}^{\gamma_1}$, $H_{\gamma_2}: \{0,1\}^*\rightarrow \{0,1\}^{\gamma_2}$.
\end{itemize}

\item Input of $\mathcal{S}$: $\vecq = (q_1,\dots,q_d)\in \mathbb{U}^d$

\item Input of $\mathcal{R}$: $\vecw = (w_1,\dots,w_d)\in \mathbb{U}^d$

\item \textbf{Protocol:}

\begin{enumerate}

\item $\mathcal{R}$ generates $(pk,sk)\leftarrow \mathsf{KeyGen}(1^\kappa)$.
\item $\mathcal{R}$ computes ${W}_k^*=\{w_{k,1}^*,\dots,w_{k,l}^*\}:=\mathsf{PrefixTrie}(w_k-\delta,w_k+\delta)$ for $k\in[d]$. Then, it computes $E:=\mathsf{Encode}(\{(H_{\gamma_1}(k||w_{k,v}^*),\mathsf{Enc}_{pk}(0))\}_{k\in[d],v\in[l]})$.

\item $\mathcal{R}$ sends $(pk,E)$ to $\mathcal{S}$.

\item $\mathcal{S}$ computes ${Q}_k^*=\{q_{k,1}^*,\dots,q_{k,l'}^*\}:=\mathsf{PrefixPath}(q_k,\delta)$ and picks random $a_k\leftarrow \mathcal{P}$ for $k\in[d]$.

\item $\mathcal{S}$ computes $u_{k,v}:=\mathsf{Decode}(E,H_{\gamma_1}(k||q_{k,v}^*)), k\in[d],v\in[l']$. Then, it computes $u_{k,v}^*:=u_{k,v}\boxplus \mathsf{Enc}_{pk}(a_k), k\in[d],v\in[l']$.

\item $\mathcal{S}$ shuffles $\{u^*_{k,v}\}$ with $v$, and sends to $\mathcal{R}$ the shuffled set. $\mathcal{S}$ also sends $r:= H_{\gamma_2}(\sum_{k\in[d]}a_k)$ to $\mathcal{R}$.

\item $\mathcal{R}$ outputs $1$ if $\exists \mathbf{v} = (v_1,\dots,v_d)\in[l']^d$ s.t. $H_{\gamma_2}(\sum_{k\in[d]}\mathsf{Dec}_{sk}(u^*_{k,v_k}))=r$. Otherwise, $\mathcal{R}$ outputs $0$.

\end{enumerate}
\end{trivlist}
\end{minipage}
\end{framed}
\caption{Fuzzy Matching Protocol for $L_\infty$ Distance in \cite{DBLP:conf/ccs/DZL25}} 
\label{fig:pifmat-dzl25}
\end{figure}

\begin{trivlist}
    \item \textbf{Security Flaw.} The primary security flaw in their protocol occurs in step~5. Here, each value $u_{k,v}$ is derived from the sender's element as $u_{k,v} = \mathsf{Decode}(E, q_{k,v}^*)$. To ensure security, each $u_{k,v}$ must be masked with an independent random value, analogous to a one-time pad encryption. However, the protocol reuses a single random value $a_k$ to mask all $\{u_{k,v}\}_{v \in [l]}$ within the same dimension. This reuse is equivalent to key reuse in a one-time pad and therefore compromises the protocol's security.
    \item \textbf{A Concrete Attack.} We present a concrete attack. In step 6, the receiver obtains all ciphertexts $\{u^*_{k,v}\}_{k\in[d],v\in[l']}$. Decrypting each yields
    \[
    m_{k,v} := \mathsf{Dec}_{sk}(u_{k,v}^*) = \mathsf{Dec}_{sk}\bigl(\mathsf{Decode}(E, H_{\gamma_1}(k||q_{k,v}^*))\bigr) + a_k .
    \]
    Since the same mask $a_k$ is reused across all $v\in[l']$, the receiver can eliminate $a_k$ by computing differences between the $m_{k,v}$, even without knowing $a_k$. For instance,
    \[
    \Delta_1 := m_{k,2} - m_{k,1} = \mathsf{Dec}_{sk}\bigl(\mathsf{Decode}(E, H_{\gamma_1}(q_{k,2}^*))\bigr) - \mathsf{Dec}_{sk}\bigl(\mathsf{Decode}(E, H_{\gamma_1}(q_{k,1}^*))\bigr).
    \]
    Now $\Delta_1$ depends only on the sender's element $\vecq$, not on $a_k$. The receiver can then guess candidate values for $q_{k,1}^*$ and $q_{k,2}^*$, compute the corresponding right-hand side of the above equation, and check whether it matches $\Delta_1$. A successful match reveals information about the sender's private input $\vecq$.
    \item \textbf{Proof Flaw.} We then examine where their security proof breaks down. In their introduction, they assert that security holds because ``only one value in $\{u_{k,v} \boxplus a_k\}_{v \in [l]}$ encrypts $a_k$ for each $k$, while others encrypt random values.''

    For instance, in the case where the receiver's output is $0$, their proof (Theorem~4) simulates each ciphertext $u^*_{k,v}$ as a random ciphertext (see Step~$3'$ of the simulator in \emph{Security against a Corrupted Receiver}). They argue that due to the randomness of the OKVS, $u^*_{k,v}$ is indistinguishable from a random ciphertext (cf. hybrids $\mathcal{H}_0$--$\mathcal{H}'_0$ in the same proof).

    This reasoning misapplies the randomness property of OKVS. The formal randomness guarantee of an OKVS states that if a key is not in the encoded set, the decoded value appears random to any party \textit{that does not know the OKVS encoding $E$} (see Section~\ref{subsec:okvs} for the formal definition). In their protocol, however, the receiver itself generates $E$ and thus knows it completely. Consequently, from the receiver’s perspective, no randomness remains in the decoding result $u^*_{k,v}$—the receiver can directly check whether a query $q$ satisfies $u^* = \mathsf{Decode}(E, H(q))$.

    The situation parallels that of a Pseudo-Random Function (PRF). A function family is a PRF if an adversary with oracle access cannot distinguish it from a truly random function, provided the adversary does \textit{not} know the PRF key. If the key is revealed, the adversary trivially distinguishes the two. Here the OKVS encoding $E$ plays the role of the key; once it is known to the receiver, the randomness condition no longer applies. Hence, their security proof is invalid on this point.
\end{trivlist}

\subsection{Security Flaws in Distributed ID Generation}
\label{app:flaw-didg}

We analyze the security flaws in their distributed ID generation protocol. We first review their dIDG protocol, which is shown in Figure~\ref{fig:pididg-dzl25}. Their construction requires invoking a Private Index Search (PIS) functionality, which we have formally defined in Figure~\ref{fig:fpis}.

\begin{figure}[!hbth]
\begin{framed}
\begin{minipage}[center]{\textwidth}
\begin{trivlist}
\item \textbf{Parameters:} Sender $\mathcal{S}$, Receiver $\mathcal{R}$, vector length $n$.

\item \textbf{Functionality:}
\begin{itemize}
\item Wait for input $b$ from the sender $\mathcal{S}$.
\item Wait for input $\veca=(a_0,\dots,a_{n-1})$ from the receiver $\mathcal{R}$.
\item If $b=a_{i}$, define $\mathsf{idx}:=i$, otherwise, pick a random $\mathsf{idx}\leftarrow [0,n-1]$.
\item Give $\mathsf{idx}$ to the receiver $\mathcal{R}$.

\end{itemize}
\end{trivlist}
\end{minipage}
\end{framed}
\caption{Private Index Search Functionality $\mathcal{F}_{\mathsf{PIS}}$}
\label{fig:fpis}
\end{figure}

\begin{figure}[!hbth]
\begin{framed}
\begin{minipage}[center]{\textwidth}
\begin{trivlist}
\item \textbf{Parameters:} 
\begin{itemize}
\item Two parties: sender $\mathcal{S}$ and receiver $\mathcal{R}$, set size $m,n$, item universe $\mathbb{U}^d$, dimension $d$, distance threshold $\delta$.
\item An OKVS scheme $(\mathsf{Encode},\mathsf{Decode})$.
\item A prefix trie scheme $(\mathsf{PrefixTrie}, \mathsf{PrefixPath})$.
\item An AHE scheme $(\mathsf{KeyGen},\mathsf{Enc},\mathsf{Dec},\boxplus)$ with plaintext space $\mathcal{P}$ and ciphertext space $\mathcal{C}$.
\item Ideal $\mathcal{F}_{\mathsf{PIS}}$ primitive specified in Figure \ref{fig:fpis}. 
\end{itemize}

\item Input of $\mathcal{S}$: $Q=\{\vecq_1,\dots,\vecq_{m}\}\subset \mathbb{U}^d$

\item Input of $\mathcal{R}$: $W=\{\vecw_1,\dots,\vecw_{n}\}\subset \mathbb{U}^d$

\item \textbf{Protocol:}

\begin{enumerate}


\item $\mathcal{R}$ generates $(pk,sk)\leftarrow \mathsf{KeyGen}(1^\kappa)$.

\item $\mathcal{R}$ computes the prefix $\{w_{i,k,v}^*\}_{v\in [l]}:=\mathsf{PrefixTrie}(w_{i,k}-\delta,w_{i,k}+\delta)$ for $i\in[n],k\in[d]$.

\item $\mathcal{R}$ picks random $\randr_{i,k}\leftarrow \mathbb{F}$ and defines $B_{i,k}:=\{(k||w_{i,k,v}^*, \mathsf{Enc}_{pk}(\randr_{i,k})||\mathsf{Enc}_{pk}(0))\}_{v\in[l]}$ for $i\in[n],k\in [d]$. For $k\in[d],i_1\in[n]$: if $\exists i_2\in[i_1],v_1,v_2\in [l]$, s.t. $w_{i_1,k,v_1}^* = w_{i_2,k,v_2}^*$, then update $B_{i_2,k}:=\{(k||w_{i_2,k,v}^*,  \mathsf{Enc}_{pk}(\randr_{i_1,k})||\mathsf{Enc}_{pk}(0))\}_{v\in[l]}$. Let $B:= \cup_{i\in[n],k\in[d]}B_{i,k}$ and pad $B$ with dummy random key-value pairs to size of $dnl$. Then, $\mathcal{R}$ computes $id_{\vecw_i}:= \sum_{k\in [d]}B[k||w_{i,k,1}^*], i\in [n]$.

\item $\mathcal{R}$ computes $E:=\mathsf{Encode}(B)$. Then, $\mathcal{R}$ sends $E$ to $\mathcal{S}$.

\item $\mathcal{S}$ computes the prefix $\{q_{j,k,v}^*\}_{v\in [l']}:=\mathsf{PrefixPath}(q_{j,k},\delta)$ and picks random $\mathsf{mask}_{j,k},\mathsf{mask}_{j,k}'\leftarrow\mathcal{P}$ for $j\in[m],k\in[d]$.

\item $\mathcal{S}$ computes $u_{j,k,v}||v_{j,k,v} = \mathsf{Decode}(E,k||q_{j,k,v}^*),j\in[m],k\in[d],v\in[l']$. Then, it computes $u_{j,k,v}':=u_{j,k,v}\boxplus \mathsf{Enc}_{pk}(\mathsf{mask}_{j,k}), v_{j,k,v}':=v_{j,k,v}\boxplus \mathsf{Enc}_{pk}(\mathsf{mask}'_{j,k}), k\in[d],v\in[l']$.

\item $\mathcal{S}$ shuffles $\{u'_{j,k,v}\}$ and $\{v'_{j,k,v}\}$ with $j,v$, and sends to $\mathcal{R}$ the shuffled set. 

\item For $j\in[m],k\in[d]$, $\mathcal{S}$ and $\mathcal{R}$ invoke the PIS functionality $\mathcal{F}_{\mathsf{PIS}}$. $\mathcal{S}$ acts as the sender in PIS with input $\mathsf{mask}_{j,k}'$ and learns nothing. $\mathcal{R}$ acts as the receiver in PIS with input $\{\mathsf{Dec}_{sk}(v'_{j,k,v}))\}_{v\in[l']}$ and learns $v_{j,k}$. 

\item $\mathcal{R}$ computes and sends $w_j:=\sum_{k\in[d]}\mathsf{Dec}_{sk}(u_{j,k,v_{j,k}}')$ to $\mathcal{S}$, $j\in[m]$.

\item $\mathcal{S}$ outputs $id_{\vecq_j}:= w_j-\sum_{k\in[d]}\mathsf{mask}_{j,k},j\in[m]$.

\end{enumerate}
\end{trivlist}
\end{minipage}
\end{framed}
\caption{Distributed ID Generation Protocol in \cite{DBLP:conf/ccs/DZL25}} 
\label{fig:pididg-dzl25}
\end{figure}

\begin{trivlist}
    \item \textbf{Security Flaws.} Their dIDG protocol suffers from two security flaws. The first, already discussed in Section~\ref{app:flaw-fmat}, arises from the sender's reuse of a single $\mathsf{mask}_{j,k}$ to mask all $\{u_{j,k,v}\}_{v\in[l']}$; we do not elaborate on it further here.

    A second problem is that the IDs generated by their protocol fail to satisfy the randomness requirement defined for fuzzy mapping (see Section~\ref{subsec:fmap}). In other words, the ID set $\{id_{\vecq_j}\}_{j\in[m]}$ depends on the receiver’s set $W$, thereby leaking additional information about receiver's set. The lack of randomness stems from the fact that their ID definition incorporates only randomness contributed by the receiver, while the receiver’s OKVS encoding may assign identical random values to overlapping intervals (see step~3). This can lead to collisions of the form $\exists j_1 \neq j_2$ such that $id_{\vecq_{j_1}} = id_{\vecq_{j_2}}$, which reveals to the sender that some receiver's element lies near either $\mathbf{q}_{j_1}$ or $\mathbf{q}_{j_2}$. 

    \item \textbf{A Concrete Attack.} We present a concrete example to illustrate the leakage in their protocol. Let $m = n = 2$, $d = 2$, $\delta = 1$, and suppose the sender holds $Q = \{\vecq_1 = (1,0), \vecq_2 = (1,3)\}$ while the receiver holds $W = \{\vecw_1 = (1,1), \vecw_2 = (4,2)\}$. It is straightforward to verify that $Q$ and $W$ satisfy the separate assumption.

    Following the receiver's encoding procedure, in the first dimension the projection interval $[0,2]$ of $\vecw_1$ is assigned a random value $\randr_{1,1}$, and the interval $[3,5]$ of $\vecw_2$ is assigned $\randr_{2,1}$. In the second dimension, $\vecw_1$'s interval $[0,2]$ is assigned $\randr_{1,2}$, and $\vecw_2$'s interval $[1,3]$ is assigned $\randr_{2,2}$. However, because the intervals $[0,2]$ and $[1,3]$ overlap in the second dimension, the random value associated with $[1,3]$ is reassigned to $\randr_{1,2}$, meaning the entire combined interval $[0,3]$ is encoded with $\randr_{1,2}$.

    Consequently, the receiver defines the ID of its first element as $id_{\vecw_1} = \randr_{1,1} + \randr_{1,2}$ and the ID of its second element as $id_{\vecw_2} = \randr_{2,1} + \randr_{1,2}$.

    Now consider the sender. Its first element $\vecq_1 = (1,0)$ decodes to $\randr_{1,1}$ in the first dimension and $\randr_{1,2}$ in the second dimension, yielding $id_{\vecq_1} = \randr_{1,1} + \randr_{1,2}$. Its second element $\vecq_2 = (1,3)$ also decodes to $\randr_{1,1}$ and $\randr_{1,2}$, giving $id_{\vecq_2} = \randr_{1,1} + \randr_{1,2} = id_{\vecq_1}$. Observing that $id_{\vecq_1} = id_{\vecq_2}$, the sender learns that there exists a receiver element within distance $\delta$ of $\vecq_1$ or $\vecq_2$. In an FPSI protocol, the sender should obtain no information about the receiver's set, so this constitutes an unintended information leakage.

    \item \textbf{Proof Flaw.} We then examine where their security proof breaks down. The security proof for their dIDG protocol (Theorem 12) is not itself flawed (apart from the issue described in Section \ref{app:flaw-fmat}), because in that proof the simulator is given the actual inputs and outputs of the parties. For instance, when the sender is corrupted, the simulator receives $(Q, \id_Q)$ and can completely simulate the sender's view of the protocol.

    The actual error lies in the proof of their full FPSI protocol, namely Theorem 14. In that proof, to invoke the dIDG simulator, the authors simulate the identifiers using randomly generated IDs (see step $2'$ of the simulator under \textit{Security for corrupted Sender}). This step implicitly assumes that the generated IDs are uniformly random. However, as we have shown, their ID construction does not satisfy this randomness property. Consequently, this assumption invalidates their proof.
\end{trivlist}

\section{Extended Experiments}
\label{app:extend-exp}

\subsection{Comparison Results of FPSI for $L_\infty$ Distances in WAN Setting}
\label{app:wan}

This section presents an experimental comparison of our FPSI protocols for the $L_\infty$ distance against related works under a WAN setting with 400 Mbps bandwidth and 80 ms RTT latency.  
We include the protocols from \cite{DBLP:conf/ccs/DZL25,AC/BGMP25} in the comparison.  
The implementation in \cite{DBLP:conf/asiacrypt/GaoQLLW24} employs an in‑memory transport, specifically `coproto::LocalAsyncSocket::makePair()', which operates independently of actual network conditions. Consequently, it is unaffected by the latency and bandwidth constraints of our WAN setup and therefore cannot be meaningfully evaluated under this configuration.  

Following the methodology used for our LAN experiments, we conducted tests in both low‑dimensional ($d \in \{2,3,4\}$) and high‑dimensional ($d \in \{6,8,10\}$) cases. The results are presented in Table \ref{tab:infty-low-wan} and Table \ref{tab:infty-high-wan}, respectively.

\begin{table}[!htbp]
\renewcommand\arraystretch{1}
	\centering
 \resizebox{0.83\linewidth}{!}{
\begin{tabular}{|c|c|c|cccccccccc|}
\hline
 &
   &
   &
  \multicolumn{10}{c|}{Threshold $\delta$} \\ \cline{4-13} 
 &
   &
   &
  \multicolumn{2}{c|}{16} &
  \multicolumn{2}{c|}{32} &
  \multicolumn{2}{c|}{64} &
  \multicolumn{2}{c|}{128} &
  \multicolumn{2}{c|}{256} \\ \cline{4-13} 
\multirow{-3}{*}{\begin{tabular}[c]{@{}c@{}}Set Size\\ $m=n$\end{tabular}} &
  \multirow{-3}{*}{\begin{tabular}[c]{@{}c@{}}Dimension\\ $d$\end{tabular}} &
  \multirow{-3}{*}{Protocol} &
  \multicolumn{1}{c|}{Comm.} &
  \multicolumn{1}{c|}{Time} &
  \multicolumn{1}{c|}{Comm.} &
  \multicolumn{1}{c|}{Time} &
  \multicolumn{1}{c|}{Comm.} &
  \multicolumn{1}{c|}{Time} &
  \multicolumn{1}{c|}{Comm.} &
  \multicolumn{1}{c|}{Time} &
  \multicolumn{1}{c|}{Comm.} &
  Time \\ \hline
 &
   &
  \cite{AC/BGMP25} &
  \multicolumn{1}{c|}{2.882} &
  \multicolumn{1}{c|}{4.269} &
  \multicolumn{1}{c|}{3.349} &
  \multicolumn{1}{c|}{4.935} &
  \multicolumn{1}{c|}{3.521} &
  \multicolumn{1}{c|}{5.586} &
  \multicolumn{1}{c|}{4.115} &
  \multicolumn{1}{c|}{5.727} &
  \multicolumn{1}{c|}{4.286} &
  5.858 \\ \cline{3-13} 
 &
   &
  \cite{DBLP:conf/ccs/DZL25}-Low &
  \multicolumn{1}{c|}{7.930} &
  \multicolumn{1}{c|}{\cellcolor[HTML]{C7ECFF}1.849} &
  \multicolumn{1}{c|}{9.346} &
  \multicolumn{1}{c|}{\cellcolor[HTML]{C7ECFF}1.975} &
  \multicolumn{1}{c|}{10.72} &
  \multicolumn{1}{c|}{\cellcolor[HTML]{C7ECFF}2.088} &
  \multicolumn{1}{c|}{11.90} &
  \multicolumn{1}{c|}{\cellcolor[HTML]{C7ECFF}2.222} &
  \multicolumn{1}{c|}{13.46} &
  \cellcolor[HTML]{C7ECFF}2.382 \\ \cline{3-13} 
 &
   &
  $\Pi_{\mathsf{OE}\text{-1}}^{\mathsf{apart}}$ &
  \multicolumn{1}{c|}{2.237} &
  \multicolumn{1}{c|}{3.292} &
  \multicolumn{1}{c|}{2.391} &
  \multicolumn{1}{c|}{3.358} &
  \multicolumn{1}{c|}{2.589} &
  \multicolumn{1}{c|}{3.389} &
  \multicolumn{1}{c|}{2.814} &
  \multicolumn{1}{c|}{3.392} &
  \multicolumn{1}{c|}{2.985} &
  3.623 \\ \cline{3-13} 
 &
   &
  $\Pi_{\mathsf{OE}\text{-2}}^{\mathsf{apart}}$ &
  \multicolumn{1}{c|}{2.673} &
  \multicolumn{1}{c|}{3.329} &
  \multicolumn{1}{c|}{2.832} &
  \multicolumn{1}{c|}{3.391} &
  \multicolumn{1}{c|}{3.024} &
  \multicolumn{1}{c|}{3.398} &
  \multicolumn{1}{c|}{3.224} &
  \multicolumn{1}{c|}{3.448} &
  \multicolumn{1}{c|}{3.415} &
  3.666 \\ \cline{3-13} 
 &
   &
  $\Pi_{\mathsf{OA}}^{\mathsf{apart}}$ &
  \multicolumn{1}{c|}{\cellcolor[HTML]{C7ECFF}1.894} &
  \multicolumn{1}{c|}{2.321} &
  \multicolumn{1}{c|}{\cellcolor[HTML]{C7ECFF}2.035} &
  \multicolumn{1}{c|}{2.339} &
  \multicolumn{1}{c|}{\cellcolor[HTML]{C7ECFF}2.224} &
  \multicolumn{1}{c|}{2.391} &
  \multicolumn{1}{c|}{\cellcolor[HTML]{C7ECFF}2.395} &
  \multicolumn{1}{c|}{2.461} &
  \multicolumn{1}{c|}{\cellcolor[HTML]{C7ECFF}2.558} &
  2.549 \\ \cline{3-13} 
 &
   &
  $\Pi_{\mathsf{SE}}^{\mathsf{apart}}$ &
  \multicolumn{1}{c|}{2.195} &
  \multicolumn{1}{c|}{2.922} &
  \multicolumn{1}{c|}{2.391} &
  \multicolumn{1}{c|}{2.941} &
  \multicolumn{1}{c|}{2.593} &
  \multicolumn{1}{c|}{2.962} &
  \multicolumn{1}{c|}{2.853} &
  \multicolumn{1}{c|}{2.968} &
  \multicolumn{1}{c|}{3.057} &
  2.969 \\ \cline{3-13} 
 &
  \multirow{-7}{*}{2} &
  $\Pi_{\mathsf{SA}}^{\mathsf{apart}}$ &
  \multicolumn{1}{c|}{\cellcolor[HTML]{99BEFF}1.421} &
  \multicolumn{1}{c|}{\cellcolor[HTML]{99BEFF}1.748} &
  \multicolumn{1}{c|}{\cellcolor[HTML]{99BEFF}1.603} &
  \multicolumn{1}{c|}{\cellcolor[HTML]{99BEFF}1.768} &
  \multicolumn{1}{c|}{\cellcolor[HTML]{99BEFF}1.804} &
  \multicolumn{1}{c|}{\cellcolor[HTML]{99BEFF}1.812} &
  \multicolumn{1}{c|}{\cellcolor[HTML]{99BEFF}1.994} &
  \multicolumn{1}{c|}{\cellcolor[HTML]{99BEFF}1.833} &
  \multicolumn{1}{c|}{\cellcolor[HTML]{99BEFF}2.191} &
  \cellcolor[HTML]{99BEFF}1.849 \\ \cline{2-13} 
 &
   &
  \cite{AC/BGMP25} &
  \multicolumn{1}{c|}{5.573} &
  \multicolumn{1}{c|}{5.741} &
  \multicolumn{1}{c|}{9.274} &
  \multicolumn{1}{c|}{6.044} &
  \multicolumn{1}{c|}{9.550} &
  \multicolumn{1}{c|}{6.252} &
  \multicolumn{1}{c|}{15.50} &
  \multicolumn{1}{c|}{6.608} &
  \multicolumn{1}{c|}{15.75} &
  7.005 \\ \cline{3-13} 
 &
   &
  \cite{DBLP:conf/ccs/DZL25}-Low &
  \multicolumn{1}{c|}{40.94} &
  \multicolumn{1}{c|}{3.082} &
  \multicolumn{1}{c|}{{\color[HTML]{FE0000} 54.39}} &
  \multicolumn{1}{c|}{{\color[HTML]{FE0000} 3.602}} &
  \multicolumn{1}{c|}{{\color[HTML]{FE0000} 38.39}} &
  \multicolumn{1}{c|}{{\color[HTML]{FE0000} 3.427}} &
  \multicolumn{1}{c|}{48.45} &
  \multicolumn{1}{c|}{3.613} &
  \multicolumn{1}{c|}{61.85} &
  4.156 \\ \cline{3-13} 
 &
   &
  $\Pi_{\mathsf{OE}\text{-1}}^{\mathsf{apart}}$ &
  \multicolumn{1}{c|}{3.784} &
  \multicolumn{1}{c|}{3.398} &
  \multicolumn{1}{c|}{4.200} &
  \multicolumn{1}{c|}{3.526} &
  \multicolumn{1}{c|}{4.897} &
  \multicolumn{1}{c|}{3.562} &
  \multicolumn{1}{c|}{5.406} &
  \multicolumn{1}{c|}{3.625} &
  \multicolumn{1}{c|}{5.902} &
  3.682 \\ \cline{3-13} 
 &
   &
  $\Pi_{\mathsf{OE}\text{-2}}^{\mathsf{apart}}$ &
  \multicolumn{1}{c|}{4.176} &
  \multicolumn{1}{c|}{3.468} &
  \multicolumn{1}{c|}{4.666} &
  \multicolumn{1}{c|}{3.642} &
  \multicolumn{1}{c|}{5.333} &
  \multicolumn{1}{c|}{3.668} &
  \multicolumn{1}{c|}{5.875} &
  \multicolumn{1}{c|}{3.692} &
  \multicolumn{1}{c|}{6.348} &
  3.711 \\ \cline{3-13} 
 &
   &
  $\Pi_{\mathsf{OA}}^{\mathsf{apart}}$ &
  \multicolumn{1}{c|}{3.359} &
  \multicolumn{1}{c|}{\cellcolor[HTML]{C7ECFF}2.459} &
  \multicolumn{1}{c|}{3.786} &
  \multicolumn{1}{c|}{\cellcolor[HTML]{C7ECFF}2.706} &
  \multicolumn{1}{c|}{4.48} &
  \multicolumn{1}{c|}{\cellcolor[HTML]{C7ECFF}2.792} &
  \multicolumn{1}{c|}{4.862} &
  \multicolumn{1}{c|}{\cellcolor[HTML]{C7ECFF}2.859} &
  \multicolumn{1}{c|}{5.374} &
  \cellcolor[HTML]{C7ECFF}2.878 \\ \cline{3-13} 
 &
   &
  $\Pi_{\mathsf{SE}}^{\mathsf{apart}}$ &
  \multicolumn{1}{c|}{\cellcolor[HTML]{C7ECFF}3.334} &
  \multicolumn{1}{c|}{2.935} &
  \multicolumn{1}{c|}{\cellcolor[HTML]{C7ECFF}3.727} &
  \multicolumn{1}{c|}{2.961} &
  \multicolumn{1}{c|}{\cellcolor[HTML]{C7ECFF}4.310} &
  \multicolumn{1}{c|}{2.971} &
  \multicolumn{1}{c|}{\cellcolor[HTML]{C7ECFF}4.812} &
  \multicolumn{1}{c|}{2.978} &
  \multicolumn{1}{c|}{\cellcolor[HTML]{C7ECFF}5.190} &
  3.019 \\ \cline{3-13} 
 &
  \multirow{-7}{*}{3} &
  $\Pi_{\mathsf{SA}}^{\mathsf{apart}}$ &
  \multicolumn{1}{c|}{\cellcolor[HTML]{99BEFF}2.497} &
  \multicolumn{1}{c|}{\cellcolor[HTML]{99BEFF}1.953} &
  \multicolumn{1}{c|}{\cellcolor[HTML]{99BEFF}2.876} &
  \multicolumn{1}{c|}{\cellcolor[HTML]{99BEFF}1.987} &
  \multicolumn{1}{c|}{\cellcolor[HTML]{99BEFF}3.441} &
  \multicolumn{1}{c|}{\cellcolor[HTML]{99BEFF}2.075} &
  \multicolumn{1}{c|}{\cellcolor[HTML]{99BEFF}3.813} &
  \multicolumn{1}{c|}{\cellcolor[HTML]{99BEFF}2.101} &
  \multicolumn{1}{c|}{\cellcolor[HTML]{99BEFF}4.227} &
  \cellcolor[HTML]{99BEFF}2.181 \\ \cline{2-13} 
 &
   &
  \cite{AC/BGMP25} &
  \multicolumn{1}{c|}{19.02} &
  \multicolumn{1}{c|}{8.806} &
  \multicolumn{1}{c|}{52.07} &
  \multicolumn{1}{c|}{11.53} &
  \multicolumn{1}{c|}{52.34} &
  \multicolumn{1}{c|}{15.13} &
  \multicolumn{1}{c|}{121.7} &
  \multicolumn{1}{c|}{18.46} &
  \multicolumn{1}{c|}{122.1} &
  23.93 \\ \cline{3-13} 
 &
   &
  \cite{DBLP:conf/ccs/DZL25}-Low &
  \multicolumn{1}{c|}{108.2} &
  \multicolumn{1}{c|}{6.178} &
  \multicolumn{1}{c|}{{\color[HTML]{FE0000} 144.0}} &
  \multicolumn{1}{c|}{{\color[HTML]{FE0000} 7.791}} &
  \multicolumn{1}{c|}{{\color[HTML]{FE0000} 100.3}} &
  \multicolumn{1}{c|}{{\color[HTML]{FE0000} 5.974}} &
  \multicolumn{1}{c|}{127.2} &
  \multicolumn{1}{c|}{7.745} &
  \multicolumn{1}{c|}{162.9} &
  9.024 \\ \cline{3-13} 
 &
   &
  $\Pi_{\mathsf{OE}\text{-1}}^{\mathsf{apart}}$ &
  \multicolumn{1}{c|}{7.530} &
  \multicolumn{1}{c|}{3.674} &
  \multicolumn{1}{c|}{8.720} &
  \multicolumn{1}{c|}{3.719} &
  \multicolumn{1}{c|}{9.830} &
  \multicolumn{1}{c|}{3.986} &
  \multicolumn{1}{c|}{11.42} &
  \multicolumn{1}{c|}{4.019} &
  \multicolumn{1}{c|}{12.51} &
  4.126 \\ \cline{3-13} 
 &
   &
  $\Pi_{\mathsf{OE}\text{-2}}^{\mathsf{apart}}$ &
  \multicolumn{1}{c|}{7.949} &
  \multicolumn{1}{c|}{3.742} &
  \multicolumn{1}{c|}{9.11} &
  \multicolumn{1}{c|}{3.799} &
  \multicolumn{1}{c|}{10.26} &
  \multicolumn{1}{c|}{3.972} &
  \multicolumn{1}{c|}{11.86} &
  \multicolumn{1}{c|}{4.091} &
  \multicolumn{1}{c|}{13.02} &
  4.154 \\ \cline{3-13} 
 &
   &
  $\Pi_{\mathsf{OA}}^{\mathsf{apart}}$ &
  \multicolumn{1}{c|}{6.976} &
  \multicolumn{1}{c|}{\cellcolor[HTML]{C7ECFF}2.919} &
  \multicolumn{1}{c|}{8.104} &
  \multicolumn{1}{c|}{\cellcolor[HTML]{C7ECFF}2.986} &
  \multicolumn{1}{c|}{9.330} &
  \multicolumn{1}{c|}{\cellcolor[HTML]{C7ECFF}3.075} &
  \multicolumn{1}{c|}{10.82} &
  \multicolumn{1}{c|}{\cellcolor[HTML]{C7ECFF}3.281} &
  \multicolumn{1}{c|}{11.88} &
  \cellcolor[HTML]{C7ECFF}3.361 \\ \cline{3-13} 
 &
   &
  $\Pi_{\mathsf{SE}}^{\mathsf{apart}}$ &
  \multicolumn{1}{c|}{\cellcolor[HTML]{C7ECFF}5.631} &
  \multicolumn{1}{c|}{3.051} &
  \multicolumn{1}{c|}{\cellcolor[HTML]{C7ECFF}6.443} &
  \multicolumn{1}{c|}{3.075} &
  \multicolumn{1}{c|}{\cellcolor[HTML]{C7ECFF}7.227} &
  \multicolumn{1}{c|}{3.099} &
  \multicolumn{1}{c|}{\cellcolor[HTML]{C7ECFF}8.175} &
  \multicolumn{1}{c|}{3.317} &
  \multicolumn{1}{c|}{\cellcolor[HTML]{C7ECFF}8.982} &
  3.381 \\ \cline{3-13} 
\multirow{-21}{*}{$2^8$} &
  \multirow{-7}{*}{4} &
  $\Pi_{\mathsf{SA}}^{\mathsf{apart}}$ &
  \multicolumn{1}{c|}{\cellcolor[HTML]{99BEFF}4.704} &
  \multicolumn{1}{c|}{\cellcolor[HTML]{99BEFF}2.166} &
  \multicolumn{1}{c|}{\cellcolor[HTML]{99BEFF}5.557} &
  \multicolumn{1}{c|}{\cellcolor[HTML]{99BEFF}2.212} &
  \multicolumn{1}{c|}{\cellcolor[HTML]{99BEFF}6.329} &
  \multicolumn{1}{c|}{\cellcolor[HTML]{99BEFF}2.252} &
  \multicolumn{1}{c|}{\cellcolor[HTML]{99BEFF}7.110} &
  \multicolumn{1}{c|}{\cellcolor[HTML]{99BEFF}2.441} &
  \multicolumn{1}{c|}{\cellcolor[HTML]{99BEFF}7.917} &
  \cellcolor[HTML]{99BEFF}2.492 \\ \hline
 &
   &
  \cite{AC/BGMP25} &
  \multicolumn{1}{c|}{27.67} &
  \multicolumn{1}{c|}{6.022} &
  \multicolumn{1}{c|}{35.55} &
  \multicolumn{1}{c|}{6.617} &
  \multicolumn{1}{c|}{38.17} &
  \multicolumn{1}{c|}{7.263} &
  \multicolumn{1}{c|}{47.55} &
  \multicolumn{1}{c|}{8.416} &
  \multicolumn{1}{c|}{50.18} &
  10.29 \\ \cline{3-13} 
 &
   &
  \cite{DBLP:conf/ccs/DZL25}-Low &
  \multicolumn{1}{c|}{126.3} &
  \multicolumn{1}{c|}{16.17} &
  \multicolumn{1}{c|}{149.0} &
  \multicolumn{1}{c|}{19.02} &
  \multicolumn{1}{c|}{171.5} &
  \multicolumn{1}{c|}{22.25} &
  \multicolumn{1}{c|}{189.1} &
  \multicolumn{1}{c|}{25.92} &
  \multicolumn{1}{c|}{207.9} &
  29.17 \\ \cline{3-13} 
 &
   &
  $\Pi_{\mathsf{OE}\text{-1}}^{\mathsf{apart}}$ &
  \multicolumn{1}{c|}{19.50} &
  \multicolumn{1}{c|}{5.347} &
  \multicolumn{1}{c|}{22.69} &
  \multicolumn{1}{c|}{5.491} &
  \multicolumn{1}{c|}{25.86} &
  \multicolumn{1}{c|}{5.878} &
  \multicolumn{1}{c|}{29.62} &
  \multicolumn{1}{c|}{6.206} &
  \multicolumn{1}{c|}{32.89} &
  6.219 \\ \cline{3-13} 
 &
   &
  $\Pi_{\mathsf{OE}\text{-2}}^{\mathsf{apart}}$ &
  \multicolumn{1}{c|}{19.97} &
  \multicolumn{1}{c|}{5.161} &
  \multicolumn{1}{c|}{23.19} &
  \multicolumn{1}{c|}{5.365} &
  \multicolumn{1}{c|}{26.33} &
  \multicolumn{1}{c|}{5.746} &
  \multicolumn{1}{c|}{30.05} &
  \multicolumn{1}{c|}{6.067} &
  \multicolumn{1}{c|}{33.33} &
  6.271 \\ \cline{3-13} 
 &
   &
  $\Pi_{\mathsf{OA}}^{\mathsf{apart}}$ &
  \multicolumn{1}{c|}{\cellcolor[HTML]{99BEFF}17.12} &
  \multicolumn{1}{c|}{\cellcolor[HTML]{C7ECFF}3.918} &
  \multicolumn{1}{c|}{\cellcolor[HTML]{99BEFF}20.20} &
  \multicolumn{1}{c|}{\cellcolor[HTML]{C7ECFF}4.318} &
  \multicolumn{1}{c|}{\cellcolor[HTML]{99BEFF}23.14} &
  \multicolumn{1}{c|}{\cellcolor[HTML]{C7ECFF}4.558} &
  \multicolumn{1}{c|}{\cellcolor[HTML]{99BEFF}26.13} &
  \multicolumn{1}{c|}{\cellcolor[HTML]{C7ECFF}4.844} &
  \multicolumn{1}{c|}{\cellcolor[HTML]{99BEFF}29.17} &
  \cellcolor[HTML]{C7ECFF}5.139 \\ \cline{3-13} 
 &
   &
  $\Pi_{\mathsf{SE}}^{\mathsf{apart}}$ &
  \multicolumn{1}{c|}{21.49} &
  \multicolumn{1}{c|}{4.699} &
  \multicolumn{1}{c|}{24.89} &
  \multicolumn{1}{c|}{4.936} &
  \multicolumn{1}{c|}{28.29} &
  \multicolumn{1}{c|}{5.218} &
  \multicolumn{1}{c|}{32.23} &
  \multicolumn{1}{c|}{5.641} &
  \multicolumn{1}{c|}{35.68} &
  5.916 \\ \cline{3-13} 
 &
  \multirow{-7}{*}{2} &
  $\Pi_{\mathsf{SA}}^{\mathsf{apart}}$ &
  \multicolumn{1}{c|}{\cellcolor[HTML]{C7ECFF}18.67} &
  \multicolumn{1}{c|}{\cellcolor[HTML]{99BEFF}3.604} &
  \multicolumn{1}{c|}{\cellcolor[HTML]{C7ECFF}21.91} &
  \multicolumn{1}{c|}{\cellcolor[HTML]{99BEFF}3.879} &
  \multicolumn{1}{c|}{\cellcolor[HTML]{C7ECFF}25.10} &
  \multicolumn{1}{c|}{\cellcolor[HTML]{99BEFF}4.185} &
  \multicolumn{1}{c|}{\cellcolor[HTML]{C7ECFF}28.27} &
  \multicolumn{1}{c|}{\cellcolor[HTML]{99BEFF}4.601} &
  \multicolumn{1}{c|}{\cellcolor[HTML]{C7ECFF}31.60} &
  \cellcolor[HTML]{99BEFF}4.891 \\ \cline{2-13} 
 &
   &
  \cite{AC/BGMP25} &
  \multicolumn{1}{c|}{71.11} &
  \multicolumn{1}{c|}{10.03} &
  \multicolumn{1}{c|}{130.6} &
  \multicolumn{1}{c|}{14.09} &
  \multicolumn{1}{c|}{134.5} &
  \multicolumn{1}{c|}{15.58} &
  \multicolumn{1}{c|}{229.9} &
  \multicolumn{1}{c|}{21.21} &
  \multicolumn{1}{c|}{233.9} &
  28.28 \\ \cline{3-13} 
 &
   &
  \cite{DBLP:conf/ccs/DZL25}-Low &
  \multicolumn{1}{c|}{655.3} &
  \multicolumn{1}{c|}{36.44} &
  \multicolumn{1}{c|}{{\color[HTML]{FE0000} 869.9}} &
  \multicolumn{1}{c|}{{\color[HTML]{FE0000} 45.82}} &
  \multicolumn{1}{c|}{{\color[HTML]{FE0000} 614.0}} &
  \multicolumn{1}{c|}{{\color[HTML]{FE0000} 39.51}} &
  \multicolumn{1}{c|}{775.2} &
  \multicolumn{1}{c|}{46.39} &
  \multicolumn{1}{c|}{989.7} &
  56.91 \\ \cline{3-13} 
 &
   &
  $\Pi_{\mathsf{OE}\text{-1}}^{\mathsf{apart}}$ &
  \multicolumn{1}{c|}{44.79} &
  \multicolumn{1}{c|}{7.185} &
  \multicolumn{1}{c|}{52.57} &
  \multicolumn{1}{c|}{8.051} &
  \multicolumn{1}{c|}{60.35} &
  \multicolumn{1}{c|}{8.954} &
  \multicolumn{1}{c|}{69.33} &
  \multicolumn{1}{c|}{9.771} &
  \multicolumn{1}{c|}{77.42} &
  10.66 \\ \cline{3-13} 
 &
   &
  $\Pi_{\mathsf{OE}\text{-2}}^{\mathsf{apart}}$ &
  \multicolumn{1}{c|}{45.33} &
  \multicolumn{1}{c|}{7.161} &
  \multicolumn{1}{c|}{53.09} &
  \multicolumn{1}{c|}{7.919} &
  \multicolumn{1}{c|}{60.86} &
  \multicolumn{1}{c|}{8.762} &
  \multicolumn{1}{c|}{69.84} &
  \multicolumn{1}{c|}{9.678} &
  \multicolumn{1}{c|}{77.75} &
  10.41 \\ \cline{3-13} 
 &
   &
  $\Pi_{\mathsf{OA}}^{\mathsf{apart}}$ &
  \multicolumn{1}{c|}{41.51} &
  \multicolumn{1}{c|}{\cellcolor[HTML]{C7ECFF}6.251} &
  \multicolumn{1}{c|}{49.06} &
  \multicolumn{1}{c|}{\cellcolor[HTML]{C7ECFF}7.054} &
  \multicolumn{1}{c|}{56.50} &
  \multicolumn{1}{c|}{\cellcolor[HTML]{C7ECFF}7.838} &
  \multicolumn{1}{c|}{64.29} &
  \multicolumn{1}{c|}{\cellcolor[HTML]{C7ECFF}8.602} &
  \multicolumn{1}{c|}{75.13} &
  \cellcolor[HTML]{C7ECFF}9.307 \\ \cline{3-13} 
 &
   &
  $\Pi_{\mathsf{SE}}^{\mathsf{apart}}$ &
  \multicolumn{1}{c|}{\cellcolor[HTML]{C7ECFF}39.70} &
  \multicolumn{1}{c|}{6.581} &
  \multicolumn{1}{c|}{\cellcolor[HTML]{C7ECFF}46.25} &
  \multicolumn{1}{c|}{7.353} &
  \multicolumn{1}{c|}{\cellcolor[HTML]{C7ECFF}52.89} &
  \multicolumn{1}{c|}{8.137} &
  \multicolumn{1}{c|}{\cellcolor[HTML]{C7ECFF}60.58} &
  \multicolumn{1}{c|}{9.039} &
  \multicolumn{1}{c|}{\cellcolor[HTML]{C7ECFF}67.34} &
  9.855 \\ \cline{3-13} 
 &
  \multirow{-7}{*}{3} &
  $\Pi_{\mathsf{SA}}^{\mathsf{apart}}$ &
  \multicolumn{1}{c|}{\cellcolor[HTML]{99BEFF}35.99} &
  \multicolumn{1}{c|}{\cellcolor[HTML]{99BEFF}5.663} &
  \multicolumn{1}{c|}{\cellcolor[HTML]{99BEFF}42.29} &
  \multicolumn{1}{c|}{\cellcolor[HTML]{99BEFF}6.438} &
  \multicolumn{1}{c|}{\cellcolor[HTML]{99BEFF}48.71} &
  \multicolumn{1}{c|}{\cellcolor[HTML]{99BEFF}7.149} &
  \multicolumn{1}{c|}{\cellcolor[HTML]{99BEFF}55.15} &
  \multicolumn{1}{c|}{\cellcolor[HTML]{99BEFF}7.871} &
  \multicolumn{1}{c|}{\cellcolor[HTML]{99BEFF}61.56} &
  \cellcolor[HTML]{99BEFF}8.626 \\ \cline{2-13} 
 &
   &
  \cite{AC/BGMP25} &
  \multicolumn{1}{c|}{284.2} &
  \multicolumn{1}{c|}{46.15} &
  \multicolumn{1}{c|}{814.3} &
  \multicolumn{1}{c|}{84.53} &
  \multicolumn{1}{c|}{819.6} &
  \multicolumn{1}{c|}{129.1} &
  \multicolumn{1}{c|}{1932} &
  \multicolumn{1}{c|}{188.7} &
  \multicolumn{1}{c|}{1937} &
  300.3 \\ \cline{3-13} 
 &
   &
  \cite{DBLP:conf/ccs/DZL25}-Low &
  \multicolumn{1}{c|}{1732} &
  \multicolumn{1}{c|}{86.11} &
  \multicolumn{1}{c|}{{\color[HTML]{FE0000} 2304}} &
  \multicolumn{1}{c|}{{\color[HTML]{FE0000} 111.8}} &
  \multicolumn{1}{c|}{{\color[HTML]{FE0000} 1602}} &
  \multicolumn{1}{c|}{{\color[HTML]{FE0000} 86.87}} &
  \multicolumn{1}{c|}{2033} &
  \multicolumn{1}{c|}{107.1} &
  \multicolumn{1}{c|}{2604} &
  132.9 \\ \cline{3-13} 
 &
   &
  $\Pi_{\mathsf{OE}\text{-1}}^{\mathsf{apart}}$ &
  \multicolumn{1}{c|}{102.2} &
  \multicolumn{1}{c|}{12.89} &
  \multicolumn{1}{c|}{121.5} &
  \multicolumn{1}{c|}{14.73} &
  \multicolumn{1}{c|}{140.1} &
  \multicolumn{1}{c|}{16.72} &
  \multicolumn{1}{c|}{160.6} &
  \multicolumn{1}{c|}{18.69} &
  \multicolumn{1}{c|}{179.9} &
  20.56 \\ \cline{3-13} 
 &
   &
  $\Pi_{\mathsf{OE}\text{-2}}^{\mathsf{apart}}$ &
  \multicolumn{1}{c|}{103.0} &
  \multicolumn{1}{c|}{12.83} &
  \multicolumn{1}{c|}{121.7} &
  \multicolumn{1}{c|}{14.82} &
  \multicolumn{1}{c|}{140.7} &
  \multicolumn{1}{c|}{16.64} &
  \multicolumn{1}{c|}{161.3} &
  \multicolumn{1}{c|}{18.59} &
  \multicolumn{1}{c|}{180.4} &
  20.73 \\ \cline{3-13} 
 &
   &
  $\Pi_{\mathsf{OA}}^{\mathsf{apart}}$ &
  \multicolumn{1}{c|}{97.95} &
  \multicolumn{1}{c|}{11.75} &
  \multicolumn{1}{c|}{116.7} &
  \multicolumn{1}{c|}{13.64} &
  \multicolumn{1}{c|}{135.1} &
  \multicolumn{1}{c|}{15.63} &
  \multicolumn{1}{c|}{154.3} &
  \multicolumn{1}{c|}{17.58} &
  \multicolumn{1}{c|}{172.8} &
  19.53 \\ \cline{3-13} 
 &
   &
  $\Pi_{\mathsf{SE}}^{\mathsf{apart}}$ &
  \multicolumn{1}{c|}{\cellcolor[HTML]{C7ECFF}74.07} &
  \multicolumn{1}{c|}{\cellcolor[HTML]{C7ECFF}11.41} &
  \multicolumn{1}{c|}{\cellcolor[HTML]{C7ECFF}87.29} &
  \multicolumn{1}{c|}{\cellcolor[HTML]{C7ECFF}13.31} &
  \multicolumn{1}{c|}{\cellcolor[HTML]{C7ECFF}100.3} &
  \multicolumn{1}{c|}{\cellcolor[HTML]{C7ECFF}15.02} &
  \multicolumn{1}{c|}{\cellcolor[HTML]{C7ECFF}114.7} &
  \multicolumn{1}{c|}{\cellcolor[HTML]{C7ECFF}16.95} &
  \multicolumn{1}{c|}{\cellcolor[HTML]{C7ECFF}127.8} &
  \cellcolor[HTML]{C7ECFF}19.41 \\ \cline{3-13} 
\multirow{-21}{*}{$2^{12}$} &
  \multirow{-7}{*}{4} &
  $\Pi_{\mathsf{SA}}^{\mathsf{apart}}$ &
  \multicolumn{1}{c|}{\cellcolor[HTML]{99BEFF}69.30} &
  \multicolumn{1}{c|}{\cellcolor[HTML]{99BEFF}10.73} &
  \multicolumn{1}{c|}{\cellcolor[HTML]{99BEFF}82.28} &
  \multicolumn{1}{c|}{\cellcolor[HTML]{99BEFF}12.15} &
  \multicolumn{1}{c|}{\cellcolor[HTML]{99BEFF}94.78} &
  \multicolumn{1}{c|}{\cellcolor[HTML]{99BEFF}14.02} &
  \multicolumn{1}{c|}{\cellcolor[HTML]{99BEFF}107.9} &
  \multicolumn{1}{c|}{\cellcolor[HTML]{99BEFF}15.72} &
  \multicolumn{1}{c|}{\cellcolor[HTML]{99BEFF}120.5} &
  \cellcolor[HTML]{99BEFF}17.49 \\ \hline
 &
   &
  \cite{AC/BGMP25} &
  \multicolumn{1}{c|}{416.0} &
  \multicolumn{1}{c|}{\cellcolor[HTML]{99BEFF}30.76} &
  \multicolumn{1}{c|}{542.1} &
  \multicolumn{1}{c|}{\cellcolor[HTML]{C7ECFF}38.52} &
  \multicolumn{1}{c|}{584.1} &
  \multicolumn{1}{c|}{50.15} &
  \multicolumn{1}{c|}{734.2} &
  \multicolumn{1}{c|}{68.63} &
  \multicolumn{1}{c|}{-} &
  - \\ \cline{3-13} 
 &
   &
  \cite{DBLP:conf/ccs/DZL25}-Low &
  \multicolumn{1}{c|}{1946} &
  \multicolumn{1}{c|}{250.6} &
  \multicolumn{1}{c|}{2262} &
  \multicolumn{1}{c|}{298.7} &
  \multicolumn{1}{c|}{2614} &
  \multicolumn{1}{c|}{348.5} &
  \multicolumn{1}{c|}{2953} &
  \multicolumn{1}{c|}{387.8} &
  \multicolumn{1}{c|}{3295} &
  438.6 \\ \cline{3-13} 
 &
   &
  $\Pi_{\mathsf{OE}\text{-1}}^{\mathsf{apart}}$ &
  \multicolumn{1}{c|}{\cellcolor[HTML]{C7ECFF}297.7} &
  \multicolumn{1}{c|}{35.62} &
  \multicolumn{1}{c|}{\cellcolor[HTML]{C7ECFF}348.3} &
  \multicolumn{1}{c|}{40.54} &
  \multicolumn{1}{c|}{\cellcolor[HTML]{C7ECFF}398.8} &
  \multicolumn{1}{c|}{47.29} &
  \multicolumn{1}{c|}{460.0} &
  \multicolumn{1}{c|}{53.87} &
  \multicolumn{1}{c|}{510.8} &
  59.69 \\ \cline{3-13} 
 &
   &
  $\Pi_{\mathsf{OE}\text{-2}}^{\mathsf{apart}}$ &
  \multicolumn{1}{c|}{298.3} &
  \multicolumn{1}{c|}{34.51} &
  \multicolumn{1}{c|}{348.9} &
  \multicolumn{1}{c|}{39.81} &
  \multicolumn{1}{c|}{399.4} &
  \multicolumn{1}{c|}{\cellcolor[HTML]{C7ECFF}46.06} &
  \multicolumn{1}{c|}{460.8} &
  \multicolumn{1}{c|}{\cellcolor[HTML]{C7ECFF}53.13} &
  \multicolumn{1}{c|}{511.6} &
  \cellcolor[HTML]{C7ECFF}58.98 \\ \cline{3-13} 
 &
   &
  $\Pi_{\mathsf{OA}}^{\mathsf{apart}}$ &
  \multicolumn{1}{c|}{\cellcolor[HTML]{99BEFF}262.0} &
  \multicolumn{1}{c|}{\cellcolor[HTML]{C7ECFF}32.35} &
  \multicolumn{1}{c|}{\cellcolor[HTML]{99BEFF}310.1} &
  \multicolumn{1}{c|}{\cellcolor[HTML]{99BEFF}37.59} &
  \multicolumn{1}{c|}{\cellcolor[HTML]{99BEFF}357.8} &
  \multicolumn{1}{c|}{\cellcolor[HTML]{99BEFF}43.09} &
  \multicolumn{1}{c|}{\cellcolor[HTML]{99BEFF}406.4} &
  \multicolumn{1}{c|}{\cellcolor[HTML]{99BEFF}49.72} &
  \multicolumn{1}{c|}{\cellcolor[HTML]{99BEFF}454.3} &
  \cellcolor[HTML]{99BEFF}55.53 \\ \cline{3-13} 
 &
   &
  $\Pi_{\mathsf{SE}}^{\mathsf{apart}}$ &
  \multicolumn{1}{c|}{334.4} &
  \multicolumn{1}{c|}{37.37} &
  \multicolumn{1}{c|}{388.8} &
  \multicolumn{1}{c|}{43.22} &
  \multicolumn{1}{c|}{443.1} &
  \multicolumn{1}{c|}{49.84} &
  \multicolumn{1}{c|}{508.3} &
  \multicolumn{1}{c|}{56.92} &
  \multicolumn{1}{c|}{562.7} &
  63.52 \\ \cline{3-13} 
 &
  \multirow{-7}{*}{2} &
  $\Pi_{\mathsf{SA}}^{\mathsf{apart}}$ &
  \multicolumn{1}{c|}{299.2} &
  \multicolumn{1}{c|}{34.77} &
  \multicolumn{1}{c|}{351.0} &
  \multicolumn{1}{c|}{40.35} &
  \multicolumn{1}{c|}{403.0} &
  \multicolumn{1}{c|}{46.95} &
  \multicolumn{1}{c|}{\cellcolor[HTML]{C7ECFF}455.0} &
  \multicolumn{1}{c|}{53.62} &
  \multicolumn{1}{c|}{\cellcolor[HTML]{C7ECFF}506.8} &
  59.63 \\ \cline{2-13} 
 &
   &
  \cite{AC/BGMP25} &
  \multicolumn{1}{c|}{1111} &
  \multicolumn{1}{c|}{93.36} &
  \multicolumn{1}{c|}{2062} &
  \multicolumn{1}{c|}{143.7} &
  \multicolumn{1}{c|}{2125} &
  \multicolumn{1}{c|}{181.5} &
  \multicolumn{1}{c|}{-} &
  \multicolumn{1}{c|}{-} &
  \multicolumn{1}{c|}{-} &
  - \\ \cline{3-13} 
 &
   &
  \cite{DBLP:conf/ccs/DZL25}-Low &
  \multicolumn{1}{c|}{-} &
  \multicolumn{1}{c|}{-} &
  \multicolumn{1}{c|}{-} &
  \multicolumn{1}{c|}{-} &
  \multicolumn{1}{c|}{-} &
  \multicolumn{1}{c|}{-} &
  \multicolumn{1}{c|}{-} &
  \multicolumn{1}{c|}{-} &
  \multicolumn{1}{c|}{-} &
  - \\ \cline{3-13} 
 &
   &
  $\Pi_{\mathsf{OE}\text{-1}}^{\mathsf{apart}}$ &
  \multicolumn{1}{c|}{704.0} &
  \multicolumn{1}{c|}{81.89} &
  \multicolumn{1}{c|}{828.5} &
  \multicolumn{1}{c|}{98.51} &
  \multicolumn{1}{c|}{953.4} &
  \multicolumn{1}{c|}{111.7} &
  \multicolumn{1}{c|}{1098} &
  \multicolumn{1}{c|}{128.3} &
  \multicolumn{1}{c|}{1226} &
  144.6 \\ \cline{3-13} 
 &
   &
  $\Pi_{\mathsf{OE}\text{-2}}^{\mathsf{apart}}$ &
  \multicolumn{1}{c|}{704.7} &
  \multicolumn{1}{c|}{80.66} &
  \multicolumn{1}{c|}{828.8} &
  \multicolumn{1}{c|}{97.38} &
  \multicolumn{1}{c|}{953.5} &
  \multicolumn{1}{c|}{110.9} &
  \multicolumn{1}{c|}{1098} &
  \multicolumn{1}{c|}{127.3} &
  \multicolumn{1}{c|}{1226} &
  143.8 \\ \cline{3-13} 
 &
   &
  $\Pi_{\mathsf{OA}}^{\mathsf{apart}}$ &
  \multicolumn{1}{c|}{652.9} &
  \multicolumn{1}{c|}{\cellcolor[HTML]{C7ECFF}77.29} &
  \multicolumn{1}{c|}{773.1} &
  \multicolumn{1}{c|}{\cellcolor[HTML]{C7ECFF}94.43} &
  \multicolumn{1}{c|}{894.2} &
  \multicolumn{1}{c|}{\cellcolor[HTML]{C7ECFF}107.4} &
  \multicolumn{1}{c|}{1019} &
  \multicolumn{1}{c|}{\cellcolor[HTML]{C7ECFF}123.5} &
  \multicolumn{1}{c|}{1144} &
  \cellcolor[HTML]{C7ECFF}139.3 \\ \cline{3-13} 
 &
   &
  $\Pi_{\mathsf{SE}}^{\mathsf{apart}}$ &
  \multicolumn{1}{c|}{\cellcolor[HTML]{C7ECFF}629.1} &
  \multicolumn{1}{c|}{79.97} &
  \multicolumn{1}{c|}{\cellcolor[HTML]{C7ECFF}734.6} &
  \multicolumn{1}{c|}{96.75} &
  \multicolumn{1}{c|}{\cellcolor[HTML]{C7ECFF}840.8} &
  \multicolumn{1}{c|}{110.4} &
  \multicolumn{1}{c|}{\cellcolor[HTML]{C7ECFF}965.1} &
  \multicolumn{1}{c|}{126.7} &
  \multicolumn{1}{c|}{\cellcolor[HTML]{C7ECFF}1073} &
  142.6 \\ \cline{3-13} 
 &
  \multirow{-7}{*}{3} &
  $\Pi_{\mathsf{SA}}^{\mathsf{apart}}$ &
  \multicolumn{1}{c|}{\cellcolor[HTML]{99BEFF}578.1} &
  \multicolumn{1}{c|}{\cellcolor[HTML]{99BEFF}76.98} &
  \multicolumn{1}{c|}{\cellcolor[HTML]{99BEFF}680.3} &
  \multicolumn{1}{c|}{\cellcolor[HTML]{99BEFF}93.53} &
  \multicolumn{1}{c|}{\cellcolor[HTML]{99BEFF}782.6} &
  \multicolumn{1}{c|}{\cellcolor[HTML]{99BEFF}106.2} &
  \multicolumn{1}{c|}{\cellcolor[HTML]{99BEFF}887.1} &
  \multicolumn{1}{c|}{\cellcolor[HTML]{99BEFF}121.6} &
  \multicolumn{1}{c|}{\cellcolor[HTML]{99BEFF}991.1} &
  \cellcolor[HTML]{99BEFF}137.1 \\ \cline{2-13} 
 &
   &
  \cite{AC/BGMP25} &
  \multicolumn{1}{c|}{4517} &
  \multicolumn{1}{c|}{701.5} &
  \multicolumn{1}{c|}{-} &
  \multicolumn{1}{c|}{-} &
  \multicolumn{1}{c|}{-} &
  \multicolumn{1}{c|}{-} &
  \multicolumn{1}{c|}{-} &
  \multicolumn{1}{c|}{-} &
  \multicolumn{1}{c|}{-} &
  - \\ \cline{3-13} 
 &
   &
  \cite{DBLP:conf/ccs/DZL25}-Low &
  \multicolumn{1}{c|}{-} &
  \multicolumn{1}{c|}{-} &
  \multicolumn{1}{c|}{-} &
  \multicolumn{1}{c|}{-} &
  \multicolumn{1}{c|}{-} &
  \multicolumn{1}{c|}{-} &
  \multicolumn{1}{c|}{-} &
  \multicolumn{1}{c|}{-} &
  \multicolumn{1}{c|}{-} &
  - \\ \cline{3-13} 
 &
   &
  $\Pi_{\mathsf{OE}\text{-1}}^{\mathsf{apart}}$ &
  \multicolumn{1}{c|}{1628} &
  \multicolumn{1}{c|}{194.6} &
  \multicolumn{1}{c|}{1930} &
  \multicolumn{1}{c|}{233.2} &
  \multicolumn{1}{c|}{2237} &
  \multicolumn{1}{c|}{274.7} &
  \multicolumn{1}{c|}{2563} &
  \multicolumn{1}{c|}{317.2} &
  \multicolumn{1}{c|}{2867} &
  361.2 \\ \cline{3-13} 
 &
   &
  $\Pi_{\mathsf{OE}\text{-2}}^{\mathsf{apart}}$ &
  \multicolumn{1}{c|}{1629} &
  \multicolumn{1}{c|}{193.6} &
  \multicolumn{1}{c|}{1931} &
  \multicolumn{1}{c|}{232.3} &
  \multicolumn{1}{c|}{2235} &
  \multicolumn{1}{c|}{273.4} &
  \multicolumn{1}{c|}{2563} &
  \multicolumn{1}{c|}{316.5} &
  \multicolumn{1}{c|}{2870} &
  361.2 \\ \cline{3-13} 
 &
   &
  $\Pi_{\mathsf{OA}}^{\mathsf{apart}}$ &
  \multicolumn{1}{c|}{1560} &
  \multicolumn{1}{c|}{190.2} &
  \multicolumn{1}{c|}{1858} &
  \multicolumn{1}{c|}{228.9} &
  \multicolumn{1}{c|}{2158} &
  \multicolumn{1}{c|}{269.6} &
  \multicolumn{1}{c|}{2460} &
  \multicolumn{1}{c|}{311.7} &
  \multicolumn{1}{c|}{2760} &
  354.5 \\ \cline{3-13} 
 &
   &
  $\Pi_{\mathsf{SE}}^{\mathsf{apart}}$ &
  \multicolumn{1}{c|}{\cellcolor[HTML]{C7ECFF}1182} &
  \multicolumn{1}{c|}{\cellcolor[HTML]{C7ECFF}181.2} &
  \multicolumn{1}{c|}{\cellcolor[HTML]{C7ECFF}1391} &
  \multicolumn{1}{c|}{\cellcolor[HTML]{C7ECFF}217.6} &
  \multicolumn{1}{c|}{\cellcolor[HTML]{C7ECFF}1602} &
  \multicolumn{1}{c|}{\cellcolor[HTML]{C7ECFF}255.3} &
  \multicolumn{1}{c|}{\cellcolor[HTML]{C7ECFF}1835} &
  \multicolumn{1}{c|}{\cellcolor[HTML]{C7ECFF}295.7} &
  \multicolumn{1}{c|}{\cellcolor[HTML]{C7ECFF}2045} &
  \cellcolor[HTML]{C7ECFF}337.1 \\ \cline{3-13} 
\multirow{-21}{*}{$2^{16}$} &
  \multirow{-7}{*}{4} &
  $\Pi_{\mathsf{SA}}^{\mathsf{apart}}$ &
  \multicolumn{1}{c|}{\cellcolor[HTML]{99BEFF}1116} &
  \multicolumn{1}{c|}{\cellcolor[HTML]{99BEFF}177.1} &
  \multicolumn{1}{c|}{\cellcolor[HTML]{99BEFF}1320} &
  \multicolumn{1}{c|}{\cellcolor[HTML]{99BEFF}213.4} &
  \multicolumn{1}{c|}{\cellcolor[HTML]{99BEFF}1526} &
  \multicolumn{1}{c|}{\cellcolor[HTML]{99BEFF}250.2} &
  \multicolumn{1}{c|}{\cellcolor[HTML]{99BEFF}1732} &
  \multicolumn{1}{c|}{\cellcolor[HTML]{99BEFF}290.8} &
  \multicolumn{1}{c|}{\cellcolor[HTML]{99BEFF}1938} &
  \cellcolor[HTML]{99BEFF}329.3 \\ \hline
\end{tabular}
}
\caption{Communication cost (in MB) and running time (in seconds) comparing our protocols to \cite{AC/BGMP25,DBLP:conf/ccs/DZL25} in WAN setting. Cells with - denote trials that ran out of memory. The best result is highlighted in \textcolor[rgb]{0,0.3,1}{blue}, the second best in \textcolor[rgb]{0,0.7,1}{cyan}, and data in \textcolor[rgb]{1,0,0}{red} font indicates abnormal values.}
\label{tab:infty-low-wan}
\end{table}

\begin{table}[!htbp]
\renewcommand\arraystretch{1}
	\centering
 \resizebox{0.83\linewidth}{!}{
\begin{tabular}{|c|c|c|cccccccccc|}
\hline
 &
   &
   &
  \multicolumn{10}{c|}{Threshold $\delta$} \\ \cline{4-13} 
 &
   &
   &
  \multicolumn{2}{c|}{16} &
  \multicolumn{2}{c|}{32} &
  \multicolumn{2}{c|}{64} &
  \multicolumn{2}{c|}{128} &
  \multicolumn{2}{c|}{256} \\ \cline{4-13} 
\multirow{-3}{*}{\begin{tabular}[c]{@{}c@{}}Set Size\\ $m=n$\end{tabular}} &
  \multirow{-3}{*}{\begin{tabular}[c]{@{}c@{}}Dimension\\ $d$\end{tabular}} &
  \multirow{-3}{*}{Protocol} &
  \multicolumn{1}{c|}{Comm.} &
  \multicolumn{1}{c|}{Time} &
  \multicolumn{1}{c|}{Comm.} &
  \multicolumn{1}{c|}{Time} &
  \multicolumn{1}{c|}{Comm.} &
  \multicolumn{1}{c|}{Time} &
  \multicolumn{1}{c|}{Comm.} &
  \multicolumn{1}{c|}{Time} &
  \multicolumn{1}{c|}{Comm.} &
  Time \\ \hline
 &
   &
  \cite{DBLP:conf/ccs/DZL25}-High &
  \multicolumn{1}{c|}{65.14} &
  \multicolumn{1}{c|}{11.71} &
  \multicolumn{1}{c|}{70.16} &
  \multicolumn{1}{c|}{11.81} &
  \multicolumn{1}{c|}{75.26} &
  \multicolumn{1}{c|}{12.52} &
  \multicolumn{1}{c|}{79.40} &
  \multicolumn{1}{c|}{14.08} &
  \multicolumn{1}{c|}{121.0} &
  21.32 \\ \cline{3-13} 
 &
   &
  $\Pi_{\mathsf{OE}\text{-1}}^{\mathsf{sep}}$ &
  \multicolumn{1}{c|}{11.71} &
  \multicolumn{1}{c|}{4.909} &
  \multicolumn{1}{c|}{13.52} &
  \multicolumn{1}{c|}{5.117} &
  \multicolumn{1}{c|}{\cellcolor[HTML]{C7ECFF}15.09} &
  \multicolumn{1}{c|}{5.321} &
  \multicolumn{1}{c|}{\cellcolor[HTML]{C7ECFF}16.83} &
  \multicolumn{1}{c|}{5.601} &
  \multicolumn{1}{c|}{\cellcolor[HTML]{C7ECFF}18.39} &
  5.666 \\ \cline{3-13} 
 &
   &
  $\Pi_{\mathsf{OE}\text{-2}}^{\mathsf{sep}}$ &
  \multicolumn{1}{c|}{12.13} &
  \multicolumn{1}{c|}{5.026} &
  \multicolumn{1}{c|}{13.98} &
  \multicolumn{1}{c|}{5.236} &
  \multicolumn{1}{c|}{15.55} &
  \multicolumn{1}{c|}{5.552} &
  \multicolumn{1}{c|}{17.29} &
  \multicolumn{1}{c|}{5.627} &
  \multicolumn{1}{c|}{18.85} &
  5.718 \\ \cline{3-13} 
 &
   &
  $\Pi_{\mathsf{OA}}^{\mathsf{sep}}$ &
  \multicolumn{1}{c|}{\cellcolor[HTML]{99BEFF}11.10} &
  \multicolumn{1}{c|}{\cellcolor[HTML]{C7ECFF}3.984} &
  \multicolumn{1}{c|}{\cellcolor[HTML]{99BEFF}12.89} &
  \multicolumn{1}{c|}{\cellcolor[HTML]{C7ECFF}4.231} &
  \multicolumn{1}{c|}{\cellcolor[HTML]{99BEFF}14.43} &
  \multicolumn{1}{c|}{\cellcolor[HTML]{C7ECFF}4.512} &
  \multicolumn{1}{c|}{\cellcolor[HTML]{99BEFF}15.94} &
  \multicolumn{1}{c|}{\cellcolor[HTML]{C7ECFF}4.576} &
  \multicolumn{1}{c|}{\cellcolor[HTML]{99BEFF}17.45} &
  \cellcolor[HTML]{C7ECFF}4.642 \\ \cline{3-13} 
 &
   &
  $\Pi_{\mathsf{SE}}^{\mathsf{sep}}$ &
  \multicolumn{1}{c|}{12.46} &
  \multicolumn{1}{c|}{4.643} &
  \multicolumn{1}{c|}{14.53} &
  \multicolumn{1}{c|}{4.761} &
  \multicolumn{1}{c|}{16.46} &
  \multicolumn{1}{c|}{4.919} &
  \multicolumn{1}{c|}{18.46} &
  \multicolumn{1}{c|}{5.145} &
  \multicolumn{1}{c|}{20.29} &
  5.302 \\ \cline{3-13} 
 &
  \multirow{-6}{*}{6} &
  $\Pi_{\mathsf{SA}}^{\mathsf{sep}}$ &
  \multicolumn{1}{c|}{\cellcolor[HTML]{C7ECFF}11.41} &
  \multicolumn{1}{c|}{\cellcolor[HTML]{99BEFF}3.558} &
  \multicolumn{1}{c|}{\cellcolor[HTML]{C7ECFF}13.44} &
  \multicolumn{1}{c|}{\cellcolor[HTML]{99BEFF}3.699} &
  \multicolumn{1}{c|}{15.35} &
  \multicolumn{1}{c|}{\cellcolor[HTML]{99BEFF}3.898} &
  \multicolumn{1}{c|}{17.16} &
  \multicolumn{1}{c|}{\cellcolor[HTML]{99BEFF}4.118} &
  \multicolumn{1}{c|}{18.94} &
  \cellcolor[HTML]{99BEFF}4.369 \\ \cline{2-13} 
 &
   &
  \cite{DBLP:conf/ccs/DZL25}-High &
  \multicolumn{1}{c|}{86.86} &
  \multicolumn{1}{c|}{14.47} &
  \multicolumn{1}{c|}{93.55} &
  \multicolumn{1}{c|}{15.69} &
  \multicolumn{1}{c|}{100.3} &
  \multicolumn{1}{c|}{16.13} &
  \multicolumn{1}{c|}{105.9} &
  \multicolumn{1}{c|}{16.85} &
  \multicolumn{1}{c|}{161.3} &
  26.89 \\ \cline{3-13} 
 &
   &
  $\Pi_{\mathsf{OE}\text{-1}}^{\mathsf{sep}}$ &
  \multicolumn{1}{c|}{\cellcolor[HTML]{C7ECFF}15.07} &
  \multicolumn{1}{c|}{5.189} &
  \multicolumn{1}{c|}{\cellcolor[HTML]{C7ECFF}17.11} &
  \multicolumn{1}{c|}{5.619} &
  \multicolumn{1}{c|}{\cellcolor[HTML]{C7ECFF}19.15} &
  \multicolumn{1}{c|}{6.017} &
  \multicolumn{1}{c|}{\cellcolor[HTML]{C7ECFF}21.76} &
  \multicolumn{1}{c|}{6.035} &
  \multicolumn{1}{c|}{\cellcolor[HTML]{C7ECFF}23.96} &
  6.082 \\ \cline{3-13} 
 &
   &
  $\Pi_{\mathsf{OE}\text{-2}}^{\mathsf{sep}}$ &
  \multicolumn{1}{c|}{15.51} &
  \multicolumn{1}{c|}{5.252} &
  \multicolumn{1}{c|}{17.51} &
  \multicolumn{1}{c|}{5.478} &
  \multicolumn{1}{c|}{19.62} &
  \multicolumn{1}{c|}{5.861} &
  \multicolumn{1}{c|}{22.18} &
  \multicolumn{1}{c|}{5.904} &
  \multicolumn{1}{c|}{24.40} &
  6.295 \\ \cline{3-13} 
 &
   &
  $\Pi_{\mathsf{OA}}^{\mathsf{sep}}$ &
  \multicolumn{1}{c|}{\cellcolor[HTML]{99BEFF}14.32} &
  \multicolumn{1}{c|}{\cellcolor[HTML]{C7ECFF}4.561} &
  \multicolumn{1}{c|}{\cellcolor[HTML]{99BEFF}16.32} &
  \multicolumn{1}{c|}{\cellcolor[HTML]{C7ECFF}4.829} &
  \multicolumn{1}{c|}{\cellcolor[HTML]{99BEFF}18.33} &
  \multicolumn{1}{c|}{\cellcolor[HTML]{C7ECFF}4.952} &
  \multicolumn{1}{c|}{\cellcolor[HTML]{99BEFF}20.68} &
  \multicolumn{1}{c|}{\cellcolor[HTML]{C7ECFF}5.087} &
  \multicolumn{1}{c|}{\cellcolor[HTML]{99BEFF}22.86} &
  \cellcolor[HTML]{C7ECFF}5.389 \\ \cline{3-13} 
 &
   &
  $\Pi_{\mathsf{SE}}^{\mathsf{sep}}$ &
  \multicolumn{1}{c|}{16.41} &
  \multicolumn{1}{c|}{5.051} &
  \multicolumn{1}{c|}{18.80} &
  \multicolumn{1}{c|}{5.103} &
  \multicolumn{1}{c|}{21.22} &
  \multicolumn{1}{c|}{5.356} &
  \multicolumn{1}{c|}{23.99} &
  \multicolumn{1}{c|}{5.642} &
  \multicolumn{1}{c|}{26.61} &
  5.899 \\ \cline{3-13} 
 &
  \multirow{-6}{*}{8} &
  $\Pi_{\mathsf{SA}}^{\mathsf{sep}}$ &
  \multicolumn{1}{c|}{15.22} &
  \multicolumn{1}{c|}{\cellcolor[HTML]{99BEFF}3.739} &
  \multicolumn{1}{c|}{17.58} &
  \multicolumn{1}{c|}{\cellcolor[HTML]{99BEFF}4.097} &
  \multicolumn{1}{c|}{19.96} &
  \multicolumn{1}{c|}{\cellcolor[HTML]{99BEFF}4.549} &
  \multicolumn{1}{c|}{22.56} &
  \multicolumn{1}{c|}{\cellcolor[HTML]{99BEFF}4.696} &
  \multicolumn{1}{c|}{25.10} &
  \cellcolor[HTML]{99BEFF}4.824 \\ \cline{2-13} 
 &
   &
  \cite{DBLP:conf/ccs/DZL25}-High &
  \multicolumn{1}{c|}{108.5} &
  \multicolumn{1}{c|}{18.83} &
  \multicolumn{1}{c|}{116.9} &
  \multicolumn{1}{c|}{19.74} &
  \multicolumn{1}{c|}{125.3} &
  \multicolumn{1}{c|}{20.07} &
  \multicolumn{1}{c|}{132.3} &
  \multicolumn{1}{c|}{20.83} &
  \multicolumn{1}{c|}{201.6} &
  30.32 \\ \cline{3-13} 
 &
   &
  $\Pi_{\mathsf{OE}\text{-1}}^{\mathsf{sep}}$ &
  \multicolumn{1}{c|}{\cellcolor[HTML]{C7ECFF}17.99} &
  \multicolumn{1}{c|}{5.284} &
  \multicolumn{1}{c|}{\cellcolor[HTML]{C7ECFF}20.70} &
  \multicolumn{1}{c|}{5.871} &
  \multicolumn{1}{c|}{\cellcolor[HTML]{C7ECFF}23.63} &
  \multicolumn{1}{c|}{6.243} &
  \multicolumn{1}{c|}{\cellcolor[HTML]{C7ECFF}26.76} &
  \multicolumn{1}{c|}{6.536} &
  \multicolumn{1}{c|}{\cellcolor[HTML]{C7ECFF}30.01} &
  6.725 \\ \cline{3-13} 
 &
   &
  $\Pi_{\mathsf{OE}\text{-2}}^{\mathsf{sep}}$ &
  \multicolumn{1}{c|}{18.45} &
  \multicolumn{1}{c|}{5.609} &
  \multicolumn{1}{c|}{21.12} &
  \multicolumn{1}{c|}{6.253} &
  \multicolumn{1}{c|}{24.09} &
  \multicolumn{1}{c|}{6.253} &
  \multicolumn{1}{c|}{27.13} &
  \multicolumn{1}{c|}{6.478} &
  \multicolumn{1}{c|}{30.50} &
  6.759 \\ \cline{3-13} 
 &
   &
  $\Pi_{\mathsf{OA}}^{\mathsf{sep}}$ &
  \multicolumn{1}{c|}{\cellcolor[HTML]{99BEFF}17.17} &
  \multicolumn{1}{c|}{\cellcolor[HTML]{C7ECFF}4.587} &
  \multicolumn{1}{c|}{\cellcolor[HTML]{99BEFF}19.79} &
  \multicolumn{1}{c|}{\cellcolor[HTML]{C7ECFF}5.087} &
  \multicolumn{1}{c|}{\cellcolor[HTML]{99BEFF}22.64} &
  \multicolumn{1}{c|}{\cellcolor[HTML]{C7ECFF}5.654} &
  \multicolumn{1}{c|}{\cellcolor[HTML]{99BEFF}25.45} &
  \multicolumn{1}{c|}{\cellcolor[HTML]{C7ECFF}5.915} &
  \multicolumn{1}{c|}{\cellcolor[HTML]{99BEFF}28.84} &
  \cellcolor[HTML]{C7ECFF}5.942 \\ \cline{3-13} 
 &
   &
  $\Pi_{\mathsf{SE}}^{\mathsf{sep}}$ &
  \multicolumn{1}{c|}{19.97} &
  \multicolumn{1}{c|}{5.185} &
  \multicolumn{1}{c|}{23.03} &
  \multicolumn{1}{c|}{5.606} &
  \multicolumn{1}{c|}{26.31} &
  \multicolumn{1}{c|}{5.828} &
  \multicolumn{1}{c|}{29.72} &
  \multicolumn{1}{c|}{6.172} &
  \multicolumn{1}{c|}{33.16} &
  6.438 \\ \cline{3-13} 
\multirow{-18}{*}{$2^8$} &
  \multirow{-6}{*}{10} &
  $\Pi_{\mathsf{SA}}^{\mathsf{sep}}$ &
  \multicolumn{1}{c|}{18.66} &
  \multicolumn{1}{c|}{\cellcolor[HTML]{99BEFF}4.035} &
  \multicolumn{1}{c|}{21.70} &
  \multicolumn{1}{c|}{\cellcolor[HTML]{99BEFF}4.578} &
  \multicolumn{1}{c|}{24.95} &
  \multicolumn{1}{c|}{\cellcolor[HTML]{99BEFF}4.873} &
  \multicolumn{1}{c|}{28.10} &
  \multicolumn{1}{c|}{\cellcolor[HTML]{99BEFF}5.351} &
  \multicolumn{1}{c|}{31.41} &
  \cellcolor[HTML]{99BEFF}5.464 \\ \hline
 &
   &
  \cite{DBLP:conf/ccs/DZL25}-High &
  \multicolumn{1}{c|}{1043} &
  \multicolumn{1}{c|}{138.5} &
  \multicolumn{1}{c|}{1123} &
  \multicolumn{1}{c|}{140.7} &
  \multicolumn{1}{c|}{1204} &
  \multicolumn{1}{c|}{150.8} &
  \multicolumn{1}{c|}{1271} &
  \multicolumn{1}{c|}{156.2} &
  \multicolumn{1}{c|}{1936} &
  257.5 \\ \cline{3-13} 
 &
   &
  $\Pi_{\mathsf{OE}\text{-1}}^{\mathsf{sep}}$ &
  \multicolumn{1}{c|}{\cellcolor[HTML]{C7ECFF}159.3} &
  \multicolumn{1}{c|}{19.58} &
  \multicolumn{1}{c|}{\cellcolor[HTML]{C7ECFF}185.5} &
  \multicolumn{1}{c|}{21.74} &
  \multicolumn{1}{c|}{\cellcolor[HTML]{C7ECFF}212.6} &
  \multicolumn{1}{c|}{24.65} &
  \multicolumn{1}{c|}{\cellcolor[HTML]{C7ECFF}241.3} &
  \multicolumn{1}{c|}{27.39} &
  \multicolumn{1}{c|}{\cellcolor[HTML]{C7ECFF}268.0} &
  30.08 \\ \cline{3-13} 
 &
   &
  $\Pi_{\mathsf{OE}\text{-2}}^{\mathsf{sep}}$ &
  \multicolumn{1}{c|}{159.7} &
  \multicolumn{1}{c|}{19.61} &
  \multicolumn{1}{c|}{186.2} &
  \multicolumn{1}{c|}{21.86} &
  \multicolumn{1}{c|}{213.0} &
  \multicolumn{1}{c|}{24.83} &
  \multicolumn{1}{c|}{241.8} &
  \multicolumn{1}{c|}{27.51} &
  \multicolumn{1}{c|}{268.5} &
  29.98 \\ \cline{3-13} 
 &
   &
  $\Pi_{\mathsf{OA}}^{\mathsf{sep}}$ &
  \multicolumn{1}{c|}{\cellcolor[HTML]{99BEFF}153.0} &
  \multicolumn{1}{c|}{\cellcolor[HTML]{99BEFF}17.92} &
  \multicolumn{1}{c|}{\cellcolor[HTML]{99BEFF}179.0} &
  \multicolumn{1}{c|}{\cellcolor[HTML]{99BEFF}21.12} &
  \multicolumn{1}{c|}{\cellcolor[HTML]{99BEFF}205.4} &
  \multicolumn{1}{c|}{\cellcolor[HTML]{99BEFF}23.42} &
  \multicolumn{1}{c|}{\cellcolor[HTML]{99BEFF}231.8} &
  \multicolumn{1}{c|}{\cellcolor[HTML]{99BEFF}25.66} &
  \multicolumn{1}{c|}{\cellcolor[HTML]{99BEFF}258.0} &
  \cellcolor[HTML]{99BEFF}28.24 \\ \cline{3-13} 
 &
   &
  $\Pi_{\mathsf{SE}}^{\mathsf{sep}}$ &
  \multicolumn{1}{c|}{177.8} &
  \multicolumn{1}{c|}{20.04} &
  \multicolumn{1}{c|}{207.4} &
  \multicolumn{1}{c|}{22.69} &
  \multicolumn{1}{c|}{237.2} &
  \multicolumn{1}{c|}{25.34} &
  \multicolumn{1}{c|}{269.0} &
  \multicolumn{1}{c|}{28.01} &
  \multicolumn{1}{c|}{298.6} &
  30.81 \\ \cline{3-13} 
 &
  \multirow{-6}{*}{6} &
  $\Pi_{\mathsf{SA}}^{\mathsf{sep}}$ &
  \multicolumn{1}{c|}{171.1} &
  \multicolumn{1}{c|}{\cellcolor[HTML]{C7ECFF}18.45} &
  \multicolumn{1}{c|}{200.1} &
  \multicolumn{1}{c|}{\cellcolor[HTML]{C7ECFF}21.45} &
  \multicolumn{1}{c|}{229.5} &
  \multicolumn{1}{c|}{\cellcolor[HTML]{C7ECFF}24.39} &
  \multicolumn{1}{c|}{259.0} &
  \multicolumn{1}{c|}{\cellcolor[HTML]{C7ECFF}26.79} &
  \multicolumn{1}{c|}{288.2} &
  \cellcolor[HTML]{C7ECFF}29.66 \\ \cline{2-13} 
 &
   &
  \cite{DBLP:conf/ccs/DZL25}-High &
  \multicolumn{1}{c|}{1390} &
  \multicolumn{1}{c|}{179.2} &
  \multicolumn{1}{c|}{1497} &
  \multicolumn{1}{c|}{186.7} &
  \multicolumn{1}{c|}{1605} &
  \multicolumn{1}{c|}{203.2} &
  \multicolumn{1}{c|}{1694} &
  \multicolumn{1}{c|}{205.2} &
  \multicolumn{1}{c|}{2581} &
  348.2 \\ \cline{3-13} 
 &
   &
  $\Pi_{\mathsf{OE}\text{-1}}^{\mathsf{sep}}$ &
  \multicolumn{1}{c|}{\cellcolor[HTML]{C7ECFF}210.6} &
  \multicolumn{1}{c|}{24.79} &
  \multicolumn{1}{c|}{\cellcolor[HTML]{C7ECFF}246.1} &
  \multicolumn{1}{c|}{27.97} &
  \multicolumn{1}{c|}{\cellcolor[HTML]{C7ECFF}281.8} &
  \multicolumn{1}{c|}{31.63} &
  \multicolumn{1}{c|}{\cellcolor[HTML]{C7ECFF}320.1} &
  \multicolumn{1}{c|}{35.78} &
  \multicolumn{1}{c|}{\cellcolor[HTML]{C7ECFF}355.9} &
  39.52 \\ \cline{3-13} 
 &
   &
  $\Pi_{\mathsf{OE}\text{-2}}^{\mathsf{sep}}$ &
  \multicolumn{1}{c|}{211.1} &
  \multicolumn{1}{c|}{25.23} &
  \multicolumn{1}{c|}{246.4} &
  \multicolumn{1}{c|}{27.73} &
  \multicolumn{1}{c|}{282.1} &
  \multicolumn{1}{c|}{31.92} &
  \multicolumn{1}{c|}{320.5} &
  \multicolumn{1}{c|}{35.14} &
  \multicolumn{1}{c|}{356.5} &
  38.58 \\ \cline{3-13} 
 &
   &
  $\Pi_{\mathsf{OA}}^{\mathsf{sep}}$ &
  \multicolumn{1}{c|}{\cellcolor[HTML]{99BEFF}202.6} &
  \multicolumn{1}{c|}{\cellcolor[HTML]{99BEFF}23.41} &
  \multicolumn{1}{c|}{\cellcolor[HTML]{99BEFF}237.1} &
  \multicolumn{1}{c|}{\cellcolor[HTML]{99BEFF}26.47} &
  \multicolumn{1}{c|}{\cellcolor[HTML]{99BEFF}272.3} &
  \multicolumn{1}{c|}{\cellcolor[HTML]{99BEFF}29.83} &
  \multicolumn{1}{c|}{\cellcolor[HTML]{99BEFF}307.4} &
  \multicolumn{1}{c|}{\cellcolor[HTML]{99BEFF}33.78} &
  \multicolumn{1}{c|}{\cellcolor[HTML]{99BEFF}342.7} &
  \cellcolor[HTML]{99BEFF}36.99 \\ \cline{3-13} 
 &
   &
  $\Pi_{\mathsf{SE}}^{\mathsf{sep}}$ &
  \multicolumn{1}{c|}{235.7} &
  \multicolumn{1}{c|}{25.25} &
  \multicolumn{1}{c|}{274.9} &
  \multicolumn{1}{c|}{28.38} &
  \multicolumn{1}{c|}{314.6} &
  \multicolumn{1}{c|}{32.78} &
  \multicolumn{1}{c|}{357.2} &
  \multicolumn{1}{c|}{36.59} &
  \multicolumn{1}{c|}{396.9} &
  40.42 \\ \cline{3-13} 
 &
  \multirow{-6}{*}{8} &
  $\Pi_{\mathsf{SA}}^{\mathsf{sep}}$ &
  \multicolumn{1}{c|}{227.0} &
  \multicolumn{1}{c|}{\cellcolor[HTML]{C7ECFF}24.16} &
  \multicolumn{1}{c|}{265.7} &
  \multicolumn{1}{c|}{\cellcolor[HTML]{C7ECFF}27.46} &
  \multicolumn{1}{c|}{304.9} &
  \multicolumn{1}{c|}{\cellcolor[HTML]{C7ECFF}31.08} &
  \multicolumn{1}{c|}{344.1} &
  \multicolumn{1}{c|}{\cellcolor[HTML]{C7ECFF}34.79} &
  \multicolumn{1}{c|}{383.1} &
  \cellcolor[HTML]{C7ECFF}38.52 \\ \cline{2-13} 
 &
   &
  \cite{DBLP:conf/ccs/DZL25}-High &
  \multicolumn{1}{c|}{1736} &
  \multicolumn{1}{c|}{223.9} &
  \multicolumn{1}{c|}{1870} &
  \multicolumn{1}{c|}{228.9} &
  \multicolumn{1}{c|}{2006} &
  \multicolumn{1}{c|}{251.4} &
  \multicolumn{1}{c|}{2117} &
  \multicolumn{1}{c|}{254.2} &
  \multicolumn{1}{c|}{3226} &
  430.6 \\ \cline{3-13} 
 &
   &
  $\Pi_{\mathsf{OE}\text{-1}}^{\mathsf{sep}}$ &
  \multicolumn{1}{c|}{\cellcolor[HTML]{C7ECFF}262.1} &
  \multicolumn{1}{c|}{\cellcolor[HTML]{C7ECFF}29.54} &
  \multicolumn{1}{c|}{\cellcolor[HTML]{C7ECFF}306.4} &
  \multicolumn{1}{c|}{\cellcolor[HTML]{C7ECFF}34.19} &
  \multicolumn{1}{c|}{\cellcolor[HTML]{C7ECFF}350.9} &
  \multicolumn{1}{c|}{\cellcolor[HTML]{C7ECFF}38.37} &
  \multicolumn{1}{c|}{\cellcolor[HTML]{C7ECFF}398.8} &
  \multicolumn{1}{c|}{43.46} &
  \multicolumn{1}{c|}{\cellcolor[HTML]{C7ECFF}443.6} &
  47.89 \\ \cline{3-13} 
 &
   &
  $\Pi_{\mathsf{OE}\text{-2}}^{\mathsf{sep}}$ &
  \multicolumn{1}{c|}{262.4} &
  \multicolumn{1}{c|}{30.17} &
  \multicolumn{1}{c|}{306.7} &
  \multicolumn{1}{c|}{34.39} &
  \multicolumn{1}{c|}{351.4} &
  \multicolumn{1}{c|}{38.44} &
  \multicolumn{1}{c|}{399.3} &
  \multicolumn{1}{c|}{\cellcolor[HTML]{C7ECFF}43.19} &
  \multicolumn{1}{c|}{444.4} &
  \cellcolor[HTML]{C7ECFF}46.94 \\ \cline{3-13} 
 &
   &
  $\Pi_{\mathsf{OA}}^{\mathsf{sep}}$ &
  \multicolumn{1}{c|}{\cellcolor[HTML]{99BEFF}252.1} &
  \multicolumn{1}{c|}{\cellcolor[HTML]{99BEFF}28.25} &
  \multicolumn{1}{c|}{\cellcolor[HTML]{99BEFF}295.5} &
  \multicolumn{1}{c|}{\cellcolor[HTML]{99BEFF}33.16} &
  \multicolumn{1}{c|}{\cellcolor[HTML]{99BEFF}339.5} &
  \multicolumn{1}{c|}{\cellcolor[HTML]{99BEFF}37.09} &
  \multicolumn{1}{c|}{\cellcolor[HTML]{99BEFF}383.4} &
  \multicolumn{1}{c|}{\cellcolor[HTML]{99BEFF}41.47} &
  \multicolumn{1}{c|}{\cellcolor[HTML]{99BEFF}427.4} &
  \cellcolor[HTML]{99BEFF}45.96 \\ \cline{3-13} 
 &
   &
  $\Pi_{\mathsf{SE}}^{\mathsf{sep}}$ &
  \multicolumn{1}{c|}{293.2} &
  \multicolumn{1}{c|}{30.83} &
  \multicolumn{1}{c|}{342.8} &
  \multicolumn{1}{c|}{35.34} &
  \multicolumn{1}{c|}{392.4} &
  \multicolumn{1}{c|}{40.08} &
  \multicolumn{1}{c|}{445.0} &
  \multicolumn{1}{c|}{44.58} &
  \multicolumn{1}{c|}{495.0} &
  49.62 \\ \cline{3-13} 
\multirow{-18}{*}{$2^{12}$} &
  \multirow{-6}{*}{10} &
  $\Pi_{\mathsf{SA}}^{\mathsf{sep}}$ &
  \multicolumn{1}{c|}{282.6} &
  \multicolumn{1}{c|}{29.67} &
  \multicolumn{1}{c|}{331.4} &
  \multicolumn{1}{c|}{34.58} &
  \multicolumn{1}{c|}{380.1} &
  \multicolumn{1}{c|}{38.93} &
  \multicolumn{1}{c|}{429.1} &
  \multicolumn{1}{c|}{43.23} &
  \multicolumn{1}{c|}{478.5} &
  47.73 \\ \hline
 &
   &
  \cite{DBLP:conf/ccs/DZL25}-High &
  \multicolumn{1}{c|}{-} &
  \multicolumn{1}{c|}{-} &
  \multicolumn{1}{c|}{-} &
  \multicolumn{1}{c|}{-} &
  \multicolumn{1}{c|}{-} &
  \multicolumn{1}{c|}{-} &
  \multicolumn{1}{c|}{-} &
  \multicolumn{1}{c|}{-} &
  \multicolumn{1}{c|}{-} &
  - \\ \cline{3-13} 
 &
   &
  $\Pi_{\mathsf{OE}\text{-1}}^{\mathsf{sep}}$ &
  \multicolumn{1}{c|}{\cellcolor[HTML]{C7ECFF}2540} &
  \multicolumn{1}{c|}{263.3} &
  \multicolumn{1}{c|}{\cellcolor[HTML]{C7ECFF}2967} &
  \multicolumn{1}{c|}{308.3} &
  \multicolumn{1}{c|}{\cellcolor[HTML]{C7ECFF}3397} &
  \multicolumn{1}{c|}{347.6} &
  \multicolumn{1}{c|}{\cellcolor[HTML]{C7ECFF}3860} &
  \multicolumn{1}{c|}{395.6} &
  \multicolumn{1}{c|}{\cellcolor[HTML]{C7ECFF}4292} &
  433.9 \\ \cline{3-13} 
 &
   &
  $\Pi_{\mathsf{OE}\text{-2}}^{\mathsf{sep}}$ &
  \multicolumn{1}{c|}{2541} &
  \multicolumn{1}{c|}{\cellcolor[HTML]{C7ECFF}262.7} &
  \multicolumn{1}{c|}{2968} &
  \multicolumn{1}{c|}{\cellcolor[HTML]{C7ECFF}307.2} &
  \multicolumn{1}{c|}{3398} &
  \multicolumn{1}{c|}{\cellcolor[HTML]{C7ECFF}346.3} &
  \multicolumn{1}{c|}{\cellcolor[HTML]{C7ECFF}3860} &
  \multicolumn{1}{c|}{\cellcolor[HTML]{C7ECFF}394.2} &
  \multicolumn{1}{c|}{4293} &
  \cellcolor[HTML]{C7ECFF}433.5 \\ \cline{3-13} 
 &
   &
  $\Pi_{\mathsf{OA}}^{\mathsf{sep}}$ &
  \multicolumn{1}{c|}{\cellcolor[HTML]{99BEFF}2442} &
  \multicolumn{1}{c|}{\cellcolor[HTML]{99BEFF}257.1} &
  \multicolumn{1}{c|}{\cellcolor[HTML]{99BEFF}2862} &
  \multicolumn{1}{c|}{\cellcolor[HTML]{99BEFF}302.3} &
  \multicolumn{1}{c|}{\cellcolor[HTML]{99BEFF}3283} &
  \multicolumn{1}{c|}{\cellcolor[HTML]{99BEFF}341.7} &
  \multicolumn{1}{c|}{\cellcolor[HTML]{99BEFF}3707} &
  \multicolumn{1}{c|}{\cellcolor[HTML]{99BEFF}387.5} &
  \multicolumn{1}{c|}{\cellcolor[HTML]{99BEFF}4131} &
  \cellcolor[HTML]{99BEFF}425.3 \\ \cline{3-13} 
 &
   &
  $\Pi_{\mathsf{SE}}^{\mathsf{sep}}$ &
  \multicolumn{1}{c|}{2848} &
  \multicolumn{1}{c|}{278.6} &
  \multicolumn{1}{c|}{3324} &
  \multicolumn{1}{c|}{325.4} &
  \multicolumn{1}{c|}{3803} &
  \multicolumn{1}{c|}{368.5} &
  \multicolumn{1}{c|}{4316} &
  \multicolumn{1}{c|}{418.6} &
  \multicolumn{1}{c|}{4797} &
  459.1 \\ \cline{3-13} 
 &
  \multirow{-6}{*}{6} &
  $\Pi_{\mathsf{SA}}^{\mathsf{sep}}$ &
  \multicolumn{1}{c|}{2750} &
  \multicolumn{1}{c|}{272.8} &
  \multicolumn{1}{c|}{3219} &
  \multicolumn{1}{c|}{320.9} &
  \multicolumn{1}{c|}{3689} &
  \multicolumn{1}{c|}{363.3} &
  \multicolumn{1}{c|}{4163} &
  \multicolumn{1}{c|}{411.5} &
  \multicolumn{1}{c|}{4638} &
  451.3 \\ \cline{2-13} 
 &
   &
  \cite{DBLP:conf/ccs/DZL25}-High &
  \multicolumn{1}{c|}{-} &
  \multicolumn{1}{c|}{-} &
  \multicolumn{1}{c|}{-} &
  \multicolumn{1}{c|}{-} &
  \multicolumn{1}{c|}{-} &
  \multicolumn{1}{c|}{-} &
  \multicolumn{1}{c|}{-} &
  \multicolumn{1}{c|}{-} &
  \multicolumn{1}{c|}{-} &
  - \\ \cline{3-13} 
 &
   &
  $\Pi_{\mathsf{OE}\text{-1}}^{\mathsf{sep}}$ &
  \multicolumn{1}{c|}{\cellcolor[HTML]{C7ECFF}3367} &
  \multicolumn{1}{c|}{\cellcolor[HTML]{C7ECFF}348.9} &
  \multicolumn{1}{c|}{\cellcolor[HTML]{C7ECFF}3937} &
  \multicolumn{1}{c|}{405.7} &
  \multicolumn{1}{c|}{\cellcolor[HTML]{C7ECFF}4511} &
  \multicolumn{1}{c|}{460.7} &
  \multicolumn{1}{c|}{\cellcolor[HTML]{C7ECFF}5128} &
  \multicolumn{1}{c|}{519.7} &
  \multicolumn{1}{c|}{\cellcolor[HTML]{C7ECFF}5704} &
  582.4 \\ \cline{3-13} 
 &
   &
  $\Pi_{\mathsf{OE}\text{-2}}^{\mathsf{sep}}$ &
  \multicolumn{1}{c|}{3369} &
  \multicolumn{1}{c|}{\cellcolor[HTML]{C7ECFF}348.9} &
  \multicolumn{1}{c|}{3938} &
  \multicolumn{1}{c|}{\cellcolor[HTML]{C7ECFF}404.8} &
  \multicolumn{1}{c|}{\cellcolor[HTML]{C7ECFF}4511} &
  \multicolumn{1}{c|}{\cellcolor[HTML]{C7ECFF}460.2} &
  \multicolumn{1}{c|}{\cellcolor[HTML]{C7ECFF}5128} &
  \multicolumn{1}{c|}{\cellcolor[HTML]{C7ECFF}518.2} &
  \multicolumn{1}{c|}{5705} &
  \cellcolor[HTML]{C7ECFF}581.8 \\ \cline{3-13} 
 &
   &
  $\Pi_{\mathsf{OA}}^{\mathsf{sep}}$ &
  \multicolumn{1}{c|}{\cellcolor[HTML]{99BEFF}3239} &
  \multicolumn{1}{c|}{\cellcolor[HTML]{99BEFF}342.2} &
  \multicolumn{1}{c|}{\cellcolor[HTML]{99BEFF}3798} &
  \multicolumn{1}{c|}{\cellcolor[HTML]{99BEFF}397.7} &
  \multicolumn{1}{c|}{\cellcolor[HTML]{99BEFF}4361} &
  \multicolumn{1}{c|}{\cellcolor[HTML]{99BEFF}452.8} &
  \multicolumn{1}{c|}{\cellcolor[HTML]{99BEFF}4926} &
  \multicolumn{1}{c|}{\cellcolor[HTML]{99BEFF}508.2} &
  \multicolumn{1}{c|}{\cellcolor[HTML]{99BEFF}5492} &
  \cellcolor[HTML]{99BEFF}572.1 \\ \cline{3-13} 
 &
   &
  $\Pi_{\mathsf{SE}}^{\mathsf{sep}}$ &
  \multicolumn{1}{c|}{3778} &
  \multicolumn{1}{c|}{369.1} &
  \multicolumn{1}{c|}{4414} &
  \multicolumn{1}{c|}{429.8} &
  \multicolumn{1}{c|}{5053} &
  \multicolumn{1}{c|}{488.8} &
  \multicolumn{1}{c|}{5737} &
  \multicolumn{1}{c|}{551.7} &
  \multicolumn{1}{c|}{6379} &
  617.8 \\ \cline{3-13} 
 &
  \multirow{-6}{*}{8} &
  $\Pi_{\mathsf{SA}}^{\mathsf{sep}}$ &
  \multicolumn{1}{c|}{3649} &
  \multicolumn{1}{c|}{364.2} &
  \multicolumn{1}{c|}{4275} &
  \multicolumn{1}{c|}{422.8} &
  \multicolumn{1}{c|}{4904} &
  \multicolumn{1}{c|}{480.9} &
  \multicolumn{1}{c|}{5535} &
  \multicolumn{1}{c|}{541.2} &
  \multicolumn{1}{c|}{6167} &
  608.7 \\ \cline{2-13} 
 &
   &
  \cite{DBLP:conf/ccs/DZL25}-High &
  \multicolumn{1}{c|}{-} &
  \multicolumn{1}{c|}{-} &
  \multicolumn{1}{c|}{-} &
  \multicolumn{1}{c|}{-} &
  \multicolumn{1}{c|}{-} &
  \multicolumn{1}{c|}{-} &
  \multicolumn{1}{c|}{-} &
  \multicolumn{1}{c|}{-} &
  \multicolumn{1}{c|}{-} &
  - \\ \cline{3-13} 
 &
   &
  $\Pi_{\mathsf{OE}\text{-1}}^{\mathsf{sep}}$ &
  \multicolumn{1}{c|}{\cellcolor[HTML]{C7ECFF}4196} &
  \multicolumn{1}{c|}{436.4} &
  \multicolumn{1}{c|}{\cellcolor[HTML]{C7ECFF}4908} &
  \multicolumn{1}{c|}{504.5} &
  \multicolumn{1}{c|}{\cellcolor[HTML]{C7ECFF}5624} &
  \multicolumn{1}{c|}{581.7} &
  \multicolumn{1}{c|}{\cellcolor[HTML]{C7ECFF}6396} &
  \multicolumn{1}{c|}{\cellcolor[HTML]{C7ECFF}647.3} &
  \multicolumn{1}{c|}{\cellcolor[HTML]{C7ECFF}7118} &
  728.4 \\ \cline{3-13} 
 &
   &
  $\Pi_{\mathsf{OE}\text{-2}}^{\mathsf{sep}}$ &
  \multicolumn{1}{c|}{\cellcolor[HTML]{C7ECFF}4196} &
  \multicolumn{1}{c|}{\cellcolor[HTML]{C7ECFF}434.7} &
  \multicolumn{1}{c|}{\cellcolor[HTML]{C7ECFF}4908} &
  \multicolumn{1}{c|}{\cellcolor[HTML]{C7ECFF}502.8} &
  \multicolumn{1}{c|}{5625} &
  \multicolumn{1}{c|}{\cellcolor[HTML]{C7ECFF}580.8} &
  \multicolumn{1}{c|}{\cellcolor[HTML]{C7ECFF}6396} &
  \multicolumn{1}{c|}{647.4} &
  \multicolumn{1}{c|}{\cellcolor[HTML]{C7ECFF}7118} &
  \cellcolor[HTML]{C7ECFF}726.9 \\ \cline{3-13} 
 &
   &
  $\Pi_{\mathsf{OA}}^{\mathsf{sep}}$ &
  \multicolumn{1}{c|}{\cellcolor[HTML]{99BEFF}4036} &
  \multicolumn{1}{c|}{\cellcolor[HTML]{99BEFF}427.1} &
  \multicolumn{1}{c|}{\cellcolor[HTML]{99BEFF}4734} &
  \multicolumn{1}{c|}{\cellcolor[HTML]{99BEFF}495.2} &
  \multicolumn{1}{c|}{\cellcolor[HTML]{99BEFF}5439} &
  \multicolumn{1}{c|}{\cellcolor[HTML]{99BEFF}573.6} &
  \multicolumn{1}{c|}{\cellcolor[HTML]{99BEFF}6144} &
  \multicolumn{1}{c|}{\cellcolor[HTML]{99BEFF}635.6} &
  \multicolumn{1}{c|}{\cellcolor[HTML]{99BEFF}6853} &
  \cellcolor[HTML]{99BEFF}714.7 \\ \cline{3-13} 
 &
   &
  $\Pi_{\mathsf{SE}}^{\mathsf{sep}}$ &
  \multicolumn{1}{c|}{4709} &
  \multicolumn{1}{c|}{460.8} &
  \multicolumn{1}{c|}{5502} &
  \multicolumn{1}{c|}{535.7} &
  \multicolumn{1}{c|}{6303} &
  \multicolumn{1}{c|}{616.8} &
  \multicolumn{1}{c|}{7157} &
  \multicolumn{1}{c|}{684.9} &
  \multicolumn{1}{c|}{7960} &
  772.8 \\ \cline{3-13} 
\multirow{-18}{*}{$2^{16}$} &
  \multirow{-6}{*}{10} &
  $\Pi_{\mathsf{SA}}^{\mathsf{sep}}$ &
  \multicolumn{1}{c|}{4549} &
  \multicolumn{1}{c|}{454.2} &
  \multicolumn{1}{c|}{5331} &
  \multicolumn{1}{c|}{525.4} &
  \multicolumn{1}{c|}{6117} &
  \multicolumn{1}{c|}{609.1} &
  \multicolumn{1}{c|}{6906} &
  \multicolumn{1}{c|}{676.4} &
  \multicolumn{1}{c|}{7696} &
  759.5 \\ \hline
\end{tabular}
}
\caption{Communication cost (in MB) and running time (in seconds) comparing our protocols to \cite{DBLP:conf/ccs/DZL25} in WAN setting. Cells with - denote trials that ran out of memory. The best result is highlighted in \textcolor[rgb]{0,0.3,1}{blue}, the second best in \textcolor[rgb]{0,0.7,1}{cyan}.}
\label{tab:infty-high-wan}
\end{table}

As shown in Table~\ref{tab:infty-low-wan} and \ref{tab:infty-high-wan}, our protocols maintain performance advantages in WAN settings similar to those observed in LAN scenarios. We find that \cite{DBLP:conf/ccs/DZL25}-Low still has data anomaly issues (marked in red) as in the LAN settings. Given that communication overhead contributes more significantly to the total runtime in WAN environments, the communication efficiency of our protocol translates into corresponding improvements in running time. 
For instance, in the low‑dimensional setting with \(n=2^{12}\), \(d=4\), and \(\delta=256\), our \(\Pi_{\mathsf{SA}}^{\mathsf{apart}}\) finishes in 17.49 seconds, compared to 300.3 seconds for \cite{AC/BGMP25} and 132.9 seconds for \cite{DBLP:conf/ccs/DZL25}-{Low}. This represents a speedup of approximately $17.2\times$ and $7.6\times$, respectively.  
In the high‑dimensional setting with \(n=2^{12}\), \(d=10\), and \(\delta=256\), our \(\Pi_{\mathsf{OA}}^{\mathsf{sep}}\) completes in 45.96 seconds, while \cite{DBLP:conf/ccs/DZL25}-{High} requires 430.6 seconds, yielding a speedup of about $9.4\times$.

\subsection{Comparison Results of FPSI for $L_1$ and $L_2$ Distances}
\label{app:l1l2}

In this section, we compare our FPSI protocols for $L_{p\in\{1,2\}}$ distance with existing works \cite{DBLP:conf/asiacrypt/GaoQLLW24,DBLP:conf/ccs/DZL25}. We omitted \cite{AC/BGMP25} as its protocol exclusively supports $L_\infty$ distance. Furthermore, we exclude \cite{DBLP:conf/ccs/piske+25} due to apparent issues in their reported results for $L_{p}$ distance. Specifically, their data exhibits an anomalous trend where the runtime decreases as the parameter $\delta$ increases. For example, in Table 2 ($L_1$ distance, $n = 2^{12}$, $d = 10$), the runtime declines from 147.94s ($\delta = 10$) to 126.94s ($\delta = 30$). This reduction lies beyond typical experimental variance and suggests potential inconsistencies in their implementation or measurement.

Similar to our $L_{\infty}$ protocols, we will name our protocol:
\begin{itemize}
    \item $\Pi_{\mathsf{OE}\text{-}\mathsf{1}}^{\mathsf{apart}}$: The FPSI construction in Figure \ref{fig:pifpsi-p-apart-opprf}, it is based on OPPRF and ECIPS;
    \item $\Pi_{\mathsf{OE}\text{-}\mathsf{2}}^{\mathsf{apart}}$: same as Figure \ref{fig:pifpsi-p-apart-opprf} but use the trade-off we discussed in section \ref{subsubsec:fpsi-apart-p-opprf}, that is, it use PS + IFmat instead of ssIfmat + PS;
    \item $\Pi_{\mathsf{SE}}^{\mathsf{apart}}$: The FPSI construction in Figure \ref{fig:pifpsi-p-apart-soprf}, it is based on sOPRF and ECIPS.
\end{itemize}

The protocols under the \textit{separate} assumption follow the same naming pattern, only with the superscript changed to $\mathsf{sep}$.

Same as our experiments for $L_\infty$ distance, our evaluation uses the following parameters: set size $m=n \in\{2^{8}, 2^{12}, 2^{16}\}$, threshold $\delta \in \{16,32,64,128,256\}$, and dimension $d$ chosen as $ \{2,3,4 \} $ for the \textit{apart} assumption and $ \{6,8,10 \} $ for the \textit{separate} assumption. We note that in our testing for $L_2$ distance, we only considered the case of $d=2$ when $m=n=2^{16}$, as we found that all protocols had memory overflow when $d\in \{3,4 \}$. We tested the performance of LAN (10Gbps bandwidth with 0.02ms RTT latency) and WAN (400 Mbps bandwidth, 80 ms RTT latency) network environments separately. For the same reason as $L_\infty$ distance, we excluded \cite{DBLP:conf/asiacrypt/GaoQLLW24} in our comparison under WAN setting.

For $L_1$ distance, experimental results are presented in Tables \ref{tab:l1-low-lan}, \ref{tab:l1-high-lan}, \ref{tab:l1-low-wan}, and \ref{tab:l1-high-wan} for the low-dimensional/LAN, high-dimensional/LAN, low-dimensional/WAN, and high-dimensional/WAN settings, respectively. Corresponding results for $L_2$ distance are provided in Tables \ref{tab:l2-low-lan}, \ref{tab:l2-high-lan}, \ref{tab:l2-low-wan}, and \ref{tab:l2-high-wan}.

Similar to the case of $L_\infty$ distance, we observe that for $d > 2$, the communication cost of \cite{DBLP:conf/ccs/DZL25}-Low still decreases when $\delta$ changes from $32$ to $64$. However, unlike in the $L_\infty$ setting, its running time increases significantly. These anomalies are marked in red in Tables \ref{tab:l1-low-lan}, \ref{tab:l1-low-wan}, \ref{tab:l2-low-lan}, and \ref{tab:l2-low-wan}.

Even accounting for the above irregularities, our protocol consistently achieves the best and second‑best performance across all tested parameter sets and network settings. For example, in the case of $L_1$ distance under LAN with $m=n=2^8$, $d=4$, and $\delta=256$, our $\Pi_{\mathsf{SE}}^{\mathsf{apart}}$ completes in only $2.823$ seconds with $34.6$ MB of communication, whereas \cite{DBLP:conf/ccs/DZL25}-Low requires $195.8$ seconds and $398.6$ MB—corresponding to a $69.4\times$ speed‑up and an $11.5\times$ reduction in communication overhead.

Similarly, for the high‑dimensional setting ($d=10$, $\delta=256$, $m=n=2^{8}$), $\Pi_{\mathsf{OE}\text{-}\mathsf{2}}^{\mathsf{sep}}$ finishes in $4.237$ seconds, yielding speed‑ups of $34\times$ over \cite{DBLP:conf/asiacrypt/GaoQLLW24} and $7.3\times$ over \cite{DBLP:conf/ccs/DZL25}-High, while reducing communication by factors of $16.4\times$ and $5.3\times$, respectively.

\begin{table}[!htbp]
\renewcommand\arraystretch{1}
	\centering
 \resizebox{0.83\linewidth}{!}{
\begin{tabular}{|c|c|c|cccccccccc|}
\hline
 &
   &
   &
  \multicolumn{10}{c|}{Threshold $\delta$} \\ \cline{4-13} 
 &
   &
   &
  \multicolumn{2}{c|}{16} &
  \multicolumn{2}{c|}{32} &
  \multicolumn{2}{c|}{64} &
  \multicolumn{2}{c|}{128} &
  \multicolumn{2}{c|}{256} \\ \cline{4-13} 
\multirow{-3}{*}{\begin{tabular}[c]{@{}c@{}}Set Size\\ $m=n$\end{tabular}} &
  \multirow{-3}{*}{\begin{tabular}[c]{@{}c@{}}Dimension\\ $d$\end{tabular}} &
  \multirow{-3}{*}{Protocol} &
  \multicolumn{1}{c|}{Comm.} &
  \multicolumn{1}{c|}{Time} &
  \multicolumn{1}{c|}{Comm.} &
  \multicolumn{1}{c|}{Time} &
  \multicolumn{1}{c|}{Comm.} &
  \multicolumn{1}{c|}{Time} &
  \multicolumn{1}{c|}{Comm.} &
  \multicolumn{1}{c|}{Time} &
  \multicolumn{1}{c|}{Comm.} &
  Time \\ \hline
 &
   &
  \cite{DBLP:conf/ccs/DZL25}-Low &
  \multicolumn{1}{c|}{15.51} &
  \multicolumn{1}{c|}{3.853} &
  \multicolumn{1}{c|}{19.55} &
  \multicolumn{1}{c|}{5.767} &
  \multicolumn{1}{c|}{23.87} &
  \multicolumn{1}{c|}{7.823} &
  \multicolumn{1}{c|}{28.50} &
  \multicolumn{1}{c|}{10.95} &
  \multicolumn{1}{c|}{33.42} &
  13.86 \\ \cline{3-13} 
 &
   &
  $\Pi_{\mathsf{OE}\text{-1}}^{\mathsf{apart}}$ &
  \multicolumn{1}{c|}{\cellcolor[HTML]{C7ECFF}7.638} &
  \multicolumn{1}{c|}{0.981} &
  \multicolumn{1}{c|}{\cellcolor[HTML]{C7ECFF}8.602} &
  \multicolumn{1}{c|}{0.986} &
  \multicolumn{1}{c|}{\cellcolor[HTML]{C7ECFF}9.197} &
  \multicolumn{1}{c|}{1.034} &
  \multicolumn{1}{c|}{\cellcolor[HTML]{99BEFF}9.883} &
  \multicolumn{1}{c|}{1.068} &
  \multicolumn{1}{c|}{\cellcolor[HTML]{99BEFF}11.24} &
  1.158 \\ \cline{3-13} 
 &
   &
  $\Pi_{\mathsf{OE}\text{-2}}^{\mathsf{apart}}$ &
  \multicolumn{1}{c|}{7.698} &
  \multicolumn{1}{c|}{\cellcolor[HTML]{C7ECFF}0.917} &
  \multicolumn{1}{c|}{8.656} &
  \multicolumn{1}{c|}{\cellcolor[HTML]{C7ECFF}0.929} &
  \multicolumn{1}{c|}{\cellcolor[HTML]{99BEFF}9.181} &
  \multicolumn{1}{c|}{\cellcolor[HTML]{C7ECFF}0.965} &
  \multicolumn{1}{c|}{\cellcolor[HTML]{C7ECFF}9.959} &
  \multicolumn{1}{c|}{\cellcolor[HTML]{C7ECFF}1.016} &
  \multicolumn{1}{c|}{\cellcolor[HTML]{C7ECFF}11.25} &
  \cellcolor[HTML]{C7ECFF}1.094 \\ \cline{3-13} 
 &
  \multirow{-4}{*}{2} &
  $\Pi_{\mathsf{SE}}^{\mathsf{apart}}$ &
  \multicolumn{1}{c|}{\cellcolor[HTML]{99BEFF}7.329} &
  \multicolumn{1}{c|}{\cellcolor[HTML]{99BEFF}0.695} &
  \multicolumn{1}{c|}{\cellcolor[HTML]{99BEFF}8.435} &
  \multicolumn{1}{c|}{\cellcolor[HTML]{99BEFF}0.719} &
  \multicolumn{1}{c|}{9.269} &
  \multicolumn{1}{c|}{\cellcolor[HTML]{99BEFF}0.804} &
  \multicolumn{1}{c|}{10.19} &
  \multicolumn{1}{c|}{\cellcolor[HTML]{99BEFF}0.887} &
  \multicolumn{1}{c|}{11.55} &
  \cellcolor[HTML]{99BEFF}0.988 \\ \cline{2-13} 
 &
   &
  \cite{DBLP:conf/ccs/DZL25}-Low &
  \multicolumn{1}{c|}{57.46} &
  \multicolumn{1}{c|}{4.731} &
  \multicolumn{1}{c|}{{\color[HTML]{FE0000} 84.30}} &
  \multicolumn{1}{c|}{{\color[HTML]{FE0000} 5.776}} &
  \multicolumn{1}{c|}{{\color[HTML]{FE0000} 81.50}} &
  \multicolumn{1}{c|}{{\color[HTML]{FE0000} 26.21}} &
  \multicolumn{1}{c|}{94.96} &
  \multicolumn{1}{c|}{{\color[HTML]{FE0000} 24.61}} &
  \multicolumn{1}{c|}{115.1} &
  25.47 \\ \cline{3-13} 
 &
   &
  $\Pi_{\mathsf{OE}\text{-1}}^{\mathsf{apart}}$ &
  \multicolumn{1}{c|}{\cellcolor[HTML]{C7ECFF}13.51} &
  \multicolumn{1}{c|}{1.119} &
  \multicolumn{1}{c|}{15.54} &
  \multicolumn{1}{c|}{1.244} &
  \multicolumn{1}{c|}{17.31} &
  \multicolumn{1}{c|}{1.338} &
  \multicolumn{1}{c|}{\cellcolor[HTML]{C7ECFF}19.62} &
  \multicolumn{1}{c|}{1.487} &
  \multicolumn{1}{c|}{\cellcolor[HTML]{C7ECFF}22.33} &
  1.674 \\ \cline{3-13} 
 &
   &
  $\Pi_{\mathsf{OE}\text{-2}}^{\mathsf{apart}}$ &
  \multicolumn{1}{c|}{\cellcolor[HTML]{C7ECFF}13.51} &
  \multicolumn{1}{c|}{\cellcolor[HTML]{C7ECFF}1.088} &
  \multicolumn{1}{c|}{\cellcolor[HTML]{C7ECFF}15.39} &
  \multicolumn{1}{c|}{\cellcolor[HTML]{C7ECFF}1.174} &
  \multicolumn{1}{c|}{\cellcolor[HTML]{C7ECFF}17.26} &
  \multicolumn{1}{c|}{\cellcolor[HTML]{C7ECFF}1.271} &
  \multicolumn{1}{c|}{19.65} &
  \multicolumn{1}{c|}{\cellcolor[HTML]{C7ECFF}1.407} &
  \multicolumn{1}{c|}{22.54} &
  \cellcolor[HTML]{C7ECFF}1.578 \\ \cline{3-13} 
 &
  \multirow{-4}{*}{3} &
  $\Pi_{\mathsf{SE}}^{\mathsf{apart}}$ &
  \multicolumn{1}{c|}{\cellcolor[HTML]{99BEFF}12.15} &
  \multicolumn{1}{c|}{\cellcolor[HTML]{99BEFF}0.967} &
  \multicolumn{1}{c|}{\cellcolor[HTML]{99BEFF}13.83} &
  \multicolumn{1}{c|}{\cellcolor[HTML]{99BEFF}1.058} &
  \multicolumn{1}{c|}{\cellcolor[HTML]{99BEFF}15.55} &
  \multicolumn{1}{c|}{\cellcolor[HTML]{99BEFF}1.203} &
  \multicolumn{1}{c|}{\cellcolor[HTML]{99BEFF}17.47} &
  \multicolumn{1}{c|}{\cellcolor[HTML]{99BEFF}1.305} &
  \multicolumn{1}{c|}{\cellcolor[HTML]{99BEFF}19.86} &
  \cellcolor[HTML]{99BEFF}1.498 \\ \cline{2-13} 
 &
   &
  \cite{DBLP:conf/ccs/DZL25}-Low &
  \multicolumn{1}{c|}{154.1} &
  \multicolumn{1}{c|}{16.21} &
  \multicolumn{1}{c|}{225.6} &
  \multicolumn{1}{c|}{{\color[HTML]{FE0000} 17.09}} &
  \multicolumn{1}{c|}{309.1} &
  \multicolumn{1}{c|}{{\color[HTML]{FE0000} 167.7}} &
  \multicolumn{1}{c|}{344.9} &
  \multicolumn{1}{c|}{194.1} &
  \multicolumn{1}{c|}{398.6} &
  195.8 \\ \cline{3-13} 
 &
   &
  $\Pi_{\mathsf{OE}\text{-1}}^{\mathsf{apart}}$ &
  \multicolumn{1}{c|}{\cellcolor[HTML]{C7ECFF}25.83} &
  \multicolumn{1}{c|}{1.716} &
  \multicolumn{1}{c|}{30.68} &
  \multicolumn{1}{c|}{2.012} &
  \multicolumn{1}{c|}{35.65} &
  \multicolumn{1}{c|}{2.256} &
  \multicolumn{1}{c|}{\cellcolor[HTML]{C7ECFF}40.04} &
  \multicolumn{1}{c|}{2.534} &
  \multicolumn{1}{c|}{46.27} &
  2.916 \\ \cline{3-13} 
 &
   &
  $\Pi_{\mathsf{OE}\text{-2}}^{\mathsf{apart}}$ &
  \multicolumn{1}{c|}{26.11} &
  \multicolumn{1}{c|}{\cellcolor[HTML]{C7ECFF}1.673} &
  \multicolumn{1}{c|}{\cellcolor[HTML]{C7ECFF}30.46} &
  \multicolumn{1}{c|}{\cellcolor[HTML]{C7ECFF}1.934} &
  \multicolumn{1}{c|}{\cellcolor[HTML]{C7ECFF}35.52} &
  \multicolumn{1}{c|}{\cellcolor[HTML]{C7ECFF}2.215} &
  \multicolumn{1}{c|}{40.08} &
  \multicolumn{1}{c|}{\cellcolor[HTML]{C7ECFF}2.475} &
  \multicolumn{1}{c|}{\cellcolor[HTML]{C7ECFF}46.09} &
  \cellcolor[HTML]{C7ECFF}2.824 \\ \cline{3-13} 
\multirow{-12}{*}{$2^8$} &
  \multirow{-4}{*}{4} &
  $\Pi_{\mathsf{SE}}^{\mathsf{apart}}$ &
  \multicolumn{1}{c|}{\cellcolor[HTML]{99BEFF}20.08} &
  \multicolumn{1}{c|}{\cellcolor[HTML]{99BEFF}1.538} &
  \multicolumn{1}{c|}{\cellcolor[HTML]{99BEFF}23.47} &
  \multicolumn{1}{c|}{\cellcolor[HTML]{99BEFF}1.821} &
  \multicolumn{1}{c|}{\cellcolor[HTML]{99BEFF}26.67} &
  \multicolumn{1}{c|}{\cellcolor[HTML]{99BEFF}2.128} &
  \multicolumn{1}{c|}{\cellcolor[HTML]{99BEFF}30.21} &
  \multicolumn{1}{c|}{\cellcolor[HTML]{99BEFF}2.464} &
  \multicolumn{1}{c|}{\cellcolor[HTML]{99BEFF}34.60} &
  \cellcolor[HTML]{99BEFF}2.823 \\ \hline
 &
   &
  \cite{DBLP:conf/ccs/DZL25}-Low &
  \multicolumn{1}{c|}{248.2} &
  \multicolumn{1}{c|}{56.23} &
  \multicolumn{1}{c|}{312.8} &
  \multicolumn{1}{c|}{79.41} &
  \multicolumn{1}{c|}{381.9} &
  \multicolumn{1}{c|}{115.7} &
  \multicolumn{1}{c|}{456.1} &
  \multicolumn{1}{c|}{153.5} &
  \multicolumn{1}{c|}{534.7} &
  194.1 \\ \cline{3-13} 
 &
   &
  $\Pi_{\mathsf{OE}\text{-1}}^{\mathsf{apart}}$ &
  \multicolumn{1}{c|}{\cellcolor[HTML]{C7ECFF}89.29} &
  \multicolumn{1}{c|}{\cellcolor[HTML]{C7ECFF}5.833} &
  \multicolumn{1}{c|}{\cellcolor[HTML]{C7ECFF}101.9} &
  \multicolumn{1}{c|}{\cellcolor[HTML]{C7ECFF}6.798} &
  \multicolumn{1}{c|}{\cellcolor[HTML]{C7ECFF}115.2} &
  \multicolumn{1}{c|}{\cellcolor[HTML]{C7ECFF}7.854} &
  \multicolumn{1}{c|}{\cellcolor[HTML]{C7ECFF}128.2} &
  \multicolumn{1}{c|}{\cellcolor[HTML]{C7ECFF}8.869} &
  \multicolumn{1}{c|}{\cellcolor[HTML]{C7ECFF}148.0} &
  \cellcolor[HTML]{C7ECFF}10.23 \\ \cline{3-13} 
 &
   &
  $\Pi_{\mathsf{OE}\text{-2}}^{\mathsf{apart}}$ &
  \multicolumn{1}{c|}{\cellcolor[HTML]{99BEFF}88.55} &
  \multicolumn{1}{c|}{\cellcolor[HTML]{99BEFF}5.026} &
  \multicolumn{1}{c|}{\cellcolor[HTML]{99BEFF}101.0} &
  \multicolumn{1}{c|}{\cellcolor[HTML]{99BEFF}5.949} &
  \multicolumn{1}{c|}{\cellcolor[HTML]{99BEFF}114.2} &
  \multicolumn{1}{c|}{\cellcolor[HTML]{99BEFF}6.561} &
  \multicolumn{1}{c|}{\cellcolor[HTML]{99BEFF}127.3} &
  \multicolumn{1}{c|}{\cellcolor[HTML]{99BEFF}7.343} &
  \multicolumn{1}{c|}{\cellcolor[HTML]{99BEFF}147.0} &
  \cellcolor[HTML]{99BEFF}8.603 \\ \cline{3-13} 
 &
  \multirow{-4}{*}{2} &
  $\Pi_{\mathsf{SE}}^{\mathsf{apart}}$ &
  \multicolumn{1}{c|}{96.44} &
  \multicolumn{1}{c|}{6.608} &
  \multicolumn{1}{c|}{110.1} &
  \multicolumn{1}{c|}{7.858} &
  \multicolumn{1}{c|}{124.2} &
  \multicolumn{1}{c|}{9.062} &
  \multicolumn{1}{c|}{138.2} &
  \multicolumn{1}{c|}{10.28} &
  \multicolumn{1}{c|}{159.1} &
  11.89 \\ \cline{2-13} 
 &
   &
  \cite{DBLP:conf/ccs/DZL25}-Low &
  \multicolumn{1}{c|}{919.3} &
  \multicolumn{1}{c|}{69.22} &
  \multicolumn{1}{c|}{{\color[HTML]{FE0000} 1349}} &
  \multicolumn{1}{c|}{{\color[HTML]{FE0000} 83.46}} &
  \multicolumn{1}{c|}{{\color[HTML]{FE0000} 1304}} &
  \multicolumn{1}{c|}{{\color[HTML]{FE0000} 364.5}} &
  \multicolumn{1}{c|}{1519} &
  \multicolumn{1}{c|}{365.5} &
  \multicolumn{1}{c|}{1842} &
  371.6 \\ \cline{3-13} 
 &
   &
  $\Pi_{\mathsf{OE}\text{-1}}^{\mathsf{apart}}$ &
  \multicolumn{1}{c|}{183.1} &
  \multicolumn{1}{c|}{\cellcolor[HTML]{C7ECFF}11.71} &
  \multicolumn{1}{c|}{214.6} &
  \multicolumn{1}{c|}{\cellcolor[HTML]{C7ECFF}14.13} &
  \multicolumn{1}{c|}{246.3} &
  \multicolumn{1}{c|}{\cellcolor[HTML]{C7ECFF}16.48} &
  \multicolumn{1}{c|}{278.4} &
  \multicolumn{1}{c|}{\cellcolor[HTML]{C7ECFF}19.27} &
  \multicolumn{1}{c|}{320.5} &
  \cellcolor[HTML]{C7ECFF}22.14 \\ \cline{3-13} 
 &
   &
  $\Pi_{\mathsf{OE}\text{-2}}^{\mathsf{apart}}$ &
  \multicolumn{1}{c|}{\cellcolor[HTML]{C7ECFF}182.3} &
  \multicolumn{1}{c|}{\cellcolor[HTML]{99BEFF}10.83} &
  \multicolumn{1}{c|}{\cellcolor[HTML]{C7ECFF}213.7} &
  \multicolumn{1}{c|}{\cellcolor[HTML]{99BEFF}13.29} &
  \multicolumn{1}{c|}{\cellcolor[HTML]{C7ECFF}245.2} &
  \multicolumn{1}{c|}{\cellcolor[HTML]{99BEFF}14.95} &
  \multicolumn{1}{c|}{\cellcolor[HTML]{C7ECFF}277.5} &
  \multicolumn{1}{c|}{\cellcolor[HTML]{99BEFF}17.33} &
  \multicolumn{1}{c|}{\cellcolor[HTML]{C7ECFF}319.7} &
  \cellcolor[HTML]{99BEFF}20.34 \\ \cline{3-13} 
 &
  \multirow{-4}{*}{3} &
  $\Pi_{\mathsf{SE}}^{\mathsf{apart}}$ &
  \multicolumn{1}{c|}{\cellcolor[HTML]{99BEFF}168.2} &
  \multicolumn{1}{c|}{12.84} &
  \multicolumn{1}{c|}{\cellcolor[HTML]{99BEFF}194.7} &
  \multicolumn{1}{c|}{15.68} &
  \multicolumn{1}{c|}{\cellcolor[HTML]{99BEFF}221.3} &
  \multicolumn{1}{c|}{18.35} &
  \multicolumn{1}{c|}{\cellcolor[HTML]{99BEFF}249.0} &
  \multicolumn{1}{c|}{21.43} &
  \multicolumn{1}{c|}{\cellcolor[HTML]{99BEFF}286.1} &
  24.45 \\ \cline{2-13} 
 &
   &
  \cite{DBLP:conf/ccs/DZL25}-Low &
  \multicolumn{1}{c|}{2465} &
  \multicolumn{1}{c|}{224.7} &
  \multicolumn{1}{c|}{3610} &
  \multicolumn{1}{c|}{251.7} &
  \multicolumn{1}{c|}{-} &
  \multicolumn{1}{c|}{-} &
  \multicolumn{1}{c|}{-} &
  \multicolumn{1}{c|}{-} &
  \multicolumn{1}{c|}{-} &
  - \\ \cline{3-13} 
 &
   &
  $\Pi_{\mathsf{OE}\text{-1}}^{\mathsf{apart}}$ &
  \multicolumn{1}{c|}{383.5} &
  \multicolumn{1}{c|}{\cellcolor[HTML]{C7ECFF}26.83} &
  \multicolumn{1}{c|}{456.5} &
  \multicolumn{1}{c|}{\cellcolor[HTML]{C7ECFF}32.81} &
  \multicolumn{1}{c|}{532.4} &
  \multicolumn{1}{c|}{\cellcolor[HTML]{C7ECFF}39.39} &
  \multicolumn{1}{c|}{610.9} &
  \multicolumn{1}{c|}{\cellcolor[HTML]{C7ECFF}46.14} &
  \multicolumn{1}{c|}{699.6} &
  \cellcolor[HTML]{99BEFF}52.91 \\ \cline{3-13} 
 &
   &
  $\Pi_{\mathsf{OE}\text{-2}}^{\mathsf{apart}}$ &
  \multicolumn{1}{c|}{\cellcolor[HTML]{C7ECFF}381.7} &
  \multicolumn{1}{c|}{\cellcolor[HTML]{99BEFF}25.87} &
  \multicolumn{1}{c|}{\cellcolor[HTML]{C7ECFF}455.2} &
  \multicolumn{1}{c|}{\cellcolor[HTML]{99BEFF}31.88} &
  \multicolumn{1}{c|}{\cellcolor[HTML]{C7ECFF}530.4} &
  \multicolumn{1}{c|}{\cellcolor[HTML]{99BEFF}38.31} &
  \multicolumn{1}{c|}{\cellcolor[HTML]{C7ECFF}607.9} &
  \multicolumn{1}{c|}{\cellcolor[HTML]{99BEFF}45.29} &
  \multicolumn{1}{c|}{\cellcolor[HTML]{C7ECFF}697.7} &
  \cellcolor[HTML]{C7ECFF}53.19 \\ \cline{3-13} 
\multirow{-12}{*}{$2^{12}$} &
  \multirow{-4}{*}{4} &
  $\Pi_{\mathsf{SE}}^{\mathsf{apart}}$ &
  \multicolumn{1}{c|}{\cellcolor[HTML]{99BEFF}292.9} &
  \multicolumn{1}{c|}{27.65} &
  \multicolumn{1}{c|}{\cellcolor[HTML]{99BEFF}343.8} &
  \multicolumn{1}{c|}{34.24} &
  \multicolumn{1}{c|}{\cellcolor[HTML]{99BEFF}397.0} &
  \multicolumn{1}{c|}{40.88} &
  \multicolumn{1}{c|}{\cellcolor[HTML]{99BEFF}449.7} &
  \multicolumn{1}{c|}{47.89} &
  \multicolumn{1}{c|}{\cellcolor[HTML]{99BEFF}515.8} &
  55.65 \\ \hline
 &
   &
  \cite{DBLP:conf/ccs/DZL25}-Low &
  \multicolumn{1}{c|}{3971} &
  \multicolumn{1}{c|}{957.4} &
  \multicolumn{1}{c|}{-} &
  \multicolumn{1}{c|}{-} &
  \multicolumn{1}{c|}{-} &
  \multicolumn{1}{c|}{-} &
  \multicolumn{1}{c|}{-} &
  \multicolumn{1}{c|}{-} &
  \multicolumn{1}{c|}{-} &
  - \\ \cline{3-13} 
 &
   &
  $\Pi_{\mathsf{OE}\text{-1}}^{\mathsf{apart}}$ &
  \multicolumn{1}{c|}{\cellcolor[HTML]{C7ECFF}1404} &
  \multicolumn{1}{c|}{\cellcolor[HTML]{C7ECFF}96.58} &
  \multicolumn{1}{c|}{\cellcolor[HTML]{C7ECFF}1612} &
  \multicolumn{1}{c|}{\cellcolor[HTML]{C7ECFF}114.4} &
  \multicolumn{1}{c|}{\cellcolor[HTML]{C7ECFF}1824} &
  \multicolumn{1}{c|}{\cellcolor[HTML]{C7ECFF}134.1} &
  \multicolumn{1}{c|}{\cellcolor[HTML]{C7ECFF}2039} &
  \multicolumn{1}{c|}{\cellcolor[HTML]{C7ECFF}153.8} &
  \multicolumn{1}{c|}{\cellcolor[HTML]{C7ECFF}2357} &
  \cellcolor[HTML]{C7ECFF}177.2 \\ \cline{3-13} 
 &
   &
  $\Pi_{\mathsf{OE}\text{-2}}^{\mathsf{apart}}$ &
  \multicolumn{1}{c|}{\cellcolor[HTML]{99BEFF}1392} &
  \multicolumn{1}{c|}{\cellcolor[HTML]{99BEFF}81.55} &
  \multicolumn{1}{c|}{\cellcolor[HTML]{99BEFF}1597} &
  \multicolumn{1}{c|}{\cellcolor[HTML]{99BEFF}97.77} &
  \multicolumn{1}{c|}{\cellcolor[HTML]{99BEFF}1807} &
  \multicolumn{1}{c|}{\cellcolor[HTML]{99BEFF}114.7} &
  \multicolumn{1}{c|}{\cellcolor[HTML]{99BEFF}2020} &
  \multicolumn{1}{c|}{\cellcolor[HTML]{99BEFF}131.9} &
  \multicolumn{1}{c|}{\cellcolor[HTML]{99BEFF}2336} &
  \cellcolor[HTML]{99BEFF}153.4 \\ \cline{3-13} 
 &
  \multirow{-4}{*}{2} &
  $\Pi_{\mathsf{SE}}^{\mathsf{apart}}$ &
  \multicolumn{1}{c|}{1536} &
  \multicolumn{1}{c|}{116.7} &
  \multicolumn{1}{c|}{1762} &
  \multicolumn{1}{c|}{140.1} &
  \multicolumn{1}{c|}{1991} &
  \multicolumn{1}{c|}{163.5} &
  \multicolumn{1}{c|}{2220} &
  \multicolumn{1}{c|}{189.1} &
  \multicolumn{1}{c|}{2552} &
  216.1 \\ \cline{2-13} 
 &
   &
  \cite{DBLP:conf/ccs/DZL25}-Low &
  \multicolumn{1}{c|}{-} &
  \multicolumn{1}{c|}{-} &
  \multicolumn{1}{c|}{-} &
  \multicolumn{1}{c|}{-} &
  \multicolumn{1}{c|}{-} &
  \multicolumn{1}{c|}{-} &
  \multicolumn{1}{c|}{-} &
  \multicolumn{1}{c|}{-} &
  \multicolumn{1}{c|}{-} &
  - \\ \cline{3-13} 
 &
   &
  $\Pi_{\mathsf{OE}\text{-1}}^{\mathsf{apart}}$ &
  \multicolumn{1}{c|}{2919} &
  \multicolumn{1}{c|}{\cellcolor[HTML]{C7ECFF}219.9} &
  \multicolumn{1}{c|}{3421} &
  \multicolumn{1}{c|}{\cellcolor[HTML]{C7ECFF}269.9} &
  \multicolumn{1}{c|}{3931} &
  \multicolumn{1}{c|}{\cellcolor[HTML]{C7ECFF}327.7} &
  \multicolumn{1}{c|}{4446} &
  \multicolumn{1}{c|}{\cellcolor[HTML]{C7ECFF}381.6} &
  \multicolumn{1}{c|}{5128} &
  \cellcolor[HTML]{C7ECFF}451.7 \\ \cline{3-13} 
 &
   &
  $\Pi_{\mathsf{OE}\text{-2}}^{\mathsf{apart}}$ &
  \multicolumn{1}{c|}{\cellcolor[HTML]{C7ECFF}2905} &
  \multicolumn{1}{c|}{\cellcolor[HTML]{99BEFF}206.1} &
  \multicolumn{1}{c|}{\cellcolor[HTML]{C7ECFF}3407} &
  \multicolumn{1}{c|}{\cellcolor[HTML]{99BEFF}256.3} &
  \multicolumn{1}{c|}{\cellcolor[HTML]{C7ECFF}3916} &
  \multicolumn{1}{c|}{\cellcolor[HTML]{99BEFF}309.3} &
  \multicolumn{1}{c|}{\cellcolor[HTML]{C7ECFF}4427} &
  \multicolumn{1}{c|}{\cellcolor[HTML]{99BEFF}360.9} &
  \multicolumn{1}{c|}{\cellcolor[HTML]{C7ECFF}5107} &
  \cellcolor[HTML]{99BEFF}426.9 \\ \cline{3-13} 
 &
  \multirow{-4}{*}{3} &
  $\Pi_{\mathsf{SE}}^{\mathsf{apart}}$ &
  \multicolumn{1}{c|}{\cellcolor[HTML]{99BEFF}2695} &
  \multicolumn{1}{c|}{253.9} &
  \multicolumn{1}{c|}{\cellcolor[HTML]{99BEFF}3127} &
  \multicolumn{1}{c|}{307.5} &
  \multicolumn{1}{c|}{\cellcolor[HTML]{99BEFF}3563} &
  \multicolumn{1}{c|}{372.2} &
  \multicolumn{1}{c|}{\cellcolor[HTML]{99BEFF}3999} &
  \multicolumn{1}{c|}{436.2} &
  \multicolumn{1}{c|}{\cellcolor[HTML]{99BEFF}4599} &
  506.7 \\ \cline{2-13} 
 &
   &
  \cite{DBLP:conf/ccs/DZL25}-Low &
  \multicolumn{1}{c|}{-} &
  \multicolumn{1}{c|}{-} &
  \multicolumn{1}{c|}{-} &
  \multicolumn{1}{c|}{-} &
  \multicolumn{1}{c|}{-} &
  \multicolumn{1}{c|}{-} &
  \multicolumn{1}{c|}{-} &
  \multicolumn{1}{c|}{-} &
  \multicolumn{1}{c|}{-} &
  - \\ \cline{3-13} 
 &
   &
  $\Pi_{\mathsf{OE}\text{-1}}^{\mathsf{apart}}$ &
  \multicolumn{1}{c|}{-} &
  \multicolumn{1}{c|}{-} &
  \multicolumn{1}{c|}{-} &
  \multicolumn{1}{c|}{-} &
  \multicolumn{1}{c|}{-} &
  \multicolumn{1}{c|}{-} &
  \multicolumn{1}{c|}{-} &
  \multicolumn{1}{c|}{-} &
  \multicolumn{1}{c|}{-} &
  - \\ \cline{3-13} 
 &
   &
  $\Pi_{\mathsf{OE}\text{-2}}^{\mathsf{apart}}$ &
  \multicolumn{1}{c|}{-} &
  \multicolumn{1}{c|}{-} &
  \multicolumn{1}{c|}{-} &
  \multicolumn{1}{c|}{-} &
  \multicolumn{1}{c|}{-} &
  \multicolumn{1}{c|}{-} &
  \multicolumn{1}{c|}{-} &
  \multicolumn{1}{c|}{-} &
  \multicolumn{1}{c|}{-} &
  - \\ \cline{3-13} 
\multirow{-12}{*}{$2^{16}$} &
  \multirow{-4}{*}{4} &
  $\Pi_{\mathsf{SE}}^{\mathsf{apart}}$ &
  \multicolumn{1}{c|}{\cellcolor[HTML]{99BEFF}4699} &
  \multicolumn{1}{c|}{\cellcolor[HTML]{99BEFF}612.9} &
  \multicolumn{1}{c|}{-} &
  \multicolumn{1}{c|}{-} &
  \multicolumn{1}{c|}{-} &
  \multicolumn{1}{c|}{-} &
  \multicolumn{1}{c|}{-} &
  \multicolumn{1}{c|}{-} &
  \multicolumn{1}{c|}{-} &
  - \\ \hline
\end{tabular}
}
\caption{Communication cost (in MB) and running time (in seconds) of our protocols compared with those of \cite{DBLP:conf/ccs/DZL25} for $L_1$ distance in a low-dimensional LAN setting. Cells with - denote trials that ran out of memory. The best result is highlighted in \textcolor[rgb]{0,0.3,1}{blue}, the second best in \textcolor[rgb]{0,0.7,1}{cyan}, and data in \textcolor[rgb]{1,0,0}{red} font indicates abnormal values.}
\label{tab:l1-low-lan}
\end{table}

\begin{table}[!htbp]
\renewcommand\arraystretch{1}
	\centering
 \resizebox{0.83\linewidth}{!}{
\begin{tabular}{|c|c|c|cccccccccc|}
\hline
 &
   &
   &
  \multicolumn{10}{c|}{Threshold $\delta$} \\ \cline{4-13} 
 &
   &
   &
  \multicolumn{2}{c|}{16} &
  \multicolumn{2}{c|}{32} &
  \multicolumn{2}{c|}{64} &
  \multicolumn{2}{c|}{128} &
  \multicolumn{2}{c|}{256} \\ \cline{4-13} 
\multirow{-3}{*}{\begin{tabular}[c]{@{}c@{}}Set Size\\ $m=n$\end{tabular}} &
  \multirow{-3}{*}{\begin{tabular}[c]{@{}c@{}}Dimension\\ $d$\end{tabular}} &
  \multirow{-3}{*}{Protocol} &
  \multicolumn{1}{c|}{Comm.} &
  \multicolumn{1}{c|}{Time} &
  \multicolumn{1}{c|}{Comm.} &
  \multicolumn{1}{c|}{Time} &
  \multicolumn{1}{c|}{Comm.} &
  \multicolumn{1}{c|}{Time} &
  \multicolumn{1}{c|}{Comm.} &
  \multicolumn{1}{c|}{Time} &
  \multicolumn{1}{c|}{Comm.} &
  Time \\ \hline
 &
   &
  \cite{DBLP:conf/asiacrypt/GaoQLLW24} &
  \multicolumn{1}{c|}{34.77} &
  \multicolumn{1}{c|}{5.871} &
  \multicolumn{1}{c|}{68.21} &
  \multicolumn{1}{c|}{11.32} &
  \multicolumn{1}{c|}{135.1} &
  \multicolumn{1}{c|}{19.92} &
  \multicolumn{1}{c|}{268.7} &
  \multicolumn{1}{c|}{39.32} &
  \multicolumn{1}{c|}{536.1} &
  76.97 \\ \cline{3-13} 
 &
   &
  \cite{DBLP:conf/ccs/DZL25}-High &
  \multicolumn{1}{c|}{91.67} &
  \multicolumn{1}{c|}{9.581} &
  \multicolumn{1}{c|}{106.7} &
  \multicolumn{1}{c|}{9.846} &
  \multicolumn{1}{c|}{116.7} &
  \multicolumn{1}{c|}{12.87} &
  \multicolumn{1}{c|}{125.0} &
  \multicolumn{1}{c|}{13.04} &
  \multicolumn{1}{c|}{173.3} &
  19.37 \\ \cline{3-13} 
 &
   &
  $\Pi_{\mathsf{OE}\text{-1}}^{\mathsf{sep}}$ &
  \multicolumn{1}{c|}{\cellcolor[HTML]{99BEFF}21.81} &
  \multicolumn{1}{c|}{2.031} &
  \multicolumn{1}{c|}{\cellcolor[HTML]{99BEFF}24.53} &
  \multicolumn{1}{c|}{2.225} &
  \multicolumn{1}{c|}{\cellcolor[HTML]{99BEFF}27.07} &
  \multicolumn{1}{c|}{2.486} &
  \multicolumn{1}{c|}{\cellcolor[HTML]{99BEFF}29.74} &
  \multicolumn{1}{c|}{\cellcolor[HTML]{C7ECFF}2.652} &
  \multicolumn{1}{c|}{\cellcolor[HTML]{C7ECFF}33.62} &
  \cellcolor[HTML]{C7ECFF}2.994 \\ \cline{3-13} 
 &
   &
  $\Pi_{\mathsf{OE}\text{-2}}^{\mathsf{sep}}$ &
  \multicolumn{1}{c|}{\cellcolor[HTML]{C7ECFF}21.87} &
  \multicolumn{1}{c|}{\cellcolor[HTML]{C7ECFF}2.003} &
  \multicolumn{1}{c|}{\cellcolor[HTML]{C7ECFF}24.57} &
  \multicolumn{1}{c|}{\cellcolor[HTML]{C7ECFF}2.178} &
  \multicolumn{1}{c|}{\cellcolor[HTML]{C7ECFF}27.09} &
  \multicolumn{1}{c|}{\cellcolor[HTML]{99BEFF}2.414} &
  \multicolumn{1}{c|}{\cellcolor[HTML]{C7ECFF}29.75} &
  \multicolumn{1}{c|}{\cellcolor[HTML]{99BEFF}2.567} &
  \multicolumn{1}{c|}{\cellcolor[HTML]{99BEFF}33.59} &
  \cellcolor[HTML]{99BEFF}2.825 \\ \cline{3-13} 
 &
  \multirow{-5}{*}{6} &
  $\Pi_{\mathsf{SE}}^{\mathsf{sep}}$ &
  \multicolumn{1}{c|}{25.45} &
  \multicolumn{1}{c|}{\cellcolor[HTML]{99BEFF}1.889} &
  \multicolumn{1}{c|}{29.14} &
  \multicolumn{1}{c|}{\cellcolor[HTML]{99BEFF}2.163} &
  \multicolumn{1}{c|}{32.48} &
  \multicolumn{1}{c|}{\cellcolor[HTML]{C7ECFF}2.423} &
  \multicolumn{1}{c|}{36.08} &
  \multicolumn{1}{c|}{2.664} &
  \multicolumn{1}{c|}{40.76} &
  3.098 \\ \cline{2-13} 
 &
   &
  \cite{DBLP:conf/asiacrypt/GaoQLLW24} &
  \multicolumn{1}{c|}{46.26} &
  \multicolumn{1}{c|}{14.06} &
  \multicolumn{1}{c|}{90.84} &
  \multicolumn{1}{c|}{23.64} &
  \multicolumn{1}{c|}{180.0} &
  \multicolumn{1}{c|}{27.18} &
  \multicolumn{1}{c|}{358.2} &
  \multicolumn{1}{c|}{52.4} &
  \multicolumn{1}{c|}{714.6} &
  106.5 \\ \cline{3-13} 
 &
   &
  \cite{DBLP:conf/ccs/DZL25}-High &
  \multicolumn{1}{c|}{122.2} &
  \multicolumn{1}{c|}{12.84} &
  \multicolumn{1}{c|}{142.2} &
  \multicolumn{1}{c|}{13.17} &
  \multicolumn{1}{c|}{155.5} &
  \multicolumn{1}{c|}{17.21} &
  \multicolumn{1}{c|}{166.6} &
  \multicolumn{1}{c|}{17.26} &
  \multicolumn{1}{c|}{231.0} &
  25.09 \\ \cline{3-13} 
 &
   &
  $\Pi_{\mathsf{OE}\text{-1}}^{\mathsf{sep}}$ &
  \multicolumn{1}{c|}{\cellcolor[HTML]{99BEFF}28.19} &
  \multicolumn{1}{c|}{\cellcolor[HTML]{99BEFF}2.376} &
  \multicolumn{1}{c|}{\cellcolor[HTML]{99BEFF}31.84} &
  \multicolumn{1}{c|}{\cellcolor[HTML]{C7ECFF}2.711} &
  \multicolumn{1}{c|}{\cellcolor[HTML]{99BEFF}35.03} &
  \multicolumn{1}{c|}{\cellcolor[HTML]{C7ECFF}2.962} &
  \multicolumn{1}{c|}{\cellcolor[HTML]{99BEFF}38.52} &
  \multicolumn{1}{c|}{\cellcolor[HTML]{C7ECFF}3.226} &
  \multicolumn{1}{c|}{\cellcolor[HTML]{C7ECFF}43.91} &
  \cellcolor[HTML]{C7ECFF}3.565 \\ \cline{3-13} 
 &
   &
  $\Pi_{\mathsf{OE}\text{-2}}^{\mathsf{sep}}$ &
  \multicolumn{1}{c|}{\cellcolor[HTML]{C7ECFF}28.29} &
  \multicolumn{1}{c|}{\cellcolor[HTML]{C7ECFF}2.379} &
  \multicolumn{1}{c|}{\cellcolor[HTML]{C7ECFF}31.86} &
  \multicolumn{1}{c|}{\cellcolor[HTML]{99BEFF}2.664} &
  \multicolumn{1}{c|}{\cellcolor[HTML]{C7ECFF}35.06} &
  \multicolumn{1}{c|}{\cellcolor[HTML]{99BEFF}2.945} &
  \multicolumn{1}{c|}{\cellcolor[HTML]{C7ECFF}38.64} &
  \multicolumn{1}{c|}{\cellcolor[HTML]{99BEFF}3.188} &
  \multicolumn{1}{c|}{\cellcolor[HTML]{99BEFF}43.88} &
  \cellcolor[HTML]{99BEFF}3.498 \\ \cline{3-13} 
 &
  \multirow{-5}{*}{8} &
  $\Pi_{\mathsf{SE}}^{\mathsf{sep}}$ &
  \multicolumn{1}{c|}{33.44} &
  \multicolumn{1}{c|}{2.408} &
  \multicolumn{1}{c|}{38.13} &
  \multicolumn{1}{c|}{2.783} &
  \multicolumn{1}{c|}{42.58} &
  \multicolumn{1}{c|}{3.093} &
  \multicolumn{1}{c|}{47.28} &
  \multicolumn{1}{c|}{3.401} &
  \multicolumn{1}{c|}{53.81} &
  3.859 \\ \cline{2-13} 
 &
   &
  \cite{DBLP:conf/asiacrypt/GaoQLLW24} &
  \multicolumn{1}{c|}{57.75} &
  \multicolumn{1}{c|}{19.81} &
  \multicolumn{1}{c|}{113.5} &
  \multicolumn{1}{c|}{33.91} &
  \multicolumn{1}{c|}{224.9} &
  \multicolumn{1}{c|}{37.28} &
  \multicolumn{1}{c|}{447.6} &
  \multicolumn{1}{c|}{68.99} &
  \multicolumn{1}{c|}{893.2} &
  144.2 \\ \cline{3-13} 
 &
   &
  \cite{DBLP:conf/ccs/DZL25}-High &
  \multicolumn{1}{c|}{152.7} &
  \multicolumn{1}{c|}{15.89} &
  \multicolumn{1}{c|}{177.7} &
  \multicolumn{1}{c|}{16.05} &
  \multicolumn{1}{c|}{194.3} &
  \multicolumn{1}{c|}{21.16} &
  \multicolumn{1}{c|}{208.2} &
  \multicolumn{1}{c|}{21.73} &
  \multicolumn{1}{c|}{288.6} &
  30.98 \\ \cline{3-13} 
 &
   &
  $\Pi_{\mathsf{OE}\text{-1}}^{\mathsf{sep}}$ &
  \multicolumn{1}{c|}{\cellcolor[HTML]{99BEFF}34.69} &
  \multicolumn{1}{c|}{\cellcolor[HTML]{99BEFF}2.825} &
  \multicolumn{1}{c|}{\cellcolor[HTML]{99BEFF}39.06} &
  \multicolumn{1}{c|}{\cellcolor[HTML]{C7ECFF}3.147} &
  \multicolumn{1}{c|}{\cellcolor[HTML]{C7ECFF}43.23} &
  \multicolumn{1}{c|}{\cellcolor[HTML]{C7ECFF}3.595} &
  \multicolumn{1}{c|}{\cellcolor[HTML]{C7ECFF}47.61} &
  \multicolumn{1}{c|}{\cellcolor[HTML]{C7ECFF}3.906} &
  \multicolumn{1}{c|}{\cellcolor[HTML]{99BEFF}54.41} &
  \cellcolor[HTML]{C7ECFF}4.363 \\ \cline{3-13} 
 &
   &
  $\Pi_{\mathsf{OE}\text{-2}}^{\mathsf{sep}}$ &
  \multicolumn{1}{c|}{\cellcolor[HTML]{C7ECFF}34.75} &
  \multicolumn{1}{c|}{\cellcolor[HTML]{C7ECFF}2.832} &
  \multicolumn{1}{c|}{\cellcolor[HTML]{C7ECFF}39.15} &
  \multicolumn{1}{c|}{\cellcolor[HTML]{99BEFF}3.115} &
  \multicolumn{1}{c|}{\cellcolor[HTML]{99BEFF}43.18} &
  \multicolumn{1}{c|}{\cellcolor[HTML]{99BEFF}3.514} &
  \multicolumn{1}{c|}{\cellcolor[HTML]{99BEFF}47.59} &
  \multicolumn{1}{c|}{\cellcolor[HTML]{99BEFF}3.805} &
  \multicolumn{1}{c|}{\cellcolor[HTML]{99BEFF}54.41} &
  \cellcolor[HTML]{99BEFF}4.237 \\ \cline{3-13} 
\multirow{-15}{*}{$2^8$} &
  \multirow{-5}{*}{10} &
  $\Pi_{\mathsf{SE}}^{\mathsf{sep}}$ &
  \multicolumn{1}{c|}{41.28} &
  \multicolumn{1}{c|}{2.899} &
  \multicolumn{1}{c|}{47.04} &
  \multicolumn{1}{c|}{3.308} &
  \multicolumn{1}{c|}{52.80} &
  \multicolumn{1}{c|}{3.794} &
  \multicolumn{1}{c|}{58.67} &
  \multicolumn{1}{c|}{4.183} &
  \multicolumn{1}{c|}{\cellcolor[HTML]{C7ECFF}66.82} &
  4.733 \\ \hline
 &
   &
  \cite{DBLP:conf/asiacrypt/GaoQLLW24} &
  \multicolumn{1}{c|}{556.3} &
  \multicolumn{1}{c|}{98.2} &
  \multicolumn{1}{c|}{1091} &
  \multicolumn{1}{c|}{247.1} &
  \multicolumn{1}{c|}{2161} &
  \multicolumn{1}{c|}{434.9} &
  \multicolumn{1}{c|}{4300} &
  \multicolumn{1}{c|}{683.9} &
  \multicolumn{1}{c|}{-} &
  - \\ \cline{3-13} 
 &
   &
  \cite{DBLP:conf/ccs/DZL25}-High &
  \multicolumn{1}{c|}{1466} &
  \multicolumn{1}{c|}{146.9} &
  \multicolumn{1}{c|}{1706} &
  \multicolumn{1}{c|}{149.5} &
  \multicolumn{1}{c|}{1866} &
  \multicolumn{1}{c|}{193.4} &
  \multicolumn{1}{c|}{2000} &
  \multicolumn{1}{c|}{197.2} &
  \multicolumn{1}{c|}{2773} &
  293.3 \\ \cline{3-13} 
 &
   &
  $\Pi_{\mathsf{OE}\text{-1}}^{\mathsf{sep}}$ &
  \multicolumn{1}{c|}{\cellcolor[HTML]{C7ECFF}304.5} &
  \multicolumn{1}{c|}{\cellcolor[HTML]{C7ECFF}21.72} &
  \multicolumn{1}{c|}{\cellcolor[HTML]{C7ECFF}345.1} &
  \multicolumn{1}{c|}{\cellcolor[HTML]{C7ECFF}25.12} &
  \multicolumn{1}{c|}{\cellcolor[HTML]{C7ECFF}386.6} &
  \multicolumn{1}{c|}{\cellcolor[HTML]{C7ECFF}28.46} &
  \multicolumn{1}{c|}{\cellcolor[HTML]{C7ECFF}427.8} &
  \multicolumn{1}{c|}{\cellcolor[HTML]{C7ECFF}31.62} &
  \multicolumn{1}{c|}{\cellcolor[HTML]{C7ECFF}488.8} &
  \cellcolor[HTML]{C7ECFF}35.33 \\ \cline{3-13} 
 &
   &
  $\Pi_{\mathsf{OE}\text{-2}}^{\mathsf{sep}}$ &
  \multicolumn{1}{c|}{\cellcolor[HTML]{99BEFF}303.8} &
  \multicolumn{1}{c|}{\cellcolor[HTML]{99BEFF}20.66} &
  \multicolumn{1}{c|}{\cellcolor[HTML]{99BEFF}344.3} &
  \multicolumn{1}{c|}{\cellcolor[HTML]{99BEFF}23.72} &
  \multicolumn{1}{c|}{\cellcolor[HTML]{99BEFF}385.5} &
  \multicolumn{1}{c|}{\cellcolor[HTML]{99BEFF}26.39} &
  \multicolumn{1}{c|}{\cellcolor[HTML]{99BEFF}426.8} &
  \multicolumn{1}{c|}{\cellcolor[HTML]{99BEFF}29.76} &
  \multicolumn{1}{c|}{\cellcolor[HTML]{99BEFF}487.4} &
  \cellcolor[HTML]{99BEFF}33.36 \\ \cline{3-13} 
 &
  \multirow{-5}{*}{6} &
  $\Pi_{\mathsf{SE}}^{\mathsf{sep}}$ &
  \multicolumn{1}{c|}{367.0} &
  \multicolumn{1}{c|}{25.25} &
  \multicolumn{1}{c|}{420.1} &
  \multicolumn{1}{c|}{29.42} &
  \multicolumn{1}{c|}{473.5} &
  \multicolumn{1}{c|}{33.68} &
  \multicolumn{1}{c|}{526.9} &
  \multicolumn{1}{c|}{37.60} &
  \multicolumn{1}{c|}{599.9} &
  42.19 \\ \cline{2-13} 
 &
   &
  \cite{DBLP:conf/asiacrypt/GaoQLLW24} &
  \multicolumn{1}{c|}{740.1} &
  \multicolumn{1}{c|}{141.9} &
  \multicolumn{1}{c|}{1453} &
  \multicolumn{1}{c|}{499.3} &
  \multicolumn{1}{c|}{2879} &
  \multicolumn{1}{c|}{548.1} &
  \multicolumn{1}{c|}{5731} &
  \multicolumn{1}{c|}{866.4} &
  \multicolumn{1}{c|}{-} &
  - \\ \cline{3-13} 
 &
   &
  \cite{DBLP:conf/ccs/DZL25}-High &
  \multicolumn{1}{c|}{1955} &
  \multicolumn{1}{c|}{197.2} &
  \multicolumn{1}{c|}{2275} &
  \multicolumn{1}{c|}{198.1} &
  \multicolumn{1}{c|}{2488} &
  \multicolumn{1}{c|}{259.9} &
  \multicolumn{1}{c|}{2666} &
  \multicolumn{1}{c|}{268.7} &
  \multicolumn{1}{c|}{3695} &
  387.8 \\ \cline{3-13} 
 &
   &
  $\Pi_{\mathsf{OE}\text{-1}}^{\mathsf{sep}}$ &
  \multicolumn{1}{c|}{\cellcolor[HTML]{C7ECFF}402.7} &
  \multicolumn{1}{c|}{\cellcolor[HTML]{C7ECFF}28.26} &
  \multicolumn{1}{c|}{\cellcolor[HTML]{C7ECFF}457.1} &
  \multicolumn{1}{c|}{\cellcolor[HTML]{C7ECFF}32.53} &
  \multicolumn{1}{c|}{\cellcolor[HTML]{C7ECFF}511.7} &
  \multicolumn{1}{c|}{\cellcolor[HTML]{C7ECFF}36.98} &
  \multicolumn{1}{c|}{\cellcolor[HTML]{C7ECFF}566.9} &
  \multicolumn{1}{c|}{\cellcolor[HTML]{C7ECFF}41.69} &
  \multicolumn{1}{c|}{\cellcolor[HTML]{C7ECFF}647.6} &
  \cellcolor[HTML]{C7ECFF}46.82 \\ \cline{3-13} 
 &
   &
  $\Pi_{\mathsf{OE}\text{-2}}^{\mathsf{sep}}$ &
  \multicolumn{1}{c|}{\cellcolor[HTML]{99BEFF}402.2} &
  \multicolumn{1}{c|}{\cellcolor[HTML]{99BEFF}27.06} &
  \multicolumn{1}{c|}{\cellcolor[HTML]{99BEFF}456.2} &
  \multicolumn{1}{c|}{\cellcolor[HTML]{99BEFF}31.29} &
  \multicolumn{1}{c|}{\cellcolor[HTML]{99BEFF}510.4} &
  \multicolumn{1}{c|}{\cellcolor[HTML]{99BEFF}35.24} &
  \multicolumn{1}{c|}{\cellcolor[HTML]{99BEFF}565.5} &
  \multicolumn{1}{c|}{\cellcolor[HTML]{99BEFF}39.52} &
  \multicolumn{1}{c|}{\cellcolor[HTML]{99BEFF}646.2} &
  \cellcolor[HTML]{99BEFF}44.34 \\ \cline{3-13} 
 &
  \multirow{-5}{*}{8} &
  $\Pi_{\mathsf{SE}}^{\mathsf{sep}}$ &
  \multicolumn{1}{c|}{486.5} &
  \multicolumn{1}{c|}{33.12} &
  \multicolumn{1}{c|}{557.3} &
  \multicolumn{1}{c|}{38.78} &
  \multicolumn{1}{c|}{628.1} &
  \multicolumn{1}{c|}{43.88} &
  \multicolumn{1}{c|}{699.0} &
  \multicolumn{1}{c|}{49.47} &
  \multicolumn{1}{c|}{796.2} &
  55.81 \\ \cline{2-13} 
 &
   &
  \cite{DBLP:conf/asiacrypt/GaoQLLW24} &
  \multicolumn{1}{c|}{923.9} &
  \multicolumn{1}{c|}{484.5} &
  \multicolumn{1}{c|}{1815} &
  \multicolumn{1}{c|}{558.3} &
  \multicolumn{1}{c|}{3598} &
  \multicolumn{1}{c|}{657.8} &
  \multicolumn{1}{c|}{7162} &
  \multicolumn{1}{c|}{1087} &
  \multicolumn{1}{c|}{-} &
  - \\ \cline{3-13} 
 &
   &
  \cite{DBLP:conf/ccs/DZL25}-High &
  \multicolumn{1}{c|}{2443} &
  \multicolumn{1}{c|}{242.1} &
  \multicolumn{1}{c|}{2843} &
  \multicolumn{1}{c|}{250.9} &
  \multicolumn{1}{c|}{3109} &
  \multicolumn{1}{c|}{324.3} &
  \multicolumn{1}{c|}{3332} &
  \multicolumn{1}{c|}{327.3} &
  \multicolumn{1}{c|}{4618} &
  485.8 \\ \cline{3-13} 
 &
   &
  $\Pi_{\mathsf{OE}\text{-1}}^{\mathsf{sep}}$ &
  \multicolumn{1}{c|}{\cellcolor[HTML]{C7ECFF}501.2} &
  \multicolumn{1}{c|}{\cellcolor[HTML]{C7ECFF}35.16} &
  \multicolumn{1}{c|}{\cellcolor[HTML]{C7ECFF}569.0} &
  \multicolumn{1}{c|}{\cellcolor[HTML]{C7ECFF}40.41} &
  \multicolumn{1}{c|}{\cellcolor[HTML]{C7ECFF}637.2} &
  \multicolumn{1}{c|}{\cellcolor[HTML]{C7ECFF}46.09} &
  \multicolumn{1}{c|}{\cellcolor[HTML]{C7ECFF}706.0} &
  \multicolumn{1}{c|}{\cellcolor[HTML]{C7ECFF}50.98} &
  \multicolumn{1}{c|}{\cellcolor[HTML]{C7ECFF}806.6} &
  \cellcolor[HTML]{C7ECFF}57.38 \\ \cline{3-13} 
 &
   &
  $\Pi_{\mathsf{OE}\text{-2}}^{\mathsf{sep}}$ &
  \multicolumn{1}{c|}{\cellcolor[HTML]{99BEFF}500.7} &
  \multicolumn{1}{c|}{\cellcolor[HTML]{99BEFF}33.61} &
  \multicolumn{1}{c|}{\cellcolor[HTML]{99BEFF}567.8} &
  \multicolumn{1}{c|}{\cellcolor[HTML]{99BEFF}39.22} &
  \multicolumn{1}{c|}{\cellcolor[HTML]{99BEFF}636.4} &
  \multicolumn{1}{c|}{\cellcolor[HTML]{99BEFF}44.34} &
  \multicolumn{1}{c|}{\cellcolor[HTML]{99BEFF}705.0} &
  \multicolumn{1}{c|}{\cellcolor[HTML]{99BEFF}49.41} &
  \multicolumn{1}{c|}{\cellcolor[HTML]{99BEFF}805.0} &
  \cellcolor[HTML]{99BEFF}55.61 \\ \cline{3-13} 
\multirow{-15}{*}{$2^{12}$} &
  \multirow{-5}{*}{10} &
  $\Pi_{\mathsf{SE}}^{\mathsf{sep}}$ &
  \multicolumn{1}{c|}{606.1} &
  \multicolumn{1}{c|}{41.05} &
  \multicolumn{1}{c|}{693.9} &
  \multicolumn{1}{c|}{48.08} &
  \multicolumn{1}{c|}{783.0} &
  \multicolumn{1}{c|}{54.88} &
  \multicolumn{1}{c|}{871.7} &
  \multicolumn{1}{c|}{61.26} &
  \multicolumn{1}{c|}{992.5} &
  69.31 \\ \hline
 &
   &
  \cite{DBLP:conf/asiacrypt/GaoQLLW24} &
  \multicolumn{1}{c|}{-} &
  \multicolumn{1}{c|}{-} &
  \multicolumn{1}{c|}{-} &
  \multicolumn{1}{c|}{-} &
  \multicolumn{1}{c|}{-} &
  \multicolumn{1}{c|}{-} &
  \multicolumn{1}{c|}{-} &
  \multicolumn{1}{c|}{-} &
  \multicolumn{1}{c|}{-} &
  - \\ \cline{3-13} 
 &
   &
  \cite{DBLP:conf/ccs/DZL25}-High &
  \multicolumn{1}{c|}{-} &
  \multicolumn{1}{c|}{-} &
  \multicolumn{1}{c|}{-} &
  \multicolumn{1}{c|}{-} &
  \multicolumn{1}{c|}{-} &
  \multicolumn{1}{c|}{-} &
  \multicolumn{1}{c|}{-} &
  \multicolumn{1}{c|}{-} &
  \multicolumn{1}{c|}{-} &
  - \\ \cline{3-13} 
 &
   &
  $\Pi_{\mathsf{OE}\text{-1}}^{\mathsf{sep}}$ &
  \multicolumn{1}{c|}{\cellcolor[HTML]{C7ECFF}4872} &
  \multicolumn{1}{c|}{\cellcolor[HTML]{C7ECFF}347.1} &
  \multicolumn{1}{c|}{\cellcolor[HTML]{C7ECFF}5532} &
  \multicolumn{1}{c|}{\cellcolor[HTML]{C7ECFF}405.2} &
  \multicolumn{1}{c|}{\cellcolor[HTML]{C7ECFF}6196} &
  \multicolumn{1}{c|}{\cellcolor[HTML]{C7ECFF}465.2} &
  \multicolumn{1}{c|}{\cellcolor[HTML]{C7ECFF}6861} &
  \multicolumn{1}{c|}{\cellcolor[HTML]{C7ECFF}522.1} &
  \multicolumn{1}{c|}{\cellcolor[HTML]{C7ECFF}7843} &
  \cellcolor[HTML]{C7ECFF}587.3 \\ \cline{3-13} 
 &
   &
  $\Pi_{\mathsf{OE}\text{-2}}^{\mathsf{sep}}$ &
  \multicolumn{1}{c|}{\cellcolor[HTML]{99BEFF}4860} &
  \multicolumn{1}{c|}{\cellcolor[HTML]{99BEFF}332.5} &
  \multicolumn{1}{c|}{\cellcolor[HTML]{99BEFF}5517} &
  \multicolumn{1}{c|}{\cellcolor[HTML]{99BEFF}388.9} &
  \multicolumn{1}{c|}{\cellcolor[HTML]{99BEFF}6180} &
  \multicolumn{1}{c|}{\cellcolor[HTML]{99BEFF}449.1} &
  \multicolumn{1}{c|}{\cellcolor[HTML]{99BEFF}6843} &
  \multicolumn{1}{c|}{\cellcolor[HTML]{99BEFF}501.5} &
  \multicolumn{1}{c|}{\cellcolor[HTML]{99BEFF}7822} &
  \cellcolor[HTML]{99BEFF}563.7 \\ \cline{3-13} 
 &
  \multirow{-5}{*}{6} &
  $\Pi_{\mathsf{SE}}^{\mathsf{sep}}$ &
  \multicolumn{1}{c|}{5904} &
  \multicolumn{1}{c|}{414.4} &
  \multicolumn{1}{c|}{6764} &
  \multicolumn{1}{c|}{486.4} &
  \multicolumn{1}{c|}{7627} &
  \multicolumn{1}{c|}{564.3} &
  \multicolumn{1}{c|}{8491} &
  \multicolumn{1}{c|}{634.6} &
  \multicolumn{1}{c|}{9669} &
  712.4 \\ \cline{2-13} 
 &
   &
  \cite{DBLP:conf/asiacrypt/GaoQLLW24} &
  \multicolumn{1}{c|}{-} &
  \multicolumn{1}{c|}{-} &
  \multicolumn{1}{c|}{-} &
  \multicolumn{1}{c|}{-} &
  \multicolumn{1}{c|}{-} &
  \multicolumn{1}{c|}{-} &
  \multicolumn{1}{c|}{-} &
  \multicolumn{1}{c|}{-} &
  \multicolumn{1}{c|}{-} &
  - \\ \cline{3-13} 
 &
   &
  \cite{DBLP:conf/ccs/DZL25}-High &
  \multicolumn{1}{c|}{-} &
  \multicolumn{1}{c|}{-} &
  \multicolumn{1}{c|}{-} &
  \multicolumn{1}{c|}{-} &
  \multicolumn{1}{c|}{-} &
  \multicolumn{1}{c|}{-} &
  \multicolumn{1}{c|}{-} &
  \multicolumn{1}{c|}{-} &
  \multicolumn{1}{c|}{-} &
  - \\ \cline{3-13} 
 &
   &
  $\Pi_{\mathsf{OE}\text{-1}}^{\mathsf{sep}}$ &
  \multicolumn{1}{c|}{\cellcolor[HTML]{C7ECFF}6460} &
  \multicolumn{1}{c|}{\cellcolor[HTML]{C7ECFF}463.8} &
  \multicolumn{1}{c|}{\cellcolor[HTML]{C7ECFF}7333} &
  \multicolumn{1}{c|}{\cellcolor[HTML]{C7ECFF}541.2} &
  \multicolumn{1}{c|}{\cellcolor[HTML]{C7ECFF}8216} &
  \multicolumn{1}{c|}{\cellcolor[HTML]{C7ECFF}615.3} &
  \multicolumn{1}{c|}{\cellcolor[HTML]{C7ECFF}9100} &
  \multicolumn{1}{c|}{\cellcolor[HTML]{C7ECFF}679.6} &
  \multicolumn{1}{c|}{\cellcolor[HTML]{C7ECFF}10407} &
  \cellcolor[HTML]{C7ECFF}763.2 \\ \cline{3-13} 
 &
   &
  $\Pi_{\mathsf{OE}\text{-2}}^{\mathsf{sep}}$ &
  \multicolumn{1}{c|}{\cellcolor[HTML]{99BEFF}6448} &
  \multicolumn{1}{c|}{\cellcolor[HTML]{99BEFF}454.9} &
  \multicolumn{1}{c|}{\cellcolor[HTML]{99BEFF}7319} &
  \multicolumn{1}{c|}{\cellcolor[HTML]{99BEFF}524.1} &
  \multicolumn{1}{c|}{\cellcolor[HTML]{99BEFF}8199} &
  \multicolumn{1}{c|}{\cellcolor[HTML]{99BEFF}598.5} &
  \multicolumn{1}{c|}{\cellcolor[HTML]{99BEFF}9082} &
  \multicolumn{1}{c|}{\cellcolor[HTML]{99BEFF}657.6} &
  \multicolumn{1}{c|}{\cellcolor[HTML]{99BEFF}10386} &
  \cellcolor[HTML]{99BEFF}738.9 \\ \cline{3-13} 
 &
  \multirow{-5}{*}{8} &
  $\Pi_{\mathsf{SE}}^{\mathsf{sep}}$ &
  \multicolumn{1}{c|}{7836} &
  \multicolumn{1}{c|}{559.4} &
  \multicolumn{1}{c|}{8978} &
  \multicolumn{1}{c|}{649.1} &
  \multicolumn{1}{c|}{10123} &
  \multicolumn{1}{c|}{740.6} &
  \multicolumn{1}{c|}{11273} &
  \multicolumn{1}{c|}{821.6} &
  \multicolumn{1}{c|}{12841} &
  930.1 \\ \cline{2-13} 
 &
   &
  \cite{DBLP:conf/asiacrypt/GaoQLLW24} &
  \multicolumn{1}{c|}{-} &
  \multicolumn{1}{c|}{-} &
  \multicolumn{1}{c|}{-} &
  \multicolumn{1}{c|}{-} &
  \multicolumn{1}{c|}{-} &
  \multicolumn{1}{c|}{-} &
  \multicolumn{1}{c|}{-} &
  \multicolumn{1}{c|}{-} &
  \multicolumn{1}{c|}{-} &
  - \\ \cline{3-13} 
 &
   &
  \cite{DBLP:conf/ccs/DZL25}-High &
  \multicolumn{1}{c|}{-} &
  \multicolumn{1}{c|}{-} &
  \multicolumn{1}{c|}{-} &
  \multicolumn{1}{c|}{-} &
  \multicolumn{1}{c|}{-} &
  \multicolumn{1}{c|}{-} &
  \multicolumn{1}{c|}{-} &
  \multicolumn{1}{c|}{-} &
  \multicolumn{1}{c|}{-} &
  - \\ \cline{3-13} 
 &
   &
  $\Pi_{\mathsf{OE}\text{-1}}^{\mathsf{sep}}$ &
  \multicolumn{1}{c|}{\cellcolor[HTML]{C7ECFF}8045} &
  \multicolumn{1}{c|}{\cellcolor[HTML]{C7ECFF}577.7} &
  \multicolumn{1}{c|}{\cellcolor[HTML]{C7ECFF}9136} &
  \multicolumn{1}{c|}{\cellcolor[HTML]{C7ECFF}657.1} &
  \multicolumn{1}{c|}{\cellcolor[HTML]{C7ECFF}10237} &
  \multicolumn{1}{c|}{\cellcolor[HTML]{C7ECFF}750.1} &
  \multicolumn{1}{c|}{\cellcolor[HTML]{C7ECFF}11340} &
  \multicolumn{1}{c|}{\cellcolor[HTML]{C7ECFF}856.3} &
  \multicolumn{1}{c|}{\cellcolor[HTML]{C7ECFF}12966} &
  \cellcolor[HTML]{C7ECFF}965.1 \\ \cline{3-13} 
 &
   &
  $\Pi_{\mathsf{OE}\text{-2}}^{\mathsf{sep}}$ &
  \multicolumn{1}{c|}{\cellcolor[HTML]{99BEFF}8032} &
  \multicolumn{1}{c|}{\cellcolor[HTML]{99BEFF}562.5} &
  \multicolumn{1}{c|}{\cellcolor[HTML]{99BEFF}9121} &
  \multicolumn{1}{c|}{\cellcolor[HTML]{99BEFF}639.6} &
  \multicolumn{1}{c|}{\cellcolor[HTML]{99BEFF}10220} &
  \multicolumn{1}{c|}{\cellcolor[HTML]{99BEFF}730.7} &
  \multicolumn{1}{c|}{\cellcolor[HTML]{99BEFF}11322} &
  \multicolumn{1}{c|}{\cellcolor[HTML]{99BEFF}838.1} &
  \multicolumn{1}{c|}{\cellcolor[HTML]{99BEFF}12944} &
  \cellcolor[HTML]{99BEFF}939.9 \\ \cline{3-13} 
\multirow{-15}{*}{$2^{16}$} &
  \multirow{-5}{*}{10} &
  $\Pi_{\mathsf{SE}}^{\mathsf{sep}}$ &
  \multicolumn{1}{c|}{9768} &
  \multicolumn{1}{c|}{692.2} &
  \multicolumn{1}{c|}{11191} &
  \multicolumn{1}{c|}{796.9} &
  \multicolumn{1}{c|}{12623} &
  \multicolumn{1}{c|}{916.9} &
  \multicolumn{1}{c|}{14056} &
  \multicolumn{1}{c|}{1036} &
  \multicolumn{1}{c|}{16010} &
  1169 \\ \hline
\end{tabular}
}
\caption{Communication cost (in MB) and running time (in seconds) of our protocols compared with those of \cite{DBLP:conf/asiacrypt/GaoQLLW24,DBLP:conf/ccs/DZL25} for $L_1$ distance in a high-dimensional LAN setting. Cells with - denote trials that ran out of memory. The best result is highlighted in \textcolor[rgb]{0,0.3,1}{blue}, the second best in \textcolor[rgb]{0,0.7,1}{cyan}.}
\label{tab:l1-high-lan}
\end{table}

\begin{table}[!htbp]
\renewcommand\arraystretch{1}
	\centering
 \resizebox{0.83\linewidth}{!}{
\begin{tabular}{|c|c|c|cccccccccc|}
\hline
 &
   &
   &
  \multicolumn{10}{c|}{Threshold $\delta$} \\ \cline{4-13} 
 &
   &
   &
  \multicolumn{2}{c|}{16} &
  \multicolumn{2}{c|}{32} &
  \multicolumn{2}{c|}{64} &
  \multicolumn{2}{c|}{128} &
  \multicolumn{2}{c|}{256} \\ \cline{4-13} 
\multirow{-3}{*}{\begin{tabular}[c]{@{}c@{}}Set Size\\ $m=n$\end{tabular}} &
  \multirow{-3}{*}{\begin{tabular}[c]{@{}c@{}}Dimension\\ $d$\end{tabular}} &
  \multirow{-3}{*}{Protocol} &
  \multicolumn{1}{c|}{Comm.} &
  \multicolumn{1}{c|}{Time} &
  \multicolumn{1}{c|}{Comm.} &
  \multicolumn{1}{c|}{Time} &
  \multicolumn{1}{c|}{Comm.} &
  \multicolumn{1}{c|}{Time} &
  \multicolumn{1}{c|}{Comm.} &
  \multicolumn{1}{c|}{Time} &
  \multicolumn{1}{c|}{Comm.} &
  Time \\ \hline
 &
   &
  \cite{DBLP:conf/ccs/DZL25}-Low &
  \multicolumn{1}{c|}{15.51} &
  \multicolumn{1}{c|}{5.344} &
  \multicolumn{1}{c|}{19.55} &
  \multicolumn{1}{c|}{6.937} &
  \multicolumn{1}{c|}{23.87} &
  \multicolumn{1}{c|}{11.62} &
  \multicolumn{1}{c|}{28.50} &
  \multicolumn{1}{c|}{11.83} &
  \multicolumn{1}{c|}{33.42} &
  13.62 \\ \cline{3-13} 
 &
   &
  $\Pi_{\mathsf{OE}\text{-1}}^{\mathsf{apart}}$ &
  \multicolumn{1}{c|}{\cellcolor[HTML]{C7ECFF}7.638} &
  \multicolumn{1}{c|}{4.327} &
  \multicolumn{1}{c|}{\cellcolor[HTML]{C7ECFF}8.602} &
  \multicolumn{1}{c|}{4.338} &
  \multicolumn{1}{c|}{\cellcolor[HTML]{C7ECFF}9.197} &
  \multicolumn{1}{c|}{4.447} &
  \multicolumn{1}{c|}{\cellcolor[HTML]{99BEFF}9.883} &
  \multicolumn{1}{c|}{\cellcolor[HTML]{C7ECFF}4.514} &
  \multicolumn{1}{c|}{\cellcolor[HTML]{99BEFF}11.24} &
  \cellcolor[HTML]{C7ECFF}4.659 \\ \cline{3-13} 
 &
   &
  $\Pi_{\mathsf{OE}\text{-2}}^{\mathsf{apart}}$ &
  \multicolumn{1}{c|}{7.698} &
  \multicolumn{1}{c|}{4.583} &
  \multicolumn{1}{c|}{8.656} &
  \multicolumn{1}{c|}{4.609} &
  \multicolumn{1}{c|}{\cellcolor[HTML]{99BEFF}9.181} &
  \multicolumn{1}{c|}{4.703} &
  \multicolumn{1}{c|}{\cellcolor[HTML]{C7ECFF}9.959} &
  \multicolumn{1}{c|}{4.716} &
  \multicolumn{1}{c|}{\cellcolor[HTML]{C7ECFF}11.25} &
  4.875 \\ \cline{3-13} 
 &
  \multirow{-4}{*}{2} &
  $\Pi_{\mathsf{SE}}^{\mathsf{apart}}$ &
  \multicolumn{1}{c|}{\cellcolor[HTML]{99BEFF}7.329} &
  \multicolumn{1}{c|}{\cellcolor[HTML]{99BEFF}3.645} &
  \multicolumn{1}{c|}{\cellcolor[HTML]{99BEFF}8.435} &
  \multicolumn{1}{c|}{\cellcolor[HTML]{99BEFF}3.675} &
  \multicolumn{1}{c|}{9.269} &
  \multicolumn{1}{c|}{\cellcolor[HTML]{99BEFF}3.724} &
  \multicolumn{1}{c|}{10.19} &
  \multicolumn{1}{c|}{\cellcolor[HTML]{99BEFF}3.889} &
  \multicolumn{1}{c|}{11.55} &
  \cellcolor[HTML]{99BEFF}3.988 \\ \cline{2-13} 
 &
   &
  \cite{DBLP:conf/ccs/DZL25}-Low &
  \multicolumn{1}{c|}{57.46} &
  \multicolumn{1}{c|}{6.805} &
  \multicolumn{1}{c|}{{\color[HTML]{FE0000} 84.30}} &
  \multicolumn{1}{c|}{{\color[HTML]{FE0000} 8.697}} &
  \multicolumn{1}{c|}{{\color[HTML]{FE0000} 81.50}} &
  \multicolumn{1}{c|}{{\color[HTML]{FE0000} 26.52}} &
  \multicolumn{1}{c|}{94.96} &
  \multicolumn{1}{c|}{29.92} &
  \multicolumn{1}{c|}{115.1} &
  31.31 \\ \cline{3-13} 
 &
   &
  $\Pi_{\mathsf{OE}\text{-1}}^{\mathsf{apart}}$ &
  \multicolumn{1}{c|}{\cellcolor[HTML]{C7ECFF}13.51} &
  \multicolumn{1}{c|}{\cellcolor[HTML]{C7ECFF}4.712} &
  \multicolumn{1}{c|}{15.54} &
  \multicolumn{1}{c|}{\cellcolor[HTML]{C7ECFF}4.842} &
  \multicolumn{1}{c|}{17.31} &
  \multicolumn{1}{c|}{\cellcolor[HTML]{C7ECFF}4.947} &
  \multicolumn{1}{c|}{\cellcolor[HTML]{C7ECFF}19.62} &
  \multicolumn{1}{c|}{\cellcolor[HTML]{C7ECFF}5.041} &
  \multicolumn{1}{c|}{\cellcolor[HTML]{C7ECFF}22.33} &
  \cellcolor[HTML]{C7ECFF}5.167 \\ \cline{3-13} 
 &
   &
  $\Pi_{\mathsf{OE}\text{-2}}^{\mathsf{apart}}$ &
  \multicolumn{1}{c|}{\cellcolor[HTML]{C7ECFF}13.51} &
  \multicolumn{1}{c|}{4.875} &
  \multicolumn{1}{c|}{\cellcolor[HTML]{C7ECFF}15.39} &
  \multicolumn{1}{c|}{5.054} &
  \multicolumn{1}{c|}{\cellcolor[HTML]{C7ECFF}17.26} &
  \multicolumn{1}{c|}{5.208} &
  \multicolumn{1}{c|}{19.65} &
  \multicolumn{1}{c|}{5.279} &
  \multicolumn{1}{c|}{22.54} &
  5.497 \\ \cline{3-13} 
 &
  \multirow{-4}{*}{3} &
  $\Pi_{\mathsf{SE}}^{\mathsf{apart}}$ &
  \multicolumn{1}{c|}{\cellcolor[HTML]{99BEFF}12.15} &
  \multicolumn{1}{c|}{\cellcolor[HTML]{99BEFF}3.919} &
  \multicolumn{1}{c|}{\cellcolor[HTML]{99BEFF}13.83} &
  \multicolumn{1}{c|}{\cellcolor[HTML]{99BEFF}4.138} &
  \multicolumn{1}{c|}{\cellcolor[HTML]{99BEFF}15.55} &
  \multicolumn{1}{c|}{\cellcolor[HTML]{99BEFF}4.314} &
  \multicolumn{1}{c|}{\cellcolor[HTML]{99BEFF}17.47} &
  \multicolumn{1}{c|}{\cellcolor[HTML]{99BEFF}4.518} &
  \multicolumn{1}{c|}{\cellcolor[HTML]{99BEFF}19.86} &
  \cellcolor[HTML]{99BEFF}4.714 \\ \cline{2-13} 
 &
   &
  \cite{DBLP:conf/ccs/DZL25}-Low &
  \multicolumn{1}{c|}{154.1} &
  \multicolumn{1}{c|}{21.16} &
  \multicolumn{1}{c|}{225.6} &
  \multicolumn{1}{c|}{{\color[HTML]{FE0000} 23.31}} &
  \multicolumn{1}{c|}{309.1} &
  \multicolumn{1}{c|}{{\color[HTML]{FE0000} 170.2}} &
  \multicolumn{1}{c|}{344.9} &
  \multicolumn{1}{c|}{171.8} &
  \multicolumn{1}{c|}{398.6} &
  174.9 \\ \cline{3-13} 
 &
   &
  $\Pi_{\mathsf{OE}\text{-1}}^{\mathsf{apart}}$ &
  \multicolumn{1}{c|}{\cellcolor[HTML]{C7ECFF}25.83} &
  \multicolumn{1}{c|}{\cellcolor[HTML]{C7ECFF}5.506} &
  \multicolumn{1}{c|}{30.68} &
  \multicolumn{1}{c|}{\cellcolor[HTML]{C7ECFF}5.864} &
  \multicolumn{1}{c|}{35.65} &
  \multicolumn{1}{c|}{\cellcolor[HTML]{C7ECFF}6.192} &
  \multicolumn{1}{c|}{\cellcolor[HTML]{C7ECFF}40.04} &
  \multicolumn{1}{c|}{\cellcolor[HTML]{C7ECFF}6.706} &
  \multicolumn{1}{c|}{46.27} &
  \cellcolor[HTML]{C7ECFF}7.266 \\ \cline{3-13} 
 &
   &
  $\Pi_{\mathsf{OE}\text{-2}}^{\mathsf{apart}}$ &
  \multicolumn{1}{c|}{26.11} &
  \multicolumn{1}{c|}{5.822} &
  \multicolumn{1}{c|}{\cellcolor[HTML]{C7ECFF}30.46} &
  \multicolumn{1}{c|}{6.156} &
  \multicolumn{1}{c|}{\cellcolor[HTML]{C7ECFF}35.52} &
  \multicolumn{1}{c|}{6.463} &
  \multicolumn{1}{c|}{40.08} &
  \multicolumn{1}{c|}{6.849} &
  \multicolumn{1}{c|}{\cellcolor[HTML]{C7ECFF}46.09} &
  7.508 \\ \cline{3-13} 
\multirow{-12}{*}{$2^8$} &
  \multirow{-4}{*}{4} &
  $\Pi_{\mathsf{SE}}^{\mathsf{apart}}$ &
  \multicolumn{1}{c|}{\cellcolor[HTML]{99BEFF}20.08} &
  \multicolumn{1}{c|}{\cellcolor[HTML]{99BEFF}4.803} &
  \multicolumn{1}{c|}{\cellcolor[HTML]{99BEFF}23.47} &
  \multicolumn{1}{c|}{\cellcolor[HTML]{99BEFF}5.052} &
  \multicolumn{1}{c|}{\cellcolor[HTML]{99BEFF}26.67} &
  \multicolumn{1}{c|}{\cellcolor[HTML]{99BEFF}5.422} &
  \multicolumn{1}{c|}{\cellcolor[HTML]{99BEFF}30.21} &
  \multicolumn{1}{c|}{\cellcolor[HTML]{99BEFF}5.857} &
  \multicolumn{1}{c|}{\cellcolor[HTML]{99BEFF}34.60} &
  \cellcolor[HTML]{99BEFF}6.293 \\ \hline
 &
   &
  \cite{DBLP:conf/ccs/DZL25}-Low &
  \multicolumn{1}{c|}{248.2} &
  \multicolumn{1}{c|}{66.21} &
  \multicolumn{1}{c|}{312.8} &
  \multicolumn{1}{c|}{93.19} &
  \multicolumn{1}{c|}{381.9} &
  \multicolumn{1}{c|}{127.9} &
  \multicolumn{1}{c|}{456.1} &
  \multicolumn{1}{c|}{155.1} &
  \multicolumn{1}{c|}{534.7} &
  206.5 \\ \cline{3-13} 
 &
   &
  $\Pi_{\mathsf{OE}\text{-1}}^{\mathsf{apart}}$ &
  \multicolumn{1}{c|}{\cellcolor[HTML]{C7ECFF}89.29} &
  \multicolumn{1}{c|}{\cellcolor[HTML]{C7ECFF}11.83} &
  \multicolumn{1}{c|}{\cellcolor[HTML]{C7ECFF}101.9} &
  \multicolumn{1}{c|}{\cellcolor[HTML]{C7ECFF}12.96} &
  \multicolumn{1}{c|}{\cellcolor[HTML]{C7ECFF}115.2} &
  \multicolumn{1}{c|}{\cellcolor[HTML]{C7ECFF}14.41} &
  \multicolumn{1}{c|}{\cellcolor[HTML]{C7ECFF}128.2} &
  \multicolumn{1}{c|}{\cellcolor[HTML]{C7ECFF}15.75} &
  \multicolumn{1}{c|}{\cellcolor[HTML]{C7ECFF}148.0} &
  \cellcolor[HTML]{C7ECFF}17.73 \\ \cline{3-13} 
 &
   &
  $\Pi_{\mathsf{OE}\text{-2}}^{\mathsf{apart}}$ &
  \multicolumn{1}{c|}{\cellcolor[HTML]{99BEFF}88.55} &
  \multicolumn{1}{c|}{\cellcolor[HTML]{99BEFF}11.14} &
  \multicolumn{1}{c|}{\cellcolor[HTML]{99BEFF}101.0} &
  \multicolumn{1}{c|}{\cellcolor[HTML]{99BEFF}12.19} &
  \multicolumn{1}{c|}{\cellcolor[HTML]{99BEFF}114.2} &
  \multicolumn{1}{c|}{\cellcolor[HTML]{99BEFF}13.39} &
  \multicolumn{1}{c|}{\cellcolor[HTML]{99BEFF}127.3} &
  \multicolumn{1}{c|}{\cellcolor[HTML]{99BEFF}14.62} &
  \multicolumn{1}{c|}{\cellcolor[HTML]{99BEFF}147.0} &
  \cellcolor[HTML]{99BEFF}16.13 \\ \cline{3-13} 
 &
  \multirow{-4}{*}{2} &
  $\Pi_{\mathsf{SE}}^{\mathsf{apart}}$ &
  \multicolumn{1}{c|}{96.44} &
  \multicolumn{1}{c|}{11.88} &
  \multicolumn{1}{c|}{110.1} &
  \multicolumn{1}{c|}{13.39} &
  \multicolumn{1}{c|}{124.2} &
  \multicolumn{1}{c|}{14.88} &
  \multicolumn{1}{c|}{138.2} &
  \multicolumn{1}{c|}{16.33} &
  \multicolumn{1}{c|}{159.1} &
  18.28 \\ \cline{2-13} 
 &
   &
  \cite{DBLP:conf/ccs/DZL25}-Low &
  \multicolumn{1}{c|}{919.3} &
  \multicolumn{1}{c|}{92.95} &
  \multicolumn{1}{c|}{{\color[HTML]{FE0000} 1349}} &
  \multicolumn{1}{c|}{{\color[HTML]{FE0000} 112.1}} &
  \multicolumn{1}{c|}{{\color[HTML]{FE0000} 1304}} &
  \multicolumn{1}{c|}{{\color[HTML]{FE0000} 378.9}} &
  \multicolumn{1}{c|}{1519} &
  \multicolumn{1}{c|}{388.3} &
  \multicolumn{1}{c|}{1842} &
  403.2 \\ \cline{3-13} 
 &
   &
  $\Pi_{\mathsf{OE}\text{-1}}^{\mathsf{apart}}$ &
  \multicolumn{1}{c|}{183.1} &
  \multicolumn{1}{c|}{20.45} &
  \multicolumn{1}{c|}{214.6} &
  \multicolumn{1}{c|}{23.74} &
  \multicolumn{1}{c|}{246.3} &
  \multicolumn{1}{c|}{27.04} &
  \multicolumn{1}{c|}{278.4} &
  \multicolumn{1}{c|}{30.57} &
  \multicolumn{1}{c|}{320.5} &
  34.71 \\ \cline{3-13} 
 &
   &
  $\Pi_{\mathsf{OE}\text{-2}}^{\mathsf{apart}}$ &
  \multicolumn{1}{c|}{\cellcolor[HTML]{C7ECFF}182.3} &
  \multicolumn{1}{c|}{\cellcolor[HTML]{99BEFF}19.79} &
  \multicolumn{1}{c|}{\cellcolor[HTML]{C7ECFF}213.7} &
  \multicolumn{1}{c|}{\cellcolor[HTML]{99BEFF}22.87} &
  \multicolumn{1}{c|}{\cellcolor[HTML]{C7ECFF}245.2} &
  \multicolumn{1}{c|}{\cellcolor[HTML]{99BEFF}26.05} &
  \multicolumn{1}{c|}{\cellcolor[HTML]{C7ECFF}277.5} &
  \multicolumn{1}{c|}{\cellcolor[HTML]{99BEFF}29.16} &
  \multicolumn{1}{c|}{\cellcolor[HTML]{C7ECFF}319.7} &
  \cellcolor[HTML]{99BEFF}33.15 \\ \cline{3-13} 
 &
  \multirow{-4}{*}{3} &
  $\Pi_{\mathsf{SE}}^{\mathsf{apart}}$ &
  \multicolumn{1}{c|}{\cellcolor[HTML]{99BEFF}168.2} &
  \multicolumn{1}{c|}{\cellcolor[HTML]{C7ECFF}19.91} &
  \multicolumn{1}{c|}{\cellcolor[HTML]{99BEFF}194.7} &
  \multicolumn{1}{c|}{\cellcolor[HTML]{C7ECFF}23.23} &
  \multicolumn{1}{c|}{\cellcolor[HTML]{99BEFF}221.3} &
  \multicolumn{1}{c|}{\cellcolor[HTML]{C7ECFF}26.67} &
  \multicolumn{1}{c|}{\cellcolor[HTML]{99BEFF}249.0} &
  \multicolumn{1}{c|}{\cellcolor[HTML]{C7ECFF}29.92} &
  \multicolumn{1}{c|}{\cellcolor[HTML]{99BEFF}286.1} &
  \cellcolor[HTML]{C7ECFF}34.29 \\ \cline{2-13} 
 &
   &
  \cite{DBLP:conf/ccs/DZL25}-Low &
  \multicolumn{1}{c|}{2465} &
  \multicolumn{1}{c|}{277.2} &
  \multicolumn{1}{c|}{3610} &
  \multicolumn{1}{c|}{326.1} &
  \multicolumn{1}{c|}{-} &
  \multicolumn{1}{c|}{-} &
  \multicolumn{1}{c|}{-} &
  \multicolumn{1}{c|}{-} &
  \multicolumn{1}{c|}{-} &
  - \\ \cline{3-13} 
 &
   &
  $\Pi_{\mathsf{OE}\text{-1}}^{\mathsf{apart}}$ &
  \multicolumn{1}{c|}{383.5} &
  \multicolumn{1}{c|}{41.67} &
  \multicolumn{1}{c|}{456.5} &
  \multicolumn{1}{c|}{49.47} &
  \multicolumn{1}{c|}{532.4} &
  \multicolumn{1}{c|}{58.11} &
  \multicolumn{1}{c|}{610.9} &
  \multicolumn{1}{c|}{67.08} &
  \multicolumn{1}{c|}{699.6} &
  77.24 \\ \cline{3-13} 
 &
   &
  $\Pi_{\mathsf{OE}\text{-2}}^{\mathsf{apart}}$ &
  \multicolumn{1}{c|}{\cellcolor[HTML]{C7ECFF}381.7} &
  \multicolumn{1}{c|}{\cellcolor[HTML]{C7ECFF}40.67} &
  \multicolumn{1}{c|}{\cellcolor[HTML]{C7ECFF}455.2} &
  \multicolumn{1}{c|}{\cellcolor[HTML]{C7ECFF}48.92} &
  \multicolumn{1}{c|}{\cellcolor[HTML]{C7ECFF}530.4} &
  \multicolumn{1}{c|}{\cellcolor[HTML]{C7ECFF}57.16} &
  \multicolumn{1}{c|}{\cellcolor[HTML]{C7ECFF}607.9} &
  \multicolumn{1}{c|}{\cellcolor[HTML]{C7ECFF}65.95} &
  \multicolumn{1}{c|}{\cellcolor[HTML]{C7ECFF}697.7} &
  \cellcolor[HTML]{C7ECFF}75.27 \\ \cline{3-13} 
\multirow{-12}{*}{$2^{12}$} &
  \multirow{-4}{*}{4} &
  $\Pi_{\mathsf{SE}}^{\mathsf{apart}}$ &
  \multicolumn{1}{c|}{\cellcolor[HTML]{99BEFF}292.9} &
  \multicolumn{1}{c|}{\cellcolor[HTML]{99BEFF}38.37} &
  \multicolumn{1}{c|}{\cellcolor[HTML]{99BEFF}343.8} &
  \multicolumn{1}{c|}{\cellcolor[HTML]{99BEFF}46.21} &
  \multicolumn{1}{c|}{\cellcolor[HTML]{99BEFF}397.0} &
  \multicolumn{1}{c|}{\cellcolor[HTML]{99BEFF}53.93} &
  \multicolumn{1}{c|}{\cellcolor[HTML]{99BEFF}449.7} &
  \multicolumn{1}{c|}{\cellcolor[HTML]{99BEFF}62.35} &
  \multicolumn{1}{c|}{\cellcolor[HTML]{99BEFF}515.8} &
  \cellcolor[HTML]{99BEFF}71.39 \\ \hline
 &
   &
  \cite{DBLP:conf/ccs/DZL25}-Low &
  \multicolumn{1}{c|}{3971} &
  \multicolumn{1}{c|}{1004} &
  \multicolumn{1}{c|}{-} &
  \multicolumn{1}{c|}{-} &
  \multicolumn{1}{c|}{-} &
  \multicolumn{1}{c|}{-} &
  \multicolumn{1}{c|}{-} &
  \multicolumn{1}{c|}{-} &
  \multicolumn{1}{c|}{-} &
  - \\ \cline{3-13} 
 &
   &
  $\Pi_{\mathsf{OE}\text{-1}}^{\mathsf{apart}}$ &
  \multicolumn{1}{c|}{\cellcolor[HTML]{C7ECFF}1404} &
  \multicolumn{1}{c|}{\cellcolor[HTML]{C7ECFF}142.9} &
  \multicolumn{1}{c|}{\cellcolor[HTML]{C7ECFF}1612} &
  \multicolumn{1}{c|}{\cellcolor[HTML]{C7ECFF}168.5} &
  \multicolumn{1}{c|}{\cellcolor[HTML]{C7ECFF}1824} &
  \multicolumn{1}{c|}{\cellcolor[HTML]{C7ECFF}194.8} &
  \multicolumn{1}{c|}{\cellcolor[HTML]{C7ECFF}2039} &
  \multicolumn{1}{c|}{\cellcolor[HTML]{C7ECFF}220.6} &
  \multicolumn{1}{c|}{\cellcolor[HTML]{C7ECFF}2357} &
  \cellcolor[HTML]{C7ECFF}254.7 \\ \cline{3-13} 
 &
   &
  $\Pi_{\mathsf{OE}\text{-2}}^{\mathsf{apart}}$ &
  \multicolumn{1}{c|}{\cellcolor[HTML]{99BEFF}1392} &
  \multicolumn{1}{c|}{\cellcolor[HTML]{99BEFF}128.3} &
  \multicolumn{1}{c|}{\cellcolor[HTML]{99BEFF}1597} &
  \multicolumn{1}{c|}{\cellcolor[HTML]{99BEFF}151.8} &
  \multicolumn{1}{c|}{\cellcolor[HTML]{99BEFF}1807} &
  \multicolumn{1}{c|}{\cellcolor[HTML]{99BEFF}174.8} &
  \multicolumn{1}{c|}{\cellcolor[HTML]{99BEFF}2020} &
  \multicolumn{1}{c|}{\cellcolor[HTML]{99BEFF}198.5} &
  \multicolumn{1}{c|}{\cellcolor[HTML]{99BEFF}2336} &
  \cellcolor[HTML]{99BEFF}228.7 \\ \cline{3-13} 
 &
  \multirow{-4}{*}{2} &
  $\Pi_{\mathsf{SE}}^{\mathsf{apart}}$ &
  \multicolumn{1}{c|}{1536} &
  \multicolumn{1}{c|}{152.7} &
  \multicolumn{1}{c|}{1762} &
  \multicolumn{1}{c|}{180.2} &
  \multicolumn{1}{c|}{1991} &
  \multicolumn{1}{c|}{207.4} &
  \multicolumn{1}{c|}{2220} &
  \multicolumn{1}{c|}{237.1} &
  \multicolumn{1}{c|}{2552} &
  272.6 \\ \cline{2-13} 
 &
   &
  \cite{DBLP:conf/ccs/DZL25}-Low &
  \multicolumn{1}{c|}{-} &
  \multicolumn{1}{c|}{-} &
  \multicolumn{1}{c|}{-} &
  \multicolumn{1}{c|}{-} &
  \multicolumn{1}{c|}{-} &
  \multicolumn{1}{c|}{-} &
  \multicolumn{1}{c|}{-} &
  \multicolumn{1}{c|}{-} &
  \multicolumn{1}{c|}{-} &
  - \\ \cline{3-13} 
 &
   &
  $\Pi_{\mathsf{OE}\text{-1}}^{\mathsf{apart}}$ &
  \multicolumn{1}{c|}{2919} &
  \multicolumn{1}{c|}{\cellcolor[HTML]{C7ECFF}315.4} &
  \multicolumn{1}{c|}{3421} &
  \multicolumn{1}{c|}{383.8} &
  \multicolumn{1}{c|}{3931} &
  \multicolumn{1}{c|}{458.4} &
  \multicolumn{1}{c|}{4446} &
  \multicolumn{1}{c|}{\cellcolor[HTML]{C7ECFF}531.2} &
  \multicolumn{1}{c|}{5128} &
  618.7 \\ \cline{3-13} 
 &
   &
  $\Pi_{\mathsf{OE}\text{-2}}^{\mathsf{apart}}$ &
  \multicolumn{1}{c|}{\cellcolor[HTML]{C7ECFF}2905} &
  \multicolumn{1}{c|}{\cellcolor[HTML]{99BEFF}302.3} &
  \multicolumn{1}{c|}{\cellcolor[HTML]{C7ECFF}3407} &
  \multicolumn{1}{c|}{\cellcolor[HTML]{99BEFF}365.6} &
  \multicolumn{1}{c|}{\cellcolor[HTML]{C7ECFF}3916} &
  \multicolumn{1}{c|}{\cellcolor[HTML]{99BEFF}437.5} &
  \multicolumn{1}{c|}{\cellcolor[HTML]{C7ECFF}4427} &
  \multicolumn{1}{c|}{\cellcolor[HTML]{99BEFF}505.1} &
  \multicolumn{1}{c|}{\cellcolor[HTML]{C7ECFF}5107} &
  \cellcolor[HTML]{99BEFF}592.3 \\ \cline{3-13} 
 &
  \multirow{-4}{*}{3} &
  $\Pi_{\mathsf{SE}}^{\mathsf{apart}}$ &
  \multicolumn{1}{c|}{\cellcolor[HTML]{99BEFF}2695} &
  \multicolumn{1}{c|}{316.6} &
  \multicolumn{1}{c|}{\cellcolor[HTML]{99BEFF}3127} &
  \multicolumn{1}{c|}{\cellcolor[HTML]{C7ECFF}383.4} &
  \multicolumn{1}{c|}{\cellcolor[HTML]{99BEFF}3563} &
  \multicolumn{1}{c|}{\cellcolor[HTML]{C7ECFF}458.1} &
  \multicolumn{1}{c|}{\cellcolor[HTML]{99BEFF}3999} &
  \multicolumn{1}{c|}{541.6} &
  \multicolumn{1}{c|}{\cellcolor[HTML]{99BEFF}4599} &
  \cellcolor[HTML]{C7ECFF}616.1 \\ \cline{2-13} 
 &
   &
  \cite{DBLP:conf/ccs/DZL25}-Low &
  \multicolumn{1}{c|}{-} &
  \multicolumn{1}{c|}{-} &
  \multicolumn{1}{c|}{-} &
  \multicolumn{1}{c|}{-} &
  \multicolumn{1}{c|}{-} &
  \multicolumn{1}{c|}{-} &
  \multicolumn{1}{c|}{-} &
  \multicolumn{1}{c|}{-} &
  \multicolumn{1}{c|}{-} &
  - \\ \cline{3-13} 
 &
   &
  $\Pi_{\mathsf{OE}\text{-1}}^{\mathsf{apart}}$ &
  \multicolumn{1}{c|}{-} &
  \multicolumn{1}{c|}{-} &
  \multicolumn{1}{c|}{-} &
  \multicolumn{1}{c|}{-} &
  \multicolumn{1}{c|}{-} &
  \multicolumn{1}{c|}{-} &
  \multicolumn{1}{c|}{-} &
  \multicolumn{1}{c|}{-} &
  \multicolumn{1}{c|}{-} &
  - \\ \cline{3-13} 
 &
   &
  $\Pi_{\mathsf{OE}\text{-2}}^{\mathsf{apart}}$ &
  \multicolumn{1}{c|}{-} &
  \multicolumn{1}{c|}{-} &
  \multicolumn{1}{c|}{-} &
  \multicolumn{1}{c|}{-} &
  \multicolumn{1}{c|}{-} &
  \multicolumn{1}{c|}{-} &
  \multicolumn{1}{c|}{-} &
  \multicolumn{1}{c|}{-} &
  \multicolumn{1}{c|}{-} &
  - \\ \cline{3-13} 
\multirow{-12}{*}{$2^{16}$} &
  \multirow{-4}{*}{4} &
  $\Pi_{\mathsf{SE}}^{\mathsf{apart}}$ &
  \multicolumn{1}{c|}{\cellcolor[HTML]{99BEFF}4699} &
  \multicolumn{1}{c|}{\cellcolor[HTML]{99BEFF}733.7} &
  \multicolumn{1}{c|}{-} &
  \multicolumn{1}{c|}{-} &
  \multicolumn{1}{c|}{-} &
  \multicolumn{1}{c|}{-} &
  \multicolumn{1}{c|}{-} &
  \multicolumn{1}{c|}{-} &
  \multicolumn{1}{c|}{-} &
  - \\ \hline
\end{tabular}
}
\caption{Communication cost (in MB) and running time (in seconds) of our protocols compared with those of \cite{DBLP:conf/ccs/DZL25} for $L_1$ distance in a low-dimensional WAN setting. Cells with - denote trials that ran out of memory. The best result is highlighted in \textcolor[rgb]{0,0.3,1}{blue}, the second best in \textcolor[rgb]{0,0.7,1}{cyan}, and data in \textcolor[rgb]{1,0,0}{red} font indicates abnormal values.}
\label{tab:l1-low-wan}
\end{table}

\begin{table}[!htbp]
\renewcommand\arraystretch{1}
	\centering
 \resizebox{0.83\linewidth}{!}{
\begin{tabular}{|c|c|c|cccccccccc|}
\hline
 &
   &
   &
  \multicolumn{10}{c|}{Threshold $\delta$} \\ \cline{4-13} 
 &
   &
   &
  \multicolumn{2}{c|}{16} &
  \multicolumn{2}{c|}{32} &
  \multicolumn{2}{c|}{64} &
  \multicolumn{2}{c|}{128} &
  \multicolumn{2}{c|}{256} \\ \cline{4-13} 
\multirow{-3}{*}{\begin{tabular}[c]{@{}c@{}}Set Size\\ $m=n$\end{tabular}} &
  \multirow{-3}{*}{\begin{tabular}[c]{@{}c@{}}Dimension\\ $d$\end{tabular}} &
  \multirow{-3}{*}{Protocol} &
  \multicolumn{1}{c|}{Comm.} &
  \multicolumn{1}{c|}{Time} &
  \multicolumn{1}{c|}{Comm.} &
  \multicolumn{1}{c|}{Time} &
  \multicolumn{1}{c|}{Comm.} &
  \multicolumn{1}{c|}{Time} &
  \multicolumn{1}{c|}{Comm.} &
  \multicolumn{1}{c|}{Time} &
  \multicolumn{1}{c|}{Comm.} &
  Time \\ \hline
 &
   &
  \cite{DBLP:conf/ccs/DZL25}-High &
  \multicolumn{1}{c|}{91.67} &
  \multicolumn{1}{c|}{13.9} &
  \multicolumn{1}{c|}{106.7} &
  \multicolumn{1}{c|}{14.49} &
  \multicolumn{1}{c|}{116.7} &
  \multicolumn{1}{c|}{17.56} &
  \multicolumn{1}{c|}{125.0} &
  \multicolumn{1}{c|}{19.81} &
  \multicolumn{1}{c|}{173.3} &
  26.49 \\ \cline{3-13} 
 &
   &
  $\Pi_{\mathsf{OE}\text{-1}}^{\mathsf{sep}}$ &
  \multicolumn{1}{c|}{\cellcolor[HTML]{99BEFF}21.81} &
  \multicolumn{1}{c|}{\cellcolor[HTML]{C7ECFF}6.098} &
  \multicolumn{1}{c|}{\cellcolor[HTML]{99BEFF}24.53} &
  \multicolumn{1}{c|}{\cellcolor[HTML]{99BEFF}6.251} &
  \multicolumn{1}{c|}{\cellcolor[HTML]{99BEFF}27.07} &
  \multicolumn{1}{c|}{\cellcolor[HTML]{C7ECFF}6.529} &
  \multicolumn{1}{c|}{\cellcolor[HTML]{99BEFF}29.74} &
  \multicolumn{1}{c|}{\cellcolor[HTML]{C7ECFF}6.872} &
  \multicolumn{1}{c|}{\cellcolor[HTML]{C7ECFF}33.62} &
  \cellcolor[HTML]{C7ECFF}7.326 \\ \cline{3-13} 
 &
   &
  $\Pi_{\mathsf{OE}\text{-2}}^{\mathsf{sep}}$ &
  \multicolumn{1}{c|}{\cellcolor[HTML]{C7ECFF}21.87} &
  \multicolumn{1}{c|}{6.248} &
  \multicolumn{1}{c|}{\cellcolor[HTML]{C7ECFF}24.57} &
  \multicolumn{1}{c|}{6.585} &
  \multicolumn{1}{c|}{\cellcolor[HTML]{C7ECFF}27.09} &
  \multicolumn{1}{c|}{7.052} &
  \multicolumn{1}{c|}{\cellcolor[HTML]{C7ECFF}29.75} &
  \multicolumn{1}{c|}{7.323} &
  \multicolumn{1}{c|}{\cellcolor[HTML]{99BEFF}33.59} &
  7.378 \\ \cline{3-13} 
 &
  \multirow{-4}{*}{6} &
  $\Pi_{\mathsf{SE}}^{\mathsf{sep}}$ &
  \multicolumn{1}{c|}{25.45} &
  \multicolumn{1}{c|}{\cellcolor[HTML]{99BEFF}5.566} &
  \multicolumn{1}{c|}{29.14} &
  \multicolumn{1}{c|}{\cellcolor[HTML]{C7ECFF}6.256} &
  \multicolumn{1}{c|}{32.48} &
  \multicolumn{1}{c|}{\cellcolor[HTML]{99BEFF}6.485} &
  \multicolumn{1}{c|}{36.08} &
  \multicolumn{1}{c|}{\cellcolor[HTML]{99BEFF}6.739} &
  \multicolumn{1}{c|}{40.76} &
  \cellcolor[HTML]{99BEFF}6.982 \\ \cline{2-13} 
 &
   &
  \cite{DBLP:conf/ccs/DZL25}-High &
  \multicolumn{1}{c|}{122.2} &
  \multicolumn{1}{c|}{17.51} &
  \multicolumn{1}{c|}{142.2} &
  \multicolumn{1}{c|}{20.55} &
  \multicolumn{1}{c|}{155.5} &
  \multicolumn{1}{c|}{24.63} &
  \multicolumn{1}{c|}{166.6} &
  \multicolumn{1}{c|}{24.78} &
  \multicolumn{1}{c|}{231.0} &
  33.91 \\ \cline{3-13} 
 &
   &
  $\Pi_{\mathsf{OE}\text{-1}}^{\mathsf{sep}}$ &
  \multicolumn{1}{c|}{\cellcolor[HTML]{99BEFF}28.19} &
  \multicolumn{1}{c|}{\cellcolor[HTML]{C7ECFF}6.489} &
  \multicolumn{1}{c|}{\cellcolor[HTML]{99BEFF}31.84} &
  \multicolumn{1}{c|}{\cellcolor[HTML]{C7ECFF}6.931} &
  \multicolumn{1}{c|}{\cellcolor[HTML]{99BEFF}35.03} &
  \multicolumn{1}{c|}{\cellcolor[HTML]{C7ECFF}7.321} &
  \multicolumn{1}{c|}{\cellcolor[HTML]{99BEFF}38.52} &
  \multicolumn{1}{c|}{\cellcolor[HTML]{C7ECFF}7.813} &
  \multicolumn{1}{c|}{\cellcolor[HTML]{C7ECFF}43.91} &
  8.511 \\ \cline{3-13} 
 &
   &
  $\Pi_{\mathsf{OE}\text{-2}}^{\mathsf{sep}}$ &
  \multicolumn{1}{c|}{\cellcolor[HTML]{C7ECFF}28.29} &
  \multicolumn{1}{c|}{6.919} &
  \multicolumn{1}{c|}{\cellcolor[HTML]{C7ECFF}31.86} &
  \multicolumn{1}{c|}{7.293} &
  \multicolumn{1}{c|}{\cellcolor[HTML]{C7ECFF}35.06} &
  \multicolumn{1}{c|}{7.505} &
  \multicolumn{1}{c|}{\cellcolor[HTML]{C7ECFF}38.64} &
  \multicolumn{1}{c|}{8.066} &
  \multicolumn{1}{c|}{\cellcolor[HTML]{99BEFF}43.88} &
  \cellcolor[HTML]{C7ECFF}8.472 \\ \cline{3-13} 
 &
  \multirow{-4}{*}{8} &
  $\Pi_{\mathsf{SE}}^{\mathsf{sep}}$ &
  \multicolumn{1}{c|}{33.44} &
  \multicolumn{1}{c|}{\cellcolor[HTML]{99BEFF}6.401} &
  \multicolumn{1}{c|}{38.13} &
  \multicolumn{1}{c|}{\cellcolor[HTML]{99BEFF}6.445} &
  \multicolumn{1}{c|}{42.58} &
  \multicolumn{1}{c|}{\cellcolor[HTML]{99BEFF}6.972} &
  \multicolumn{1}{c|}{47.28} &
  \multicolumn{1}{c|}{\cellcolor[HTML]{99BEFF}7.376} &
  \multicolumn{1}{c|}{53.81} &
  \cellcolor[HTML]{99BEFF}8.069 \\ \cline{2-13} 
 &
   &
  \cite{DBLP:conf/ccs/DZL25}-High &
  \multicolumn{1}{c|}{152.7} &
  \multicolumn{1}{c|}{23.24} &
  \multicolumn{1}{c|}{177.7} &
  \multicolumn{1}{c|}{23.83} &
  \multicolumn{1}{c|}{194.3} &
  \multicolumn{1}{c|}{29.22} &
  \multicolumn{1}{c|}{208.2} &
  \multicolumn{1}{c|}{29.62} &
  \multicolumn{1}{c|}{288.6} &
  41.05 \\ \cline{3-13} 
 &
   &
  $\Pi_{\mathsf{OE}\text{-1}}^{\mathsf{sep}}$ &
  \multicolumn{1}{c|}{\cellcolor[HTML]{99BEFF}34.69} &
  \multicolumn{1}{c|}{\cellcolor[HTML]{C7ECFF}7.228} &
  \multicolumn{1}{c|}{\cellcolor[HTML]{99BEFF}39.06} &
  \multicolumn{1}{c|}{\cellcolor[HTML]{C7ECFF}7.391} &
  \multicolumn{1}{c|}{\cellcolor[HTML]{C7ECFF}43.23} &
  \multicolumn{1}{c|}{\cellcolor[HTML]{C7ECFF}8.389} &
  \multicolumn{1}{c|}{\cellcolor[HTML]{C7ECFF}47.61} &
  \multicolumn{1}{c|}{\cellcolor[HTML]{C7ECFF}8.624} &
  \multicolumn{1}{c|}{\cellcolor[HTML]{99BEFF}54.41} &
  \cellcolor[HTML]{C7ECFF}9.252 \\ \cline{3-13} 
 &
   &
  $\Pi_{\mathsf{OE}\text{-2}}^{\mathsf{sep}}$ &
  \multicolumn{1}{c|}{\cellcolor[HTML]{C7ECFF}34.75} &
  \multicolumn{1}{c|}{7.394} &
  \multicolumn{1}{c|}{\cellcolor[HTML]{C7ECFF}39.15} &
  \multicolumn{1}{c|}{7.745} &
  \multicolumn{1}{c|}{\cellcolor[HTML]{99BEFF}43.18} &
  \multicolumn{1}{c|}{8.525} &
  \multicolumn{1}{c|}{\cellcolor[HTML]{99BEFF}47.59} &
  \multicolumn{1}{c|}{9.112} &
  \multicolumn{1}{c|}{\cellcolor[HTML]{99BEFF}54.41} &
  9.371 \\ \cline{3-13} 
\multirow{-12}{*}{$2^8$} &
  \multirow{-4}{*}{10} &
  $\Pi_{\mathsf{SE}}^{\mathsf{sep}}$ &
  \multicolumn{1}{c|}{41.28} &
  \multicolumn{1}{c|}{\cellcolor[HTML]{99BEFF}6.932} &
  \multicolumn{1}{c|}{47.04} &
  \multicolumn{1}{c|}{\cellcolor[HTML]{99BEFF}7.331} &
  \multicolumn{1}{c|}{52.80} &
  \multicolumn{1}{c|}{\cellcolor[HTML]{99BEFF}8.041} &
  \multicolumn{1}{c|}{58.67} &
  \multicolumn{1}{c|}{\cellcolor[HTML]{99BEFF}8.621} &
  \multicolumn{1}{c|}{\cellcolor[HTML]{C7ECFF}66.82} &
  \cellcolor[HTML]{99BEFF}8.998 \\ \hline
 &
   &
  \cite{DBLP:conf/ccs/DZL25}-High &
  \multicolumn{1}{c|}{1466} &
  \multicolumn{1}{c|}{182.1} &
  \multicolumn{1}{c|}{1706} &
  \multicolumn{1}{c|}{191.2} &
  \multicolumn{1}{c|}{1866} &
  \multicolumn{1}{c|}{245.1} &
  \multicolumn{1}{c|}{2000} &
  \multicolumn{1}{c|}{250.3} &
  \multicolumn{1}{c|}{2773} &
  355.7 \\ \cline{3-13} 
 &
   &
  $\Pi_{\mathsf{OE}\text{-1}}^{\mathsf{sep}}$ &
  \multicolumn{1}{c|}{\cellcolor[HTML]{C7ECFF}304.5} &
  \multicolumn{1}{c|}{\cellcolor[HTML]{C7ECFF}30.99} &
  \multicolumn{1}{c|}{\cellcolor[HTML]{C7ECFF}345.1} &
  \multicolumn{1}{c|}{\cellcolor[HTML]{C7ECFF}34.83} &
  \multicolumn{1}{c|}{\cellcolor[HTML]{C7ECFF}386.6} &
  \multicolumn{1}{c|}{\cellcolor[HTML]{C7ECFF}38.75} &
  \multicolumn{1}{c|}{\cellcolor[HTML]{C7ECFF}427.8} &
  \multicolumn{1}{c|}{\cellcolor[HTML]{C7ECFF}43.24} &
  \multicolumn{1}{c|}{\cellcolor[HTML]{C7ECFF}488.8} &
  \cellcolor[HTML]{C7ECFF}48.11 \\ \cline{3-13} 
 &
   &
  $\Pi_{\mathsf{OE}\text{-2}}^{\mathsf{sep}}$ &
  \multicolumn{1}{c|}{\cellcolor[HTML]{99BEFF}303.8} &
  \multicolumn{1}{c|}{\cellcolor[HTML]{99BEFF}30.51} &
  \multicolumn{1}{c|}{\cellcolor[HTML]{99BEFF}344.3} &
  \multicolumn{1}{c|}{\cellcolor[HTML]{99BEFF}33.87} &
  \multicolumn{1}{c|}{\cellcolor[HTML]{99BEFF}385.5} &
  \multicolumn{1}{c|}{\cellcolor[HTML]{99BEFF}37.54} &
  \multicolumn{1}{c|}{\cellcolor[HTML]{99BEFF}426.8} &
  \multicolumn{1}{c|}{\cellcolor[HTML]{99BEFF}41.46} &
  \multicolumn{1}{c|}{\cellcolor[HTML]{99BEFF}487.4} &
  \cellcolor[HTML]{99BEFF}46.04 \\ \cline{3-13} 
 &
  \multirow{-4}{*}{6} &
  $\Pi_{\mathsf{SE}}^{\mathsf{sep}}$ &
  \multicolumn{1}{c|}{367.0} &
  \multicolumn{1}{c|}{34.26} &
  \multicolumn{1}{c|}{420.1} &
  \multicolumn{1}{c|}{41.18} &
  \multicolumn{1}{c|}{473.5} &
  \multicolumn{1}{c|}{46.21} &
  \multicolumn{1}{c|}{526.9} &
  \multicolumn{1}{c|}{49.64} &
  \multicolumn{1}{c|}{599.9} &
  56.52 \\ \cline{2-13} 
 &
   &
  \cite{DBLP:conf/ccs/DZL25}-High &
  \multicolumn{1}{c|}{1955} &
  \multicolumn{1}{c|}{242.1} &
  \multicolumn{1}{c|}{2275} &
  \multicolumn{1}{c|}{254.6} &
  \multicolumn{1}{c|}{2488} &
  \multicolumn{1}{c|}{318.4} &
  \multicolumn{1}{c|}{2666} &
  \multicolumn{1}{c|}{329.8} &
  \multicolumn{1}{c|}{3695} &
  476.2 \\ \cline{3-13} 
 &
   &
  $\Pi_{\mathsf{OE}\text{-1}}^{\mathsf{sep}}$ &
  \multicolumn{1}{c|}{\cellcolor[HTML]{C7ECFF}402.7} &
  \multicolumn{1}{c|}{\cellcolor[HTML]{C7ECFF}39.59} &
  \multicolumn{1}{c|}{\cellcolor[HTML]{C7ECFF}457.1} &
  \multicolumn{1}{c|}{\cellcolor[HTML]{C7ECFF}44.22} &
  \multicolumn{1}{c|}{\cellcolor[HTML]{C7ECFF}511.7} &
  \multicolumn{1}{c|}{\cellcolor[HTML]{C7ECFF}49.87} &
  \multicolumn{1}{c|}{\cellcolor[HTML]{C7ECFF}566.9} &
  \multicolumn{1}{c|}{\cellcolor[HTML]{C7ECFF}54.99} &
  \multicolumn{1}{c|}{\cellcolor[HTML]{C7ECFF}647.6} &
  \cellcolor[HTML]{C7ECFF}61.48 \\ \cline{3-13} 
 &
   &
  $\Pi_{\mathsf{OE}\text{-2}}^{\mathsf{sep}}$ &
  \multicolumn{1}{c|}{\cellcolor[HTML]{99BEFF}402.2} &
  \multicolumn{1}{c|}{\cellcolor[HTML]{99BEFF}38.28} &
  \multicolumn{1}{c|}{\cellcolor[HTML]{99BEFF}456.2} &
  \multicolumn{1}{c|}{\cellcolor[HTML]{99BEFF}43.66} &
  \multicolumn{1}{c|}{\cellcolor[HTML]{99BEFF}510.4} &
  \multicolumn{1}{c|}{\cellcolor[HTML]{99BEFF}48.52} &
  \multicolumn{1}{c|}{\cellcolor[HTML]{99BEFF}565.5} &
  \multicolumn{1}{c|}{\cellcolor[HTML]{99BEFF}53.31} &
  \multicolumn{1}{c|}{\cellcolor[HTML]{99BEFF}646.2} &
  \cellcolor[HTML]{99BEFF}59.62 \\ \cline{3-13} 
 &
  \multirow{-4}{*}{8} &
  $\Pi_{\mathsf{SE}}^{\mathsf{sep}}$ &
  \multicolumn{1}{c|}{486.5} &
  \multicolumn{1}{c|}{46.09} &
  \multicolumn{1}{c|}{557.3} &
  \multicolumn{1}{c|}{51.39} &
  \multicolumn{1}{c|}{628.1} &
  \multicolumn{1}{c|}{58.71} &
  \multicolumn{1}{c|}{699.0} &
  \multicolumn{1}{c|}{64.46} &
  \multicolumn{1}{c|}{796.2} &
  71.32 \\ \cline{2-13} 
 &
   &
  \cite{DBLP:conf/ccs/DZL25}-High &
  \multicolumn{1}{c|}{2443} &
  \multicolumn{1}{c|}{297.1} &
  \multicolumn{1}{c|}{2843} &
  \multicolumn{1}{c|}{319.6} &
  \multicolumn{1}{c|}{3109} &
  \multicolumn{1}{c|}{395.8} &
  \multicolumn{1}{c|}{3332} &
  \multicolumn{1}{c|}{408.3} &
  \multicolumn{1}{c|}{4618} &
  592.4 \\ \cline{3-13} 
 &
   &
  $\Pi_{\mathsf{OE}\text{-1}}^{\mathsf{sep}}$ &
  \multicolumn{1}{c|}{\cellcolor[HTML]{C7ECFF}501.2} &
  \multicolumn{1}{c|}{\cellcolor[HTML]{C7ECFF}47.84} &
  \multicolumn{1}{c|}{\cellcolor[HTML]{C7ECFF}569.0} &
  \multicolumn{1}{c|}{\cellcolor[HTML]{C7ECFF}54.52} &
  \multicolumn{1}{c|}{\cellcolor[HTML]{C7ECFF}637.2} &
  \multicolumn{1}{c|}{\cellcolor[HTML]{C7ECFF}60.17} &
  \multicolumn{1}{c|}{\cellcolor[HTML]{C7ECFF}706.0} &
  \multicolumn{1}{c|}{\cellcolor[HTML]{C7ECFF}66.99} &
  \multicolumn{1}{c|}{\cellcolor[HTML]{C7ECFF}806.6} &
  \cellcolor[HTML]{C7ECFF}75.21 \\ \cline{3-13} 
 &
   &
  $\Pi_{\mathsf{OE}\text{-2}}^{\mathsf{sep}}$ &
  \multicolumn{1}{c|}{\cellcolor[HTML]{99BEFF}500.7} &
  \multicolumn{1}{c|}{\cellcolor[HTML]{99BEFF}46.71} &
  \multicolumn{1}{c|}{\cellcolor[HTML]{99BEFF}567.8} &
  \multicolumn{1}{c|}{\cellcolor[HTML]{99BEFF}53.01} &
  \multicolumn{1}{c|}{\cellcolor[HTML]{99BEFF}636.4} &
  \multicolumn{1}{c|}{\cellcolor[HTML]{99BEFF}59.26} &
  \multicolumn{1}{c|}{\cellcolor[HTML]{99BEFF}705.0} &
  \multicolumn{1}{c|}{\cellcolor[HTML]{99BEFF}65.15} &
  \multicolumn{1}{c|}{\cellcolor[HTML]{99BEFF}805.0} &
  \cellcolor[HTML]{99BEFF}73.65 \\ \cline{3-13} 
\multirow{-12}{*}{$2^{12}$} &
  \multirow{-4}{*}{10} &
  $\Pi_{\mathsf{SE}}^{\mathsf{sep}}$ &
  \multicolumn{1}{c|}{606.1} &
  \multicolumn{1}{c|}{54.85} &
  \multicolumn{1}{c|}{693.9} &
  \multicolumn{1}{c|}{63.73} &
  \multicolumn{1}{c|}{783.0} &
  \multicolumn{1}{c|}{70.58} &
  \multicolumn{1}{c|}{871.7} &
  \multicolumn{1}{c|}{75.21} &
  \multicolumn{1}{c|}{992.5} &
  89.49 \\ \hline
 &
   &
  \cite{DBLP:conf/ccs/DZL25}-High &
  \multicolumn{1}{c|}{-} &
  \multicolumn{1}{c|}{-} &
  \multicolumn{1}{c|}{-} &
  \multicolumn{1}{c|}{-} &
  \multicolumn{1}{c|}{-} &
  \multicolumn{1}{c|}{-} &
  \multicolumn{1}{c|}{-} &
  \multicolumn{1}{c|}{-} &
  \multicolumn{1}{c|}{-} &
  - \\ \cline{3-13} 
 &
   &
  $\Pi_{\mathsf{OE}\text{-1}}^{\mathsf{sep}}$ &
  \multicolumn{1}{c|}{\cellcolor[HTML]{C7ECFF}4872} &
  \multicolumn{1}{c|}{\cellcolor[HTML]{C7ECFF}442.9} &
  \multicolumn{1}{c|}{\cellcolor[HTML]{C7ECFF}5532} &
  \multicolumn{1}{c|}{\cellcolor[HTML]{C7ECFF}508.8} &
  \multicolumn{1}{c|}{\cellcolor[HTML]{C7ECFF}6196} &
  \multicolumn{1}{c|}{\cellcolor[HTML]{C7ECFF}581.8} &
  \multicolumn{1}{c|}{\cellcolor[HTML]{C7ECFF}6861} &
  \multicolumn{1}{c|}{\cellcolor[HTML]{C7ECFF}649.7} &
  \multicolumn{1}{c|}{\cellcolor[HTML]{C7ECFF}7843} &
  \cellcolor[HTML]{C7ECFF}735.8 \\ \cline{3-13} 
 &
   &
  $\Pi_{\mathsf{OE}\text{-2}}^{\mathsf{sep}}$ &
  \multicolumn{1}{c|}{\cellcolor[HTML]{99BEFF}4860} &
  \multicolumn{1}{c|}{\cellcolor[HTML]{99BEFF}429.1} &
  \multicolumn{1}{c|}{\cellcolor[HTML]{99BEFF}5517} &
  \multicolumn{1}{c|}{\cellcolor[HTML]{99BEFF}491.7} &
  \multicolumn{1}{c|}{\cellcolor[HTML]{99BEFF}6180} &
  \multicolumn{1}{c|}{\cellcolor[HTML]{99BEFF}562.3} &
  \multicolumn{1}{c|}{\cellcolor[HTML]{99BEFF}6843} &
  \multicolumn{1}{c|}{\cellcolor[HTML]{99BEFF}627.2} &
  \multicolumn{1}{c|}{\cellcolor[HTML]{99BEFF}7822} &
  \cellcolor[HTML]{99BEFF}710.8 \\ \cline{3-13} 
 &
  \multirow{-4}{*}{6} &
  $\Pi_{\mathsf{SE}}^{\mathsf{sep}}$ &
  \multicolumn{1}{c|}{5904} &
  \multicolumn{1}{c|}{499.1} &
  \multicolumn{1}{c|}{6764} &
  \multicolumn{1}{c|}{573.4} &
  \multicolumn{1}{c|}{7627} &
  \multicolumn{1}{c|}{657.1} &
  \multicolumn{1}{c|}{8491} &
  \multicolumn{1}{c|}{735.9} &
  \multicolumn{1}{c|}{9669} &
  833.5 \\ \cline{2-13} 
 &
   &
  \cite{DBLP:conf/ccs/DZL25}-High &
  \multicolumn{1}{c|}{-} &
  \multicolumn{1}{c|}{-} &
  \multicolumn{1}{c|}{-} &
  \multicolumn{1}{c|}{-} &
  \multicolumn{1}{c|}{-} &
  \multicolumn{1}{c|}{-} &
  \multicolumn{1}{c|}{-} &
  \multicolumn{1}{c|}{-} &
  \multicolumn{1}{c|}{-} &
  - \\ \cline{3-13} 
 &
   &
  $\Pi_{\mathsf{OE}\text{-1}}^{\mathsf{sep}}$ &
  \multicolumn{1}{c|}{\cellcolor[HTML]{C7ECFF}6460} &
  \multicolumn{1}{c|}{\cellcolor[HTML]{C7ECFF}588.2} &
  \multicolumn{1}{c|}{\cellcolor[HTML]{C7ECFF}7333} &
  \multicolumn{1}{c|}{\cellcolor[HTML]{C7ECFF}677.3} &
  \multicolumn{1}{c|}{\cellcolor[HTML]{C7ECFF}8216} &
  \multicolumn{1}{c|}{\cellcolor[HTML]{C7ECFF}766.9} &
  \multicolumn{1}{c|}{\cellcolor[HTML]{C7ECFF}9100} &
  \multicolumn{1}{c|}{\cellcolor[HTML]{C7ECFF}843.4} &
  \multicolumn{1}{c|}{\cellcolor[HTML]{C7ECFF}10407} &
  \cellcolor[HTML]{C7ECFF}957.6 \\ \cline{3-13} 
 &
   &
  $\Pi_{\mathsf{OE}\text{-2}}^{\mathsf{sep}}$ &
  \multicolumn{1}{c|}{\cellcolor[HTML]{99BEFF}6448} &
  \multicolumn{1}{c|}{\cellcolor[HTML]{99BEFF}575.8} &
  \multicolumn{1}{c|}{\cellcolor[HTML]{99BEFF}7319} &
  \multicolumn{1}{c|}{\cellcolor[HTML]{99BEFF}662.5} &
  \multicolumn{1}{c|}{\cellcolor[HTML]{99BEFF}8199} &
  \multicolumn{1}{c|}{\cellcolor[HTML]{99BEFF}747.4} &
  \multicolumn{1}{c|}{\cellcolor[HTML]{99BEFF}9082} &
  \multicolumn{1}{c|}{\cellcolor[HTML]{99BEFF}822.9} &
  \multicolumn{1}{c|}{\cellcolor[HTML]{99BEFF}10386} &
  \cellcolor[HTML]{99BEFF}938.9 \\ \cline{3-13} 
 &
  \multirow{-4}{*}{8} &
  $\Pi_{\mathsf{SE}}^{\mathsf{sep}}$ &
  \multicolumn{1}{c|}{7836} &
  \multicolumn{1}{c|}{663.1} &
  \multicolumn{1}{c|}{8978} &
  \multicolumn{1}{c|}{766.7} &
  \multicolumn{1}{c|}{10123} &
  \multicolumn{1}{c|}{868.3} &
  \multicolumn{1}{c|}{11273} &
  \multicolumn{1}{c|}{959.6} &
  \multicolumn{1}{c|}{12841} &
  1088 \\ \cline{2-13} 
 &
   &
  \cite{DBLP:conf/ccs/DZL25}-High &
  \multicolumn{1}{c|}{-} &
  \multicolumn{1}{c|}{-} &
  \multicolumn{1}{c|}{-} &
  \multicolumn{1}{c|}{-} &
  \multicolumn{1}{c|}{-} &
  \multicolumn{1}{c|}{-} &
  \multicolumn{1}{c|}{-} &
  \multicolumn{1}{c|}{-} &
  \multicolumn{1}{c|}{-} &
  - \\ \cline{3-13} 
 &
   &
  $\Pi_{\mathsf{OE}\text{-1}}^{\mathsf{sep}}$ &
  \multicolumn{1}{c|}{\cellcolor[HTML]{C7ECFF}8045} &
  \multicolumn{1}{c|}{\cellcolor[HTML]{C7ECFF}732.2} &
  \multicolumn{1}{c|}{\cellcolor[HTML]{C7ECFF}9136} &
  \multicolumn{1}{c|}{\cellcolor[HTML]{C7ECFF}831.6} &
  \multicolumn{1}{c|}{\cellcolor[HTML]{C7ECFF}10237} &
  \multicolumn{1}{c|}{\cellcolor[HTML]{C7ECFF}938.1} &
  \multicolumn{1}{c|}{\cellcolor[HTML]{C7ECFF}11340} &
  \multicolumn{1}{c|}{\cellcolor[HTML]{C7ECFF}1064} &
  \multicolumn{1}{c|}{\cellcolor[HTML]{C7ECFF}12966} &
  \cellcolor[HTML]{C7ECFF}1207 \\ \cline{3-13} 
 &
   &
  $\Pi_{\mathsf{OE}\text{-2}}^{\mathsf{sep}}$ &
  \multicolumn{1}{c|}{\cellcolor[HTML]{99BEFF}8032} &
  \multicolumn{1}{c|}{\cellcolor[HTML]{99BEFF}720.9} &
  \multicolumn{1}{c|}{\cellcolor[HTML]{99BEFF}9121} &
  \multicolumn{1}{c|}{\cellcolor[HTML]{99BEFF}815.5} &
  \multicolumn{1}{c|}{\cellcolor[HTML]{99BEFF}10220} &
  \multicolumn{1}{c|}{\cellcolor[HTML]{99BEFF}921.6} &
  \multicolumn{1}{c|}{\cellcolor[HTML]{99BEFF}11322} &
  \multicolumn{1}{c|}{\cellcolor[HTML]{99BEFF}1041} &
  \multicolumn{1}{c|}{\cellcolor[HTML]{99BEFF}12944} &
  \cellcolor[HTML]{99BEFF}1183 \\ \cline{3-13} 
\multirow{-12}{*}{$2^{16}$} &
  \multirow{-4}{*}{10} &
  $\Pi_{\mathsf{SE}}^{\mathsf{sep}}$ &
  \multicolumn{1}{c|}{9768} &
  \multicolumn{1}{c|}{824.4} &
  \multicolumn{1}{c|}{11191} &
  \multicolumn{1}{c|}{941.8} &
  \multicolumn{1}{c|}{12623} &
  \multicolumn{1}{c|}{1069} &
  \multicolumn{1}{c|}{14056} &
  \multicolumn{1}{c|}{1211} &
  \multicolumn{1}{c|}{16010} &
  1369 \\ \hline
\end{tabular}
}
\caption{Communication cost (in MB) and running time (in seconds) of our protocols compared with those of \cite{DBLP:conf/ccs/DZL25} for $L_1$ distance in a high-dimensional WAN setting. Cells with - denote trials that ran out of memory. The best result is highlighted in \textcolor[rgb]{0,0.3,1}{blue}, the second best in \textcolor[rgb]{0,0.7,1}{cyan}.}
\label{tab:l1-high-wan}
\end{table}

\begin{table}[!htbp]
\renewcommand\arraystretch{1}
	\centering
 \resizebox{0.83\linewidth}{!}{
\begin{tabular}{|c|c|c|cccccccccc|}
\hline
 &
   &
   &
  \multicolumn{10}{c|}{Threshold $\delta$} \\ \cline{4-13} 
 &
   &
   &
  \multicolumn{2}{c|}{16} &
  \multicolumn{2}{c|}{32} &
  \multicolumn{2}{c|}{64} &
  \multicolumn{2}{c|}{128} &
  \multicolumn{2}{c|}{256} \\ \cline{4-13} 
\multirow{-3}{*}{\begin{tabular}[c]{@{}c@{}}Set Size\\ $m=n$\end{tabular}} &
  \multirow{-3}{*}{\begin{tabular}[c]{@{}c@{}}Dimension\\ $d$\end{tabular}} &
  \multirow{-3}{*}{Protocol} &
  \multicolumn{1}{c|}{Comm.} &
  \multicolumn{1}{c|}{Time} &
  \multicolumn{1}{c|}{Comm.} &
  \multicolumn{1}{c|}{Time} &
  \multicolumn{1}{c|}{Comm.} &
  \multicolumn{1}{c|}{Time} &
  \multicolumn{1}{c|}{Comm.} &
  \multicolumn{1}{c|}{Time} &
  \multicolumn{1}{c|}{Comm.} &
  Time \\ \hline
 &
   &
  \cite{DBLP:conf/ccs/DZL25}-Low &
  \multicolumn{1}{c|}{26.70} &
  \multicolumn{1}{c|}{3.912} &
  \multicolumn{1}{c|}{33.15} &
  \multicolumn{1}{c|}{5.464} &
  \multicolumn{1}{c|}{40.14} &
  \multicolumn{1}{c|}{7.343} &
  \multicolumn{1}{c|}{47.92} &
  \multicolumn{1}{c|}{9.841} &
  \multicolumn{1}{c|}{56.97} &
  12.21 \\ \cline{3-13} 
 &
   &
  $\Pi_{\mathsf{OE}\text{-1}}^{\mathsf{apart}}$ &
  \multicolumn{1}{c|}{\cellcolor[HTML]{99BEFF}10.03} &
  \multicolumn{1}{c|}{1.089} &
  \multicolumn{1}{c|}{\cellcolor[HTML]{99BEFF}11.36} &
  \multicolumn{1}{c|}{1.182} &
  \multicolumn{1}{c|}{\cellcolor[HTML]{C7ECFF}12.58} &
  \multicolumn{1}{c|}{1.234} &
  \multicolumn{1}{c|}{\cellcolor[HTML]{C7ECFF}13.82} &
  \multicolumn{1}{c|}{1.331} &
  \multicolumn{1}{c|}{15.79} &
  1.418 \\ \cline{3-13} 
 &
   &
  $\Pi_{\mathsf{OE}\text{-2}}^{\mathsf{apart}}$ &
  \multicolumn{1}{c|}{\cellcolor[HTML]{C7ECFF}10.07} &
  \multicolumn{1}{c|}{\cellcolor[HTML]{C7ECFF}1.028} &
  \multicolumn{1}{c|}{\cellcolor[HTML]{99BEFF}11.36} &
  \multicolumn{1}{c|}{\cellcolor[HTML]{C7ECFF}1.039} &
  \multicolumn{1}{c|}{\cellcolor[HTML]{99BEFF}12.53} &
  \multicolumn{1}{c|}{\cellcolor[HTML]{C7ECFF}1.102} &
  \multicolumn{1}{c|}{\cellcolor[HTML]{99BEFF}13.74} &
  \multicolumn{1}{c|}{\cellcolor[HTML]{C7ECFF}1.199} &
  \multicolumn{1}{c|}{\cellcolor[HTML]{C7ECFF}15.74} &
  \cellcolor[HTML]{C7ECFF}1.318 \\ \cline{3-13} 
 &
  \multirow{-4}{*}{2} &
  $\Pi_{\mathsf{SE}}^{\mathsf{apart}}$ &
  \multicolumn{1}{c|}{10.28} &
  \multicolumn{1}{c|}{\cellcolor[HTML]{99BEFF}0.869} &
  \multicolumn{1}{c|}{11.80} &
  \multicolumn{1}{c|}{\cellcolor[HTML]{99BEFF}0.969} &
  \multicolumn{1}{c|}{13.09} &
  \multicolumn{1}{c|}{\cellcolor[HTML]{99BEFF}1.049} &
  \multicolumn{1}{c|}{14.51} &
  \multicolumn{1}{c|}{\cellcolor[HTML]{99BEFF}1.185} &
  \multicolumn{1}{c|}{16.46} &
  \cellcolor[HTML]{99BEFF}1.278 \\ \cline{2-13} 
 &
   &
  \cite{DBLP:conf/ccs/DZL25}-Low &
  \multicolumn{1}{c|}{110.4} &
  \multicolumn{1}{c|}{4.447} &
  \multicolumn{1}{c|}{{\color[HTML]{FE0000} 163.9}} &
  \multicolumn{1}{c|}{{\color[HTML]{FE0000} 5.681}} &
  \multicolumn{1}{c|}{{\color[HTML]{FE0000} 141.8}} &
  \multicolumn{1}{c|}{{\color[HTML]{FE0000} 22.34}} &
  \multicolumn{1}{c|}{169.4} &
  \multicolumn{1}{c|}{22.86} &
  \multicolumn{1}{c|}{211.3} &
  23.37 \\ \cline{3-13} 
 &
   &
  $\Pi_{\mathsf{OE}\text{-1}}^{\mathsf{apart}}$ &
  \multicolumn{1}{c|}{\cellcolor[HTML]{C7ECFF}18.88} &
  \multicolumn{1}{c|}{1.447} &
  \multicolumn{1}{c|}{22.00} &
  \multicolumn{1}{c|}{1.615} &
  \multicolumn{1}{c|}{24.92} &
  \multicolumn{1}{c|}{1.767} &
  \multicolumn{1}{c|}{27.88} &
  \multicolumn{1}{c|}{1.934} &
  \multicolumn{1}{c|}{31.97} &
  \cellcolor[HTML]{C7ECFF}2.125 \\ \cline{3-13} 
 &
   &
  $\Pi_{\mathsf{OE}\text{-2}}^{\mathsf{apart}}$ &
  \multicolumn{1}{c|}{18.96} &
  \multicolumn{1}{c|}{\cellcolor[HTML]{C7ECFF}1.372} &
  \multicolumn{1}{c|}{\cellcolor[HTML]{C7ECFF}21.96} &
  \multicolumn{1}{c|}{\cellcolor[HTML]{C7ECFF}1.522} &
  \multicolumn{1}{c|}{\cellcolor[HTML]{C7ECFF}24.55} &
  \multicolumn{1}{c|}{\cellcolor[HTML]{C7ECFF}1.723} &
  \multicolumn{1}{c|}{\cellcolor[HTML]{C7ECFF}27.70} &
  \multicolumn{1}{c|}{\cellcolor[HTML]{99BEFF}1.783} &
  \multicolumn{1}{c|}{\cellcolor[HTML]{C7ECFF}31.80} &
  \cellcolor[HTML]{99BEFF}2.034 \\ \cline{3-13} 
 &
  \multirow{-4}{*}{3} &
  $\Pi_{\mathsf{SE}}^{\mathsf{apart}}$ &
  \multicolumn{1}{c|}{\cellcolor[HTML]{99BEFF}17.37} &
  \multicolumn{1}{c|}{\cellcolor[HTML]{99BEFF}1.256} &
  \multicolumn{1}{c|}{\cellcolor[HTML]{99BEFF}19.79} &
  \multicolumn{1}{c|}{\cellcolor[HTML]{99BEFF}1.472} &
  \multicolumn{1}{c|}{\cellcolor[HTML]{99BEFF}22.35} &
  \multicolumn{1}{c|}{\cellcolor[HTML]{99BEFF}1.657} &
  \multicolumn{1}{c|}{\cellcolor[HTML]{99BEFF}25.12} &
  \multicolumn{1}{c|}{\cellcolor[HTML]{C7ECFF}1.865} &
  \multicolumn{1}{c|}{\cellcolor[HTML]{99BEFF}28.58} &
  2.138 \\ \cline{2-13} 
 &
   &
  \cite{DBLP:conf/ccs/DZL25}-Low &
  \multicolumn{1}{c|}{295.1} &
  \multicolumn{1}{c|}{14.12} &
  \multicolumn{1}{c|}{437.2} &
  \multicolumn{1}{c|}{{\color[HTML]{FE0000} 15.86}} &
  \multicolumn{1}{c|}{468.3} &
  \multicolumn{1}{c|}{{\color[HTML]{FE0000} 164.1}} &
  \multicolumn{1}{c|}{540.3} &
  \multicolumn{1}{c|}{164.8} &
  \multicolumn{1}{c|}{648.7} &
  167.3 \\ \cline{3-13} 
 &
   &
  $\Pi_{\mathsf{OE}\text{-1}}^{\mathsf{apart}}$ &
  \multicolumn{1}{c|}{\cellcolor[HTML]{C7ECFF}37.40} &
  \multicolumn{1}{c|}{\cellcolor[HTML]{C7ECFF}2.319} &
  \multicolumn{1}{c|}{\cellcolor[HTML]{C7ECFF}44.39} &
  \multicolumn{1}{c|}{\cellcolor[HTML]{C7ECFF}2.758} &
  \multicolumn{1}{c|}{\cellcolor[HTML]{C7ECFF}51.35} &
  \multicolumn{1}{c|}{\cellcolor[HTML]{99BEFF}3.187} &
  \multicolumn{1}{c|}{\cellcolor[HTML]{C7ECFF}58.61} &
  \multicolumn{1}{c|}{\cellcolor[HTML]{C7ECFF}3.685} &
  \multicolumn{1}{c|}{67.17} &
  \cellcolor[HTML]{99BEFF}4.139 \\ \cline{3-13} 
 &
   &
  $\Pi_{\mathsf{OE}\text{-2}}^{\mathsf{apart}}$ &
  \multicolumn{1}{c|}{37.78} &
  \multicolumn{1}{c|}{2.402} &
  \multicolumn{1}{c|}{44.56} &
  \multicolumn{1}{c|}{2.774} &
  \multicolumn{1}{c|}{51.93} &
  \multicolumn{1}{c|}{3.312} &
  \multicolumn{1}{c|}{58.78} &
  \multicolumn{1}{c|}{3.728} &
  \multicolumn{1}{c|}{\cellcolor[HTML]{C7ECFF}66.96} &
  4.226 \\ \cline{3-13} 
\multirow{-12}{*}{$2^8$} &
  \multirow{-4}{*}{4} &
  $\Pi_{\mathsf{SE}}^{\mathsf{apart}}$ &
  \multicolumn{1}{c|}{\cellcolor[HTML]{99BEFF}29.00} &
  \multicolumn{1}{c|}{\cellcolor[HTML]{99BEFF}2.256} &
  \multicolumn{1}{c|}{\cellcolor[HTML]{99BEFF}33.98} &
  \multicolumn{1}{c|}{\cellcolor[HTML]{99BEFF}2.653} &
  \multicolumn{1}{c|}{\cellcolor[HTML]{99BEFF}39.03} &
  \multicolumn{1}{c|}{\cellcolor[HTML]{C7ECFF}3.194} &
  \multicolumn{1}{c|}{\cellcolor[HTML]{99BEFF}43.94} &
  \multicolumn{1}{c|}{\cellcolor[HTML]{99BEFF}3.663} &
  \multicolumn{1}{c|}{\cellcolor[HTML]{99BEFF}50.25} &
  \cellcolor[HTML]{C7ECFF}4.148 \\ \hline
 &
   &
  \cite{DBLP:conf/ccs/DZL25}-Low &
  \multicolumn{1}{c|}{427.2} &
  \multicolumn{1}{c|}{60.38} &
  \multicolumn{1}{c|}{530.5} &
  \multicolumn{1}{c|}{86.35} &
  \multicolumn{1}{c|}{642.3} &
  \multicolumn{1}{c|}{117.4} &
  \multicolumn{1}{c|}{766.7} &
  \multicolumn{1}{c|}{153.5} &
  \multicolumn{1}{c|}{911.5} &
  195.3 \\ \cline{3-13} 
 &
   &
  $\Pi_{\mathsf{OE}\text{-1}}^{\mathsf{apart}}$ &
  \multicolumn{1}{c|}{\cellcolor[HTML]{C7ECFF}129.3} &
  \multicolumn{1}{c|}{\cellcolor[HTML]{C7ECFF}7.996} &
  \multicolumn{1}{c|}{\cellcolor[HTML]{C7ECFF}148.4} &
  \multicolumn{1}{c|}{\cellcolor[HTML]{C7ECFF}9.487} &
  \multicolumn{1}{c|}{\cellcolor[HTML]{C7ECFF}167.8} &
  \multicolumn{1}{c|}{\cellcolor[HTML]{C7ECFF}10.93} &
  \multicolumn{1}{c|}{\cellcolor[HTML]{C7ECFF}187.3} &
  \multicolumn{1}{c|}{\cellcolor[HTML]{C7ECFF}12.55} &
  \multicolumn{1}{c|}{\cellcolor[HTML]{C7ECFF}216.1} &
  \cellcolor[HTML]{C7ECFF}14.38 \\ \cline{3-13} 
 &
   &
  $\Pi_{\mathsf{OE}\text{-2}}^{\mathsf{apart}}$ &
  \multicolumn{1}{c|}{\cellcolor[HTML]{99BEFF}127.8} &
  \multicolumn{1}{c|}{\cellcolor[HTML]{99BEFF}6.668} &
  \multicolumn{1}{c|}{\cellcolor[HTML]{99BEFF}147.0} &
  \multicolumn{1}{c|}{\cellcolor[HTML]{99BEFF}7.654} &
  \multicolumn{1}{c|}{\cellcolor[HTML]{99BEFF}166.3} &
  \multicolumn{1}{c|}{\cellcolor[HTML]{99BEFF}8.818} &
  \multicolumn{1}{c|}{\cellcolor[HTML]{99BEFF}184.7} &
  \multicolumn{1}{c|}{\cellcolor[HTML]{99BEFF}10.04} &
  \multicolumn{1}{c|}{\cellcolor[HTML]{99BEFF}213.9} &
  \cellcolor[HTML]{99BEFF}11.75 \\ \cline{3-13} 
 &
  \multirow{-4}{*}{2} &
  $\Pi_{\mathsf{SE}}^{\mathsf{apart}}$ &
  \multicolumn{1}{c|}{140.4} &
  \multicolumn{1}{c|}{9.635} &
  \multicolumn{1}{c|}{161.3} &
  \multicolumn{1}{c|}{11.58} &
  \multicolumn{1}{c|}{182.0} &
  \multicolumn{1}{c|}{13.46} &
  \multicolumn{1}{c|}{202.5} &
  \multicolumn{1}{c|}{15.38} &
  \multicolumn{1}{c|}{233.0} &
  17.62 \\ \cline{2-13} 
 &
   &
  \cite{DBLP:conf/ccs/DZL25}-Low &
  \multicolumn{1}{c|}{1767} &
  \multicolumn{1}{c|}{70.64} &
  \multicolumn{1}{c|}{{\color[HTML]{FE0000} 2622}} &
  \multicolumn{1}{c|}{{\color[HTML]{FE0000} 89.98}} &
  \multicolumn{1}{c|}{{\color[HTML]{FE0000} 2268}} &
  \multicolumn{1}{c|}{{\color[HTML]{FE0000} 363.9}} &
  \multicolumn{1}{c|}{2710} &
  \multicolumn{1}{c|}{366.6} &
  \multicolumn{1}{c|}{3381} &
  375.4 \\ \cline{3-13} 
 &
   &
  $\Pi_{\mathsf{OE}\text{-1}}^{\mathsf{apart}}$ &
  \multicolumn{1}{c|}{267.9} &
  \multicolumn{1}{c|}{\cellcolor[HTML]{C7ECFF}16.51} &
  \multicolumn{1}{c|}{315.8} &
  \multicolumn{1}{c|}{\cellcolor[HTML]{C7ECFF}20.16} &
  \multicolumn{1}{c|}{362.3} &
  \multicolumn{1}{c|}{\cellcolor[HTML]{C7ECFF}23.91} &
  \multicolumn{1}{c|}{410.2} &
  \multicolumn{1}{c|}{\cellcolor[HTML]{C7ECFF}27.77} &
  \multicolumn{1}{c|}{471.3} &
  \cellcolor[HTML]{C7ECFF}31.75 \\ \cline{3-13} 
 &
   &
  $\Pi_{\mathsf{OE}\text{-2}}^{\mathsf{apart}}$ &
  \multicolumn{1}{c|}{\cellcolor[HTML]{C7ECFF}266.9} &
  \multicolumn{1}{c|}{\cellcolor[HTML]{99BEFF}15.09} &
  \multicolumn{1}{c|}{\cellcolor[HTML]{C7ECFF}313.9} &
  \multicolumn{1}{c|}{\cellcolor[HTML]{99BEFF}18.29} &
  \multicolumn{1}{c|}{\cellcolor[HTML]{C7ECFF}360.6} &
  \multicolumn{1}{c|}{\cellcolor[HTML]{99BEFF}21.76} &
  \multicolumn{1}{c|}{\cellcolor[HTML]{C7ECFF}407.2} &
  \multicolumn{1}{c|}{\cellcolor[HTML]{99BEFF}24.94} &
  \multicolumn{1}{c|}{\cellcolor[HTML]{C7ECFF}468.4} &
  \cellcolor[HTML]{99BEFF}28.71 \\ \cline{3-13} 
 &
  \multirow{-4}{*}{3} &
  $\Pi_{\mathsf{SE}}^{\mathsf{apart}}$ &
  \multicolumn{1}{c|}{\cellcolor[HTML]{99BEFF}246.1} &
  \multicolumn{1}{c|}{19.13} &
  \multicolumn{1}{c|}{\cellcolor[HTML]{99BEFF}285.9} &
  \multicolumn{1}{c|}{23.16} &
  \multicolumn{1}{c|}{\cellcolor[HTML]{99BEFF}325.8} &
  \multicolumn{1}{c|}{27.47} &
  \multicolumn{1}{c|}{\cellcolor[HTML]{99BEFF}365.6} &
  \multicolumn{1}{c|}{31.93} &
  \multicolumn{1}{c|}{\cellcolor[HTML]{99BEFF}419.7} &
  36.57 \\ \cline{2-13} 
 &
   &
  \cite{DBLP:conf/ccs/DZL25}-Low &
  \multicolumn{1}{c|}{4721} &
  \multicolumn{1}{c|}{229.6} &
  \multicolumn{1}{c|}{6996} &
  \multicolumn{1}{c|}{262.7} &
  \multicolumn{1}{c|}{-} &
  \multicolumn{1}{c|}{-} &
  \multicolumn{1}{c|}{-} &
  \multicolumn{1}{c|}{-} &
  \multicolumn{1}{c|}{-} &
  - \\ \cline{3-13} 
 &
   &
  $\Pi_{\mathsf{OE}\text{-1}}^{\mathsf{apart}}$ &
  \multicolumn{1}{c|}{567.1} &
  \multicolumn{1}{c|}{\cellcolor[HTML]{C7ECFF}38.61} &
  \multicolumn{1}{c|}{676.5} &
  \multicolumn{1}{c|}{\cellcolor[HTML]{C7ECFF}48.09} &
  \multicolumn{1}{c|}{789.3} &
  \multicolumn{1}{c|}{\cellcolor[HTML]{C7ECFF}57.28} &
  \multicolumn{1}{c|}{900.1} &
  \multicolumn{1}{c|}{\cellcolor[HTML]{C7ECFF}66.89} &
  \multicolumn{1}{c|}{1035} &
  \cellcolor[HTML]{C7ECFF}77.86 \\ \cline{3-13} 
 &
   &
  $\Pi_{\mathsf{OE}\text{-2}}^{\mathsf{apart}}$ &
  \multicolumn{1}{c|}{\cellcolor[HTML]{C7ECFF}563.6} &
  \multicolumn{1}{c|}{\cellcolor[HTML]{99BEFF}37.25} &
  \multicolumn{1}{c|}{\cellcolor[HTML]{C7ECFF}674.2} &
  \multicolumn{1}{c|}{\cellcolor[HTML]{99BEFF}46.07} &
  \multicolumn{1}{c|}{\cellcolor[HTML]{C7ECFF}786.6} &
  \multicolumn{1}{c|}{\cellcolor[HTML]{99BEFF}55.19} &
  \multicolumn{1}{c|}{\cellcolor[HTML]{C7ECFF}898.9} &
  \multicolumn{1}{c|}{\cellcolor[HTML]{99BEFF}64.81} &
  \multicolumn{1}{c|}{\cellcolor[HTML]{C7ECFF}1031} &
  \cellcolor[HTML]{99BEFF}74.79 \\ \cline{3-13} 
\multirow{-12}{*}{$2^{12}$} &
  \multirow{-4}{*}{4} &
  $\Pi_{\mathsf{SE}}^{\mathsf{apart}}$ &
  \multicolumn{1}{c|}{\cellcolor[HTML]{99BEFF}431.3} &
  \multicolumn{1}{c|}{41.92} &
  \multicolumn{1}{c|}{\cellcolor[HTML]{99BEFF}507.4} &
  \multicolumn{1}{c|}{51.77} &
  \multicolumn{1}{c|}{\cellcolor[HTML]{99BEFF}585.5} &
  \multicolumn{1}{c|}{61.97} &
  \multicolumn{1}{c|}{\cellcolor[HTML]{99BEFF}662.5} &
  \multicolumn{1}{c|}{72.65} &
  \multicolumn{1}{c|}{\cellcolor[HTML]{99BEFF}760.7} &
  84.09 \\ \hline
 &
   &
  \cite{DBLP:conf/ccs/DZL25}-Low &
  \multicolumn{1}{c|}{6835} &
  \multicolumn{1}{c|}{971.7} &
  \multicolumn{1}{c|}{-} &
  \multicolumn{1}{c|}{-} &
  \multicolumn{1}{c|}{-} &
  \multicolumn{1}{c|}{-} &
  \multicolumn{1}{c|}{-} &
  \multicolumn{1}{c|}{-} &
  \multicolumn{1}{c|}{-} &
  - \\ \cline{3-13} 
 &
   &
  $\Pi_{\mathsf{OE}\text{-1}}^{\mathsf{apart}}$ &
  \multicolumn{1}{c|}{\cellcolor[HTML]{C7ECFF}2052} &
  \multicolumn{1}{c|}{\cellcolor[HTML]{C7ECFF}141.6} &
  \multicolumn{1}{c|}{\cellcolor[HTML]{C7ECFF}2361} &
  \multicolumn{1}{c|}{\cellcolor[HTML]{C7ECFF}170.8} &
  \multicolumn{1}{c|}{\cellcolor[HTML]{C7ECFF}2674} &
  \multicolumn{1}{c|}{\cellcolor[HTML]{C7ECFF}201.9} &
  \multicolumn{1}{c|}{\cellcolor[HTML]{C7ECFF}2986} &
  \multicolumn{1}{c|}{\cellcolor[HTML]{C7ECFF}233.7} &
  \multicolumn{1}{c|}{\cellcolor[HTML]{C7ECFF}3452} &
  \cellcolor[HTML]{C7ECFF}269.1 \\ \cline{3-13} 
 &
   &
  $\Pi_{\mathsf{OE}\text{-2}}^{\mathsf{apart}}$ &
  \multicolumn{1}{c|}{\cellcolor[HTML]{99BEFF}2032} &
  \multicolumn{1}{c|}{\cellcolor[HTML]{99BEFF}116.7} &
  \multicolumn{1}{c|}{\cellcolor[HTML]{99BEFF}2336} &
  \multicolumn{1}{c|}{\cellcolor[HTML]{99BEFF}141.5} &
  \multicolumn{1}{c|}{\cellcolor[HTML]{99BEFF}2643} &
  \multicolumn{1}{c|}{\cellcolor[HTML]{99BEFF}166.2} &
  \multicolumn{1}{c|}{\cellcolor[HTML]{99BEFF}2950} &
  \multicolumn{1}{c|}{\cellcolor[HTML]{99BEFF}192.8} &
  \multicolumn{1}{c|}{\cellcolor[HTML]{99BEFF}3412} &
  \cellcolor[HTML]{99BEFF}223.2 \\ \cline{3-13} 
\multirow{-4}{*}{$2^{16}$} &
  \multirow{-4}{*}{2} &
  $\Pi_{\mathsf{SE}}^{\mathsf{apart}}$ &
  \multicolumn{1}{c|}{2251} &
  \multicolumn{1}{c|}{173.9} &
  \multicolumn{1}{c|}{2586} &
  \multicolumn{1}{c|}{209.7} &
  \multicolumn{1}{c|}{2923} &
  \multicolumn{1}{c|}{247.5} &
  \multicolumn{1}{c|}{3260} &
  \multicolumn{1}{c|}{285.8} &
  \multicolumn{1}{c|}{3745} &
  334.1 \\ \hline
\end{tabular}
}
\caption{Communication cost (in MB) and running time (in seconds) of our protocols compared with those of \cite{DBLP:conf/ccs/DZL25} for $L_2$ distance in a low-dimensional LAN setting. Cells with - denote trials that ran out of memory. The best result is highlighted in \textcolor[rgb]{0,0.3,1}{blue}, the second best in \textcolor[rgb]{0,0.7,1}{cyan}, and data in \textcolor[rgb]{1,0,0}{red} font indicates abnormal values.}
\label{tab:l2-low-lan}
\end{table}

\begin{table}[!htbp]
\renewcommand\arraystretch{1}
	\centering
 \resizebox{0.83\linewidth}{!}{
\begin{tabular}{|c|c|c|cccccccccc|}
\hline
 &
   &
   &
  \multicolumn{10}{c|}{Threshold $\delta$} \\ \cline{4-13} 
 &
   &
   &
  \multicolumn{2}{c|}{16} &
  \multicolumn{2}{c|}{32} &
  \multicolumn{2}{c|}{64} &
  \multicolumn{2}{c|}{128} &
  \multicolumn{2}{c|}{256} \\ \cline{4-13} 
\multirow{-3}{*}{\begin{tabular}[c]{@{}c@{}}Set Size\\ $m=n$\end{tabular}} &
  \multirow{-3}{*}{\begin{tabular}[c]{@{}c@{}}Dimension\\ $d$\end{tabular}} &
  \multirow{-3}{*}{Protocol} &
  \multicolumn{1}{c|}{Comm.} &
  \multicolumn{1}{c|}{Time} &
  \multicolumn{1}{c|}{Comm.} &
  \multicolumn{1}{c|}{Time} &
  \multicolumn{1}{c|}{Comm.} &
  \multicolumn{1}{c|}{Time} &
  \multicolumn{1}{c|}{Comm.} &
  \multicolumn{1}{c|}{Time} &
  \multicolumn{1}{c|}{Comm.} &
  Time \\ \hline
 &
   &
  \cite{DBLP:conf/asiacrypt/GaoQLLW24} &
  \multicolumn{1}{c|}{34.87} &
  \multicolumn{1}{c|}{7.812} &
  \multicolumn{1}{c|}{68.33} &
  \multicolumn{1}{c|}{15.77} &
  \multicolumn{1}{c|}{135.2} &
  \multicolumn{1}{c|}{20.08} &
  \multicolumn{1}{c|}{268.9} &
  \multicolumn{1}{c|}{39.31} &
  \multicolumn{1}{c|}{536.2} &
  81.12 \\ \cline{3-13} 
 &
   &
  \cite{DBLP:conf/ccs/DZL25}-High &
  \multicolumn{1}{c|}{105.1} &
  \multicolumn{1}{c|}{9.669} &
  \multicolumn{1}{c|}{126.9} &
  \multicolumn{1}{c|}{9.976} &
  \multicolumn{1}{c|}{132.4} &
  \multicolumn{1}{c|}{13.33} &
  \multicolumn{1}{c|}{145.0} &
  \multicolumn{1}{c|}{14.55} &
  \multicolumn{1}{c|}{200.2} &
  22.18 \\ \cline{3-13} 
 &
   &
  $\Pi_{\mathsf{OE}\text{-1}}^{\mathsf{sep}}$ &
  \multicolumn{1}{c|}{\cellcolor[HTML]{99BEFF}26.67} &
  \multicolumn{1}{c|}{\cellcolor[HTML]{C7ECFF}2.208} &
  \multicolumn{1}{c|}{\cellcolor[HTML]{C7ECFF}30.06} &
  \multicolumn{1}{c|}{\cellcolor[HTML]{C7ECFF}2.447} &
  \multicolumn{1}{c|}{\cellcolor[HTML]{C7ECFF}33.14} &
  \multicolumn{1}{c|}{\cellcolor[HTML]{C7ECFF}2.685} &
  \multicolumn{1}{c|}{\cellcolor[HTML]{C7ECFF}36.30} &
  \multicolumn{1}{c|}{\cellcolor[HTML]{C7ECFF}2.902} &
  \multicolumn{1}{c|}{\cellcolor[HTML]{C7ECFF}41.17} &
  \cellcolor[HTML]{C7ECFF}3.187 \\ \cline{3-13} 
 &
   &
  $\Pi_{\mathsf{OE}\text{-2}}^{\mathsf{sep}}$ &
  \multicolumn{1}{c|}{\cellcolor[HTML]{C7ECFF}26.74} &
  \multicolumn{1}{c|}{\cellcolor[HTML]{99BEFF}2.182} &
  \multicolumn{1}{c|}{\cellcolor[HTML]{99BEFF}30.04} &
  \multicolumn{1}{c|}{\cellcolor[HTML]{99BEFF}2.317} &
  \multicolumn{1}{c|}{\cellcolor[HTML]{99BEFF}32.99} &
  \multicolumn{1}{c|}{\cellcolor[HTML]{99BEFF}2.596} &
  \multicolumn{1}{c|}{\cellcolor[HTML]{99BEFF}36.16} &
  \multicolumn{1}{c|}{\cellcolor[HTML]{99BEFF}2.735} &
  \multicolumn{1}{c|}{\cellcolor[HTML]{99BEFF}41.06} &
  \cellcolor[HTML]{99BEFF}3.035 \\ \cline{3-13} 
 &
  \multirow{-5}{*}{6} &
  $\Pi_{\mathsf{SE}}^{\mathsf{sep}}$ &
  \multicolumn{1}{c|}{32.67} &
  \multicolumn{1}{c|}{2.217} &
  \multicolumn{1}{c|}{37.28} &
  \multicolumn{1}{c|}{2.571} &
  \multicolumn{1}{c|}{41.58} &
  \multicolumn{1}{c|}{2.827} &
  \multicolumn{1}{c|}{46.07} &
  \multicolumn{1}{c|}{3.145} &
  \multicolumn{1}{c|}{52.27} &
  3.531 \\ \cline{2-13} 
 &
   &
  \cite{DBLP:conf/asiacrypt/GaoQLLW24} &
  \multicolumn{1}{c|}{46.36} &
  \multicolumn{1}{c|}{8.954} &
  \multicolumn{1}{c|}{90.95} &
  \multicolumn{1}{c|}{16.41} &
  \multicolumn{1}{c|}{180.1} &
  \multicolumn{1}{c|}{27.11} &
  \multicolumn{1}{c|}{358.3} &
  \multicolumn{1}{c|}{52.19} &
  \multicolumn{1}{c|}{714.8} &
  121.4 \\ \cline{3-13} 
 &
   &
  \cite{DBLP:conf/ccs/DZL25}-High &
  \multicolumn{1}{c|}{140.0} &
  \multicolumn{1}{c|}{12.84} &
  \multicolumn{1}{c|}{169.0} &
  \multicolumn{1}{c|}{13.28} &
  \multicolumn{1}{c|}{176.2} &
  \multicolumn{1}{c|}{17.46} &
  \multicolumn{1}{c|}{192.6} &
  \multicolumn{1}{c|}{18.57} &
  \multicolumn{1}{c|}{265.5} &
  28.57 \\ \cline{3-13} 
 &
   &
  $\Pi_{\mathsf{OE}\text{-1}}^{\mathsf{sep}}$ &
  \multicolumn{1}{c|}{\cellcolor[HTML]{C7ECFF}35.09} &
  \multicolumn{1}{c|}{\cellcolor[HTML]{C7ECFF}2.661} &
  \multicolumn{1}{c|}{\cellcolor[HTML]{C7ECFF}38.95} &
  \multicolumn{1}{c|}{\cellcolor[HTML]{C7ECFF}2.944} &
  \multicolumn{1}{c|}{\cellcolor[HTML]{C7ECFF}42.97} &
  \multicolumn{1}{c|}{\cellcolor[HTML]{C7ECFF}3.226} &
  \multicolumn{1}{c|}{\cellcolor[HTML]{C7ECFF}47.26} &
  \multicolumn{1}{c|}{\cellcolor[HTML]{C7ECFF}3.569} &
  \multicolumn{1}{c|}{\cellcolor[HTML]{C7ECFF}53.86} &
  \cellcolor[HTML]{C7ECFF}3.965 \\ \cline{3-13} 
 &
   &
  $\Pi_{\mathsf{OE}\text{-2}}^{\mathsf{sep}}$ &
  \multicolumn{1}{c|}{\cellcolor[HTML]{99BEFF}34.99} &
  \multicolumn{1}{c|}{\cellcolor[HTML]{99BEFF}2.592} &
  \multicolumn{1}{c|}{\cellcolor[HTML]{99BEFF}38.93} &
  \multicolumn{1}{c|}{\cellcolor[HTML]{99BEFF}2.852} &
  \multicolumn{1}{c|}{\cellcolor[HTML]{99BEFF}42.86} &
  \multicolumn{1}{c|}{\cellcolor[HTML]{99BEFF}3.122} &
  \multicolumn{1}{c|}{\cellcolor[HTML]{99BEFF}47.17} &
  \multicolumn{1}{c|}{\cellcolor[HTML]{99BEFF}3.451} &
  \multicolumn{1}{c|}{\cellcolor[HTML]{99BEFF}53.84} &
  \cellcolor[HTML]{99BEFF}3.792 \\ \cline{3-13} 
 &
  \multirow{-5}{*}{8} &
  $\Pi_{\mathsf{SE}}^{\mathsf{sep}}$ &
  \multicolumn{1}{c|}{43.04} &
  \multicolumn{1}{c|}{2.789} &
  \multicolumn{1}{c|}{48.91} &
  \multicolumn{1}{c|}{3.199} &
  \multicolumn{1}{c|}{54.62} &
  \multicolumn{1}{c|}{3.701} &
  \multicolumn{1}{c|}{60.61} &
  \multicolumn{1}{c|}{4.037} &
  \multicolumn{1}{c|}{69.01} &
  4.506 \\ \cline{2-13} 
 &
   &
  \cite{DBLP:conf/asiacrypt/GaoQLLW24} &
  \multicolumn{1}{c|}{57.84} &
  \multicolumn{1}{c|}{10.04} &
  \multicolumn{1}{c|}{113.6} &
  \multicolumn{1}{c|}{17.55} &
  \multicolumn{1}{c|}{225.0} &
  \multicolumn{1}{c|}{33.83} &
  \multicolumn{1}{c|}{447.8} &
  \multicolumn{1}{c|}{69.04} &
  \multicolumn{1}{c|}{893.3} &
  136.8 \\ \cline{3-13} 
 &
   &
  \cite{DBLP:conf/ccs/DZL25}-High &
  \multicolumn{1}{c|}{174.9} &
  \multicolumn{1}{c|}{15.54} &
  \multicolumn{1}{c|}{211.1} &
  \multicolumn{1}{c|}{16.20} &
  \multicolumn{1}{c|}{219.9} &
  \multicolumn{1}{c|}{21.47} &
  \multicolumn{1}{c|}{240.3} &
  \multicolumn{1}{c|}{22.49} &
  \multicolumn{1}{c|}{330.9} &
  34.07 \\ \cline{3-13} 
 &
   &
  $\Pi_{\mathsf{OE}\text{-1}}^{\mathsf{sep}}$ &
  \multicolumn{1}{c|}{\cellcolor[HTML]{99BEFF}42.74} &
  \multicolumn{1}{c|}{\cellcolor[HTML]{C7ECFF}3.079} &
  \multicolumn{1}{c|}{\cellcolor[HTML]{C7ECFF}47.94} &
  \multicolumn{1}{c|}{\cellcolor[HTML]{C7ECFF}3.528} &
  \multicolumn{1}{c|}{\cellcolor[HTML]{99BEFF}53.00} &
  \multicolumn{1}{c|}{\cellcolor[HTML]{C7ECFF}3.976} &
  \multicolumn{1}{c|}{\cellcolor[HTML]{C7ECFF}58.22} &
  \multicolumn{1}{c|}{\cellcolor[HTML]{C7ECFF}4.252} &
  \multicolumn{1}{c|}{\cellcolor[HTML]{99BEFF}66.58} &
  \cellcolor[HTML]{C7ECFF}4.678 \\ \cline{3-13} 
 &
   &
  $\Pi_{\mathsf{OE}\text{-2}}^{\mathsf{sep}}$ &
  \multicolumn{1}{c|}{\cellcolor[HTML]{C7ECFF}42.76} &
  \multicolumn{1}{c|}{\cellcolor[HTML]{99BEFF}3.025} &
  \multicolumn{1}{c|}{\cellcolor[HTML]{99BEFF}47.92} &
  \multicolumn{1}{c|}{\cellcolor[HTML]{99BEFF}3.384} &
  \multicolumn{1}{c|}{\cellcolor[HTML]{C7ECFF}53.02} &
  \multicolumn{1}{c|}{\cellcolor[HTML]{99BEFF}3.804} &
  \multicolumn{1}{c|}{\cellcolor[HTML]{99BEFF}58.15} &
  \multicolumn{1}{c|}{\cellcolor[HTML]{99BEFF}4.079} &
  \multicolumn{1}{c|}{\cellcolor[HTML]{C7ECFF}66.63} &
  \cellcolor[HTML]{99BEFF}4.537 \\ \cline{3-13} 
\multirow{-15}{*}{$2^8$} &
  \multirow{-5}{*}{10} &
  $\Pi_{\mathsf{SE}}^{\mathsf{sep}}$ &
  \multicolumn{1}{c|}{53.19} &
  \multicolumn{1}{c|}{3.418} &
  \multicolumn{1}{c|}{60.54} &
  \multicolumn{1}{c|}{3.868} &
  \multicolumn{1}{c|}{67.79} &
  \multicolumn{1}{c|}{4.467} &
  \multicolumn{1}{c|}{75.19} &
  \multicolumn{1}{c|}{4.895} &
  \multicolumn{1}{c|}{85.74} &
  5.481 \\ \hline
 &
   &
  \cite{DBLP:conf/asiacrypt/GaoQLLW24} &
  \multicolumn{1}{c|}{557.8} &
  \multicolumn{1}{c|}{99.78} &
  \multicolumn{1}{c|}{1093} &
  \multicolumn{1}{c|}{314.8} &
  \multicolumn{1}{c|}{2163} &
  \multicolumn{1}{c|}{438.2} &
  \multicolumn{1}{c|}{4302} &
  \multicolumn{1}{c|}{687.9} &
  \multicolumn{1}{c|}{-} &
  - \\ \cline{3-13} 
 &
   &
  \cite{DBLP:conf/ccs/DZL25}-High &
  \multicolumn{1}{c|}{1681} &
  \multicolumn{1}{c|}{150.3} &
  \multicolumn{1}{c|}{2030} &
  \multicolumn{1}{c|}{156.5} &
  \multicolumn{1}{c|}{2118} &
  \multicolumn{1}{c|}{209.9} &
  \multicolumn{1}{c|}{2320} &
  \multicolumn{1}{c|}{228.5} &
  \multicolumn{1}{c|}{3202} &
  351.8 \\ \cline{3-13} 
 &
   &
  $\Pi_{\mathsf{OE}\text{-1}}^{\mathsf{sep}}$ &
  \multicolumn{1}{c|}{\cellcolor[HTML]{C7ECFF}379.7} &
  \multicolumn{1}{c|}{\cellcolor[HTML]{C7ECFF}24.36} &
  \multicolumn{1}{c|}{\cellcolor[HTML]{C7ECFF}428.7} &
  \multicolumn{1}{c|}{\cellcolor[HTML]{C7ECFF}27.99} &
  \multicolumn{1}{c|}{\cellcolor[HTML]{C7ECFF}477.7} &
  \multicolumn{1}{c|}{\cellcolor[HTML]{C7ECFF}31.52} &
  \multicolumn{1}{c|}{\cellcolor[HTML]{C7ECFF}527.3} &
  \multicolumn{1}{c|}{\cellcolor[HTML]{C7ECFF}35.51} &
  \multicolumn{1}{c|}{\cellcolor[HTML]{C7ECFF}603.4} &
  \cellcolor[HTML]{C7ECFF}40.19 \\ \cline{3-13} 
 &
   &
  $\Pi_{\mathsf{OE}\text{-2}}^{\mathsf{sep}}$ &
  \multicolumn{1}{c|}{\cellcolor[HTML]{99BEFF}378.4} &
  \multicolumn{1}{c|}{\cellcolor[HTML]{99BEFF}22.96} &
  \multicolumn{1}{c|}{\cellcolor[HTML]{99BEFF}427.3} &
  \multicolumn{1}{c|}{\cellcolor[HTML]{99BEFF}26.46} &
  \multicolumn{1}{c|}{\cellcolor[HTML]{99BEFF}476.0} &
  \multicolumn{1}{c|}{\cellcolor[HTML]{99BEFF}29.48} &
  \multicolumn{1}{c|}{\cellcolor[HTML]{99BEFF}524.8} &
  \multicolumn{1}{c|}{\cellcolor[HTML]{99BEFF}32.78} &
  \multicolumn{1}{c|}{\cellcolor[HTML]{99BEFF}601.3} &
  \cellcolor[HTML]{99BEFF}37.53 \\ \cline{3-13} 
 &
  \multirow{-5}{*}{6} &
  $\Pi_{\mathsf{SE}}^{\mathsf{sep}}$ &
  \multicolumn{1}{c|}{473.9} &
  \multicolumn{1}{c|}{30.63} &
  \multicolumn{1}{c|}{541.2} &
  \multicolumn{1}{c|}{35.35} &
  \multicolumn{1}{c|}{608.7} &
  \multicolumn{1}{c|}{41.37} &
  \multicolumn{1}{c|}{676.3} &
  \multicolumn{1}{c|}{46.19} &
  \multicolumn{1}{c|}{770.8} &
  51.29 \\ \cline{2-13} 
 &
   &
  \cite{DBLP:conf/asiacrypt/GaoQLLW24} &
  \multicolumn{1}{c|}{741.6} &
  \multicolumn{1}{c|}{143.6} &
  \multicolumn{1}{c|}{1455} &
  \multicolumn{1}{c|}{455.2} &
  \multicolumn{1}{c|}{2882} &
  \multicolumn{1}{c|}{485.8} &
  \multicolumn{1}{c|}{5733} &
  \multicolumn{1}{c|}{1053} &
  \multicolumn{1}{c|}{-} &
  - \\ \cline{3-13} 
 &
   &
  \cite{DBLP:conf/ccs/DZL25}-High &
  \multicolumn{1}{c|}{2239} &
  \multicolumn{1}{c|}{198.5} &
  \multicolumn{1}{c|}{2703} &
  \multicolumn{1}{c|}{204.9} &
  \multicolumn{1}{c|}{2818} &
  \multicolumn{1}{c|}{275.4} &
  \multicolumn{1}{c|}{3082} &
  \multicolumn{1}{c|}{291.9} &
  \multicolumn{1}{c|}{4248} &
  448.1 \\ \cline{3-13} 
 &
   &
  $\Pi_{\mathsf{OE}\text{-1}}^{\mathsf{sep}}$ &
  \multicolumn{1}{c|}{\cellcolor[HTML]{C7ECFF}501.8} &
  \multicolumn{1}{c|}{\cellcolor[HTML]{C7ECFF}31.46} &
  \multicolumn{1}{c|}{\cellcolor[HTML]{C7ECFF}567.0} &
  \multicolumn{1}{c|}{\cellcolor[HTML]{C7ECFF}36.28} &
  \multicolumn{1}{c|}{\cellcolor[HTML]{C7ECFF}632.1} &
  \multicolumn{1}{c|}{\cellcolor[HTML]{C7ECFF}41.14} &
  \multicolumn{1}{c|}{\cellcolor[HTML]{C7ECFF}697.1} &
  \multicolumn{1}{c|}{\cellcolor[HTML]{C7ECFF}46.17} &
  \multicolumn{1}{c|}{\cellcolor[HTML]{C7ECFF}798.8} &
  \cellcolor[HTML]{C7ECFF}52.06 \\ \cline{3-13} 
 &
   &
  $\Pi_{\mathsf{OE}\text{-2}}^{\mathsf{sep}}$ &
  \multicolumn{1}{c|}{\cellcolor[HTML]{99BEFF}500.7} &
  \multicolumn{1}{c|}{\cellcolor[HTML]{99BEFF}30.03} &
  \multicolumn{1}{c|}{\cellcolor[HTML]{99BEFF}565.5} &
  \multicolumn{1}{c|}{\cellcolor[HTML]{99BEFF}34.58} &
  \multicolumn{1}{c|}{\cellcolor[HTML]{99BEFF}630.2} &
  \multicolumn{1}{c|}{\cellcolor[HTML]{99BEFF}39.09} &
  \multicolumn{1}{c|}{\cellcolor[HTML]{99BEFF}694.9} &
  \multicolumn{1}{c|}{\cellcolor[HTML]{99BEFF}43.78} &
  \multicolumn{1}{c|}{\cellcolor[HTML]{99BEFF}796.5} &
  \cellcolor[HTML]{99BEFF}49.17 \\ \cline{3-13} 
 &
  \multirow{-5}{*}{8} &
  $\Pi_{\mathsf{SE}}^{\mathsf{sep}}$ &
  \multicolumn{1}{c|}{628.2} &
  \multicolumn{1}{c|}{39.73} &
  \multicolumn{1}{c|}{717.7} &
  \multicolumn{1}{c|}{46.22} &
  \multicolumn{1}{c|}{806.7} &
  \multicolumn{1}{c|}{52.69} &
  \multicolumn{1}{c|}{896.3} &
  \multicolumn{1}{c|}{62.42} &
  \multicolumn{1}{c|}{1022} &
  67.08 \\ \cline{2-13} 
 &
   &
  \cite{DBLP:conf/asiacrypt/GaoQLLW24} &
  \multicolumn{1}{c|}{925.4} &
  \multicolumn{1}{c|}{241.9} &
  \multicolumn{1}{c|}{1817} &
  \multicolumn{1}{c|}{559.4} &
  \multicolumn{1}{c|}{3600} &
  \multicolumn{1}{c|}{725.1} &
  \multicolumn{1}{c|}{7165} &
  \multicolumn{1}{c|}{1281} &
  \multicolumn{1}{c|}{-} &
  - \\ \cline{3-13} 
 &
   &
  \cite{DBLP:conf/ccs/DZL25}-High &
  \multicolumn{1}{c|}{2798} &
  \multicolumn{1}{c|}{246.7} &
  \multicolumn{1}{c|}{3377} &
  \multicolumn{1}{c|}{255.6} &
  \multicolumn{1}{c|}{3519} &
  \multicolumn{1}{c|}{341.5} &
  \multicolumn{1}{c|}{3845} &
  \multicolumn{1}{c|}{356.1} &
  \multicolumn{1}{c|}{5294} &
  549.6 \\ \cline{3-13} 
 &
   &
  $\Pi_{\mathsf{OE}\text{-1}}^{\mathsf{sep}}$ &
  \multicolumn{1}{c|}{\cellcolor[HTML]{C7ECFF}625.1} &
  \multicolumn{1}{c|}{\cellcolor[HTML]{C7ECFF}38.49} &
  \multicolumn{1}{c|}{\cellcolor[HTML]{C7ECFF}705.6} &
  \multicolumn{1}{c|}{\cellcolor[HTML]{C7ECFF}44.97} &
  \multicolumn{1}{c|}{\cellcolor[HTML]{C7ECFF}786.1} &
  \multicolumn{1}{c|}{\cellcolor[HTML]{C7ECFF}50.91} &
  \multicolumn{1}{c|}{\cellcolor[HTML]{C7ECFF}867.6} &
  \multicolumn{1}{c|}{\cellcolor[HTML]{C7ECFF}57.02} &
  \multicolumn{1}{c|}{\cellcolor[HTML]{C7ECFF}994.2} &
  \cellcolor[HTML]{C7ECFF}64.44 \\ \cline{3-13} 
 &
   &
  $\Pi_{\mathsf{OE}\text{-2}}^{\mathsf{sep}}$ &
  \multicolumn{1}{c|}{\cellcolor[HTML]{99BEFF}623.6} &
  \multicolumn{1}{c|}{\cellcolor[HTML]{99BEFF}37.55} &
  \multicolumn{1}{c|}{\cellcolor[HTML]{99BEFF}703.9} &
  \multicolumn{1}{c|}{\cellcolor[HTML]{99BEFF}43.24} &
  \multicolumn{1}{c|}{\cellcolor[HTML]{99BEFF}784.3} &
  \multicolumn{1}{c|}{\cellcolor[HTML]{99BEFF}48.68} &
  \multicolumn{1}{c|}{\cellcolor[HTML]{99BEFF}865.5} &
  \multicolumn{1}{c|}{\cellcolor[HTML]{99BEFF}54.57} &
  \multicolumn{1}{c|}{\cellcolor[HTML]{99BEFF}991.9} &
  \cellcolor[HTML]{99BEFF}61.57 \\ \cline{3-13} 
\multirow{-15}{*}{$2^{12}$} &
  \multirow{-5}{*}{10} &
  $\Pi_{\mathsf{SE}}^{\mathsf{sep}}$ &
  \multicolumn{1}{c|}{782.4} &
  \multicolumn{1}{c|}{48.67} &
  \multicolumn{1}{c|}{893.4} &
  \multicolumn{1}{c|}{57.25} &
  \multicolumn{1}{c|}{1005} &
  \multicolumn{1}{c|}{65.37} &
  \multicolumn{1}{c|}{1117} &
  \multicolumn{1}{c|}{73.48} &
  \multicolumn{1}{c|}{1273} &
  83.18 \\ \hline
 &
   &
  \cite{DBLP:conf/asiacrypt/GaoQLLW24} &
  \multicolumn{1}{c|}{-} &
  \multicolumn{1}{c|}{-} &
  \multicolumn{1}{c|}{-} &
  \multicolumn{1}{c|}{-} &
  \multicolumn{1}{c|}{-} &
  \multicolumn{1}{c|}{-} &
  \multicolumn{1}{c|}{-} &
  \multicolumn{1}{c|}{-} &
  \multicolumn{1}{c|}{-} &
  - \\ \cline{3-13} 
 &
   &
  \cite{DBLP:conf/ccs/DZL25}-High &
  \multicolumn{1}{c|}{-} &
  \multicolumn{1}{c|}{-} &
  \multicolumn{1}{c|}{-} &
  \multicolumn{1}{c|}{-} &
  \multicolumn{1}{c|}{-} &
  \multicolumn{1}{c|}{-} &
  \multicolumn{1}{c|}{-} &
  \multicolumn{1}{c|}{-} &
  \multicolumn{1}{c|}{-} &
  - \\ \cline{3-13} 
 &
   &
  $\Pi_{\mathsf{OE}\text{-1}}^{\mathsf{sep}}$ &
  \multicolumn{1}{c|}{\cellcolor[HTML]{C7ECFF}6090} &
  \multicolumn{1}{c|}{\cellcolor[HTML]{C7ECFF}402.9} &
  \multicolumn{1}{c|}{\cellcolor[HTML]{C7ECFF}6875} &
  \multicolumn{1}{c|}{\cellcolor[HTML]{C7ECFF}463.5} &
  \multicolumn{1}{c|}{\cellcolor[HTML]{C7ECFF}7667} &
  \multicolumn{1}{c|}{\cellcolor[HTML]{C7ECFF}537.2} &
  \multicolumn{1}{c|}{\cellcolor[HTML]{C7ECFF}8464} &
  \multicolumn{1}{c|}{\cellcolor[HTML]{C7ECFF}602.4} &
  \multicolumn{1}{c|}{\cellcolor[HTML]{C7ECFF}9699} &
  \cellcolor[HTML]{C7ECFF}682.5 \\ \cline{3-13} 
 &
   &
  $\Pi_{\mathsf{OE}\text{-2}}^{\mathsf{sep}}$ &
  \multicolumn{1}{c|}{\cellcolor[HTML]{99BEFF}6069} &
  \multicolumn{1}{c|}{\cellcolor[HTML]{99BEFF}376.2} &
  \multicolumn{1}{c|}{\cellcolor[HTML]{99BEFF}6850} &
  \multicolumn{1}{c|}{\cellcolor[HTML]{99BEFF}434.4} &
  \multicolumn{1}{c|}{\cellcolor[HTML]{99BEFF}7638} &
  \multicolumn{1}{c|}{\cellcolor[HTML]{99BEFF}504.1} &
  \multicolumn{1}{c|}{\cellcolor[HTML]{99BEFF}8430} &
  \multicolumn{1}{c|}{\cellcolor[HTML]{99BEFF}561.2} &
  \multicolumn{1}{c|}{\cellcolor[HTML]{99BEFF}9659} &
  \cellcolor[HTML]{99BEFF}640.7 \\ \cline{3-13} 
 &
  \multirow{-5}{*}{6} &
  $\Pi_{\mathsf{SE}}^{\mathsf{sep}}$ &
  \multicolumn{1}{c|}{7640} &
  \multicolumn{1}{c|}{502.3} &
  \multicolumn{1}{c|}{8725} &
  \multicolumn{1}{c|}{588.4} &
  \multicolumn{1}{c|}{9815} &
  \multicolumn{1}{c|}{682.4} &
  \multicolumn{1}{c|}{10908} &
  \multicolumn{1}{c|}{768.4} &
  \multicolumn{1}{c|}{12438} &
  867.9 \\ \cline{2-13} 
 &
   &
  \cite{DBLP:conf/asiacrypt/GaoQLLW24} &
  \multicolumn{1}{c|}{-} &
  \multicolumn{1}{c|}{-} &
  \multicolumn{1}{c|}{-} &
  \multicolumn{1}{c|}{-} &
  \multicolumn{1}{c|}{-} &
  \multicolumn{1}{c|}{-} &
  \multicolumn{1}{c|}{-} &
  \multicolumn{1}{c|}{-} &
  \multicolumn{1}{c|}{-} &
  - \\ \cline{3-13} 
 &
   &
  \cite{DBLP:conf/ccs/DZL25}-High &
  \multicolumn{1}{c|}{-} &
  \multicolumn{1}{c|}{-} &
  \multicolumn{1}{c|}{-} &
  \multicolumn{1}{c|}{-} &
  \multicolumn{1}{c|}{-} &
  \multicolumn{1}{c|}{-} &
  \multicolumn{1}{c|}{-} &
  \multicolumn{1}{c|}{-} &
  \multicolumn{1}{c|}{-} &
  - \\ \cline{3-13} 
 &
   &
  $\Pi_{\mathsf{OE}\text{-1}}^{\mathsf{sep}}$ &
  \multicolumn{1}{c|}{\cellcolor[HTML]{C7ECFF}8066} &
  \multicolumn{1}{c|}{\cellcolor[HTML]{C7ECFF}537.1} &
  \multicolumn{1}{c|}{\cellcolor[HTML]{C7ECFF}9109} &
  \multicolumn{1}{c|}{\cellcolor[HTML]{C7ECFF}621.5} &
  \multicolumn{1}{c|}{\cellcolor[HTML]{C7ECFF}10157} &
  \multicolumn{1}{c|}{\cellcolor[HTML]{C7ECFF}707.1} &
  \multicolumn{1}{c|}{\cellcolor[HTML]{C7ECFF}11206} &
  \multicolumn{1}{c|}{\cellcolor[HTML]{C7ECFF}786.7} &
  \multicolumn{1}{c|}{\cellcolor[HTML]{C7ECFF}12846} &
  \cellcolor[HTML]{C7ECFF}888.5 \\ \cline{3-13} 
 &
   &
  $\Pi_{\mathsf{OE}\text{-2}}^{\mathsf{sep}}$ &
  \multicolumn{1}{c|}{\cellcolor[HTML]{99BEFF}8046} &
  \multicolumn{1}{c|}{\cellcolor[HTML]{99BEFF}512.7} &
  \multicolumn{1}{c|}{\cellcolor[HTML]{99BEFF}9082} &
  \multicolumn{1}{c|}{\cellcolor[HTML]{99BEFF}589.7} &
  \multicolumn{1}{c|}{\cellcolor[HTML]{99BEFF}10125} &
  \multicolumn{1}{c|}{\cellcolor[HTML]{99BEFF}669.9} &
  \multicolumn{1}{c|}{\cellcolor[HTML]{99BEFF}11170} &
  \multicolumn{1}{c|}{\cellcolor[HTML]{99BEFF}741.2} &
  \multicolumn{1}{c|}{\cellcolor[HTML]{99BEFF}12808} &
  \cellcolor[HTML]{99BEFF}844.2 \\ \cline{3-13} 
 &
  \multirow{-5}{*}{8} &
  $\Pi_{\mathsf{SE}}^{\mathsf{sep}}$ &
  \multicolumn{1}{c|}{10133} &
  \multicolumn{1}{c|}{674.9} &
  \multicolumn{1}{c|}{11572} &
  \multicolumn{1}{c|}{784.4} &
  \multicolumn{1}{c|}{13018} &
  \multicolumn{1}{c|}{897.1} &
  \multicolumn{1}{c|}{14466} &
  \multicolumn{1}{c|}{998.9} &
  \multicolumn{1}{c|}{16501} &
  1130 \\ \cline{2-13} 
 &
   &
  \cite{DBLP:conf/asiacrypt/GaoQLLW24} &
  \multicolumn{1}{c|}{-} &
  \multicolumn{1}{c|}{-} &
  \multicolumn{1}{c|}{-} &
  \multicolumn{1}{c|}{-} &
  \multicolumn{1}{c|}{-} &
  \multicolumn{1}{c|}{-} &
  \multicolumn{1}{c|}{-} &
  \multicolumn{1}{c|}{-} &
  \multicolumn{1}{c|}{-} &
  - \\ \cline{3-13} 
 &
   &
  \cite{DBLP:conf/ccs/DZL25}-High &
  \multicolumn{1}{c|}{-} &
  \multicolumn{1}{c|}{-} &
  \multicolumn{1}{c|}{-} &
  \multicolumn{1}{c|}{-} &
  \multicolumn{1}{c|}{-} &
  \multicolumn{1}{c|}{-} &
  \multicolumn{1}{c|}{-} &
  \multicolumn{1}{c|}{-} &
  \multicolumn{1}{c|}{-} &
  - \\ \cline{3-13} 
 &
   &
  $\Pi_{\mathsf{OE}\text{-1}}^{\mathsf{sep}}$ &
  \multicolumn{1}{c|}{\cellcolor[HTML]{C7ECFF}10045} &
  \multicolumn{1}{c|}{\cellcolor[HTML]{C7ECFF}659.9} &
  \multicolumn{1}{c|}{\cellcolor[HTML]{C7ECFF}11338} &
  \multicolumn{1}{c|}{\cellcolor[HTML]{C7ECFF}755.3} &
  \multicolumn{1}{c|}{\cellcolor[HTML]{C7ECFF}12644} &
  \multicolumn{1}{c|}{\cellcolor[HTML]{C7ECFF}866.1} &
  \multicolumn{1}{c|}{\cellcolor[HTML]{C7ECFF}13953} &
  \multicolumn{1}{c|}{\cellcolor[HTML]{C7ECFF}980.9} &
  \multicolumn{1}{c|}{\cellcolor[HTML]{C7ECFF}15997} &
  \cellcolor[HTML]{C7ECFF}1115 \\ \cline{3-13} 
 &
   &
  $\Pi_{\mathsf{OE}\text{-2}}^{\mathsf{sep}}$ &
  \multicolumn{1}{c|}{\cellcolor[HTML]{99BEFF}10024} &
  \multicolumn{1}{c|}{\cellcolor[HTML]{99BEFF}634.1} &
  \multicolumn{1}{c|}{\cellcolor[HTML]{99BEFF}11313} &
  \multicolumn{1}{c|}{\cellcolor[HTML]{99BEFF}729.8} &
  \multicolumn{1}{c|}{\cellcolor[HTML]{99BEFF}12613} &
  \multicolumn{1}{c|}{\cellcolor[HTML]{99BEFF}827.1} &
  \multicolumn{1}{c|}{\cellcolor[HTML]{99BEFF}13917} &
  \multicolumn{1}{c|}{\cellcolor[HTML]{99BEFF}943.5} &
  \multicolumn{1}{c|}{\cellcolor[HTML]{99BEFF}15955} &
  \cellcolor[HTML]{99BEFF}1067 \\ \cline{3-13} 
\multirow{-15}{*}{$2^{16}$} &
  \multirow{-5}{*}{10} &
  $\Pi_{\mathsf{SE}}^{\mathsf{sep}}$ &
  \multicolumn{1}{c|}{12628} &
  \multicolumn{1}{c|}{827.6} &
  \multicolumn{1}{c|}{14420} &
  \multicolumn{1}{c|}{957.4} &
  \multicolumn{1}{c|}{16221} &
  \multicolumn{1}{c|}{1100} &
  \multicolumn{1}{c|}{18026} &
  \multicolumn{1}{c|}{1254} &
  \multicolumn{1}{c|}{20562} &
  1420 \\ \hline
\end{tabular}
}
\caption{Communication cost (in MB) and running time (in seconds) of our protocols compared with those of \cite{DBLP:conf/asiacrypt/GaoQLLW24,DBLP:conf/ccs/DZL25} for $L_2$ distance in a high-dimensional LAN setting. Cells with - denote trials that ran out of memory. The best result is highlighted in \textcolor[rgb]{0,0.3,1}{blue}, the second best in \textcolor[rgb]{0,0.7,1}{cyan}.}
\label{tab:l2-high-lan}
\end{table}

\begin{table}[!htbp]
\renewcommand\arraystretch{1}
	\centering
 \resizebox{0.83\linewidth}{!}{
\begin{tabular}{|c|c|c|cccccccccc|}
\hline
 &
   &
   &
  \multicolumn{10}{c|}{Threshold $\delta$} \\ \cline{4-13} 
 &
   &
   &
  \multicolumn{2}{c|}{16} &
  \multicolumn{2}{c|}{32} &
  \multicolumn{2}{c|}{64} &
  \multicolumn{2}{c|}{128} &
  \multicolumn{2}{c|}{256} \\ \cline{4-13} 
\multirow{-3}{*}{\begin{tabular}[c]{@{}c@{}}Set Size\\ $m=n$\end{tabular}} &
  \multirow{-3}{*}{\begin{tabular}[c]{@{}c@{}}Dimension\\ $d$\end{tabular}} &
  \multirow{-3}{*}{Protocol} &
  \multicolumn{1}{c|}{Comm.} &
  \multicolumn{1}{c|}{Time} &
  \multicolumn{1}{c|}{Comm.} &
  \multicolumn{1}{c|}{Time} &
  \multicolumn{1}{c|}{Comm.} &
  \multicolumn{1}{c|}{Time} &
  \multicolumn{1}{c|}{Comm.} &
  \multicolumn{1}{c|}{Time} &
  \multicolumn{1}{c|}{Comm.} &
  Time \\ \hline
 &
   &
  \cite{DBLP:conf/ccs/DZL25}-Low &
  \multicolumn{1}{c|}{26.70} &
  \multicolumn{1}{c|}{7.109} &
  \multicolumn{1}{c|}{33.15} &
  \multicolumn{1}{c|}{10.48} &
  \multicolumn{1}{c|}{40.14} &
  \multicolumn{1}{c|}{12.55} &
  \multicolumn{1}{c|}{47.91} &
  \multicolumn{1}{c|}{15.31} &
  \multicolumn{1}{c|}{56.96} &
  19.37 \\ \cline{3-13} 
 &
   &
  $\Pi_{\mathsf{OE}\text{-1}}^{\mathsf{apart}}$ &
  \multicolumn{1}{c|}{\cellcolor[HTML]{99BEFF}10.03} &
  \multicolumn{1}{c|}{\cellcolor[HTML]{C7ECFF}4.574} &
  \multicolumn{1}{c|}{\cellcolor[HTML]{99BEFF}11.36} &
  \multicolumn{1}{c|}{\cellcolor[HTML]{C7ECFF}4.645} &
  \multicolumn{1}{c|}{\cellcolor[HTML]{C7ECFF}12.58} &
  \multicolumn{1}{c|}{\cellcolor[HTML]{C7ECFF}4.681} &
  \multicolumn{1}{c|}{\cellcolor[HTML]{C7ECFF}13.82} &
  \multicolumn{1}{c|}{\cellcolor[HTML]{C7ECFF}4.773} &
  \multicolumn{1}{c|}{\cellcolor[HTML]{C7ECFF}15.79} &
  \cellcolor[HTML]{C7ECFF}4.779 \\ \cline{3-13} 
 &
   &
  $\Pi_{\mathsf{OE}\text{-2}}^{\mathsf{apart}}$ &
  \multicolumn{1}{c|}{\cellcolor[HTML]{C7ECFF}10.07} &
  \multicolumn{1}{c|}{4.792} &
  \multicolumn{1}{c|}{\cellcolor[HTML]{99BEFF}11.36} &
  \multicolumn{1}{c|}{4.794} &
  \multicolumn{1}{c|}{\cellcolor[HTML]{99BEFF}12.53} &
  \multicolumn{1}{c|}{4.953} &
  \multicolumn{1}{c|}{\cellcolor[HTML]{99BEFF}13.74} &
  \multicolumn{1}{c|}{4.987} &
  \multicolumn{1}{c|}{\cellcolor[HTML]{C7ECFF}15.74} &
  4.989 \\ \cline{3-13} 
 &
  \multirow{-4}{*}{2} &
  $\Pi_{\mathsf{SE}}^{\mathsf{apart}}$ &
  \multicolumn{1}{c|}{10.28} &
  \multicolumn{1}{c|}{\cellcolor[HTML]{99BEFF}3.913} &
  \multicolumn{1}{c|}{\cellcolor[HTML]{C7ECFF}11.80} &
  \multicolumn{1}{c|}{\cellcolor[HTML]{99BEFF}4.008} &
  \multicolumn{1}{c|}{13.09} &
  \multicolumn{1}{c|}{\cellcolor[HTML]{99BEFF}4.155} &
  \multicolumn{1}{c|}{14.51} &
  \multicolumn{1}{c|}{\cellcolor[HTML]{99BEFF}4.192} &
  \multicolumn{1}{c|}{16.46} &
  \cellcolor[HTML]{99BEFF}4.321 \\ \cline{2-13} 
 &
   &
  \cite{DBLP:conf/ccs/DZL25}-Low &
  \multicolumn{1}{c|}{110.4} &
  \multicolumn{1}{c|}{10.42} &
  \multicolumn{1}{c|}{{\color[HTML]{FE0000} 163.9}} &
  \multicolumn{1}{c|}{{\color[HTML]{FE0000} 12.34}} &
  \multicolumn{1}{c|}{{\color[HTML]{FE0000} 141.8}} &
  \multicolumn{1}{c|}{{\color[HTML]{FE0000} 32.17}} &
  \multicolumn{1}{c|}{169.4} &
  \multicolumn{1}{c|}{33.65} &
  \multicolumn{1}{c|}{211.3} &
  30.65 \\ \cline{3-13} 
 &
   &
  $\Pi_{\mathsf{OE}\text{-1}}^{\mathsf{apart}}$ &
  \multicolumn{1}{c|}{\cellcolor[HTML]{C7ECFF}18.88} &
  \multicolumn{1}{c|}{\cellcolor[HTML]{C7ECFF}4.943} &
  \multicolumn{1}{c|}{22.00} &
  \multicolumn{1}{c|}{\cellcolor[HTML]{C7ECFF}5.089} &
  \multicolumn{1}{c|}{24.92} &
  \multicolumn{1}{c|}{\cellcolor[HTML]{C7ECFF}5.401} &
  \multicolumn{1}{c|}{27.88} &
  \multicolumn{1}{c|}{\cellcolor[HTML]{C7ECFF}5.755} &
  \multicolumn{1}{c|}{31.97} &
  \cellcolor[HTML]{C7ECFF}6.103 \\ \cline{3-13} 
 &
   &
  $\Pi_{\mathsf{OE}\text{-2}}^{\mathsf{apart}}$ &
  \multicolumn{1}{c|}{18.96} &
  \multicolumn{1}{c|}{5.248} &
  \multicolumn{1}{c|}{\cellcolor[HTML]{C7ECFF}21.96} &
  \multicolumn{1}{c|}{5.388} &
  \multicolumn{1}{c|}{\cellcolor[HTML]{C7ECFF}24.55} &
  \multicolumn{1}{c|}{5.642} &
  \multicolumn{1}{c|}{\cellcolor[HTML]{C7ECFF}27.70} &
  \multicolumn{1}{c|}{6.012} &
  \multicolumn{1}{c|}{\cellcolor[HTML]{C7ECFF}31.80} &
  6.204 \\ \cline{3-13} 
 &
  \multirow{-4}{*}{3} &
  $\Pi_{\mathsf{SE}}^{\mathsf{apart}}$ &
  \multicolumn{1}{c|}{\cellcolor[HTML]{99BEFF}17.37} &
  \multicolumn{1}{c|}{\cellcolor[HTML]{99BEFF}4.476} &
  \multicolumn{1}{c|}{\cellcolor[HTML]{99BEFF}19.79} &
  \multicolumn{1}{c|}{\cellcolor[HTML]{99BEFF}4.704} &
  \multicolumn{1}{c|}{\cellcolor[HTML]{99BEFF}22.35} &
  \multicolumn{1}{c|}{\cellcolor[HTML]{99BEFF}4.988} &
  \multicolumn{1}{c|}{\cellcolor[HTML]{99BEFF}25.12} &
  \multicolumn{1}{c|}{\cellcolor[HTML]{99BEFF}5.337} &
  \multicolumn{1}{c|}{\cellcolor[HTML]{99BEFF}28.58} &
  \cellcolor[HTML]{99BEFF}5.545 \\ \cline{2-13} 
 &
   &
  \cite{DBLP:conf/ccs/DZL25}-Low &
  \multicolumn{1}{c|}{295.1} &
  \multicolumn{1}{c|}{23.96} &
  \multicolumn{1}{c|}{437.2} &
  \multicolumn{1}{c|}{{\color[HTML]{FE0000} 29.89}} &
  \multicolumn{1}{c|}{468.3} &
  \multicolumn{1}{c|}{{\color[HTML]{FE0000} 176.9}} &
  \multicolumn{1}{c|}{540.3} &
  \multicolumn{1}{c|}{182.3} &
  \multicolumn{1}{c|}{648.7} &
  183.6 \\ \cline{3-13} 
 &
   &
  $\Pi_{\mathsf{OE}\text{-1}}^{\mathsf{apart}}$ &
  \multicolumn{1}{c|}{\cellcolor[HTML]{C7ECFF}37.40} &
  \multicolumn{1}{c|}{\cellcolor[HTML]{C7ECFF}6.259} &
  \multicolumn{1}{c|}{\cellcolor[HTML]{C7ECFF}44.39} &
  \multicolumn{1}{c|}{\cellcolor[HTML]{C7ECFF}7.053} &
  \multicolumn{1}{c|}{\cellcolor[HTML]{C7ECFF}51.35} &
  \multicolumn{1}{c|}{\cellcolor[HTML]{C7ECFF}7.585} &
  \multicolumn{1}{c|}{\cellcolor[HTML]{C7ECFF}58.61} &
  \multicolumn{1}{c|}{\cellcolor[HTML]{C7ECFF}8.382} &
  \multicolumn{1}{c|}{67.17} &
  \cellcolor[HTML]{C7ECFF}9.154 \\ \cline{3-13} 
 &
   &
  $\Pi_{\mathsf{OE}\text{-2}}^{\mathsf{apart}}$ &
  \multicolumn{1}{c|}{37.78} &
  \multicolumn{1}{c|}{6.633} &
  \multicolumn{1}{c|}{44.56} &
  \multicolumn{1}{c|}{7.444} &
  \multicolumn{1}{c|}{51.93} &
  \multicolumn{1}{c|}{7.818} &
  \multicolumn{1}{c|}{58.78} &
  \multicolumn{1}{c|}{8.501} &
  \multicolumn{1}{c|}{\cellcolor[HTML]{C7ECFF}66.96} &
  9.363 \\ \cline{3-13} 
\multirow{-12}{*}{$2^8$} &
  \multirow{-4}{*}{4} &
  $\Pi_{\mathsf{SE}}^{\mathsf{apart}}$ &
  \multicolumn{1}{c|}{\cellcolor[HTML]{99BEFF}29.00} &
  \multicolumn{1}{c|}{\cellcolor[HTML]{99BEFF}5.587} &
  \multicolumn{1}{c|}{\cellcolor[HTML]{99BEFF}33.98} &
  \multicolumn{1}{c|}{\cellcolor[HTML]{99BEFF}6.229} &
  \multicolumn{1}{c|}{\cellcolor[HTML]{99BEFF}39.03} &
  \multicolumn{1}{c|}{\cellcolor[HTML]{99BEFF}6.682} &
  \multicolumn{1}{c|}{\cellcolor[HTML]{99BEFF}43.94} &
  \multicolumn{1}{c|}{\cellcolor[HTML]{99BEFF}7.533} &
  \multicolumn{1}{c|}{\cellcolor[HTML]{99BEFF}50.25} &
  \cellcolor[HTML]{99BEFF}8.219 \\ \hline
 &
   &
  \cite{DBLP:conf/ccs/DZL25}-Low &
  \multicolumn{1}{c|}{427.2} &
  \multicolumn{1}{c|}{69.12} &
  \multicolumn{1}{c|}{530.5} &
  \multicolumn{1}{c|}{103.1} &
  \multicolumn{1}{c|}{642.3} &
  \multicolumn{1}{c|}{138.9} &
  \multicolumn{1}{c|}{766.6} &
  \multicolumn{1}{c|}{177.8} &
  \multicolumn{1}{c|}{911.5} &
  217.9 \\ \cline{3-13} 
 &
   &
  $\Pi_{\mathsf{OE}\text{-1}}^{\mathsf{apart}}$ &
  \multicolumn{1}{c|}{\cellcolor[HTML]{C7ECFF}129.3} &
  \multicolumn{1}{c|}{\cellcolor[HTML]{C7ECFF}15.54} &
  \multicolumn{1}{c|}{\cellcolor[HTML]{C7ECFF}148.4} &
  \multicolumn{1}{c|}{\cellcolor[HTML]{C7ECFF}17.71} &
  \multicolumn{1}{c|}{\cellcolor[HTML]{C7ECFF}167.8} &
  \multicolumn{1}{c|}{\cellcolor[HTML]{C7ECFF}19.97} &
  \multicolumn{1}{c|}{\cellcolor[HTML]{C7ECFF}187.3} &
  \multicolumn{1}{c|}{\cellcolor[HTML]{C7ECFF}21.93} &
  \multicolumn{1}{c|}{\cellcolor[HTML]{C7ECFF}216.1} &
  \cellcolor[HTML]{C7ECFF}24.71 \\ \cline{3-13} 
 &
   &
  $\Pi_{\mathsf{OE}\text{-2}}^{\mathsf{apart}}$ &
  \multicolumn{1}{c|}{\cellcolor[HTML]{99BEFF}127.8} &
  \multicolumn{1}{c|}{\cellcolor[HTML]{99BEFF}13.92} &
  \multicolumn{1}{c|}{\cellcolor[HTML]{99BEFF}147.0} &
  \multicolumn{1}{c|}{\cellcolor[HTML]{99BEFF}15.84} &
  \multicolumn{1}{c|}{\cellcolor[HTML]{99BEFF}166.3} &
  \multicolumn{1}{c|}{\cellcolor[HTML]{99BEFF}17.73} &
  \multicolumn{1}{c|}{\cellcolor[HTML]{99BEFF}184.7} &
  \multicolumn{1}{c|}{\cellcolor[HTML]{99BEFF}19.36} &
  \multicolumn{1}{c|}{\cellcolor[HTML]{99BEFF}213.9} &
  \cellcolor[HTML]{99BEFF}21.99 \\ \cline{3-13} 
 &
  \multirow{-4}{*}{2} &
  $\Pi_{\mathsf{SE}}^{\mathsf{apart}}$ &
  \multicolumn{1}{c|}{140.4} &
  \multicolumn{1}{c|}{16.02} &
  \multicolumn{1}{c|}{161.3} &
  \multicolumn{1}{c|}{18.33} &
  \multicolumn{1}{c|}{182.0} &
  \multicolumn{1}{c|}{20.73} &
  \multicolumn{1}{c|}{202.5} &
  \multicolumn{1}{c|}{23.07} &
  \multicolumn{1}{c|}{233.0} &
  26.05 \\ \cline{2-13} 
 &
   &
  \cite{DBLP:conf/ccs/DZL25}-Low &
  \multicolumn{1}{c|}{1767} &
  \multicolumn{1}{c|}{129.6} &
  \multicolumn{1}{c|}{{\color[HTML]{FE0000} 2622}} &
  \multicolumn{1}{c|}{{\color[HTML]{FE0000} 166.1}} &
  \multicolumn{1}{c|}{{\color[HTML]{FE0000} 2268}} &
  \multicolumn{1}{c|}{{\color[HTML]{FE0000} 414.4}} &
  \multicolumn{1}{c|}{2710} &
  \multicolumn{1}{c|}{440.1} &
  \multicolumn{1}{c|}{3381} &
  472.4 \\ \cline{3-13} 
 &
   &
  $\Pi_{\mathsf{OE}\text{-1}}^{\mathsf{apart}}$ &
  \multicolumn{1}{c|}{267.9} &
  \multicolumn{1}{c|}{28.74} &
  \multicolumn{1}{c|}{315.8} &
  \multicolumn{1}{c|}{33.63} &
  \multicolumn{1}{c|}{362.3} &
  \multicolumn{1}{c|}{39.01} &
  \multicolumn{1}{c|}{410.2} &
  \multicolumn{1}{c|}{44.49} &
  \multicolumn{1}{c|}{471.3} &
  50.45 \\ \cline{3-13} 
 &
   &
  $\Pi_{\mathsf{OE}\text{-2}}^{\mathsf{apart}}$ &
  \multicolumn{1}{c|}{\cellcolor[HTML]{C7ECFF}266.9} &
  \multicolumn{1}{c|}{\cellcolor[HTML]{99BEFF}26.99} &
  \multicolumn{1}{c|}{\cellcolor[HTML]{C7ECFF}313.9} &
  \multicolumn{1}{c|}{\cellcolor[HTML]{99BEFF}31.93} &
  \multicolumn{1}{c|}{\cellcolor[HTML]{C7ECFF}360.6} &
  \multicolumn{1}{c|}{\cellcolor[HTML]{99BEFF}36.75} &
  \multicolumn{1}{c|}{\cellcolor[HTML]{C7ECFF}407.2} &
  \multicolumn{1}{c|}{\cellcolor[HTML]{99BEFF}41.98} &
  \multicolumn{1}{c|}{\cellcolor[HTML]{C7ECFF}468.4} &
  \cellcolor[HTML]{99BEFF}47.62 \\ \cline{3-13} 
 &
  \multirow{-4}{*}{3} &
  $\Pi_{\mathsf{SE}}^{\mathsf{apart}}$ &
  \multicolumn{1}{c|}{\cellcolor[HTML]{99BEFF}246.1} &
  \multicolumn{1}{c|}{\cellcolor[HTML]{C7ECFF}28.21} &
  \multicolumn{1}{c|}{\cellcolor[HTML]{99BEFF}285.9} &
  \multicolumn{1}{c|}{\cellcolor[HTML]{C7ECFF}33.38} &
  \multicolumn{1}{c|}{\cellcolor[HTML]{99BEFF}325.8} &
  \multicolumn{1}{c|}{\cellcolor[HTML]{C7ECFF}38.35} &
  \multicolumn{1}{c|}{\cellcolor[HTML]{99BEFF}365.6} &
  \multicolumn{1}{c|}{\cellcolor[HTML]{C7ECFF}43.78} &
  \multicolumn{1}{c|}{\cellcolor[HTML]{99BEFF}419.7} &
  \cellcolor[HTML]{C7ECFF}49.83 \\ \cline{2-13} 
 &
   &
  \cite{DBLP:conf/ccs/DZL25}-Low &
  \multicolumn{1}{c|}{4721} &
  \multicolumn{1}{c|}{370.6} &
  \multicolumn{1}{c|}{6996} &
  \multicolumn{1}{c|}{474.4} &
  \multicolumn{1}{c|}{-} &
  \multicolumn{1}{c|}{-} &
  \multicolumn{1}{c|}{-} &
  \multicolumn{1}{c|}{-} &
  \multicolumn{1}{c|}{-} &
  - \\ \cline{3-13} 
 &
   &
  $\Pi_{\mathsf{OE}\text{-1}}^{\mathsf{apart}}$ &
  \multicolumn{1}{c|}{567.1} &
  \multicolumn{1}{c|}{60.32} &
  \multicolumn{1}{c|}{676.5} &
  \multicolumn{1}{c|}{73.15} &
  \multicolumn{1}{c|}{789.3} &
  \multicolumn{1}{c|}{86.95} &
  \multicolumn{1}{c|}{900.1} &
  \multicolumn{1}{c|}{100.2} &
  \multicolumn{1}{c|}{1035} &
  115.6 \\ \cline{3-13} 
 &
   &
  $\Pi_{\mathsf{OE}\text{-2}}^{\mathsf{apart}}$ &
  \multicolumn{1}{c|}{\cellcolor[HTML]{C7ECFF}563.6} &
  \multicolumn{1}{c|}{\cellcolor[HTML]{C7ECFF}59.03} &
  \multicolumn{1}{c|}{\cellcolor[HTML]{C7ECFF}674.2} &
  \multicolumn{1}{c|}{\cellcolor[HTML]{C7ECFF}71.24} &
  \multicolumn{1}{c|}{\cellcolor[HTML]{C7ECFF}786.6} &
  \multicolumn{1}{c|}{\cellcolor[HTML]{C7ECFF}83.93} &
  \multicolumn{1}{c|}{\cellcolor[HTML]{C7ECFF}898.9} &
  \multicolumn{1}{c|}{\cellcolor[HTML]{C7ECFF}97.91} &
  \multicolumn{1}{c|}{\cellcolor[HTML]{C7ECFF}1031} &
  \cellcolor[HTML]{C7ECFF}112.5 \\ \cline{3-13} 
\multirow{-12}{*}{$2^{12}$} &
  \multirow{-4}{*}{4} &
  $\Pi_{\mathsf{SE}}^{\mathsf{apart}}$ &
  \multicolumn{1}{c|}{\cellcolor[HTML]{99BEFF}431.3} &
  \multicolumn{1}{c|}{\cellcolor[HTML]{99BEFF}56.45} &
  \multicolumn{1}{c|}{\cellcolor[HTML]{99BEFF}507.4} &
  \multicolumn{1}{c|}{\cellcolor[HTML]{99BEFF}68.35} &
  \multicolumn{1}{c|}{\cellcolor[HTML]{99BEFF}585.5} &
  \multicolumn{1}{c|}{\cellcolor[HTML]{99BEFF}80.55} &
  \multicolumn{1}{c|}{\cellcolor[HTML]{99BEFF}662.5} &
  \multicolumn{1}{c|}{\cellcolor[HTML]{99BEFF}93.18} &
  \multicolumn{1}{c|}{\cellcolor[HTML]{99BEFF}760.7} &
  \cellcolor[HTML]{99BEFF}106.9 \\ \hline
 &
   &
  \cite{DBLP:conf/ccs/DZL25}-Low &
  \multicolumn{1}{c|}{6835} &
  \multicolumn{1}{c|}{1136} &
  \multicolumn{1}{c|}{-} &
  \multicolumn{1}{c|}{-} &
  \multicolumn{1}{c|}{-} &
  \multicolumn{1}{c|}{-} &
  \multicolumn{1}{c|}{-} &
  \multicolumn{1}{c|}{-} &
  \multicolumn{1}{c|}{-} &
  - \\ \cline{3-13} 
 &
   &
  $\Pi_{\mathsf{OE}\text{-1}}^{\mathsf{apart}}$ &
  \multicolumn{1}{c|}{\cellcolor[HTML]{C7ECFF}2052} &
  \multicolumn{1}{c|}{\cellcolor[HTML]{C7ECFF}210.4} &
  \multicolumn{1}{c|}{\cellcolor[HTML]{C7ECFF}2361} &
  \multicolumn{1}{c|}{\cellcolor[HTML]{C7ECFF}249.7} &
  \multicolumn{1}{c|}{\cellcolor[HTML]{C7ECFF}2674} &
  \multicolumn{1}{c|}{\cellcolor[HTML]{C7ECFF}290.2} &
  \multicolumn{1}{c|}{\cellcolor[HTML]{C7ECFF}2986} &
  \multicolumn{1}{c|}{\cellcolor[HTML]{C7ECFF}332.6} &
  \multicolumn{1}{c|}{\cellcolor[HTML]{C7ECFF}3452} &
  \cellcolor[HTML]{C7ECFF}382.8 \\ \cline{3-13} 
 &
   &
  $\Pi_{\mathsf{OE}\text{-2}}^{\mathsf{apart}}$ &
  \multicolumn{1}{c|}{\cellcolor[HTML]{99BEFF}2032} &
  \multicolumn{1}{c|}{\cellcolor[HTML]{99BEFF}185.6} &
  \multicolumn{1}{c|}{\cellcolor[HTML]{99BEFF}2336} &
  \multicolumn{1}{c|}{\cellcolor[HTML]{99BEFF}217.7} &
  \multicolumn{1}{c|}{\cellcolor[HTML]{99BEFF}2643} &
  \multicolumn{1}{c|}{\cellcolor[HTML]{99BEFF}252.2} &
  \multicolumn{1}{c|}{\cellcolor[HTML]{99BEFF}2950} &
  \multicolumn{1}{c|}{\cellcolor[HTML]{99BEFF}289.1} &
  \multicolumn{1}{c|}{\cellcolor[HTML]{99BEFF}3412} &
  \cellcolor[HTML]{99BEFF}335.3 \\ \cline{3-13} 
\multirow{-4}{*}{$2^{16}$} &
  \multirow{-4}{*}{2} &
  $\Pi_{\mathsf{SE}}^{\mathsf{apart}}$ &
  \multicolumn{1}{c|}{2251} &
  \multicolumn{1}{c|}{226.2} &
  \multicolumn{1}{c|}{2586} &
  \multicolumn{1}{c|}{267.8} &
  \multicolumn{1}{c|}{2923} &
  \multicolumn{1}{c|}{312.1} &
  \multicolumn{1}{c|}{3260} &
  \multicolumn{1}{c|}{355.4} &
  \multicolumn{1}{c|}{3745} &
  410.5 \\ \hline
\end{tabular}
}
\caption{Communication cost (in MB) and running time (in seconds) of our protocols compared with those of \cite{DBLP:conf/ccs/DZL25} for $L_2$ distance in a low-dimensional WAN setting. Cells with - denote trials that ran out of memory. The best result is highlighted in \textcolor[rgb]{0,0.3,1}{blue}, the second best in \textcolor[rgb]{0,0.7,1}{cyan}, and data in \textcolor[rgb]{1,0,0}{red} font indicates abnormal values.}
\label{tab:l2-low-wan}
\end{table}

\begin{table}[!htbp]
\renewcommand\arraystretch{1}
	\centering
 \resizebox{0.83\linewidth}{!}{
\begin{tabular}{|c|c|c|cccccccccc|}
\hline
 &
   &
   &
  \multicolumn{10}{c|}{Threshold $\delta$} \\ \cline{4-13} 
 &
   &
   &
  \multicolumn{2}{c|}{16} &
  \multicolumn{2}{c|}{32} &
  \multicolumn{2}{c|}{64} &
  \multicolumn{2}{c|}{128} &
  \multicolumn{2}{c|}{256} \\ \cline{4-13} 
\multirow{-3}{*}{\begin{tabular}[c]{@{}c@{}}Set Size\\ $m=n$\end{tabular}} &
  \multirow{-3}{*}{\begin{tabular}[c]{@{}c@{}}Dimension\\ $d$\end{tabular}} &
  \multirow{-3}{*}{Protocol} &
  \multicolumn{1}{c|}{Comm.} &
  \multicolumn{1}{c|}{Time} &
  \multicolumn{1}{c|}{Comm.} &
  \multicolumn{1}{c|}{Time} &
  \multicolumn{1}{c|}{Comm.} &
  \multicolumn{1}{c|}{Time} &
  \multicolumn{1}{c|}{Comm.} &
  \multicolumn{1}{c|}{Time} &
  \multicolumn{1}{c|}{Comm.} &
  Time \\ \hline
 &
   &
  \cite{DBLP:conf/ccs/DZL25}-High &
  \multicolumn{1}{c|}{105.1} &
  \multicolumn{1}{c|}{14.58} &
  \multicolumn{1}{c|}{126.9} &
  \multicolumn{1}{c|}{15.39} &
  \multicolumn{1}{c|}{132.4} &
  \multicolumn{1}{c|}{18.84} &
  \multicolumn{1}{c|}{145.0} &
  \multicolumn{1}{c|}{19.95} &
  \multicolumn{1}{c|}{200.2} &
  29.34 \\ \cline{3-13} 
 &
   &
  $\Pi_{\mathsf{OE}\text{-1}}^{\mathsf{sep}}$ &
  \multicolumn{1}{c|}{\cellcolor[HTML]{99BEFF}26.67} &
  \multicolumn{1}{c|}{\cellcolor[HTML]{99BEFF}6.298} &
  \multicolumn{1}{c|}{\cellcolor[HTML]{C7ECFF}30.06} &
  \multicolumn{1}{c|}{\cellcolor[HTML]{99BEFF}6.627} &
  \multicolumn{1}{c|}{\cellcolor[HTML]{C7ECFF}33.14} &
  \multicolumn{1}{c|}{7.489} &
  \multicolumn{1}{c|}{\cellcolor[HTML]{C7ECFF}36.30} &
  \multicolumn{1}{c|}{\cellcolor[HTML]{C7ECFF}7.642} &
  \multicolumn{1}{c|}{\cellcolor[HTML]{C7ECFF}41.17} &
  \cellcolor[HTML]{C7ECFF}7.964 \\ \cline{3-13} 
 &
   &
  $\Pi_{\mathsf{OE}\text{-2}}^{\mathsf{sep}}$ &
  \multicolumn{1}{c|}{\cellcolor[HTML]{C7ECFF}26.74} &
  \multicolumn{1}{c|}{6.494} &
  \multicolumn{1}{c|}{\cellcolor[HTML]{99BEFF}30.04} &
  \multicolumn{1}{c|}{6.813} &
  \multicolumn{1}{c|}{\cellcolor[HTML]{99BEFF}32.99} &
  \multicolumn{1}{c|}{\cellcolor[HTML]{C7ECFF}7.264} &
  \multicolumn{1}{c|}{\cellcolor[HTML]{99BEFF}36.16} &
  \multicolumn{1}{c|}{7.725} &
  \multicolumn{1}{c|}{\cellcolor[HTML]{99BEFF}41.06} &
  8.099 \\ \cline{3-13} 
 &
  \multirow{-4}{*}{6} &
  $\Pi_{\mathsf{SE}}^{\mathsf{sep}}$ &
  \multicolumn{1}{c|}{32.67} &
  \multicolumn{1}{c|}{\cellcolor[HTML]{C7ECFF}6.351} &
  \multicolumn{1}{c|}{37.28} &
  \multicolumn{1}{c|}{\cellcolor[HTML]{C7ECFF}6.653} &
  \multicolumn{1}{c|}{41.58} &
  \multicolumn{1}{c|}{\cellcolor[HTML]{99BEFF}6.701} &
  \multicolumn{1}{c|}{46.07} &
  \multicolumn{1}{c|}{\cellcolor[HTML]{99BEFF}7.078} &
  \multicolumn{1}{c|}{52.27} &
  \cellcolor[HTML]{99BEFF}7.782 \\ \cline{2-13} 
 &
   &
  \cite{DBLP:conf/ccs/DZL25}-High &
  \multicolumn{1}{c|}{140.0} &
  \multicolumn{1}{c|}{18.28} &
  \multicolumn{1}{c|}{169.0} &
  \multicolumn{1}{c|}{19.85} &
  \multicolumn{1}{c|}{176.2} &
  \multicolumn{1}{c|}{26.13} &
  \multicolumn{1}{c|}{192.6} &
  \multicolumn{1}{c|}{27.39} &
  \multicolumn{1}{c|}{265.5} &
  38.67 \\ \cline{3-13} 
 &
   &
  $\Pi_{\mathsf{OE}\text{-1}}^{\mathsf{sep}}$ &
  \multicolumn{1}{c|}{\cellcolor[HTML]{C7ECFF}35.09} &
  \multicolumn{1}{c|}{\cellcolor[HTML]{C7ECFF}7.537} &
  \multicolumn{1}{c|}{\cellcolor[HTML]{C7ECFF}38.95} &
  \multicolumn{1}{c|}{\cellcolor[HTML]{C7ECFF}7.875} &
  \multicolumn{1}{c|}{\cellcolor[HTML]{C7ECFF}42.97} &
  \multicolumn{1}{c|}{\cellcolor[HTML]{C7ECFF}8.028} &
  \multicolumn{1}{c|}{\cellcolor[HTML]{C7ECFF}47.26} &
  \multicolumn{1}{c|}{\cellcolor[HTML]{C7ECFF}8.378} &
  \multicolumn{1}{c|}{\cellcolor[HTML]{C7ECFF}53.86} &
  9.188 \\ \cline{3-13} 
 &
   &
  $\Pi_{\mathsf{OE}\text{-2}}^{\mathsf{sep}}$ &
  \multicolumn{1}{c|}{\cellcolor[HTML]{99BEFF}34.99} &
  \multicolumn{1}{c|}{7.854} &
  \multicolumn{1}{c|}{\cellcolor[HTML]{99BEFF}38.93} &
  \multicolumn{1}{c|}{7.981} &
  \multicolumn{1}{c|}{\cellcolor[HTML]{99BEFF}42.86} &
  \multicolumn{1}{c|}{8.213} &
  \multicolumn{1}{c|}{\cellcolor[HTML]{99BEFF}47.17} &
  \multicolumn{1}{c|}{8.881} &
  \multicolumn{1}{c|}{\cellcolor[HTML]{99BEFF}53.84} &
  \cellcolor[HTML]{99BEFF}9.113 \\ \cline{3-13} 
 &
  \multirow{-4}{*}{8} &
  $\Pi_{\mathsf{SE}}^{\mathsf{sep}}$ &
  \multicolumn{1}{c|}{43.04} &
  \multicolumn{1}{c|}{\cellcolor[HTML]{99BEFF}6.844} &
  \multicolumn{1}{c|}{48.91} &
  \multicolumn{1}{c|}{\cellcolor[HTML]{99BEFF}7.124} &
  \multicolumn{1}{c|}{54.62} &
  \multicolumn{1}{c|}{\cellcolor[HTML]{99BEFF}7.757} &
  \multicolumn{1}{c|}{60.61} &
  \multicolumn{1}{c|}{\cellcolor[HTML]{99BEFF}8.089} &
  \multicolumn{1}{c|}{69.01} &
  \cellcolor[HTML]{C7ECFF}9.154 \\ \cline{2-13} 
 &
   &
  \cite{DBLP:conf/ccs/DZL25}-High &
  \multicolumn{1}{c|}{174.9} &
  \multicolumn{1}{c|}{24.35} &
  \multicolumn{1}{c|}{211.1} &
  \multicolumn{1}{c|}{24.85} &
  \multicolumn{1}{c|}{219.9} &
  \multicolumn{1}{c|}{31.29} &
  \multicolumn{1}{c|}{240.3} &
  \multicolumn{1}{c|}{33.11} &
  \multicolumn{1}{c|}{330.9} &
  47.25 \\ \cline{3-13} 
 &
   &
  $\Pi_{\mathsf{OE}\text{-1}}^{\mathsf{sep}}$ &
  \multicolumn{1}{c|}{\cellcolor[HTML]{99BEFF}42.74} &
  \multicolumn{1}{c|}{\cellcolor[HTML]{C7ECFF}8.145} &
  \multicolumn{1}{c|}{\cellcolor[HTML]{C7ECFF}47.94} &
  \multicolumn{1}{c|}{\cellcolor[HTML]{C7ECFF}8.284} &
  \multicolumn{1}{c|}{\cellcolor[HTML]{99BEFF}53.00} &
  \multicolumn{1}{c|}{9.223} &
  \multicolumn{1}{c|}{\cellcolor[HTML]{C7ECFF}58.22} &
  \multicolumn{1}{c|}{\cellcolor[HTML]{99BEFF}9.326} &
  \multicolumn{1}{c|}{\cellcolor[HTML]{99BEFF}66.58} &
  \cellcolor[HTML]{99BEFF}9.824 \\ \cline{3-13} 
 &
   &
  $\Pi_{\mathsf{OE}\text{-2}}^{\mathsf{sep}}$ &
  \multicolumn{1}{c|}{\cellcolor[HTML]{C7ECFF}42.76} &
  \multicolumn{1}{c|}{8.447} &
  \multicolumn{1}{c|}{\cellcolor[HTML]{99BEFF}47.92} &
  \multicolumn{1}{c|}{8.627} &
  \multicolumn{1}{c|}{\cellcolor[HTML]{C7ECFF}53.02} &
  \multicolumn{1}{c|}{\cellcolor[HTML]{C7ECFF}8.991} &
  \multicolumn{1}{c|}{\cellcolor[HTML]{99BEFF}58.15} &
  \multicolumn{1}{c|}{9.486} &
  \multicolumn{1}{c|}{\cellcolor[HTML]{C7ECFF}66.63} &
  \cellcolor[HTML]{C7ECFF}10.11 \\ \cline{3-13} 
\multirow{-12}{*}{$2^8$} &
  \multirow{-4}{*}{10} &
  $\Pi_{\mathsf{SE}}^{\mathsf{sep}}$ &
  \multicolumn{1}{c|}{53.19} &
  \multicolumn{1}{c|}{\cellcolor[HTML]{99BEFF}7.731} &
  \multicolumn{1}{c|}{60.54} &
  \multicolumn{1}{c|}{\cellcolor[HTML]{99BEFF}8.019} &
  \multicolumn{1}{c|}{67.79} &
  \multicolumn{1}{c|}{\cellcolor[HTML]{99BEFF}8.987} &
  \multicolumn{1}{c|}{75.19} &
  \multicolumn{1}{c|}{\cellcolor[HTML]{C7ECFF}9.341} &
  \multicolumn{1}{c|}{85.74} &
  10.25 \\ \hline
 &
   &
  \cite{DBLP:conf/ccs/DZL25}-High &
  \multicolumn{1}{c|}{1681} &
  \multicolumn{1}{c|}{194.4} &
  \multicolumn{1}{c|}{2030} &
  \multicolumn{1}{c|}{211.7} &
  \multicolumn{1}{c|}{2118} &
  \multicolumn{1}{c|}{266.6} &
  \multicolumn{1}{c|}{2320} &
  \multicolumn{1}{c|}{290.5} &
  \multicolumn{1}{c|}{3202} &
  437.9 \\ \cline{3-13} 
 &
   &
  $\Pi_{\mathsf{OE}\text{-1}}^{\mathsf{sep}}$ &
  \multicolumn{1}{c|}{\cellcolor[HTML]{C7ECFF}379.7} &
  \multicolumn{1}{c|}{\cellcolor[HTML]{C7ECFF}36.57} &
  \multicolumn{1}{c|}{\cellcolor[HTML]{C7ECFF}428.7} &
  \multicolumn{1}{c|}{\cellcolor[HTML]{C7ECFF}41.79} &
  \multicolumn{1}{c|}{\cellcolor[HTML]{C7ECFF}477.7} &
  \multicolumn{1}{c|}{\cellcolor[HTML]{C7ECFF}45.56} &
  \multicolumn{1}{c|}{\cellcolor[HTML]{C7ECFF}527.3} &
  \multicolumn{1}{c|}{\cellcolor[HTML]{C7ECFF}50.61} &
  \multicolumn{1}{c|}{\cellcolor[HTML]{C7ECFF}603.4} &
  \cellcolor[HTML]{C7ECFF}56.24 \\ \cline{3-13} 
 &
   &
  $\Pi_{\mathsf{OE}\text{-2}}^{\mathsf{sep}}$ &
  \multicolumn{1}{c|}{\cellcolor[HTML]{99BEFF}378.4} &
  \multicolumn{1}{c|}{\cellcolor[HTML]{99BEFF}35.33} &
  \multicolumn{1}{c|}{\cellcolor[HTML]{99BEFF}427.3} &
  \multicolumn{1}{c|}{\cellcolor[HTML]{99BEFF}39.11} &
  \multicolumn{1}{c|}{\cellcolor[HTML]{99BEFF}476.0} &
  \multicolumn{1}{c|}{\cellcolor[HTML]{99BEFF}43.43} &
  \multicolumn{1}{c|}{\cellcolor[HTML]{99BEFF}524.8} &
  \multicolumn{1}{c|}{\cellcolor[HTML]{99BEFF}47.67} &
  \multicolumn{1}{c|}{\cellcolor[HTML]{99BEFF}601.3} &
  \cellcolor[HTML]{99BEFF}53.62 \\ \cline{3-13} 
 &
  \multirow{-4}{*}{6} &
  $\Pi_{\mathsf{SE}}^{\mathsf{sep}}$ &
  \multicolumn{1}{c|}{473.9} &
  \multicolumn{1}{c|}{43.75} &
  \multicolumn{1}{c|}{541.2} &
  \multicolumn{1}{c|}{50.14} &
  \multicolumn{1}{c|}{608.7} &
  \multicolumn{1}{c|}{55.03} &
  \multicolumn{1}{c|}{676.3} &
  \multicolumn{1}{c|}{60.89} &
  \multicolumn{1}{c|}{770.8} &
  69.67 \\ \cline{2-13} 
 &
   &
  \cite{DBLP:conf/ccs/DZL25}-High &
  \multicolumn{1}{c|}{2239} &
  \multicolumn{1}{c|}{258.7} &
  \multicolumn{1}{c|}{2703} &
  \multicolumn{1}{c|}{280.5} &
  \multicolumn{1}{c|}{2818} &
  \multicolumn{1}{c|}{348.9} &
  \multicolumn{1}{c|}{3082} &
  \multicolumn{1}{c|}{371.9} &
  \multicolumn{1}{c|}{4248} &
  559.1 \\ \cline{3-13} 
 &
   &
  $\Pi_{\mathsf{OE}\text{-1}}^{\mathsf{sep}}$ &
  \multicolumn{1}{c|}{\cellcolor[HTML]{C7ECFF}501.8} &
  \multicolumn{1}{c|}{\cellcolor[HTML]{C7ECFF}46.18} &
  \multicolumn{1}{c|}{\cellcolor[HTML]{C7ECFF}567.0} &
  \multicolumn{1}{c|}{\cellcolor[HTML]{C7ECFF}51.88} &
  \multicolumn{1}{c|}{\cellcolor[HTML]{C7ECFF}632.1} &
  \multicolumn{1}{c|}{\cellcolor[HTML]{C7ECFF}58.58} &
  \multicolumn{1}{c|}{\cellcolor[HTML]{C7ECFF}697.1} &
  \multicolumn{1}{c|}{\cellcolor[HTML]{C7ECFF}64.52} &
  \multicolumn{1}{c|}{\cellcolor[HTML]{C7ECFF}798.8} &
  \cellcolor[HTML]{C7ECFF}72.65 \\ \cline{3-13} 
 &
   &
  $\Pi_{\mathsf{OE}\text{-2}}^{\mathsf{sep}}$ &
  \multicolumn{1}{c|}{\cellcolor[HTML]{99BEFF}500.7} &
  \multicolumn{1}{c|}{\cellcolor[HTML]{99BEFF}44.71} &
  \multicolumn{1}{c|}{\cellcolor[HTML]{99BEFF}565.5} &
  \multicolumn{1}{c|}{\cellcolor[HTML]{99BEFF}49.98} &
  \multicolumn{1}{c|}{\cellcolor[HTML]{99BEFF}630.2} &
  \multicolumn{1}{c|}{\cellcolor[HTML]{99BEFF}56.34} &
  \multicolumn{1}{c|}{\cellcolor[HTML]{99BEFF}694.9} &
  \multicolumn{1}{c|}{\cellcolor[HTML]{99BEFF}62.45} &
  \multicolumn{1}{c|}{\cellcolor[HTML]{99BEFF}796.5} &
  \cellcolor[HTML]{99BEFF}70.01 \\ \cline{3-13} 
 &
  \multirow{-4}{*}{8} &
  $\Pi_{\mathsf{SE}}^{\mathsf{sep}}$ &
  \multicolumn{1}{c|}{628.2} &
  \multicolumn{1}{c|}{56.14} &
  \multicolumn{1}{c|}{717.7} &
  \multicolumn{1}{c|}{62.55} &
  \multicolumn{1}{c|}{806.7} &
  \multicolumn{1}{c|}{71.66} &
  \multicolumn{1}{c|}{896.3} &
  \multicolumn{1}{c|}{78.52} &
  \multicolumn{1}{c|}{1022} &
  84.36 \\ \cline{2-13} 
 &
   &
  \cite{DBLP:conf/ccs/DZL25}-High &
  \multicolumn{1}{c|}{2798} &
  \multicolumn{1}{c|}{322.3} &
  \multicolumn{1}{c|}{3377} &
  \multicolumn{1}{c|}{347.7} &
  \multicolumn{1}{c|}{3519} &
  \multicolumn{1}{c|}{429.3} &
  \multicolumn{1}{c|}{3845} &
  \multicolumn{1}{c|}{459.4} &
  \multicolumn{1}{c|}{5294} &
  676.7 \\ \cline{3-13} 
 &
   &
  $\Pi_{\mathsf{OE}\text{-1}}^{\mathsf{sep}}$ &
  \multicolumn{1}{c|}{\cellcolor[HTML]{C7ECFF}625.1} &
  \multicolumn{1}{c|}{\cellcolor[HTML]{C7ECFF}55.56} &
  \multicolumn{1}{c|}{\cellcolor[HTML]{C7ECFF}705.6} &
  \multicolumn{1}{c|}{\cellcolor[HTML]{C7ECFF}63.87} &
  \multicolumn{1}{c|}{\cellcolor[HTML]{C7ECFF}786.1} &
  \multicolumn{1}{c|}{\cellcolor[HTML]{C7ECFF}71.04} &
  \multicolumn{1}{c|}{\cellcolor[HTML]{C7ECFF}867.6} &
  \multicolumn{1}{c|}{\cellcolor[HTML]{C7ECFF}79.12} &
  \multicolumn{1}{c|}{\cellcolor[HTML]{C7ECFF}994.2} &
  \cellcolor[HTML]{C7ECFF}88.96 \\ \cline{3-13} 
 &
   &
  $\Pi_{\mathsf{OE}\text{-2}}^{\mathsf{sep}}$ &
  \multicolumn{1}{c|}{\cellcolor[HTML]{99BEFF}623.6} &
  \multicolumn{1}{c|}{\cellcolor[HTML]{99BEFF}54.47} &
  \multicolumn{1}{c|}{\cellcolor[HTML]{99BEFF}703.9} &
  \multicolumn{1}{c|}{\cellcolor[HTML]{99BEFF}61.85} &
  \multicolumn{1}{c|}{\cellcolor[HTML]{99BEFF}784.3} &
  \multicolumn{1}{c|}{\cellcolor[HTML]{99BEFF}69.59} &
  \multicolumn{1}{c|}{\cellcolor[HTML]{99BEFF}865.5} &
  \multicolumn{1}{c|}{\cellcolor[HTML]{99BEFF}76.69} &
  \multicolumn{1}{c|}{\cellcolor[HTML]{99BEFF}991.9} &
  \cellcolor[HTML]{99BEFF}86.55 \\ \cline{3-13} 
\multirow{-12}{*}{$2^{12}$} &
  \multirow{-4}{*}{10} &
  $\Pi_{\mathsf{SE}}^{\mathsf{sep}}$ &
  \multicolumn{1}{c|}{782.4} &
  \multicolumn{1}{c|}{66.35} &
  \multicolumn{1}{c|}{893.4} &
  \multicolumn{1}{c|}{77.83} &
  \multicolumn{1}{c|}{1005} &
  \multicolumn{1}{c|}{86.01} &
  \multicolumn{1}{c|}{1117} &
  \multicolumn{1}{c|}{100.6} &
  \multicolumn{1}{c|}{1273} &
  111.7 \\ \hline
 &
   &
  \cite{DBLP:conf/ccs/DZL25}-High &
  \multicolumn{1}{c|}{-} &
  \multicolumn{1}{c|}{-} &
  \multicolumn{1}{c|}{-} &
  \multicolumn{1}{c|}{-} &
  \multicolumn{1}{c|}{-} &
  \multicolumn{1}{c|}{-} &
  \multicolumn{1}{c|}{-} &
  \multicolumn{1}{c|}{-} &
  \multicolumn{1}{c|}{-} &
  - \\ \cline{3-13} 
 &
   &
  $\Pi_{\mathsf{OE}\text{-1}}^{\mathsf{sep}}$ &
  \multicolumn{1}{c|}{\cellcolor[HTML]{C7ECFF}6090} &
  \multicolumn{1}{c|}{\cellcolor[HTML]{C7ECFF}535.9} &
  \multicolumn{1}{c|}{\cellcolor[HTML]{C7ECFF}6875} &
  \multicolumn{1}{c|}{\cellcolor[HTML]{C7ECFF}616.1} &
  \multicolumn{1}{c|}{\cellcolor[HTML]{C7ECFF}7667} &
  \multicolumn{1}{c|}{\cellcolor[HTML]{C7ECFF}702.9} &
  \multicolumn{1}{c|}{\cellcolor[HTML]{C7ECFF}8464} &
  \multicolumn{1}{c|}{\cellcolor[HTML]{C7ECFF}783.9} &
  \multicolumn{1}{c|}{\cellcolor[HTML]{C7ECFF}9699} &
  \cellcolor[HTML]{C7ECFF}892.1 \\ \cline{3-13} 
 &
   &
  $\Pi_{\mathsf{OE}\text{-2}}^{\mathsf{sep}}$ &
  \multicolumn{1}{c|}{\cellcolor[HTML]{99BEFF}6069} &
  \multicolumn{1}{c|}{\cellcolor[HTML]{99BEFF}511.9} &
  \multicolumn{1}{c|}{\cellcolor[HTML]{99BEFF}6850} &
  \multicolumn{1}{c|}{\cellcolor[HTML]{99BEFF}581.6} &
  \multicolumn{1}{c|}{\cellcolor[HTML]{99BEFF}7638} &
  \multicolumn{1}{c|}{\cellcolor[HTML]{99BEFF}664.9} &
  \multicolumn{1}{c|}{\cellcolor[HTML]{99BEFF}8430} &
  \multicolumn{1}{c|}{\cellcolor[HTML]{99BEFF}741.3} &
  \multicolumn{1}{c|}{\cellcolor[HTML]{99BEFF}9659} &
  \cellcolor[HTML]{99BEFF}842.7 \\ \cline{3-13} 
 &
  \multirow{-4}{*}{6} &
  $\Pi_{\mathsf{SE}}^{\mathsf{sep}}$ &
  \multicolumn{1}{c|}{7640} &
  \multicolumn{1}{c|}{623.8} &
  \multicolumn{1}{c|}{8725} &
  \multicolumn{1}{c|}{720.8} &
  \multicolumn{1}{c|}{9815} &
  \multicolumn{1}{c|}{820.2} &
  \multicolumn{1}{c|}{10908} &
  \multicolumn{1}{c|}{920.7} &
  \multicolumn{1}{c|}{12438} &
  1048 \\ \cline{2-13} 
 &
   &
  \cite{DBLP:conf/ccs/DZL25}-High &
  \multicolumn{1}{c|}{-} &
  \multicolumn{1}{c|}{-} &
  \multicolumn{1}{c|}{-} &
  \multicolumn{1}{c|}{-} &
  \multicolumn{1}{c|}{-} &
  \multicolumn{1}{c|}{-} &
  \multicolumn{1}{c|}{-} &
  \multicolumn{1}{c|}{-} &
  \multicolumn{1}{c|}{-} &
  - \\ \cline{3-13} 
 &
   &
  $\Pi_{\mathsf{OE}\text{-1}}^{\mathsf{sep}}$ &
  \multicolumn{1}{c|}{\cellcolor[HTML]{C7ECFF}8066} &
  \multicolumn{1}{c|}{\cellcolor[HTML]{C7ECFF}711.3} &
  \multicolumn{1}{c|}{\cellcolor[HTML]{C7ECFF}9109} &
  \multicolumn{1}{c|}{\cellcolor[HTML]{C7ECFF}816.1} &
  \multicolumn{1}{c|}{\cellcolor[HTML]{C7ECFF}10157} &
  \multicolumn{1}{c|}{\cellcolor[HTML]{C7ECFF}925.1} &
  \multicolumn{1}{c|}{\cellcolor[HTML]{C7ECFF}11206} &
  \multicolumn{1}{c|}{\cellcolor[HTML]{C7ECFF}1025} &
  \multicolumn{1}{c|}{\cellcolor[HTML]{C7ECFF}12846} &
  \cellcolor[HTML]{C7ECFF}1160 \\ \cline{3-13} 
 &
   &
  $\Pi_{\mathsf{OE}\text{-2}}^{\mathsf{sep}}$ &
  \multicolumn{1}{c|}{\cellcolor[HTML]{99BEFF}8046} &
  \multicolumn{1}{c|}{\cellcolor[HTML]{99BEFF}685.7} &
  \multicolumn{1}{c|}{\cellcolor[HTML]{99BEFF}9082} &
  \multicolumn{1}{c|}{\cellcolor[HTML]{99BEFF}788.5} &
  \multicolumn{1}{c|}{\cellcolor[HTML]{99BEFF}10125} &
  \multicolumn{1}{c|}{\cellcolor[HTML]{99BEFF}890.2} &
  \multicolumn{1}{c|}{\cellcolor[HTML]{99BEFF}11170} &
  \multicolumn{1}{c|}{\cellcolor[HTML]{99BEFF}986.1} &
  \multicolumn{1}{c|}{\cellcolor[HTML]{99BEFF}12808} &
  \cellcolor[HTML]{99BEFF}1128 \\ \cline{3-13} 
 &
  \multirow{-4}{*}{8} &
  $\Pi_{\mathsf{SE}}^{\mathsf{sep}}$ &
  \multicolumn{1}{c|}{10133} &
  \multicolumn{1}{c|}{820.7} &
  \multicolumn{1}{c|}{11572} &
  \multicolumn{1}{c|}{947.5} &
  \multicolumn{1}{c|}{13018} &
  \multicolumn{1}{c|}{1075} &
  \multicolumn{1}{c|}{14466} &
  \multicolumn{1}{c|}{1201} &
  \multicolumn{1}{c|}{16501} &
  1355 \\ \cline{2-13} 
 &
   &
  \cite{DBLP:conf/ccs/DZL25}-High &
  \multicolumn{1}{c|}{-} &
  \multicolumn{1}{c|}{-} &
  \multicolumn{1}{c|}{-} &
  \multicolumn{1}{c|}{-} &
  \multicolumn{1}{c|}{-} &
  \multicolumn{1}{c|}{-} &
  \multicolumn{1}{c|}{-} &
  \multicolumn{1}{c|}{-} &
  \multicolumn{1}{c|}{-} &
  - \\ \cline{3-13} 
 &
   &
  $\Pi_{\mathsf{OE}\text{-1}}^{\mathsf{sep}}$ &
  \multicolumn{1}{c|}{\cellcolor[HTML]{C7ECFF}10045} &
  \multicolumn{1}{c|}{\cellcolor[HTML]{C7ECFF}880.6} &
  \multicolumn{1}{c|}{\cellcolor[HTML]{C7ECFF}11338} &
  \multicolumn{1}{c|}{\cellcolor[HTML]{C7ECFF}1003} &
  \multicolumn{1}{c|}{\cellcolor[HTML]{C7ECFF}12644} &
  \multicolumn{1}{c|}{\cellcolor[HTML]{C7ECFF}1130} &
  \multicolumn{1}{c|}{\cellcolor[HTML]{C7ECFF}13953} &
  \multicolumn{1}{c|}{\cellcolor[HTML]{C7ECFF}1274} &
  \multicolumn{1}{c|}{\cellcolor[HTML]{C7ECFF}15997} &
  \cellcolor[HTML]{C7ECFF}1456 \\ \cline{3-13} 
 &
   &
  $\Pi_{\mathsf{OE}\text{-2}}^{\mathsf{sep}}$ &
  \multicolumn{1}{c|}{\cellcolor[HTML]{99BEFF}10024} &
  \multicolumn{1}{c|}{\cellcolor[HTML]{99BEFF}866.1} &
  \multicolumn{1}{c|}{\cellcolor[HTML]{99BEFF}11313} &
  \multicolumn{1}{c|}{\cellcolor[HTML]{99BEFF}987.3} &
  \multicolumn{1}{c|}{\cellcolor[HTML]{99BEFF}12613} &
  \multicolumn{1}{c|}{\cellcolor[HTML]{99BEFF}1095} &
  \multicolumn{1}{c|}{\cellcolor[HTML]{99BEFF}13917} &
  \multicolumn{1}{c|}{\cellcolor[HTML]{99BEFF}1237} &
  \multicolumn{1}{c|}{\cellcolor[HTML]{99BEFF}15955} &
  \cellcolor[HTML]{99BEFF}1420 \\ \cline{3-13} 
\multirow{-12}{*}{$2^{16}$} &
  \multirow{-4}{*}{10} &
  $\Pi_{\mathsf{SE}}^{\mathsf{sep}}$ &
  \multicolumn{1}{c|}{12628} &
  \multicolumn{1}{c|}{1020} &
  \multicolumn{1}{c|}{14420} &
  \multicolumn{1}{c|}{1167} &
  \multicolumn{1}{c|}{16221} &
  \multicolumn{1}{c|}{1327} &
  \multicolumn{1}{c|}{18026} &
  \multicolumn{1}{c|}{1494} &
  \multicolumn{1}{c|}{20562} &
  1699 \\ \hline
\end{tabular}
}
\caption{Communication cost (in MB) and running time (in seconds) of our protocols compared with those of \cite{DBLP:conf/ccs/DZL25} for $L_2$ distance in a high-dimensional WAN setting. Cells with - denote trials that ran out of memory. The best result is highlighted in \textcolor[rgb]{0,0.3,1}{blue}, the second best in \textcolor[rgb]{0,0.7,1}{cyan}.}
\label{tab:l2-high-wan}
\end{table}

\end{document}